\documentclass[12pt]{article} 
\usepackage[paper=letterpaper,margin=.8in]{geometry}

\usepackage{amsmath}
\usepackage{amssymb}
\usepackage{tensor}
\usepackage{graphicx}
\usepackage{caption}
\usepackage{color}
\usepackage{subfigure}
\usepackage{tikz}
\usepackage[numbers,sort&compress]{natbib}
\usepackage[utf8]{inputenc}
\usepackage{braket}
\usepackage{upgreek}
\usepackage[normalem]{ulem}
\usepackage[colorlinks = true, linkcolor = blue, citecolor = red, urlcolor  = blue, anchorcolor = blue]{hyperref}

\newcommand{\qt}{\tilde{q}}

\newcommand{\calo}{\mathcal{O}}

\newcommand{\cald}{\mathcal{D}}
\newcommand{\calx}{\mathcal{X}}
\newcommand{\cals}{\mathcal{S}}
\newcommand{\calb}{\mathcal{B}}
\newcommand{\calp}{\mathcal{P}}
\newcommand{\cala}{\mathcal{A}}
\newcommand{\calf}{\mathcal{F}}
\newcommand{\calr}{\mathcal{R}}
\newcommand{\calc}{\mathcal{C}}

\newcommand{\disk}{\text{disk}}
\newcommand{\cyl}{\text{cyl}}
\newcommand{\e}{\epsilon}
\newcommand{\p}{\partial}
\newcommand{\la}{\langle}
\newcommand{\ra}{\rangle}
\newcommand{\wt}{\widetilde}

\newcommand{\nn}{\nonumber}

\DeclareMathOperator{\tr}{Tr}
\DeclareMathOperator{\re}{Re}

\DeclareMathOperator{\sgn}{sgn}

\numberwithin{equation}{section}

\begin{document}

\thispagestyle{empty}

\begin{center}

~\vspace{3mm}

{\LARGE \bf {JT gravity with matter, generalized ETH,  \\ ~ \\ and Random Matrices }}

\vspace{0.5in}

{\bf Daniel Louis Jafferis,$^{1}$ David K. Kolchmeyer,$^{1,2}$ }
\vskip1em
{\bf Baur Mukhametzhanov,$^{3,4}$ and Julian Sonner$^{5}$}

\vspace{0.3in}

$^1$ Department of Physics, Harvard University, Cambridge, MA 02138, USA

\vskip1em

$^2$ Center for Theoretical Physics,

Massachusetts Institute of Technology, Cambridge, MA 02139, USA

\vskip1em

$^3$ Institute for Advanced Study, Princeton, NJ 08540, USA

\vskip1em

$^4$ Department of Physics, Cornell University, Ithaca, NY 14853, USA

\vskip1em

$^5$ Department of Theoretical Physics, University of Geneva, Geneva, Switzerland
    
    \vspace{0.3in}

\end{center}

\begin{abstract}

\vspace{0.1in}

We present evidence for a duality between Jackiw–Teitelboim gravity minimally coupled to a free massive scalar field and a single-trace two-matrix model. One matrix is the Hamiltonian $H$ of a holographic disorder-averaged quantum mechanics, while the other matrix is the light operator $\cal O$ dual to the bulk scalar field. The single-boundary observables of interest are thermal correlation functions of $\cal O$. We study the matching of the genus zero one- and two-boundary expectation values in the matrix model to the disk and cylinder Euclidean path integrals. The non-Gaussian statistics of the matrix elements of $\cal O$ correspond to a generalization of the ETH ansatz.

We describe multiple ways to construct double-scaled matrix models that reproduce the gravitational disk correlators. One method involves imposing an operator equation obeyed by $H$ and $\cal O$ as a constraint on the two matrices. Separately, we design a model that reproduces certain double-scaled SYK correlators that may be scaled once more to obtain the disk correlators.

We show that in any single-trace, two-matrix model, the genus zero two-boundary expectation value, with up to one $\cal O$ insertion on each boundary, can be computed directly from all of the genus zero one-boundary correlators. Applied to the models of interest, we find that these cylinder observables depend on the details of the double-scaling limit. To the extent we have checked, it is possible to reproduce the gravitational double-trumpet, which is UV divergent, from a systematic classification of matrix model `t Hooft diagrams. The UV divergence indicates that the matrix integral saddle of interest is perturbatively unstable. A non-perturbative treatment of the matrix models discussed in this work is left for future investigations.

\end{abstract}

\pagebreak

{
\hypersetup{linkcolor=black}
\tableofcontents
}


\pagebreak

\section{Introduction}

The Eigenstate Thermalization Hypothesis (ETH) is an ansatz for the statistical properties of the matrix elements of simple operators between high energy microstates in quantum chaotic and thermalizing systems \cite{Deutsch,Srednicki1}. In a given system that exhibits level repulsion of an exponentially dense spectrum, such matrix elements are expected to be pseudo-random, and an ensemble can be formed by averaging within narrow energy bands. Alternatively, the randomness may be understood as referring to an ensemble of systems sharing the same simple thermal correlation functions. 

We will embed the ETH in a matrix model framework. This will make the ansatz more precise, extend its regime of validity, and clarify how to compute thermal higher point functions. In the examples we will consider,
the matrix model will reproduce correlation functions at time scales ranging from pre-thermalization to post-scrambling.  Matrix models of the type we define would not be expected to reproduce behavior at ultra short times in theories with a free UV, or at exponentially long times that depend on the specific fine grained microstates.

Perhaps surprisingly, the matrix models that describe thermal mean field theory are strongly non-Gaussian. Thus the standard Gaussian ETH ansatz does not agree with higher point functions at leading order, even in systems for which the thermal correlators factorize into one and two point functions. This observation resonates well with the non-Gaussian generalizations of the ETH ansatz put forward in \cite{Foini:2018sdb} capable to accommodate non-trivial OTOCs \cite{Sonner:2017hxc,Murthy:2019fgs,Anous:2019yku,Nayak:2019evx,Wang:2021mtp}. Despite these complications, one of the models we consider will turn out to be integrable in a certain sense.

A two-matrix model provides a unified framework for the ETH and the Wigner-type matrix model for the Hamiltonian $H$ that defines the spectrum. In this paper, we will demonstrate this for the quantum mechanics dual to JT gravity with propagating matter. With $\calo$ the operator dual to a free massive scalar field in the bulk, we will define double-scaled two-matrix models for $H$ and $\calo$, generalizing the SSS matrix model \cite{Saad:2019lba} and its deformations \cite{Witten:2020wvy}, \cite{Maxfield:2020ale}. The genus expansion will now be decorated by $\calo$ graphs, which are associated to geodesic worldlines.

Unlike pure JT gravity, which is exactly solvable and finite, JT gravity with propagating matter can at best be regarded as an effective field theory due to its UV divergences. As was mentioned in section 6.1 of \cite{Saad:2019lba} and section 5 of \cite{Penington:2019kki}, these divergences arise from long, thin wormholes. We find that it is possible to reproduce these divergences using matrix model perturbation theory. Because a nonperturbative treatment of the matrix integral is not necessary to study the bulk effective field theory, we may refer to our matrix models as ``effective matrix models.''

To be precise, we are interested in two-matrix models with a single-trace potential. These matrix integrals take the form
\begin{equation}
\label{eq:theansatz}
\int \, d\calo \, dH \, e^{- \text{Tr } V(H,\calo)}
\end{equation}
where $V(H,\calo)$ is an arbitrary (and generically infinite) linear combination of words made with the letters $\calo$ and $H$. In contrast, the SSS matrix model is a single-trace model of a single matrix, $H$. Single-trace models (with any number of matrices) admit a genus expansion where each order in the expansion corresponds to a sum over `t Hooft double-line graphs with a given topology. Furthermore, each trace present in an expectation value corresponds to a boundary. In the SSS model, the leading order result for the partition function of $H$ is given by a sum over `t Hooft graphs with disk topology:\footnote{$\braket{\cdot}$ refers to the expectation value defined using the matrix integral. The ``disk'' subscript refers to the topology of the `t Hooft diagrams.}
\begin{equation}
    \braket{\text{Tr } e^{-\beta H}}_{\text{disk}} = \raisebox{-0.45in}{\includegraphics[scale=0.40]{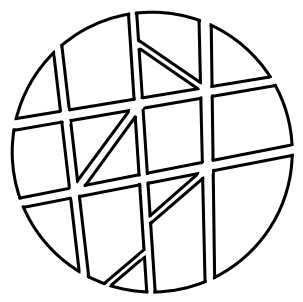}} + \cdots.
\end{equation}
This matrix model computation is equivalent to a disk gravitational path integral in pure JT gravity. In our two-matrix models, we can construct thermal $n$-point correlators using the matrices $\calo$ and $H$. To leading order in the genus expansion, the thermal two-point function in the matrix model is again given by a sum over planar diagrams:
\begin{equation}
\braket{\text{Tr } e^{-\beta_1 H} \calo e^{-\beta_2 H} \calo }_{\text{disk}} = \raisebox{-0.56in}{\includegraphics[scale=0.20]{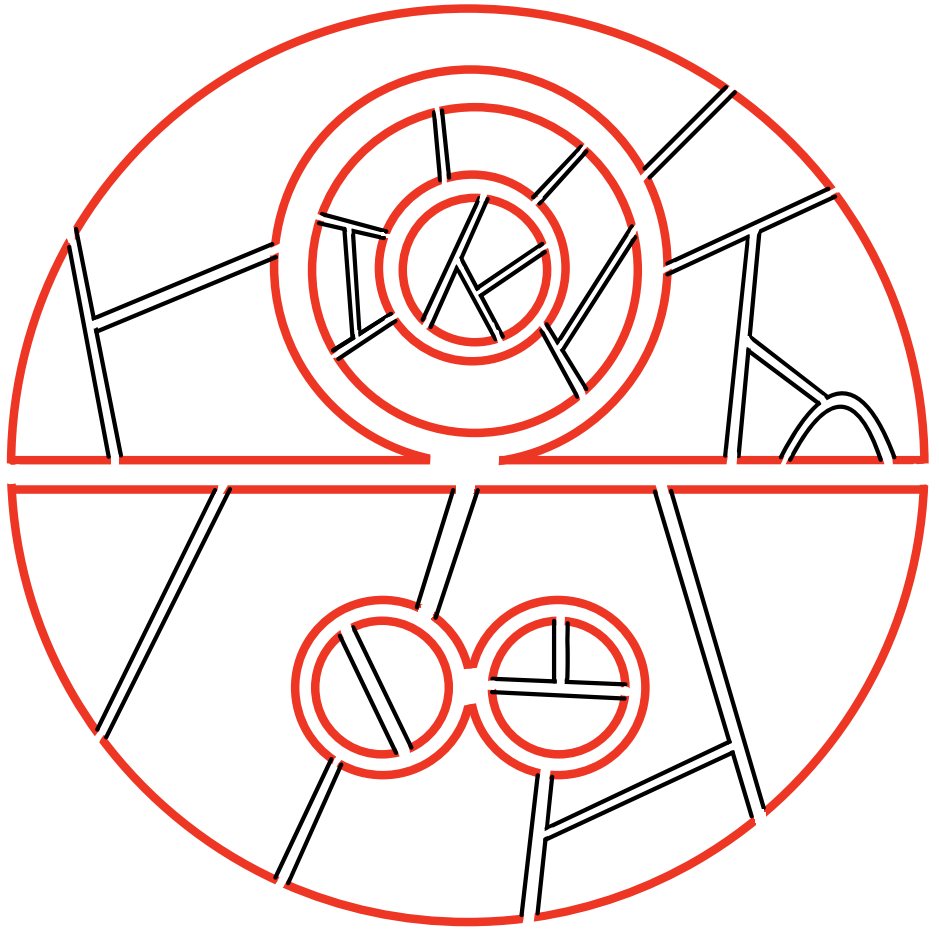}} + \cdots.
\label{eq:bigmess}
\end{equation}
The red double-line is a propagator of the $\calo$ matrix, while the black double-line is a propagator of the $H$ matrix. In JT gravity minimally coupled to a free massive scalar field,\footnote{Going forward, we may refer to this theory as ``JT gravity'' or ``JT gravity with matter'' out of convenience. The theory without matter will always be called ``pure JT gravity.''} the corresponding disk two-point function is computed using the extrapolate dictionary applied to a Euclidean black hole background.

The main technical objective of this paper is to explore the duality between single-trace, two-matrix models and JT gravity with matter. We are interested in correlators with additional $\calo$ insertions, additional boundaries, and additional handles. Our technical results may be summarized as follows:
\begin{itemize}
    \item In section \ref{sec:corrections}, we provide an algorithm for constructing a potential $V(H,\calo)$ for which the planar matrix model $n$-point functions (namely the higher-point analogues of \eqref{eq:bigmess}) equal the disk $n$-point functions of $\calo$ in JT gravity. This establishes the duality at the level of the disk. Explicitly determining $V(H,\calo)$ is tedious in practice. The algorithm introduces a fictitious parameter $\epsilon$ and weights each of the various terms that contribute to a gravitational $n$-point correlation function by some power of $\epsilon$, such that (roughly speaking) higher-point correlators are weighted by higher powers of $\epsilon$. The potential $V(H,\calo)$ is organized as a series expansion in $\epsilon$. The last step in the algorithm is to set $\epsilon \rightarrow 1$. We consider two specific schemes for $\epsilon$-deforming the gravitational correlators, which we call the ``Selberg regulator'' and the ``$q$-deformed regulator''. We have not proven that the series representation of $V(H,\calo)$ is convergent, so strictly speaking the existence of these matrix models is conjectural.
    \item In section \ref{sec:regtwomatrixmodel}, we argue that the matrix model defined using the $q$-deformed regulator can be generalized into a three-parameter model (where $\epsilon$ is one of the parameters) for which the disk correlators equal the correlators that were studied in \cite{Berkooz:2018jqr} in the double-scaled SYK model. It was shown in \cite{Berkooz:2018jqr} that these correlators can be scaled once more to obtain those of JT gravity with matter. To the extent that we checked, the three-parameter matrix model has a symmetry that may be predicted using the framework of \cite{Berkooz:2018jqr}. In our calculations, the appearance of this symmetry is highly non-trivial and provides further support for the claim that the model is well-defined.
    \item In section \ref{sec:ethasamatrixmodel}, we provide another argument for the existence of two-matrix models that reproduce the desired gravitational disk correlators. This argument is more explicit about the form that $V(H,\calo)$ takes. We first point out that in any ensemble-averaged theory whose correlators are exactly given by the disk correlators of JT gravity with matter (such as the SYK model in the appropriate scaling regime), the operators $\calo$ and $H$ obey an operator equation. In the semiclassical (or high-energy) limit,\footnote{By ``semiclassical limit,'' we are always referring to the $G_N \rightarrow 0$ limit. Equivalently, this is the limit in which the coefficient of the Schwarzian action, which we may call $\phi_b$, goes to infinity. By dimensional analysis, $\phi_b$ has units of length, and is the only scale in the Schwarzian theory. Because all inverse temperatures are naturally measured in units of $\phi_b$, the semiclassical limit may be thought of as a high energy limit. Furthermore, in this limit, the wiggly AdS boundaries become rigid, so the conformal invariance of boundary correlation functions is restored.} where the correlators become conformally invariant (with conformal group $PSL(2,\mathbb{R})$), this operator equation becomes the condition that the scaling dimensions of the primary operators appearing in the $\calo \calo$ OPE are in the set $\{1 \} \cup \{2 \Delta + 2 n \, : \, n \in \mathbb{Z}_{\ge 0} \}$. Using the 1D CFT bootstrap, one may prove\footnote{See appendix \ref{sec:gffproof} for the proof. See \cite{Mazac:2018mdx,Mazac:2018ycv} for further 1D CFT bootstrap results.} that this condition, together with associativity of the OPE and conformal invariance, guarantees that all of the $n$-point correlators of $\calo$ agree with those of a bosonic generalized free field (GFF). Although we are generally interested in JT gravity away from the semiclassical limit, this result nonetheless motivates us to construct a two-matrix ensemble by squaring our operator equation and adding it to the matrix potential with a large coefficient such that it is enforced as a constraint. Note that associativity of the OPE is guaranteed in our model because we are representing the operators $\calo$ and $H$ using matrices, and matrix multiplication is associative. This model is a single-trace matrix model and we conjecture that it correctly computes all of the disk correlators in JT gravity with matter. In section \ref{sec:crossingsymmetric}, we support this conjecture by showing that a large class of Schwinger-Dyson equations in this model are solved by the gravitational $n$-point correlators. To verify the Schwinger-Dyson equations, we need to use certain integrability relations. One of these is only available in the double-scaling limit.
    \item Our next results concern the two-boundary, genus zero correlators in the matrix model. These are computed by summing over `t Hooft diagrams with cylinder topology. In section \ref{sec:doubletrumpet}, we show that in any single-trace, two-matrix model, the connected two-boundary correlators \begin{equation}
    \label{eq:cyl1}
    \braket{\text{Tr } e^{- \beta_L H} \text{Tr } e^{- \beta_R H}}_{\text{cylinder}}
    \end{equation}
    and
    \begin{equation}
        \label{eq:cyl2}
        \braket{\text{Tr } \calo e^{- \beta_L H} \text{Tr } \calo e^{- \beta_R H}}_{\text{cylinder}}
        \end{equation}
    may be determined directly from the disk correlators even if the matrix potential $V(H,\calo)$ is unknown. Roughly speaking, our strategy is to cut into pieces the `t Hooft diagrams that contribute to the disk amplitudes and reassemble the pieces to form diagrams with cylinder topology, which we systematically classify. Our result is reminiscent of recursion in one-matrix models, where higher-genus and higher-boundary correlators may be recursively computed from lower-genus and lower-boundary correlators.\footnote{We are hopeful that there exists a general formulation of two-matrix model recursion, but finding it is beyond the scope of this work.} For a `t Hooft-scaled model (where the matrix potential is proportional to $N$, the number of eigenvalues of each matrix), our algorithm for computing the cylinder correlators yields unambiguous results. However, for the double-scaled matrix models that are dual to JT gravity at the level of the disk, our algorithm only returns an unambiguous answer once the scheme for taking the double-scaling limit is specified. We obtain results using the aforementioned Selberg and $q$-deformed regulators. As explained in the first point above, these regulators are part of the definition of the matrix models.
    
    Using the Selberg regulator, we match the matrix-model cylinder computations to their analogous double-trumpet calculations in JT gravity (either the empty double-trumpet or a double-trumpet with one $\calo$ inserted on each boundary). In JT gravity, these quantities receive contributions from the partition function of the scalar field on the rigid double-trumpet, which is given by
    \begin{align}
        Z_{\text{scalar}}(b) &= \sum_{n = 0}^\infty \frac{e^{- n \Delta b}}{(1-e^{-b})(1-e^{-2b}) \dots (1-e^{-nb})} \\
        &= {1\over (e^{-\Delta b}~ ;e^{-b})_\infty}
        \ ,
    \end{align}
    where $b$ is the length of the closed geodesic that wraps the double-trumpet and $\Delta$ encodes the mass of the scalar. We have explicitly matched the contributions from the first four terms above to the first four classes of `t Hooft diagrams that appear in our infinite classification of cylindrical `t Hooft diagrams. We conjecture that the sum over all the `t Hooft diagrams reproduces the full double-trumpet result. We further conjecture that our method for computing the sum over cylinder diagrams can be extended to compute the sum over diagrams with any topology, and we expect that the result will match the gravitational path integral with the same topology (including the matter determinant factor).
    
    Using the $q$-deformed regulator, the results for the cylinder correlators are the same as in the Selberg model, except the partition function is replaced by
    \begin{align}
        Z(b) &= \sum_{n=0}^\infty  \frac{e^{- n \Delta b}}{(1-e^{-b})^n} \\
        &= \frac{1 - e^{-b}}{1 - e^{-b} - e^{-\Delta b}} \ .
    \end{align}
    Curiously, this partition function has a Hagedorn temperature because the denominator goes to zero for sufficiently small $b$. Finding an independently-defined bulk theory that reproduces this result is an interesting question which is beyond the scope of this paper.
    \item Note that the gravitational duals of \eqref{eq:cyl1} and \eqref{eq:cyl2} are ill-defined because the gravitational path integral includes an integral over $b$, and this integral is either non-convergent (in the Selberg model) or simply ill-defined (in the $q$-deformed model) due to the small-$b$ behavior of the above partition functions. As mentioned above, we can still reproduce $Z_{\text{scalar}}(b)$ for all $b$ using the Selberg matrix model by writing the result as an infinite series of terms that are individually well-defined. In section \ref{sec:uvdivergences}, we explain that our two-matrix models are necessarily non-perturbatively unstable. To illustrate this point, we consider a single-matrix multi-trace matrix model for $H$ (which could arise by integrating out $\calo$ in \eqref{eq:theansatz}). For simplicity, we consider a `t Hooft scaled matrix model, where the $n$-trace\footnote{``1-trace'' corresponds to single-trace, ``2-trace'' corresponds to double-trace, etc.}  part of the potential is weighted by $N^{2 - n}$. We review that the partition function $\braket{\text{Tr } e^{- \beta H}}$ to leading order in $N$ is determined by the saddle-point of the integral over the eigenvalues of $H$. We then show that the connected two-boundary correlator $\braket{\text{Tr } e^{- \beta_L H} \text{Tr } e^{- \beta_R H}}_c$ to leading order in $N$ is directly determined by the Hessian of the matrix potential evaluated at the saddle. The bad behavior of the double-trumpet is directly linked to a perturbative instability of the saddle in the eigenvalue integral.
    \item Our final main result pertains only to JT gravity with matter. It was shown in \cite{Mertens:2017mtv,Suh:2020lco} that the disk $n$-point correlators may be conveniently computed using a set of Feynman rules. A graph drawn on a disk represents a set of Wick contractions of the external operators. We show in section \ref{sec:JTgravitywithmatter} that these Feynman rules may be naturally extended to other topologies. In particular, we draw graphs on genus zero surfaces with two or three boundaries and show that the naively-extended Feynman rules yield sensible results. This result supports many of the other results in this paper but may also be of independent interest to JT gravity experts.
\end{itemize}

We now outline the structure of the remainder of this paper. In section \ref{sec:JTgravitywithmatter}, we review JT gravity with matter with an emphasis on the disk and double-trumpet calculations that will be compared against the matrix models. We also conjecture how the gravitational Feynman rules which were originally developed for disk computations can be extended to arbitrary topologies. In section \ref{sec:ethasamatrixmodel}, we briefly step away from JT gravity to discuss the ETH and its relevance to holography in full generality. We then return to JT gravity to discuss the operator equation relating $\calo$ and $H$ that is used to construct a constrained matrix ensemble. In section \ref{sec:toymodel}, we introduce a toy model where $V(H,\calo)$ is Gaussian in $\calo$. This model succeeds in computing the holographic two-point function but fails to correctly compute higher-point functions. The purpose of introducing this model is to make the reader comfortable with two-matrix models, and also to introduce notations and ideas that will be useful later in the paper. Section \ref{sec:summary} summarizes the most important points of the remainder of the paper while avoiding most of the technical details. In particular, we outline the definitions of the Selberg and $q$-deformed matrix models, and we review the basic strategy of our double-trumpet computations. The reader who has no time can skip to the discussion section after section \ref{sec:summary}. In section \ref{sec:crossingsymmetric}, we show how in the constrained matrix ensemble, a large class of Schwinger-Dyson equations is solved by the disk correlators of JT gravity with matter. In section \ref{sec:corrections}, we explain in more detail our algorithm that allows one to start from the answers for the holographic disk $n$-point functions and work backwards to determine the potential $V(H,\calo)$. In section \ref{sec:regtwomatrixmodel} we carefully define the Selberg and $q$-deformed matrix models and explain the connection between the $q$-deformed model and the double-scaled SYK model, using the results of \cite{Berkooz:2018jqr}. In section \ref{sec:doubletrumpet} and appendix \ref{sec:dtw=3}, we discuss our computation of \eqref{eq:cyl1} and \eqref{eq:cyl2} in full detail. We compare the matrix model results against their gravitational counterparts. We also discuss the perturbative instability in the matrix model that is dual to the UV divergence in the double-trumpet. In the discussion section we speculate on how our results may be generalized to higher dimensions, and we compare our results with other works that used the ETH to study JT gravity with matter. The appendices mostly contain derivations of various special function identities as well as other technical calculations.


\section{JT gravity with matter}

\label{sec:JTgravitywithmatter}

In this section we review JT gravity \cite{Maldacena:2016upp, Saad:2019lba} minimally coupled to a free massive scalar. Correlation functions on the disk are determined by a set of Feynman rules \cite{Mertens:2017mtv}. Using the boundary particle formulation of JT gravity \cite{Yang:2018gdb, Kitaev:2018wpr, Suh:2020lco}, we also compute the 2-point function on the double-trumpet, including the matter 1-loop determinant. The result is again described by the Feynman rules of \cite{Mertens:2017mtv},  suggesting that they generalize to correlation functions on all genus $g$ Riemann surfaces with $n$ boundaries ${\cal M}_{g,n}$.

We consider JT gravity coupled to a scalar with the Euclidean action
\begin{align}
	I[g,\phi,\varphi] &= - S_0 \chi  + I_{JT}[g,\phi] + I_{m}[g,\varphi] \ ,  
	\\
	I_{JT}[g,\phi] &= - \frac{1}{2} \int_{\mathcal{M}} \sqrt{g} \phi (R + 2) - \int_{\partial \mathcal{M}} \sqrt{h} \phi (K-1) \ ,
	\\
	I_{m}[g,\varphi] &= \frac{1}{2} \int_{\mathcal{M}} \sqrt{g} \left(g^{ab} \partial_a \varphi \partial_b \varphi  + m^2 \varphi^2\right),
\end{align}
where $\chi$ is the Euler characteristic of the two-dimensional manifold, $S_0$ is the entropy of an extremal black hole, $\phi$ is the dilaton and $\varphi$ is a scalar field with mass $m$. To compute the path integral on ${\cal M}_{g,n}$, interpreted as $\langle Z(\beta_1) \dots Z(\beta_n) \rangle_{g,n}$ in the dual matrix model \cite{Saad:2019lba}, we fix the regulated boundary lengths of $n$ asymptotically AdS regions to be $\beta_1/\e, \dots \beta_n/\e$ and the dilaton $\phi|_{\p M} = {\gamma/\e}$. We follow the conventions of \cite{Saad:2019lba} and set $\gamma = 1/2$.\footnote{We use a different convention for the overall normalization of the density of states, see below.} For the scalar field $\varphi$ we take either Dirichlet of Neumann boundary conditions, such that the dual boundary operator $\calo$ has the scaling dimension
\begin{equation}
	\Delta = \frac{1}{2} \pm \sqrt{\frac{1}{4} + m^2} \ .
	\label{eq:deltam2}
\end{equation}
The choice of sign depends on which of the two standard boundary conditions we choose for $\varphi$. Given any $\Delta > 0$, there is a unique choice of the sign and $m^2> -{1\over 4}$ such that \eqref{eq:deltam2} holds.

\subsection{Correlation functions on the disk}
\label{sec:disk}

Disk correlators of $\calo$ have been computed in \cite{BAGRETS2016191,Mertens:2017mtv, Yang:2018gdb, Kitaev:2018wpr,Suh:2020lco}, which we now review. In the absence of gravity, the correlators of $\calo$ are those of a generalized free field (GFF). The 2-point function at zero temperature is $\la \calo(\tau) \calo(0) \ra_{\text{GFF}} = \tau^{-2\Delta}$ and higher correlators are computed by Wick contractions. 

To include the gravitational corrections, one must reparameterize the GFF correlators and then integrate over all reparameterizations with the Schwarzian action,
\begin{align}
\label{diskPI}
\begin{split}
	&\braket{\tr e^{-\beta H}\calo(\tau_1) \dots \calo(\tau_n)}_{\disk} 
\\
&=e^{S_0}\int \frac{\mathcal{D} f}{SL(2,\mathbb{R})} \quad  (f^\prime(\tau_1))^\Delta \dots (f^\prime(\tau_n))^\Delta \braket{\calo(f(\tau_1)) \dots \calo(f(\tau_n))  }_{\text{GFF}} ~ e^{-I_{Sch}[f]} \ ,
\end{split}
\end{align}
where $f(\tau)$ is a reparameterization field that determines the embedding of the boundary curve into the hyperbolic disk. The notation on the left hand side of \eqref{diskPI} indicates that the gravitational correlator is not normalized by the partition function $\la \tr e^{-\beta H} \ra$.

\begin{figure}
	\centering
	\includegraphics[scale=.5]{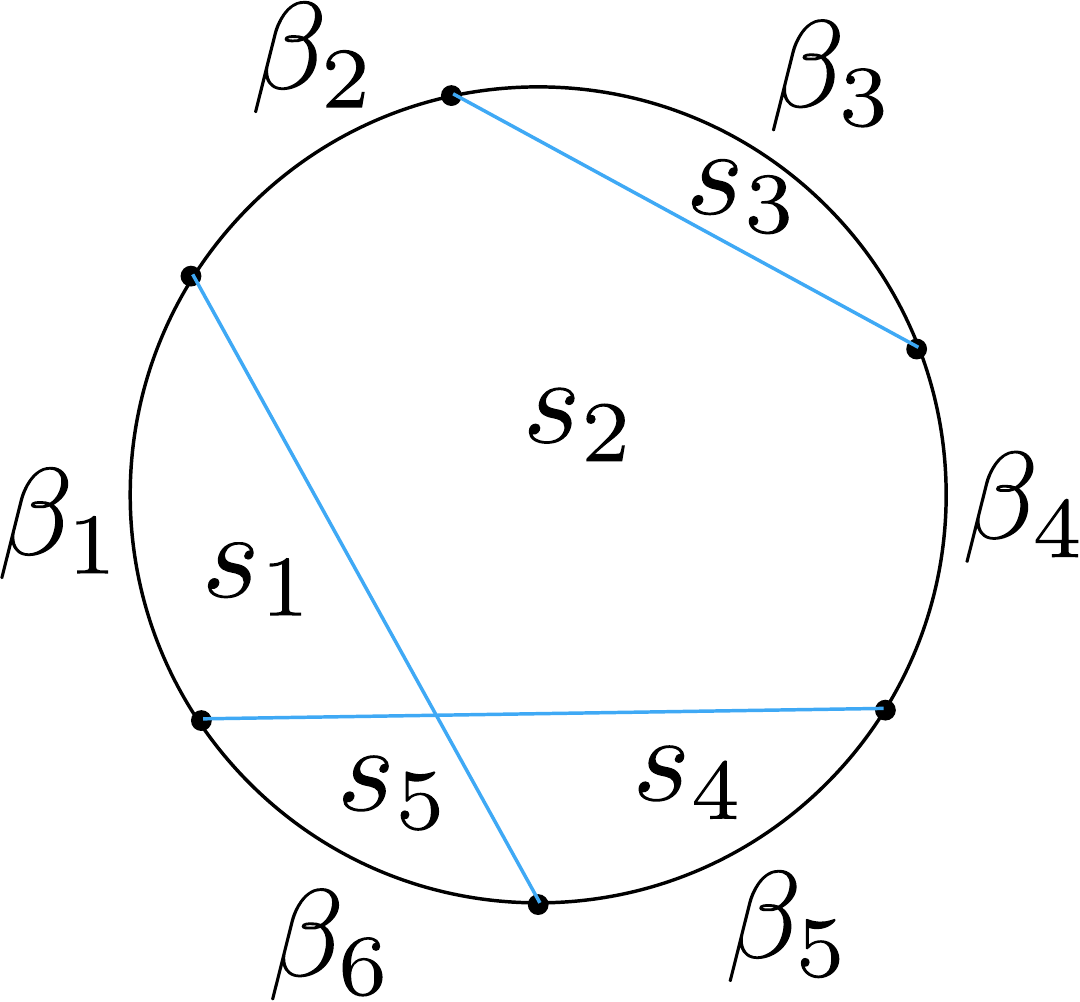}
	\caption{An example of a Feynman diagram that contributes to the thermal six-point $\calo$ correlator. Each blue line is a Wick-contraction through the bulk of two boundary operators with scaling dimension $\Delta$. The correlator is a sum over all ways of contracting the operators. Each disk-shaped region is labeled by an $s$ parameter that is integrated in the range $s \in (0,\infty)$ to obtain the value of the correlator. The Euclidean-time separation between the external operators is indicated by the $\beta_1,\cdots,\beta_6$ parameters. We adopt a convention where a blue bulk line always has scaling dimension $\Delta$.
}
	\label{fig:feyndiag}
\end{figure}

The disk correlator of an arbitrary number of $\calo$ operators can be computed exactly, and the result resums all the perturbative gravitational corrections. In particular, \cite{Mertens:2017mtv} devised an elegant set of Feynman rules that assigns a finite-dimensional integral to each Wick contraction of the $\calo$ operators.\footnote{An alternative derivation of these rules was given in \cite{Suh:2020lco}, which is based on a formalism developed in \cite{Kitaev:2018wpr,Yang:2018gdb}.}
For example, a Feynman diagram that represents a particular six-point Wick-contraction is depicted in Figure \ref{fig:feyndiag}.

\vskip .3in

The value of each Feynman diagram is set by the following rules:
\begin{itemize}
	\item For each boundary segment of length $\beta$, we include a factor of $e^{- \beta s^2}$, where $s$ labels the disk-shaped region that is adjacent to the boundary segment.
	\item For each $\calo$ insertion, we include a factor of
\begin{equation} 
	\raisebox{-.4in}{\includegraphics[scale=.35]{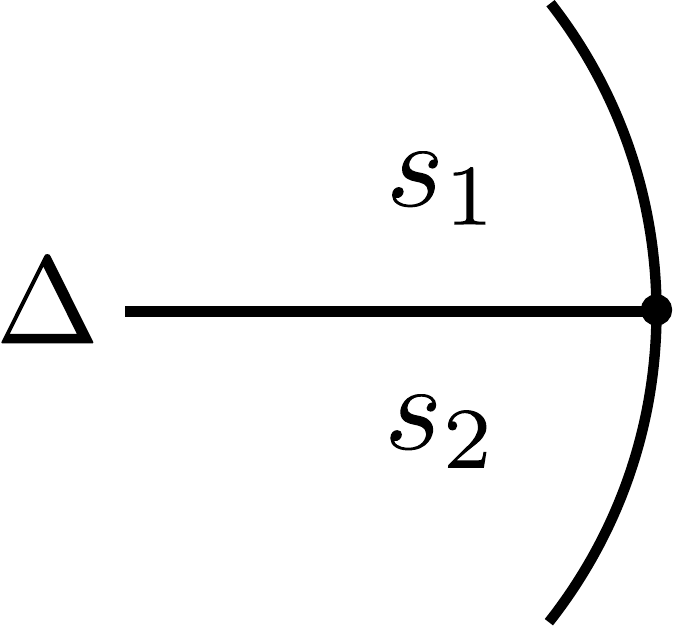}}
	~~~=~~~
\left( \Gamma(\Delta \pm i s_1 \pm i s_2) \over \Gamma(2\Delta) \right)^{1/2}
\equiv \left(\Gamma^{\Delta}_{12}  \right)^{1/2}
\ ,
\label{eq:feynruleoboundary}	
\end{equation}
	 where $s_1$ and $s_2$ are associated to the two regions adjacent to the operator, and $\pm$ means that we take a product of gamma functions for all four choices of the signs\footnote{That is $\Gamma(\Delta \pm i s_2 \pm i s_2) = \Gamma(\Delta +i s_1 +is_2)\Gamma(\Delta +i s_1 -is_2)\Gamma(\Delta -i s_1 +is_2)\Gamma(\Delta -i s_1 -is_2)$.}. Here, the normalization of operators is chosen to be such that at short distances the two-point function is $\langle \tr e^{-\beta H} \calo(\tau) \calo(0) \rangle_{\disk} \approx \tau^{-2\Delta} \, \la \tr e^{- \beta H} \ra_{\disk} \ , \tau \to 0 $.
	 \item For each crossing of two
	 lines, we include a factor of 
	 \begin{align}
	 	\raisebox{-.35in}{\includegraphics[scale=.35]{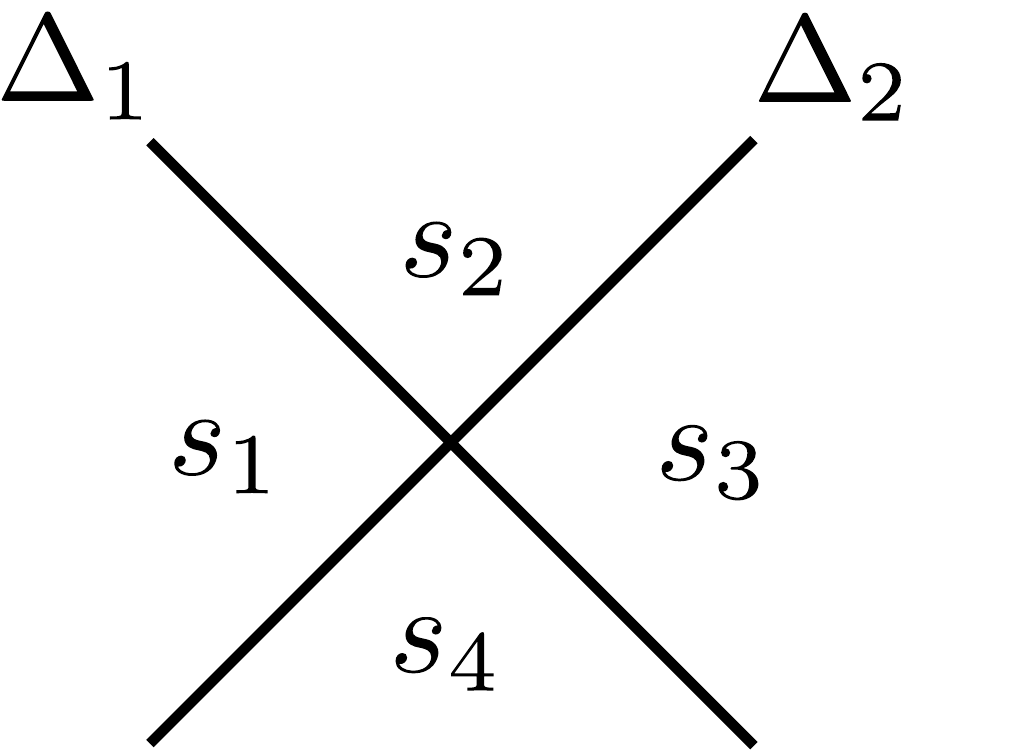}}
	 	~~~ = ~~~
	 	\left\{\begin{array}{ccc}
	 		\Delta_1 & s_1 & s_2 \\
	 		\Delta_2 & s_3 & s_4
	 	\end{array}\right\} 
	 	\ ,
 	\label{eq:sixjsymbol}
	 \end{align}
 	which is the 6j-symbol of the $\mathfrak{sl}(2,\mathbb{R})$ algebra.\footnote{6j symbols also appear in the computation of AdS amplitudes in general dimensions, as was shown in \cite{Liu:2018jhs}.} The parameters $\Delta_1$ and $\Delta_2$ represent the scaling dimensions associated to the two crossing lines. The parameters $s_1,\cdots,s_4$ represent the four disk-shaped regions that surround the crossing. In the theory with one scalar operator we set $\Delta_1 = \Delta_2=\Delta$. More generally, if we had two different free scalar fields in the bulk, we would use \eqref{eq:sixjsymbol}. The symmetries of the 6j-symbol are the symmetries of its graphical representation, e.g. reflections across $\Delta_1$ or $\Delta_2$ lines and reflections across the horizontal or vertical axes. The definition of the 6j-symbol and its properties are reviewed in Appendix \ref{sec:specialfunctions}. 
 	\item After including all of the appropriate factors as specified above, we integrate over each $s$ parameter with the Schwarzian density of states $\int_0^\infty ds \, \rho(s)$, where\footnote{Our convention for the Schwarzian density of states is related to \cite{Saad:2019lba} by $\rho(s) = 2\rho_{there}(s)$. Equivalently, we use conventions of \cite{Saad:2019lba} with a rescaling $e^{S_0} \to 2 e^{S_0}$. We find this normalization more convenient for computations involving $\mathfrak{sl}(2,\mathbb{R})$ 6j-symbols. }
 	
 \begin{equation}
 \rho(s) = {1\over 2\pi \Gamma(\pm 2i s)}= \frac{s}{\pi^2}  \sinh (2 \pi s) \ . \label{eq:rhos} 
 \end{equation}
 Sometimes we use the energy basis $E=s^2$ and the corresponding density of states 
 \begin{align} \label{rhoE}
\rho_0(E) = {1\over 2\pi^2} \sinh (2\pi \sqrt{E})
 \end{align}
 such that $\rho(s) ds = \rho_0(E) dE$.

\end{itemize}

\subsubsection*{Examples}

To demonstrate how the rules described above work in practice and also for later use, we explicitly write down a few correlation functions.

We start with the 2-point function 
\begin{align}\label{2ptGrav}
    \la \tr e^{-\beta H} \calo(\tau) \calo(0) \ra_{\disk}
    &=      	\raisebox{-.5in}{\includegraphics[scale=.2]{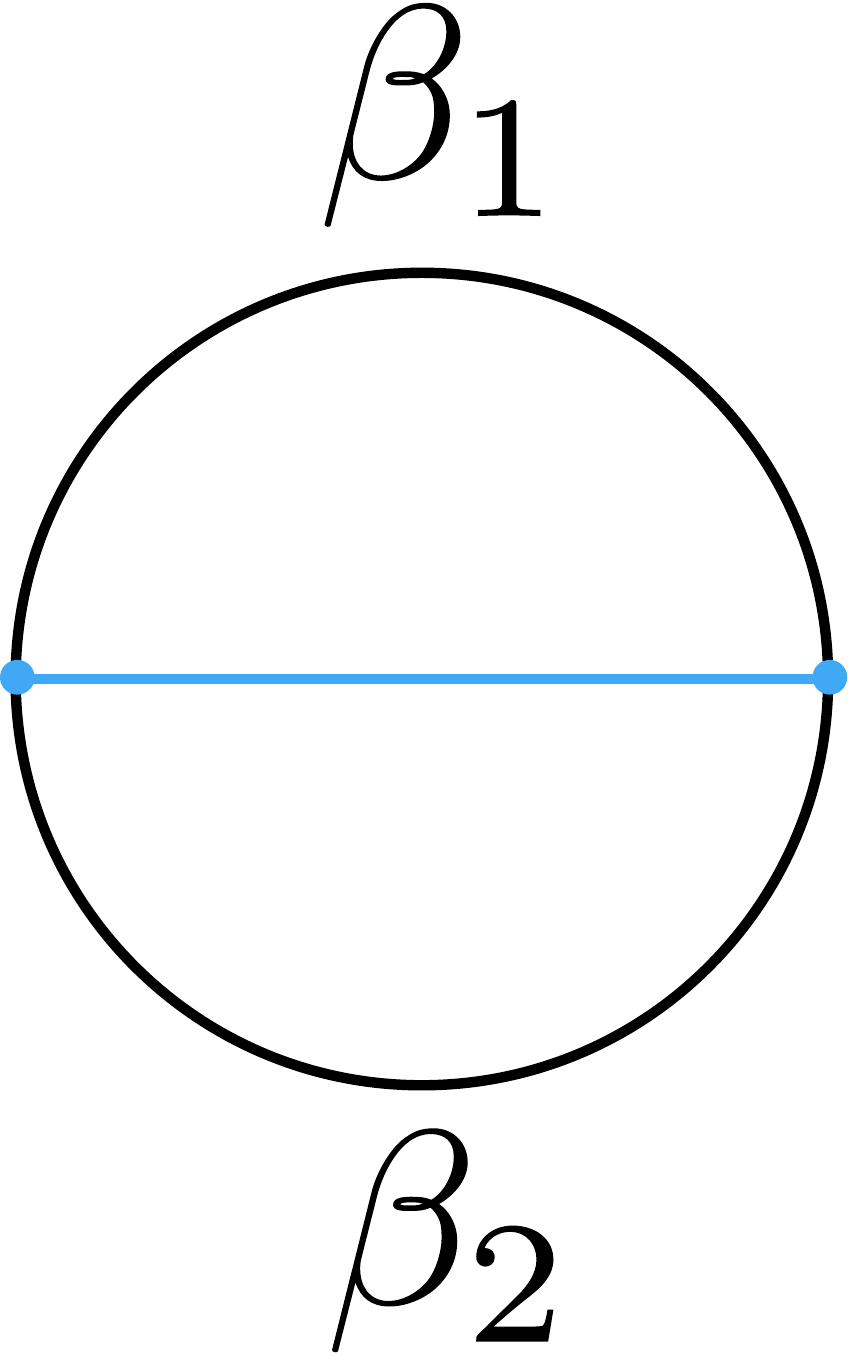}}    	
    = e^{S_0} \int_0^\infty ds_1 ds_2~
    \rho(s_1) \rho(s_2) 
    ~ e^{-(\beta_1 s_1^2 + \beta_2 s_2^2)} 
    {\Gamma(\Delta \pm i s_1 \pm i s_2) \over \Gamma(2\Delta)}
    \ ,
\end{align}
where $\beta_1 = \tau, \beta_2 = \beta - \tau$. The 4-point function has 3 terms corresponding to 3 different GFF Wick contractions
\begin{align}
\label{4pt diag}
\la \tr e^{-\beta H} \calo(\tau_1) \dots \calo(\tau_4) \ra_{\disk}
&=  
  \raisebox{-.4in}{\includegraphics[scale=.3]{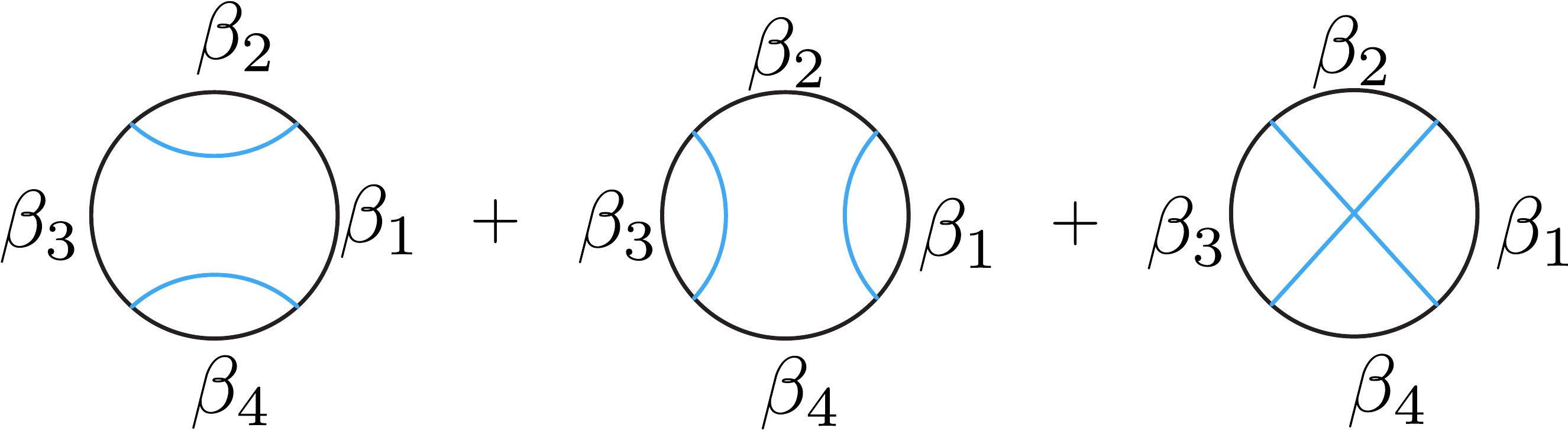}}    	\\
    &= e^{S_0} \int_0^\infty 
     \prod_{i=1}^4 \left( ds_i ~\rho(s_i) e^{-\beta_i s_i^2} \right)
     {\left( \Gamma_{12}^\Delta \Gamma_{23}^\Delta \Gamma_{34}^\Delta \Gamma_{41}^\Delta \right)^{1/2}  } \\
    &\qquad \qquad \times \left( 
     {\delta(s_1 -s_3) \over \rho(s_1)} 
     + {\delta(s_2 -s_4) \over \rho(s_2)}
     + \left\{\begin{array}{ccc}
	 		\Delta & s_1 & s_2 \\
	 		\Delta & s_3 & s_4
	 	\end{array}\right\} 
     \right)
    \ ,
\label{4pt}
\end{align}
where $\beta_i$ denote the time differences between operators. Assuming $0<\tau_4 < \dots < \tau_1 < \beta$, we defined $\beta_1 =\beta - (\tau_1 - \tau_4), \beta_2 = \tau_1-\tau_2, \beta_3 = \tau_2-\tau_3, \beta_4 = \tau_3 - \tau_4$. For convenience, in the first two terms we introduced two energies for the same region and included corresponding delta-functions to remove one of the energy integrals. In Appendix \ref{4ptCoeffCheck} we check that the relative coefficient between the first two terms and the third term is indeed as given in \eqref{4pt}. Later, we will also need diagrams contributing to the 6-point function and we will discuss them in due course.

\vskip .3in

For some Feynman diagrams we have a choice of where to put the intersections of the bulk lines. Once we fix a Wick-contraction, the result is independent of where we put the intersections. This is guaranteed by the orthogonality and Yang-Baxter equations. We express them pictorially as
\begin{align}
    &\raisebox{-.2in}{\includegraphics[scale=.2]{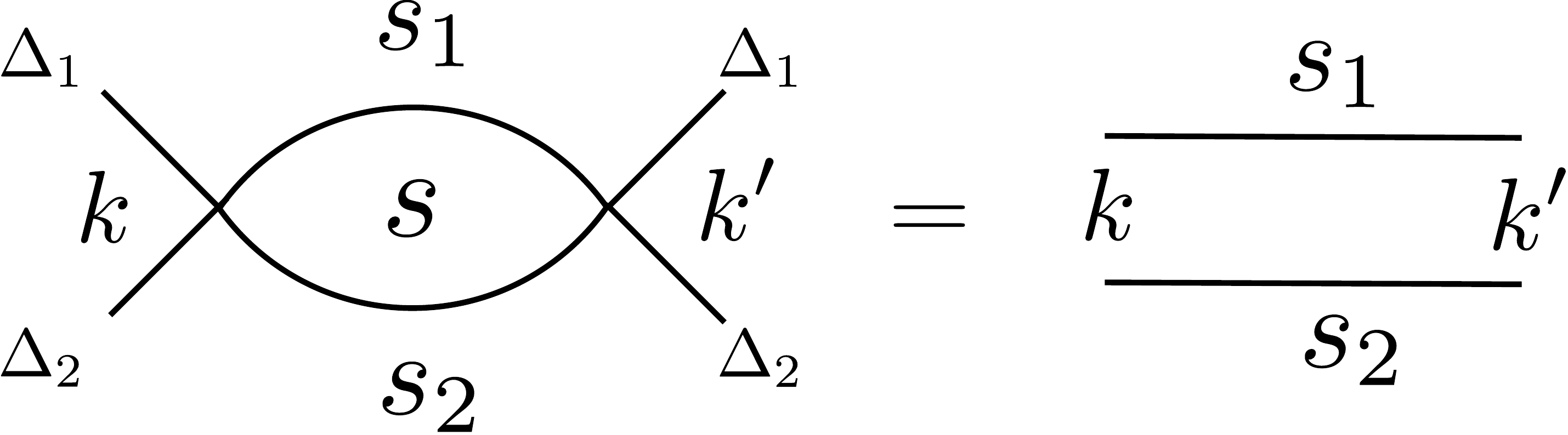} \label{eq:6jorthog}} \ \  , \\
    &\raisebox{-.2in}{\includegraphics[scale = .21]{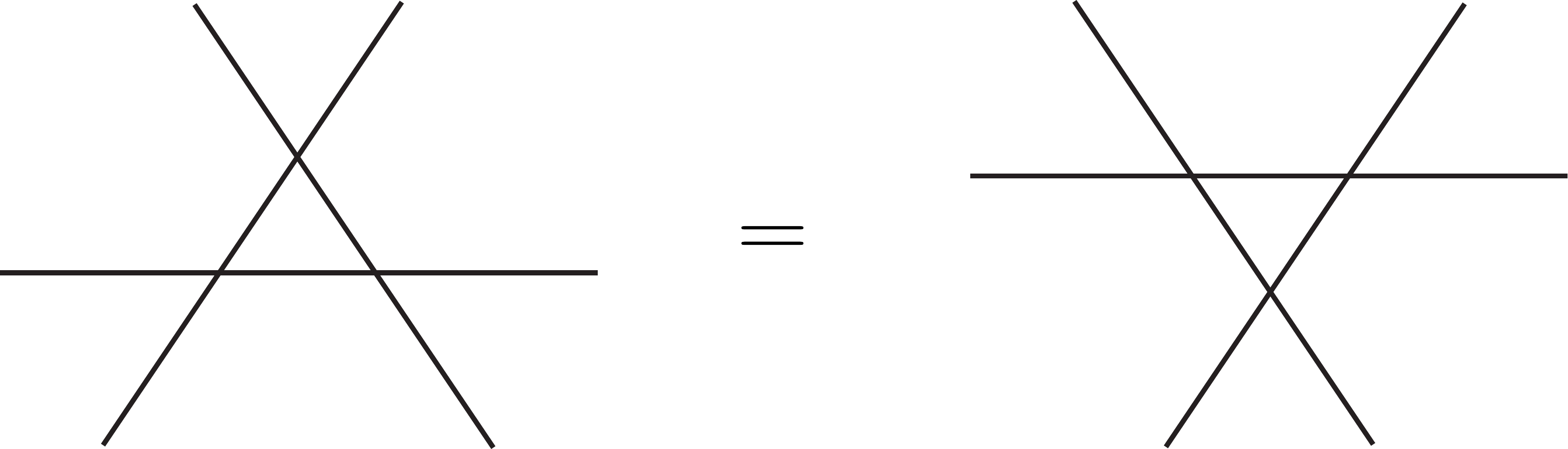} \label{eq:YB}} \ ,
\end{align}
respectively.
The expression for the orthogonality relation is
\begin{align}
    \int_0^\infty ds ~ \rho(s)  
        \left\{\begin{array}{ccc}
			\Delta_1 & s_1 & k \\
			\Delta_2 & s_2 & s
		\end{array}
		\right\} 
		\left\{\begin{array}{ccc}
			\Delta_1 & s_1 & k' \\
			\Delta_2 & s_2 & s
		\end{array}
		\right\} = \frac{\delta(k - k')}{\rho(k)} \ .
		\label{eq:2.16}
\end{align}
Similarly, one can write the Yang-Baxter equation from \eqref{eq:YB}. If we impose the condition that the bulk lines should not have any voluntary crossings, then only the Yang-Baxter equation is needed to ensure that the Feynman rules are unambiguous. More identities satisfied by the 6j-symbol are described in Appendix \ref{sec:specialfunctions}. In section \ref{sec:regtwomatrixmodel} we will encounter a $q$-deformed set of Feynman rules where the 6j-symbol obeys the Yang-Baxter equation, but not the orthogonality equation.

\subsection{Correlation functions on the double-trumpet}

\label{sec:doubletrumpetreview}

We are also interested in correlation functions on the double-trumpet. Here, we discuss the two simplest cases: the double-trumpet with no $\calo$ insertions, and the 2-point function on the double-trumpet, with one $\calo$ on each boundary. In both cases, we include the matter 1-loop determinant

\subsubsection*{Double-trumpet with 1-loop determinant}

The path integral on the double-trumpet without $\calo$ insertions is
\begin{align}	\label{eq:dttoy}
\braket{\tr e^{-\beta_L H} \tr e^{-\beta_R H}}_{\cyl}
	&= \int_0^\infty db ~b \, Z_{\text{tr}}(\beta_L,b) Z_{\text{tr}}(\beta_R,b) \, Z_{\text{scalar}}(b) \\
	& = \raisebox{-.35in}{\includegraphics[scale=.18]{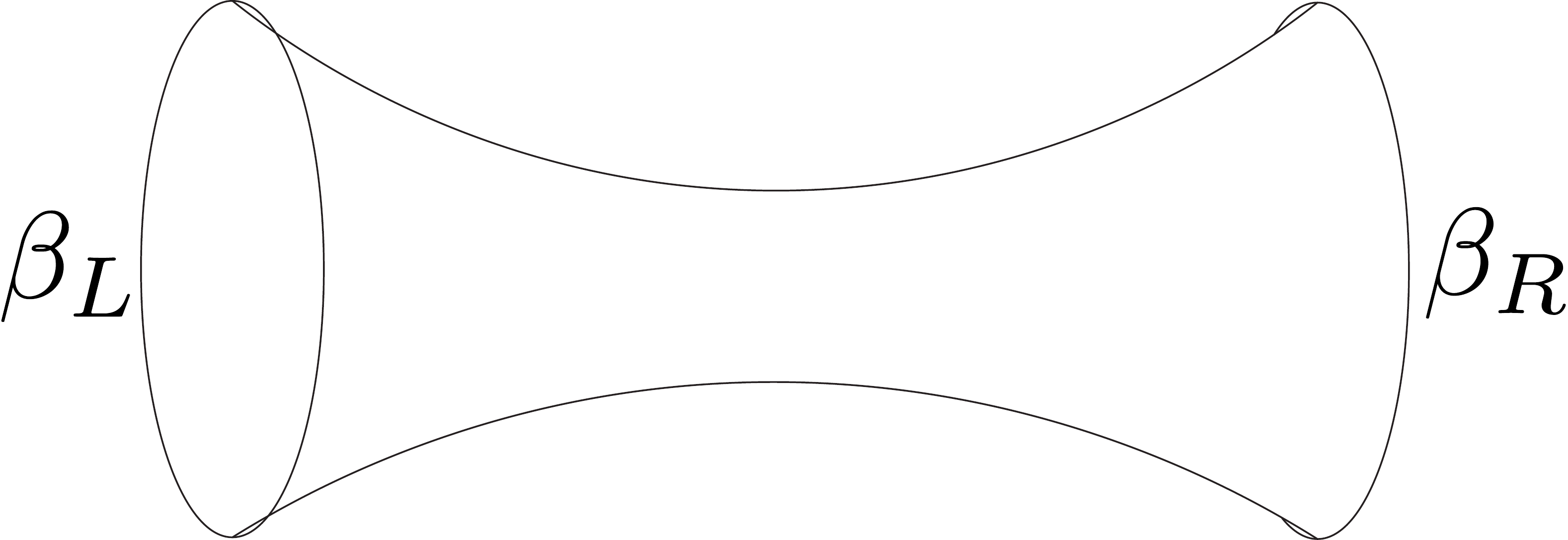} }
	\ ,
\end{align}
where $\text{cyl}$ indicates the cylinder topology. The partition function of the scalar field on the double-trumpet is $Z_{\text{scalar}}$, which we explicitly define in \eqref{eq:first}.
If we omit the $Z_{\text{scalar}}$ term, we obtain the answer computed in \cite{Saad:2019lba} for pure JT gravity. It is given as an integral over the moduli space of hyperbolic manifolds with two boundaries, parameterized by the length of the closed geodesic $b$. The ``trumpet'' partition function $Z_{\text{tr}}(\beta,b)$ represents the integral over the extrinsic curvatures of a ``wiggly'' AdS boundary and is given by \cite{Saad:2019lba}
\begin{align}
	Z_{\text{tr}}(\beta,b) 
	&=\raisebox{-.35in}{\includegraphics[scale=.2]{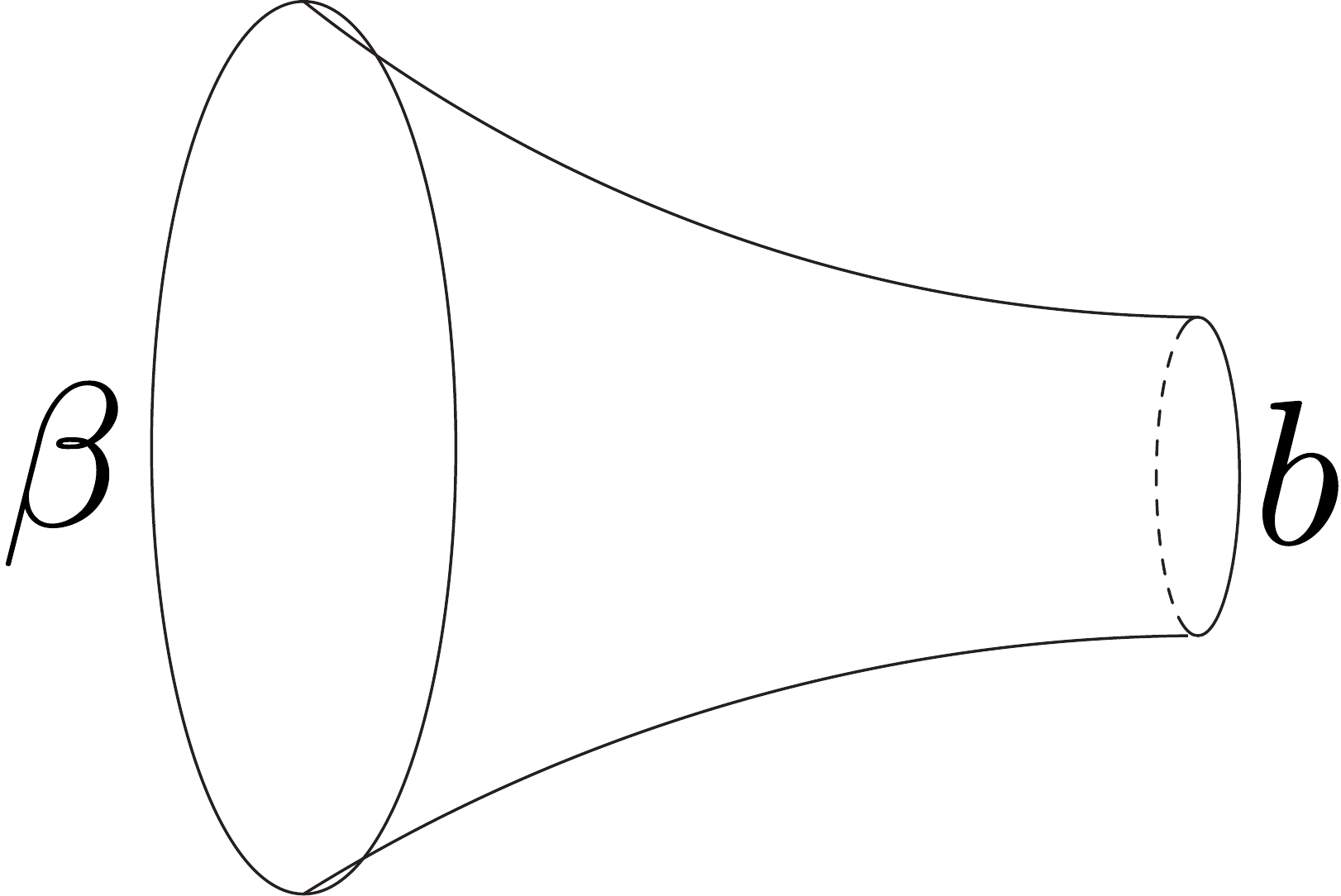} } \\
	&= \int_0^\infty ds \, {\cos(b s) \over \pi} e^{- \beta s^2} \\
    &= {1\over 2\sqrt{\pi }}  \beta^{-1/2}e^{-{b^2\over 4\beta}} \ .
	\label{eq:ztr}
\end{align}

\noindent
To compute $Z_{\text{scalar}}$, we may canonically quantize the scalar field on the analytic continuation of the double-trumpet to Lorentzian signature with the metric
\begin{equation}
	ds^2 = d\rho^2 -  \, \cosh^2 \rho ~  dt^2 \ ,
\end{equation}
where $\rho \in (-\infty,\infty)$ is the spatial coordinate, and $t$ is time. Each positive-frequency mode is labeled by $n \in \mathbb{Z}_{\ge 0}$ and has energy $\Delta + n$. The partition function, computed with periodicity $b$ in the imaginary time direction, is thus
\begin{align}
\label{eq:first}
	Z_{\text{scalar}}(b) &= \prod_{n = 0}^\infty \frac{1}{1 - e^{-b(\Delta + n) }} 
	\\ &= \sum_{n = 0}^\infty \frac{e^{- n \Delta b}}{(1-e^{-b})(1-e^{-2b}) \dots (1-e^{-nb})} \label{eq:second}
	\\ &= \exp\left(\sum_{w = 1}^\infty \frac{e^{- w b \Delta }}{w(1 - e^{- w b})}\right),\label{eq:third}
\end{align}
where we provided three equivalent expressions with different interpretations. The first formula makes the relation with the single particle spectrum manifest. In the second formula \eqref{eq:second}, we can separate out the counting of conformal primaries and descendants. For example, the $n=1$ term $${e^{-\Delta b } \over 1-e^{-b}} = \sum_{k=0}^\infty e^{-(\Delta +k)b}$$ counts the primary state $\Delta$ together with its descendants. The $n=2$ term \begin{equation}{e^{-2\Delta b } \over (1-e^{-b}) (1-e^{-2b}) }=\sum_{m=0}^\infty {e^{-(2\Delta + 2m)b} \over 1-e^{-b}} \label{eq:ne2term}\end{equation} counts the double-trace operators (states) $2\Delta + 2m$, where the factor $${1\over 1-e^{-b}}=\sum_{k=0}^\infty e^{-k b}$$ again accounts for the descendants. The $n=3$ term counts the primary states with dimensions $3\Delta + 2m_1 + 3 m_2$ (for $m_1, m_2 \in \mathbb{Z}_{\ge 0}$) and their descendants. The pattern continues for higher-trace operators. Finally, the third expression \eqref{eq:third} connects with the Selberg trace formula which we discuss next.\footnote{To relate \eqref{eq:third} to \eqref{eq:first} we first expand ${1\over 1-e^{-w b}} = \sum_{n=0}^\infty e^{- n w b}$ and then sum over $w$.}

\vskip .3in

An alternative way to compute the 1-loop determinant, which is simpler to generalize to an arbitrary hyperbolic manifold, is to use the Selberg trace formula. It relates the spectrum of the Laplace operator on a hyperbolic manifold ${\cal M}$ to the spectrum of closed geodesics (e.g. see \cite{Buser})
\begin{align}\label{Selberg}
\text{tr } e^{{\tau\over 2} \nabla^2} = f(\tau)A  + {e^{-\tau/8} \over \sqrt{2\pi \tau}} \sum_\gamma \sum_{w=1}^\infty {b_\gamma \over \sinh{wb_\gamma \over 2}} e^{- {(w b_\gamma)^2 \over 2\tau}} \ .
\end{align}
In the RHS the sum is over all closed geodesics $\gamma$ and $b_\gamma$ is the length of this geodesic.\footnote{Here, our convention is that we sum over unoriented primitive geodesics.} The sum over $w$ can be thought of as a sum over multiple windings of the ``primitive'' (traversed once) geodesic $\gamma$. In the first term in the RHS $A$ is the area of the manifold and $f(\tau)$ does not depend on the particular manifold and will not play a role in our discussion.\footnote{$f(\tau) = {e^{-\tau/8}\over (2\pi \tau)^{3/2}} \int_0^\infty dr {r e^{-r^2/2\tau} \over \sinh {r\over 2}}$}.

To compute the determinant we use the standard trick to relate it to the heat kernel
\begin{align}
Z_{scalar}({\cal M})
&=  \det(-\nabla^2 + m^2)^{-1/2} \\
&= \exp {-1\over 2}\text{tr}\log(-\nabla^2 + m^2) \\
&=\exp {1\over 2} \text{tr} \int_{\e}^\infty {d\tau \over \tau} e^{-{\tau\over 2} (-\nabla^2 + m^2)} 
\end{align}
up to a constant that diverges as we take $\e \to 0$. After inserting the Selberg trace formula \eqref{Selberg} and integrating over $\tau$ we have
\begin{align}\label{detSel}
Z_{scalar}({\cal M})
= \exp\left( 
\# A + 
{1\over 2} \sum_\gamma \sum_{w=1}^\infty {e^{-w b_\gamma (\Delta  - {1\over 2}) } \over w \sinh {w b_\gamma \over 2}} 
\right) \ .
\end{align}
Strictly speaking, the Selberg trace formula applies to compact hyperbolic manifolds. On the other hand, we are interested in manifolds with infinite area. To deal with this we use Gauss-Bonnet theorem to write $A= -2\pi \chi + \int_{\p M} K$ and absorb this term into the renormalization of $S_0$ and $\gamma$. 

On the double-trumpet there is a single primitive geodesic and \eqref{detSel} agrees with \eqref{eq:third}. While on the disk there are no closed geodesics and the 1-loop determinant contributes only the area term that renormalizes $S_0,\gamma$.

\vskip .3in

In the presence of matter the integral \eqref{eq:dttoy} has a UV divergence at small\footnote{This is analogous to the ``tachyon'' divergence in string theory.} $b$ \cite{Saad:2019lba}. This can be seen by expanding the Pochhammer symbol $(e^{-\Delta b}, e^{-b})_\infty = \prod_{n=0}^\infty (1-e^{-b(\Delta +n)})$ in \eqref{eq:first} at small $b$. Alternatively we can note that in the UV limit $b \to 0$ the mass is not important and we can approximate by a massless free scalar on a strip with the partition function dominated by the vacuum $\sim e^{{1\over 24} {4\pi^2 \over b}} = e^{\pi^2 \over 6b}$.

 We consider this divergence as a reflection of the fact that JT with matter should be considered as a low energy effective field theory. We will match this divergence in the matrix model. In other words, we consider correlation functions at fixed $b$ and match them with a computation in the matrix model.

\subsubsection*{2-point correlator}

The next observable we consider is the double-trumpet with one $\calo$ inserted on each boundary 
\begin{align}
\braket{\tr e^{-\beta_L H} \calo \tr e^{-\beta_R H} \calo}_{\cyl}
= ~
\raisebox{-.35in}
{\includegraphics[scale=.2]{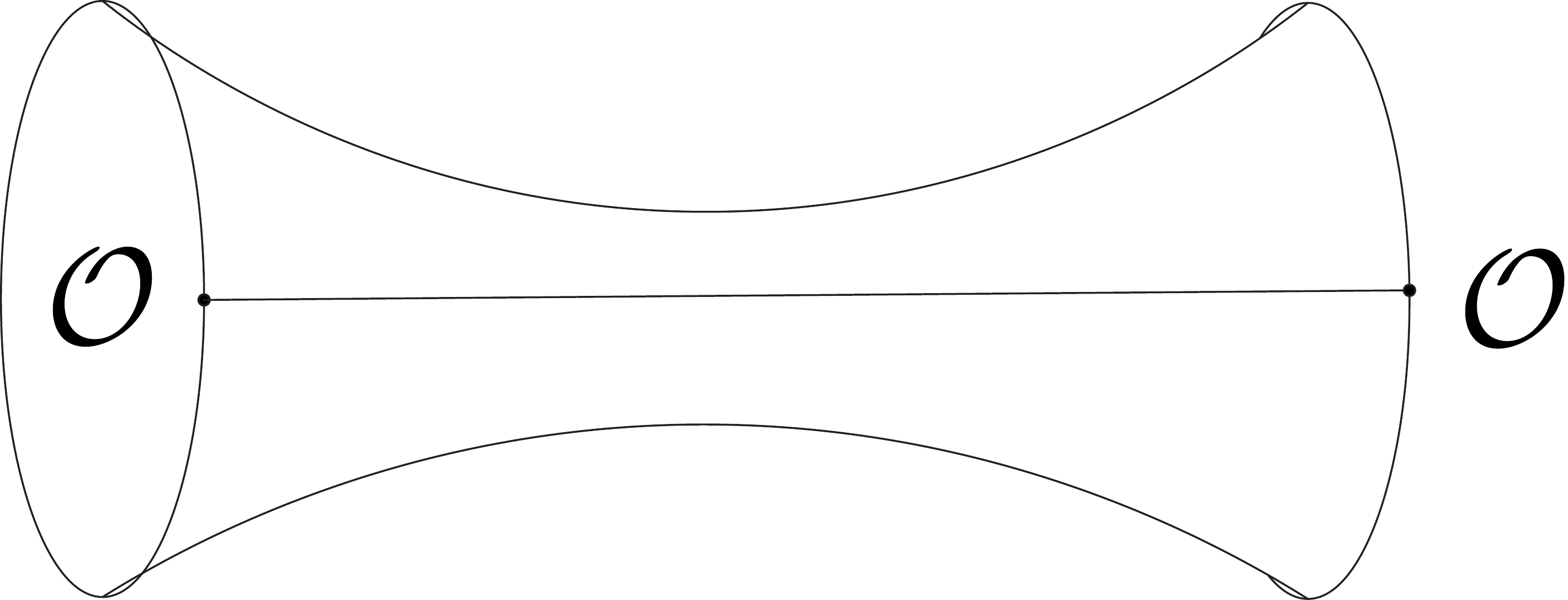} } \ .
\end{align}
To compute this quantity, we first compute the two-point function on the rigid double-trumpet as a function of $b$, integrate over boundary reparameterizations using the Schwarzian action, multiply by $Z_{\text{scalar}}(b)$, and finally integrate over $b$. The result may be expressed as an infinite sum \eqref{2pt-dt2}, where each term in the sum captures the contribution to $Z_{\text{scalar}}(b)$ from a single term in \eqref{eq:second}. We use the boundary particle formalism \cite{Yang:2018gdb} to do this computation in Appendix \ref{2pt-dt}. The upshot is that the Feynman rules of \cite{Mertens:2017mtv} discussed in \ref{sec:disk} still apply, but they are generalized to allow for closed loops of a bulk line. The 6j-symbol governs the intersection of all bulk lines, regardless of whether they form closed loops or extend to the boundaries. For the 2-point correlator on the double-trumpet we have
\begin{align}
&\braket{\tr e^{-\beta_L H} \calo \tr e^{-\beta_R H} \calo}_{\cyl} \\
&= 
\int_0^\infty ds_L ds_R ~\rho(s_L)\rho(s_R) 
e^{- (\beta_L s_L^2 + \beta_R s_R^2) }
\left(\Gamma^\Delta_{LL} \Gamma^{\Delta}_{RR}
\right)^{1/2}
\\
&\times  
\left(
{\delta(s_L - s_R) \over \rho(s_L)} 
+ 
\begin{Bmatrix}
   \Delta & s_L & s_R \\
   \Delta & s_R & s_L
\end{Bmatrix}
+ 
\sum_{m=0}^\infty  
\begin{Bmatrix}
   2\Delta + 2m & s_L & s_R \\
   \Delta & s_R & s_L
\end{Bmatrix}
+ \dots 
\right) 
\\
&=
\raisebox{-.3in}
{\includegraphics[scale=.15]{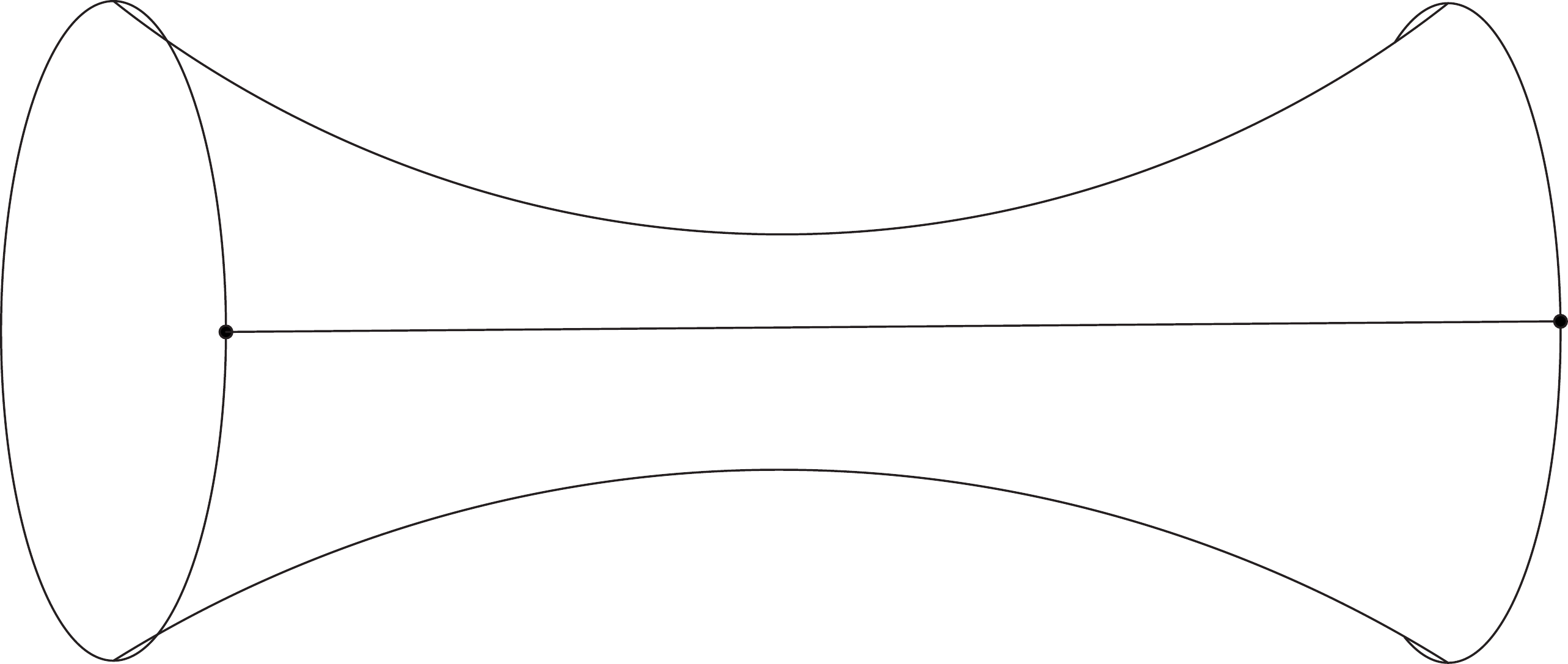} }
+ 
\raisebox{-.35in}
{\includegraphics[scale=.15]{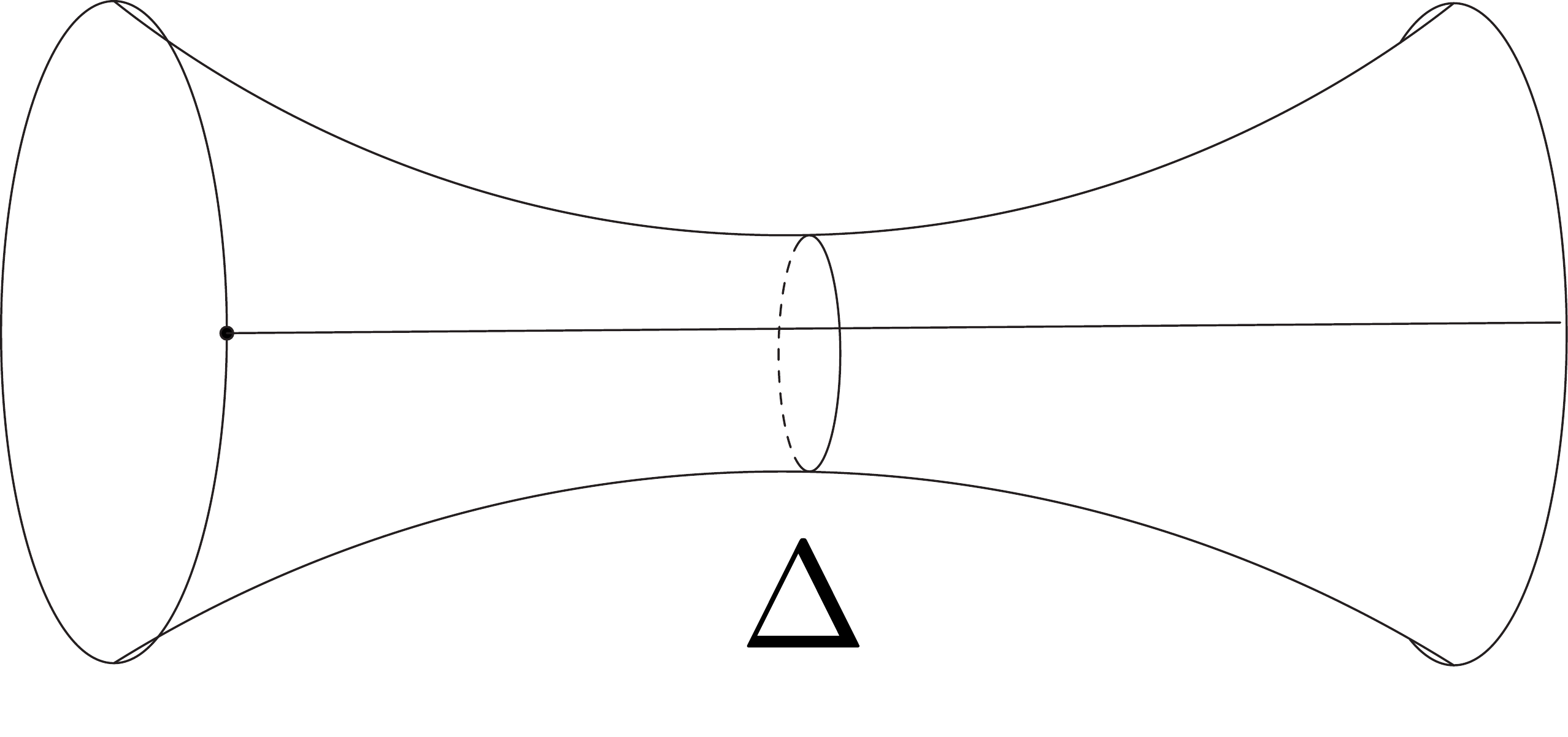} }
+ \sum_{m=0}^\infty 
\raisebox{-.35in}
{\includegraphics[scale=.15]{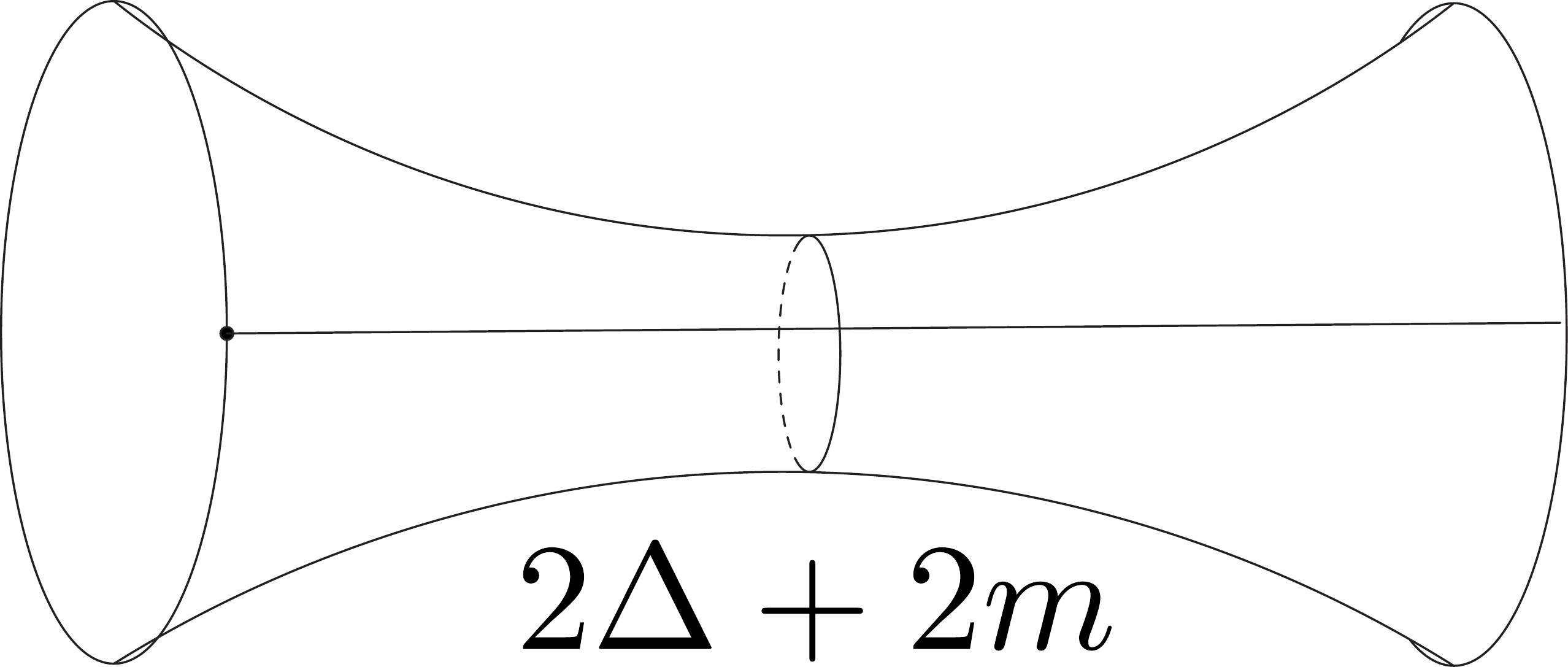} }
+ \dots \ .
\label{2pt-dt2}
\end{align}
Here, the gamma functions are as prescribed by the rule \eqref{eq:feynruleoboundary} with $\Gamma_{L L}^\Delta = {\Gamma(\Delta)^2 \over \Gamma(2 \Delta)} \Gamma(\Delta \pm 2i s_L)  $ and similarly for $s_R$. 
The three explicitly shown terms in \eqref{2pt-dt2} correspond to the $n=0,1,2$ terms in \eqref{eq:second}. In the first term, the identity operator (or ground state) propagates on the closed geodesic and the energies $s_L, s_R$ are the same. In the second term, a primary operator $\calo$ and its descendants propagate around the closed geodesic. The closed geodesic is assigned a label equal to the scaling dimension of $\calo$ (which is $\Delta$). The intersection of the closed geodesic with the geodesic connecting the AdS boundaries implies that we include a 6j-symbol. The third term corresponds to the double-trace operators propagating around the closed geodesic. They have scaling dimensions $2\Delta + 2m$ for $m \in \mathbb{Z}_{\ge 0}$. The general rule is that for each primary operator (aside from the identity, which is treated as a special case), we assign its scaling dimension to the closed geodesic line and include a 6j symbol for the intersection. We then sum over the terms we get for all the primary operators. The second term above captures the lone single-trace operator, while the third term captures the double-trace operators. In particular, the sum over $m$ in the third term directly corresponds to the sum over $m$ in \eqref{eq:ne2term}. Similarly, we can include triple-trace and higher operators, which we denoted by ellipsis in \eqref{2pt-dt2}.

\subsection{Pair of pants and beyond}

\label{sec:PairOfPants}

In the previous subsection we saw that the gravitational Feynman rules that compute disk correlators may be extended to compute the two-point function on the double-trumpet. Here, we compute an amplitude involving the pair of pants and find evidence that the Feynman rules may be extended even further.

\begin{figure}
	\centering
	\includegraphics[scale = .2]{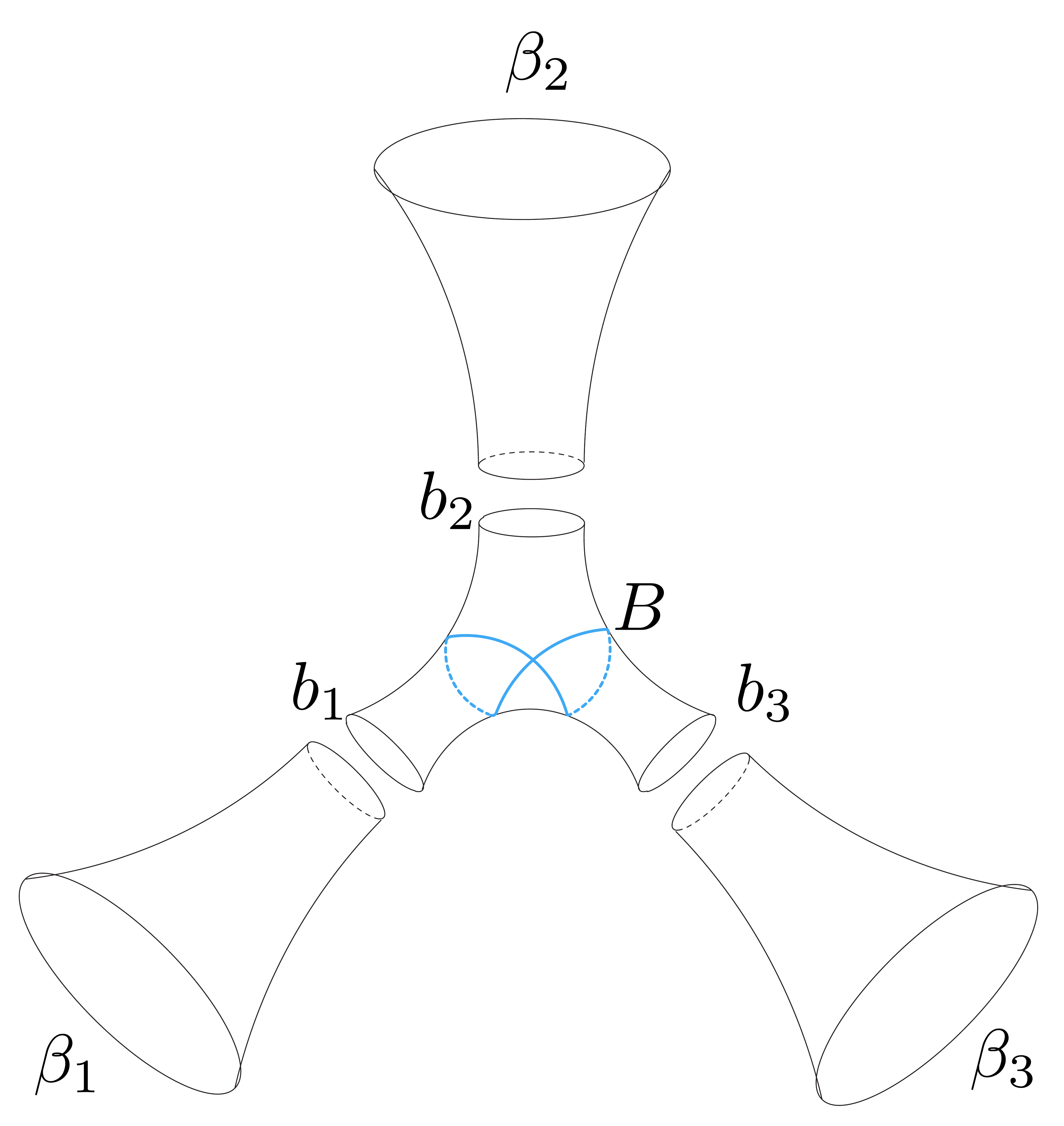}
	\caption{The blue geodesic divides the pair of pants into three regions with cylinder topology. The length of the blue geodesic $B$ is given in \eqref{eq:Bstar}.} 
	\label{fig:pairofpants}
\end{figure}

\bigskip 

We are interested in the pair of pants geometry with three AdS boundaries and a geodesic that winds in a figure-8 pattern contributing to the 1-loop determinant, see Figure \ref{fig:pairofpants}. In the previous examples we considered, the bulk lines divided the geometry into disk-shaped regions. Here, the bulk lines divide the geometry into three cylinders. We may extend the gravitational Feynman rules by declaring that for each cylinder we define two parameters $s, s'$, one near each boundary of the cylinder. And include a factor of $\braket{\rho(s) \rho(s')}_{\cyl}$ in the integrand, the zero genus two-boundary density-density correlator in the SSS model \cite{Saad:2019lba}. The self-intersection of the bulk line corresponds to the 6j symbol as usual. Using these extended Feynman rules, the full amplitude for Figure \ref{fig:pairofpants} is

\begin{align}\label{figure8}
e^{-S_0}\int_0^\infty \prod_{j=1}^3 \left(  ds_j \braket{Z(\beta_j) \rho(s_j)}_\cyl \right) 
\left\{
\begin{array}{ccc}
			\Delta & s_1 & s_2 \\
			\Delta & s_3 & s_2
\end{array}
\right\} \ .
\end{align}
To show this is correct we relate it to the contribution of the closed geodesic to the 1-loop determinant \eqref{detSel}. This is done using an integral representation of the 6j-symbol \eqref{6jb-int} that we derive in Appendix \ref{sec:6ju-int} 
\begin{align}
    \left\{
\begin{array}{ccc}
			\Delta & s_1 & s_2 \\
			\Delta & s_3 & s_2
\end{array}
\right\} 
=
{1\over 2} \int_0^\infty \prod_{j=1}^3 \left( db_j ~ b_j ~ {2 \cos(b_j s_j)\over b_j} \right) ~ {e^{-\Delta B} \over 1- e^{-B}} \ ,
\end{align}
where $B = B(b_1, b_2, b_3)$ is determined by
\begin{equation}
    \cosh \frac{B}{2} =\cosh \frac{b_2}{2} +  2 \cosh \frac{b_1}{2}  \cosh \frac{b_3}{2} \ .
    \label{eq:Bstar}
\end{equation}
It turns out that $B$ is also the length of the blue geodesic in Figure \ref{fig:pairofpants} on the pair of pants with geodesic boundaries of lengths $b_1, b_2, b_3$. Inserting this into \eqref{figure8} and exchanging integrals over $b_j$ and $s_j$ we have
\begin{align}
{1\over 2}e^{-S_0} \int_0^\infty \prod_{j=1}^3 \left( 
db_j ~ b_j \braket{Z(\beta_j) x(b_j)}_\cyl ~
\right)
{e^{-\Delta B} \over 1- e^{-B}} \ , 
\end{align}
where we defined 
\begin{align}
x(b) = {2\over b} \tr \cos(b \sqrt{H}) \ .
\end{align}
Each factor $\braket{Z(\beta) x(b)}_\cyl = Z_{\text{tr}}(\beta, b)$ is the trumpet partition function. We discuss this fact in more detail in section \ref{sec:double-trumpet}. Without the factor $ \frac{e^{-\Delta B  }}{ 1 - e^{- B}}$, this is the path integral on the pair of pants $\braket{Z(\beta_1) Z(\beta_2) Z(\beta_3)}_{g=0,n=3}$ computed in \cite{Saad:2019lba}.\footnote{We have an extra $1\over 2$ because our conventions for the density of states correspond to \cite{Saad:2019lba} with a rescaling $e^{S_0} \to 2 e^{S_0}$, also see \eqref{eq:rhos}.} Then the factor $ \frac{e^{-\Delta B  }}{ 1 - e^{- B}}$ corresponds to the $w = 1$ contribution to the determinant \eqref{detSel} on the pair of pants. We therefore showed that \eqref{figure8} is indeed the contribution of the ``figure-8'' geodesic in Figure \ref{fig:pairofpants}. Finally, it is easy to generalize the above computation to any multi-trace operator propagating on the figure-8 geodesic. We simply substitute $\Delta$ by the dimension of the desired the multi-trace operator.

\bigskip

It is natural to conjecture that the gravitational Feynman rules can be extended to compute amplitudes for arbitrary genus topologies with an arbitrary number of boundaries and $\calo$ insertions. A general diagram consists of bulk lines drawn on a Riemann surface with boundaries. The bulk lines can end on AdS boundaries where $\calo$ operators are inserted, or the bulk lines can form closed geodesics, which may represent contributions to the 1-loop determinant. These lines divide the surface into subregions, and each subregion is characterized by the number of boundaries $n$ and the genus $g$. For every boundary of every subregion, we assign an $s$ parameter. For each subregion, we get a factor of $\braket{\rho(s_1) \ldots \rho(s_n)}_{g,n}$, which is the inverse Laplace transform of $Z_{g,n}(\beta_1,\ldots,\beta_n)$ defined in equation (127) of \cite{Saad:2019lba}. Wherever two bulk lines intersect, we get a 6j symbol involving the four adjacent $s$ parameters. Furthermore, wherever there is a closed geodesic that does not intersect other geodesics (including itself), we get a function of the two adjacent $s$ parameters:
\begin{equation}
 \int_0^\infty db ~ b ~ {e^{-\Delta' b} \over 1-e^{-b}} ~ {2\over b} \cos(b s_1) ~ {2\over b} \cos(b s_2),
 \end{equation}
 where $\Delta'$ is the dimension of the primary operator propagating on the closed geodesic. This ensures that the closed geodesic is weighted by ${e^{-\Delta' b} \over 1-e^{-b}}$ in the moduli space integral. After integrating over all of the $s$ parameters, the result is equal to the sum/integral over all the gravitational configurations (including the metric and the choice of non-homotopic geodesics) that have the topology of the diagram. 

Another check of this conjecture comes from the two-point function on the disk with a handle where the geodesic connecting the two $\calo$ insertions is non-self-intersecting. This was computed in \cite{Saad:2019pqd, Iliesiu:2021ari, Blommaert:2020seb}.

\subsection{Gravitationally dressed OPE}

The GFF correlators have $SL(2,\mathbb{R})$ conformal symmetry and can be decomposed into conformal blocks. For example, the 4-point function is
\begin{align}
\braket{\calo(\tau_1)\calo(\tau_2)\calo(\tau_3)\calo(\tau_4)}_{\text{GFF}}
&=
\tau_{12}^{-2\Delta}\tau_{34}^{-2\Delta}
+\tau_{14}^{-2\Delta}\tau_{23}^{-2\Delta}
+\tau_{13}^{-2\Delta}\tau_{24}^{-2\Delta} \\
&=\tau_{12}^{-2\Delta}\tau_{34}^{-2\Delta}
\left( 
1 + \left( z\over 1-z \right)^{2\Delta} + z^{2\Delta} 
\right) \ ,
\label{4ptGFF}
\end{align}
where $\tau_{ij} = \tau_i - \tau_j$ and the cross-ratio is $z = {\tau_{12} \tau_{34} \over \tau_{13} \tau_{24}}$. In the OPE channel $\calo(\tau_1) \calo(\tau_2)$, the first term in \eqref{4ptGFF} corresponds to the exchange of the unit operator, while the other two terms are due to double-trace operators. To see the latter, we decompose (e.g. see equations (4.14), (4.15) in \cite{Gaiotto:2013nva})
\begin{align}
    \left( z\over 1-z \right)^{2\Delta} &= \sum_{n = 0}^\infty  \frac{(2 \Delta)_n^2}{n! \, ( 4 \Delta + n - 1)_n} g_{2\Delta +n}(z),
\label{eq:2.47}
    \\
     z^{2\Delta}&= \sum_{n = 0}^\infty (-1)^n \frac{(2 \Delta)_n^2}{n! \, ( 4 \Delta + n - 1)_n} g_{2\Delta+n}(z),
    \label{eq:2.48}
\end{align}
where the conformal block is defined by
\begin{equation}
    g_h(z) \equiv z^h \, _2F_1(h,h,2 h,z) \ .
\end{equation}

Let us assume $\tau_1 > \dots > \tau_4$ as in \eqref{4pt}. Now, if we dress \eqref{4ptGFF} with the Schwarzian mode as in \eqref{diskPI}, the result is the three gravitational Feynman diagrams \eqref{4pt}. Each of the three terms in \eqref{4pt} correspond to the three terms in \eqref{4ptGFF}. The first two are the uncrossed diagrams, while the last one is the crossed diagram.

One might ask what happens with the OPE expansion \eqref{eq:2.47}, \eqref{eq:2.48} after we dress it with the Schwarzian mode. It turns out that explicit expressions may be derived for each of the individual terms on the RHS of \eqref{eq:2.47}, \eqref{eq:2.48} after dressing with the Schwarzian. These are the last two terms in \eqref{4pt} expanded as
\begin{align}\label{OPE1}
{\delta(s_2-s_4) \over \rho(s_2)} = \sum_{n=0}^\infty P_n^{\Delta,\Delta}(s_2;s_1,s_3)P_n^{\Delta,\Delta}(s_4;s_1,s_3) \ , \\
\left\{ 
\begin{matrix}
\Delta & s_1 & s_2 \\
\Delta & s_3 & s_4
\end{matrix}
\right\}
=
\sum_{n=0}^\infty (-1)^n P_n^{\Delta,\Delta}(s_2;s_1,s_3)P_n^{\Delta,\Delta}(s_4;s_1,s_3) \ ,
\label{OPE2}
\end{align}
where $P_n$ is proportional to a Wilson polynomial. The definition of $P_n$ and a more detailed discussion of these identities can be found in appendix \ref{sec:specialfunctions}. These expressions can be further integrated over $s_j$ as in \eqref{4pt} to obtain formulas in Euclidean time $\tau_j$. 

An important point that we will use later is that a system of orthonormal polynomials $P_n$ diagonalizes the 6j symbol as a map between $s_2, s_4$. In Appendix \ref{sec:confblocks} we take the semiclassical limit of \eqref{OPE1}, \eqref{OPE2} and show that it reduces to \eqref{eq:2.47}, \eqref{eq:2.48}.

\section{The ETH as a matrix model}

\label{sec:ethasamatrixmodel}

Holographic CFTs have a large $N$ expansion and a sparse spectrum of light operators \cite{Heemskerk:2009pn}. By definition, the scaling dimensions of and OPE coefficients involving the light operators only are well-defined in the large $N$ limit. In contrast, the OPE coefficients that involve heavy operators do not have a large $N$ limit. In particular, the spectrum of black hole microstates becomes arbitrarily dense for large $N$. For this reason, computing the exact spectrum and OPE coefficients at some fixed, large value of $N$ is in general a hopeless task.

It is much easier to make a well-motivated guess for the CFT data. That is, it is natural to conjecture that the scaling dimensions of and OPE coefficients involving the heavy operators look like a typical draw from some ensemble.\footnote{To be clear, for the present discussion we are considering a single, non-disorder averaged theory.} We will refer to this ensemble as the ``ETH ensemble.'' This conjecture amounts to applying the ETH to holographic theories, and it was previously studied in connection to wormholes and AdS/CFT in \cite{Pollack:2020gfa}. A key point is that self-averaging observables (such as the black hole partition function, or thermal correlators of an order one number of light operators at sufficiently early times) are only sensitive to the choice of the ensemble itself at large $N$. Thus, for a judiciously chosen ensemble, the ensemble's predictions for self-averaging observables should be comparable to the results of gravitational path integral calculations. The ensemble should be chosen to correctly reproduce the results of all the gravitational calculations that we know how to perform at some given energy scale and some given level of precision.\footnote{Different calculations in the bulk should correspond to different ensembles. For example, gravitational calculations that incorporate the effects of irrelevant operators in the Lagrangian should be more sensitive to the exact CFT spectrum. Hence, the corresponding ensemble should be more fine-grained. Of course, the effects of irrelevant operators are more important at greater energy scales and higher levels of precision.} Note that there are certain calculations involving wormholes that we do not know how to perform.

\bigskip 

To illustrate these points, consider the expectation value of a light operator in a black hole microstate, which we will write as $\braket{E|\calo|E}$. Let $\overline{\braket{E|\calo|E}}$ refer to the average value of this matrix element in a microcanonical window of width $\delta E$ centered around $E$, in the limit of large $N$ and small $\delta E$,
\begin{equation}
    \overline{\braket{E|\calo|E}} \equiv \lim_{\delta E \rightarrow 0} \lim_{N \rightarrow \infty} e^{- S(E,\delta E)} \sum_{|E_a - E| < \frac{\delta E}{2}} \braket{E_a|\calo|E_a},
    \label{eq:mean}
\end{equation}
where $e^{S(E,\delta E)}$ is the number of states in the microcanonical window. Note that when $N$ is large and $\delta E$ is small, we have that $e^{S(E,\delta E)} = \delta E \rho(E)$, where $\rho(E)$ is the large $N$ density of black hole microstates, which may be computed from the black hole saddle in the Euclidean path integral. We can holographically determine $\overline{\braket{E|\calo|E}}$ by taking the inverse Laplace transform of the thermal one-point function $\tr e^{- \beta H} \calo$, again computed from the black hole saddle (we always assume that the temperature is above the Hawking-Page transition). The result is a smooth function of $E$. In JT gravity coupled to a free massive scalar, this smooth function is zero. For the remainder of this discussion, let us for simplicity assume that $\overline{\braket{E|\calo|E}}$ vanishes (this can be achieved by subtracting a multiple of the identity from the bulk operator that $\calo$ is dual to).

Computing the thermal two-point function $\tr e^{- \beta_1 H} \calo e^{- \beta_2 H} \calo$ in the same way, we can deduce another self-averaging quantity:
\begin{equation}
    \overline{|\braket{E_1 | \calo | E_2}|^2} \equiv \lim_{\delta E \rightarrow 0} \lim_{N \rightarrow \infty} e^{-S(E_1,\delta E) - S(E_2,\delta E)}
    \sum_{\substack{|E_a - E_1| < \frac{\delta E}{2}\\ |E_b - E_2| < \frac{\delta E}{2}}} |\braket{E_a | \calo | E_b}|^2.
    \label{eq:twopoint}
\end{equation}
Again, one may compute \eqref{eq:twopoint} by taking the inverse Laplace transform with respect to $\beta_1$ and $\beta_2$ of the two-point function evaluated from the black hole saddle, which takes the form
\begin{equation}
    \int_{E_0}^\infty dE_1 \, \rho(E_1) \, dE_2 \, \rho(E_2) \, \overline{|\braket{E_1 | \calo | E_2}|^2} \, e^{- \beta_1 E_1 - \beta_2 E_2},
    \label{eq:3.3}
\end{equation}
where $E_0$ is the black hole threshold.

Another statistical quantity of interest is the variance,
\begin{equation}
    \overline{\braket{E|\calo|E}^2} \equiv \lim_{\delta E \rightarrow 0} \lim_{N \rightarrow \infty} e^{- S(E,\delta E)} \sum_{|E_a - E| < \frac{\delta E}{2}} \braket{E_a|\calo|E_a}^2.
    \label{eq:variance}
\end{equation}
Equation \eqref{eq:variance} is not to be confused with \eqref{eq:twopoint} evaluated for $E_1 = E_2 = E$, which we will denote by \begin{equation}\left. \overline{|\braket{E_1 | \calo | E_2}|^2}\right|_{E_1 = E_2 = E}.\label{eq:twopoint2} \end{equation} If the matrix elements of $\calo$ in a given microcanonical Hilbert space are i.i.d. random variables (up to the Hermiticity condition for $\calo$), then \eqref{eq:variance} and \eqref{eq:twopoint2} will be equal. More generally, \eqref{eq:variance} and \eqref{eq:twopoint2} may differ, perhaps at a subleading order in $N$, due to nontrivial correlations among these variables. Because \eqref{eq:twopoint2} and its finite $N$ corrections should be computable from the bulk black hole two-point function, \eqref{eq:variance} and its subleading corrections must be computable from some other bulk observable. Starting from the black hole two-point function, one could try to obtain \eqref{eq:variance} by setting $\beta_1 = \beta + i T$ and $\beta_2 = \beta - i T$, integrating over $T \in (-\infty,\infty)$ to set the two energies equal, and then taking an inverse Laplace transform on $\beta$, but the result will be the same smooth function that one would have obtained by performing separate inverse Laplace transforms on $\beta_1$ and $\beta_2$. The black hole two-point function does not have enough information for us to determine both \eqref{eq:twopoint2} and \eqref{eq:variance} to all orders in $N$.  The only other candidate for the holographic dual to \eqref{eq:variance} is a two-point function on a wormhole geometry that connects two asymptotically AdS boundaries, where one $\calo$ operator is inserted on each boundary.\footnote{See the discussion section of \cite{Saad:2019pqd} for further comments.}

Euclidean wormhole amplitudes are challenging because the size of the wormhole can become small in the off-shell path integral. Where the wormhole is small, the local temperature of the bulk fields is large, leading to a UV divergence in the amplitude. For example, in JT gravity minimally coupled to a CFT, the double-trumpet matter partition function diverges as $e^{\frac{c \pi^2}{6 b}}$ for small $b$. Hence, the wormhole amplitude is undefined. This issue can be sidestepped when the path integral has a saddle. For instance, if the dimension of $\calo$ scales as $\frac{1}{G_N}$, then by inserting an $\calo$ on each AdS boundary we can stabilize the wormhole at a finite size. However, when the dimension is of order one, there is no saddle. Hence, wormhole amplitudes reflect a limitation on what the gravitational Euclidean path integral can teach us about the ETH ensemble.

\bigskip

As illustrated above, the Euclidean path integral alone is not able to entirely determine the ETH ensemble. Thus, some additional principle is needed to motivate which ensemble to use. The simplest guess is to follow the original ETH literature \cite{Deutsch, Srednicki1, Srednicki2} and draw the matrix elements of $\calo$ from a Gaussian ensemble:
\begin{equation}
     \braket{E_a|\calo|E_b} = \overline{\braket{E_a|\calo|E_a}} \delta_{ab} + \sqrt{\overline{|\braket{E_a|\calo|E_b}|^2}} R_{ab},
     \label{eq:gaussansatz}
\end{equation}
where $R_{ab}$ is a random GUE matrix with zero mean and unit variance. Unfortunately, a Gaussian ETH ensemble does not correctly compute all of the thermal correlators. For concreteness, let us specialize to the case of JT gravity coupled to a massive scalar and use the conventions of section \ref{sec:JTgravitywithmatter}. In the next section, we will find that the ansatz \eqref{eq:gaussansatz} computes the correct two-point function only. For the four-point function, the Gaussian ETH reproduces the two gravitational Feynman diagrams where the bulk lines do not cross (see equation \eqref{4pt diag}). In the semiclassical limit at zero temperature, the corresponding correlator is
\begin{equation}
    \braket{\calo(\tau_1) \calo(\tau_2) \calo(\tau_3) \calo(\tau_4)} = (\tau_{12}\tau_{34})^{-2 \Delta} 
 +  (\tau_{14}\tau_{23})^{-2 \Delta}
\ .
\label{eq:wrongGFFcorrelator}
\end{equation}
The main issue with \eqref{eq:wrongGFFcorrelator} is that it is inconsistent with crossing symmetry, which is the requirement that multipoint $\calo$ correlators ought to be invariant under permutations of the $\tau_j$ variables.\footnote{Strictly speaking, this definition is correct when $\Delta$ is an integer. More generally, there are branch cuts in the complex $\tau$ plane that introduce an ambiguity when continuing one operator past another. A more general definition of crossing symmetry is that the correlators should be invariant under permutations up to the phases associated with continuing the operators past one another.}
The crossing-symmetric four-point function that we wish to reproduce using the ETH ensemble is
\begin{equation}
    \braket{\calo(\tau_1) \calo(\tau_2) \calo(\tau_3) \calo(\tau_4)} = (\tau_{12}\tau_{34})^{-2 \Delta} 
 +  (\tau_{14}\tau_{23})^{-2 \Delta}
+ (\tau_{13}\tau_{24})^{-2 \Delta} \ .
\label{eq:rightGFFcorrelator}
\end{equation}
This is the four-point function of a generalized free field (GFF). In 1D CFT, a simple condition that fixes the correlation functions of $\calo$ to be those of a GFF is that the only primary operators that appear in the $\calo \calo$ OPE aside from the identity have dimensions $2 \Delta + 2n$ for $n \in  \mathbb{Z}_{\ge 0}$.\footnote{See appendix \ref{sec:gffproof} for an explanation.} In this case, the OPE looks like
\begin{equation}
    \calo(\tau) \calo(0) = \tau^{-2 \Delta} + \sum_{n = 0}^\infty \tau^{2 n} [\calo \calo]_n + \text{descendants}
    \label{eq:3.8}
\end{equation}
where $[\calo \calo]_n$ refers to a double-trace primary. Equation \eqref{eq:3.8} is equivalent to the following operator equation:
\begin{equation}
			[\calo(t_1),\calo(t_2)] = \lim_{\epsilon \rightarrow 0} \left[ \frac{1}{( i t_{12} + \epsilon)^{2 \Delta}} - \frac{1}{( i t_{12} - \epsilon  )^{2 \Delta}}\right] 
			= \left\{
			\begin{array}{cc}
	-\frac{2 i \sin(\pi \Delta)}{|t_1 - t_2|^{2 \Delta}} \, \text{sign } (t_1 - t_2), & \Delta \notin \mathbb{Z}_{\ge 1} \\
	\frac{2 \pi i}{(2 \Delta - 1)!} (-1)^{\Delta -1} \delta^{(2 \Delta - 1)}(t_{12}), & \Delta \in \mathbb{Z}_{\ge 1}
\end{array}\right. .
\label{eq:GFFconstraint}
\end{equation}
That is, the commutator of two $\calo$ operators separated in Lorentzian time is proportional to the identity operator. It is straightforward to check that \eqref{eq:3.8} implies \eqref{eq:GFFconstraint}. To see that \eqref{eq:GFFconstraint} implies \eqref{eq:3.8}, note that if any primary operators with dimensions not in the set $\{1\} \cup \{ 2 \Delta + 2 n \, : \, n \in \mathbb{Z}_{\ge 0} \}$ appeared on the RHS of \eqref{eq:3.8}, then they would make additional contributions to the RHS of \eqref{eq:GFFconstraint}. Furthermore, it is simple to show that \eqref{eq:GFFconstraint} is obeyed by the GFF correlators.

\bigskip

In a typical instance of the Gaussian ETH ensemble, the matrix $\calo$ does not obey any simple operator equations. In an ETH ensemble that reproduces the disk correlators of JT gravity with matter, $\calo$ should be constrained to obey \eqref{eq:GFFconstraint} in the semiclassical limit. We propose an ETH ensemble where the analog of \eqref{eq:GFFconstraint} away from the semiclassical limit is manifestly obeyed. First, we must find the operator equation that generalizes \eqref{eq:GFFconstraint} away from the semiclassical limit. We propose the following expression:
\begin{equation}
e^{-S_0} \sum_{b}    \left\{\begin{array}{ccc}
	 		\Delta & s_a & s_b \\
	 		\Delta & s_c & s_d
	 	\end{array}\right\} \left[\frac{\calo_{a b} \calo_{bc}}{e^{-S_0} \sqrt{\Gamma_{ab}^\Delta \Gamma_{bc}^\Delta}} - \delta_{ac} \right] = \left[\frac{\calo_{a d} \calo_{dc}}{e^{-S_0}\sqrt{\Gamma_{ad}^\Delta \Gamma_{dc}^\Delta}} - \delta_{ac} \right], \label{eq:constraint}
\end{equation}
where $\calo_{ab} \equiv \braket{E_a|\calo|E_b}$ refers to matrix elements of $\calo$ in the energy eigenbasis, and $s^2 = E$ relates the $s$ parameters to the energies. The sum is over all of the energy eigenvalues. Equation \eqref{eq:constraint} holds within any correlator (both at the disk level and, conjecturally, at higher genus) in any ensemble-averaged theory that is dual to JT gravity minimally coupled to a scalar.

\bigskip 

For example, let us insert \eqref{eq:constraint} into a correlator with two other $\calo$ insertions. The ensemble average of
\begin{equation}
    \sum_{a,b,c,d,e} e^{-\beta_1 E_a} e^{- \beta_2 E_e}e^{-S_0} \left\{\begin{array}{ccc}
	 		\Delta & s_a & s_b \\
	 		\Delta & s_c & s_e
	 	\end{array}\right\}\left[\frac{\calo_{a b}  \calo_{bc}}{e^{-S_0} \sqrt{\Gamma_{ab}^\Delta \Gamma_{bc}^\Delta}} -  \delta_{ac} \right] e^{- \beta_3 E_c} \calo_{cd} e^{- \beta_4 E_d} \calo_{da}
\end{equation}
is, at disk level,
\begin{align}
        & \int_0^\infty ds_e  \rho(s_e) e^{- \beta_2 s_e^2} ~ e^{2 S_0} \int_0^\infty 
       ds_a \rho(s_a) ~ ds_b \rho(s_b) ~ ds_c \rho(s_c) ~ ds_d \rho(s_d)    e^{-\beta_1 s_a^2}   e^{-\beta_3 s_c^2} e^{-\beta_4 s_d^2} 
     {\left(  \Gamma_{cd}^\Delta \Gamma_{da}^\Delta \right)^{1/2}  } \\
    &\qquad \qquad \times \left\{\begin{array}{ccc}
	 		\Delta & s_a & s_b \\
	 		\Delta & s_c & s_e
	 	\end{array}\right\} \left( 
      {\delta(s_b -s_d) \over {\rho(s_b)}}
     + \left\{\begin{array}{ccc}
	 		\Delta & s_a & s_b \\
	 		\Delta & s_c & s_d
	 	\end{array}\right\} 
     \right)
    \ ,
    \\
            &=  e^{2 S_0} \int_0^\infty 
       ds_a \rho(s_a) ~ds_e  \rho(s_e)  ~ ds_c \rho(s_c) ~ ds_d \rho(s_d)    e^{-\beta_1 s_a^2} e^{- \beta_2 s_e^2}  e^{-\beta_3 s_c^2} e^{-\beta_4 s_d^2} 
     {\left(  \Gamma_{cd}^\Delta \Gamma_{da}^\Delta \right)^{1/2}  } \\
    &\qquad \qquad \times  \left( 
      \left\{\begin{array}{ccc}
	 		\Delta & s_a & s_d \\
	 		\Delta & s_c & s_e
	 	\end{array}\right\}
     + {\delta(s_d - s_e) \over \rho(s_e)} 
     \right)
    \ ,
\end{align}
where we have used the orthogonality of 6j-symbols \eqref{eq:6jorthog}. This is the disk level expression in the ensemble-averaged theory for
\begin{equation}
    \sum_{a,c,d,e} e^{-\beta_1 E_a} e^{-\beta_2 E_e} \left[\frac{\calo_{a e}  \calo_{ec}}{e^{-S_0}\sqrt{\Gamma_{ae}^\Delta \Gamma_{ec}^\Delta}} -  \delta_{ac} \right] e^{- \beta_3 E_c} \calo_{cd} e^{- \beta_4 E_d} \calo_{da}.
\end{equation}
It is straightforward to check that \eqref{eq:constraint} holds in correlators with more $\calo$ insertions. Equation \eqref{eq:constraint} is a natural generalization of \eqref{eq:GFFconstraint} away from the semiclassical limit because it is the only operator equation that we are aware of that is quadratic in $\calo$. Heuristically, the 6j-symbol exchanges the order of the operators.

\bigskip 

If we integrate both sides of \eqref{eq:constraint} with respect to $P_n^{\Delta \Delta}(s_d; s_a, s_c)$ defined in \ref{Pndef}, we may derive
\begin{equation}
\sum_{b}    P_n^{\Delta \Delta}(s_b; s_a, s_c) \left[\frac{\calo_{a b} \calo_{bc}}{e^{-S_0}\sqrt{\Gamma_{ab}^\Delta \Gamma_{bc}^\Delta}} - \delta_{ac} \right] = 0, \quad n \in 2 \mathbb{Z}_{\ge 0} + 1 \ .
\end{equation}
This can be interpreted as the statement that primaries with weights $2 \Delta + n$ for odd $n$ do not appear in the $\calo \calo$ OPE. Thus, in every member of the ensemble, $\calo_{ab} \calo_{bc}$ should be equal to\footnote{after averaging the $E_b$ energy in a microcanonical window} $\delta_{ac} e^{-S_0} (\Gamma^{\Delta}_{ab} \Gamma^{\Delta}_{bc})^{1/2}$, which represents the identity operator contribution in the OPE, plus a linear combination of $(\Gamma^{\Delta}_{ab} \Gamma^{\Delta}_{bc})^{1/2} P_n^{\Delta \Delta}(s_b;s_a,s_c)$ for even $n$, which represent the blocks associated to primaries with dimensions $2 \Delta + n$ for $n$ even.\footnote{Note that the $P^{\Delta \Delta}_n(s_b;s_a,s_c)$ functions obey a completeness relation. See \eqref{IdRes}.}

\bigskip

We will construct a new ETH ensemble by imposing \eqref{eq:constraint} as a constraint. That is, let us define
\begin{equation}
    M_{ac}^d = \frac{1}{2}\sum_{b}    
    \left(
    e^{-S_0} 
    \left\{\begin{array}{ccc}
	 		\Delta & s_a & s_b \\
	 		\Delta & s_c & s_d
	 	\end{array}\right\} - \delta_{bd}
	 \right) 
	 \left(
	 \frac{\calo_{a b} \calo_{bc}}{e^{-S_0} \sqrt{\Gamma_{ab}^\Delta \Gamma_{bc}^\Delta}} - \delta_{ac} 
	 \right).
\end{equation}
The ensemble is then defined by the following matrix integral:
\begin{equation}
    \int dH d\calo \, \exp\left(- \tr V(H) - \frac{\Lambda}{2} \sum_{a,c,d}|M_{ac}^d|^2\right),
    \label{eq:constraintsquared}
\end{equation}
where $\Lambda$ is a large parameter that enforces $M_{ac}^d = 0$, or \eqref{eq:constraint}, as a constraint. $\calo$ and $H$ are Hermitian matrices with the usual measure. The role of $V(H)$ is to ensure that to leading order in $S_0$, the spectrum of the Hamiltonian $H$ agrees with $\rho(s)$. This is a two-matrix model with a single-trace potential. One can define a genus expansion in terms of `t Hooft ribbon diagrams (the diagrammatic rules are given in \eqref{eq:constraintsquaredpropagator} and \eqref{eq:constraintsquaredvertex}). We propose that the disk correlators of this matrix model agree with the disk correlators of JT gravity minimally coupled to a scalar field. In fact, \eqref{eq:constraintsquared} is a solvable model in the sense that the disk correlators of JT gravity solve its planar Schwinger-Dyson equations (to the extent that we checked). Furthermore, the analytic expressions for `t Hooft diagrams greatly simplify thanks to the unlacing rules in \eqref{eq:6jorthog} and \eqref{eq:YB}. We will explain this in more detail in section \ref{sec:crossingsymmetric}. Note that the size of the matrices in \eqref{eq:constraintsquared} is formally infinite, so \eqref{eq:constraintsquared} represents the end product of a double-scaling limit. We discuss how one can back away from the double-scaling limit in section \ref{sec:backingaway}. The manipulations in section \ref{sec:crossingsymmetric} will suggest that any matrix model that looks like \eqref{eq:constraintsquared} in a double-scaling limit will compute the JT disk correlators. Away from the double-scaling limit, the model is not solvable because generically the unlacing rules would no longer hold.\footnote{In a regulated model, the spectrum of $H$ will have compact support, so \eqref{eq:2.16} will be replaced by an integral over a finite domain. It is not possible for an integral over a finite domain to produce a delta function.
}

\bigskip

Even though \eqref{eq:constraintsquared} is not solvable away from the double-scaling limit, we can argue that in the double-scaling limit it reproduces the desired gravitational correlators at disk level. It is natural to then ask what the multiboundary and higher genus correlators are. One of the technical results of this paper is that the double-trumpet correlators\footnote{Strictly speaking, in this work we only explicitly investigate the empty double-trumpet and the double-trumpet two-point function with one $\calo$ on each AdS boundary.} (and most likely all multi-boundary correlators for any genus) of a single-trace model of two Hermitian matrices can be directly determined from the disk correlators without knowledge of the matrix potential itself. However, the result depends on how the double-scaling limit is taken. In particular, one needs to have regulated expressions for the gravitational correlators. 
Although in section \ref{sec:backingaway} we write an explicit matrix potential that represents a regulated version of \eqref{eq:constraintsquared}, we have not determined the corresponding regulated expressions for the six- and higher-point disk correlators.\footnote{For the two- and four-point disk correlators, we conjecture explicit expressions that obey a Schwinger-Dyson equation in the double-scaling limit.} Thus, beginning in section \ref{sec:corrections}, we consider different ways of regulating the disk gravitational Feynman rules that {\it a priori} are independent from the regulator in section \ref{sec:backingaway}, and we argue that one can work backwards to determine the matrix potential. We will consider two specific regulators that lead to two specific models, and we will see in section \ref{sec:doubletrumpet} (at the level of the empty double-trumpet and the double-trumpet two point function with one $\calo$ on each boundary) that the two bulk theories have the same correlators up to the matter determinant factor (which for a massive scalar on the double-trumpet is $Z_{\text{scalar}}(b)$, defined in \eqref{eq:first}, \eqref{eq:second}, \eqref{eq:third}). We will also see that the bulk 1-loop determinant is sensitive to how the matrix model is defined slightly away from the double-scaling limit.

Having an ETH ensemble allows us in principle to make sense of the otherwise ill-defined wormhole amplitudes. In section \ref{sec:uvdivergences}, we show that the empty double-trumpet in the matrix model is directly determined by the Hessian of the matrix potential evaluated at the saddle-point that defines the perturbative expansion.\footnote{To be more precise, we integrate out $\calo$ and then work with the effective potential for $H$, which is multi-trace.} The double-trumpet becomes ill-defined precisely when the Hessian is not positive-definite. Hence, the problematic UV behavior of wormholes is linked to a perturbative instability in the matrix model. To make the wormholes well-defined, it would be interesting to find a stable saddle that the unstable saddle could decay to. We leave this question to future work.

\section{Warming up with a solvable two-matrix model}

\label{sec:toymodel}

To gain intuition for how a two-matrix model can compute the disk correlators of an arbitrary number of $\calo$ insertions, it is useful to first consider a solvable toy model that only correctly computes the disk two-point function. Later, we will generalize the model described in this section to a more complicated model that correctly computes all of the disk correlators.

\bigskip

Consider a two-matrix model with Hermitian matrices $H$ and $\calo$ and a single-trace matrix potential $V(H,\calo)$. The potential has a complicated dependence on $H$ but is quadratic in $\calo$. We use the standard flat measure for the real and imaginary parts of the matrix elements of $H$ and $\calo$. The matrix integral of this toy model is given by
\begin{align}
\label{eq:1.21}	\mathcal{Z}_{\text{toy}} &= \int dH \, d\calo \, e^{-V_{\text{toy}}(H,\calo)}, \\
	V_{\text{toy}}(H,\calo) &= \sum_a (V_{SSS}(E_a) + V_{c.t.}(E_a)) + {1\over 2} \sum_{a,b} F(E_a,E_b)\calo_{ab} \calo_{ba} 
	\label{eq:1.22},
\end{align}
where $F(E_a,E_b) = F(E_b,E_a)$ is a smooth function and $E_a$ refers to an eigenvalue of $H$. We have chosen to write this single-trace potential in the eigenbasis of $H$, so the sums are over all of the eigenvalues of $H$. The $V_{SSS}(H)$ term is the matrix potential of the SSS model \cite{Saad:2019lba}, which is dual to pure JT gravity.

We have included a counterterm potential $V_{c.t.}(H)$ that is chosen to ensure that after integrating out $\calo$ the disk density of states for $H$ is still $e^{S_0} \rho_{0}(E)$, given in \eqref{rhoE}. This is because in JT gravity coupled to the free scalar, integrating out the scalar does not affect the disk partition function, except for the renormalization of $S_0, \gamma$. This was discussed after \eqref{detSel}. We will determine $V_{c.t.}(H)$ momentarily.

The last term in \eqref{eq:1.22} is single-trace because we can write an arbitrary function of two energies as $F(E_a,E_b) = \sum_{nm} c_{nm} E_a^n E_b^m$, with each term being single-trace $\tr H^n \calo H^m \calo$.

\bigskip 

Note that \cite{Saad:2019lba} did not provide an explicit formula for the matrix potential in their model, because the details of how the double-scaling limit is taken do not affect their results. These details are also irrelevant for this section. In this section, the number of eigenvalues is infinity and we are working directly in the double-scaling limit. A more rigorous treatment is provided in section \ref{sec:regtwomatrixmodel}.

\bigskip

\begin{figure}
	\centering
	\includegraphics[scale = .5]{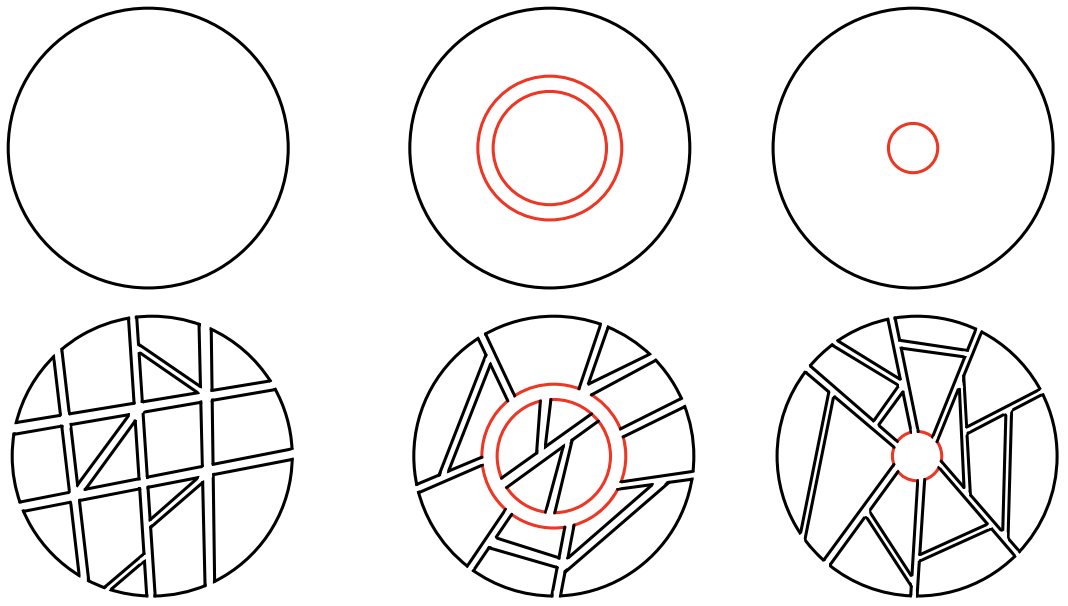}
	\caption{Top row: The left disk represents the disk computation of $\braket{\text{Tr } e^{- \beta H}}$ in the SSS model \cite{Saad:2019lba}. One should imagine filling in the disk with all possible planar 't Hooft diagrams of $H$. The middle disk represents a correction from a single insertion of the double-trace term in $\tilde{V}(H)$ (in general, there could be arbitrarily many insertions, which are all summed over). One should imagine filling in the regions inside and outside the red double-line loop with the t' Hooft diagrams of $H$ in the SSS model of disk and cylinder topology, respectively. In our terminology, the red double-line loop in the center is an example of an ``$\calo$ bubble diagram.'' The right disk represents a correction from the single-trace counterterm potential that is designed to cancel the contribution from the $\calo$ bubble diagram. The counterterm is a single-trace term in the potential and hence is represented by a single loop. Bottom row: we provide an example of one of the infinitely many ways the diagrams in the top row can be filled in with planar `t Hooft diagrams involving the $H$ matrix. A black double-line represents the propagator of the $H$ matrix.
	} 
	\label{fig:threedisks}
\end{figure}

To determine $V_{c.t.}(H)$, we should first integrate out $\calo$. The result is
\begin{align}
	\mathcal{Z}_{\text{toy}} &= \int dH \, e^{-\tr  V_{SSS}(H) - \wt{V}(H)}, \label{eq:29}\\
	\wt{V}(H) &= \sum_a V_{c.t.}(E_a) + \frac{1}{2}\sum_{a,b} \log F(E_a,E_b). \label{eq:1.23}
\end{align}
The last term in \eqref{eq:1.23} is a double-trace term, and it is represented by a double-line loop in `t Hooft diagrams. After expanding the $e^{-\wt{V}(H)}$ term, we may diagramatically compute corrections to the disk density of states for $H$, as shown in Figure \ref{fig:threedisks}. We should pick $V_{c.t.}$ such that $\wt{V}(H)$ becomes
\begin{equation}
	\wt{V}(H) = \frac{1}{2} \int dE_1 dE_2 \left(\rho_H(E_1) - e^{S_0}\rho_{0}(E_1)\right) \left(\rho_H(E_2) - e^{S_0}\rho_{0}(E_2)\right) \log F(E_1,E_2),
	\label{eq:tildeV}
\end{equation}
where
\begin{align}
\label{eq:rhoe}
	\rho_H(E) &= \tr \delta (E-H)
	= \sum_a \delta(E - E_a).
\end{align}
The variation of \eqref{eq:tildeV} with respect to $\rho_H(E)$ vanishes to first order, when evaluated for $\rho_H(E) = e^{S_0} \rho_{0}(E)$. Hence, the addition of $\wt{V}$ to $V_{SSS}$ does not change the saddle-point density of states, as desired. Note that adding a single-trace counterterm $\tr V_{c.t.}(H)$ is enough to ensure this.

\subsection{Correlation functions on the disk}

We next consider the disk two-point function of $\calo$ in our toy model. It is given by
\begin{align}
		\braket{\text{Tr } e^{- \beta_1 H} \calo e^{-\beta_2 H} \calo }_{\disk} 
		= e^{2S_0}  \int_0^\infty dE_1 dE_2 ~ e^{- \beta_1 E_1} e^{-\beta_2 E_2}
		~ \rho_{0}(E_1) \rho_{0}(E_2) ~F(E_1, E_2)^{-1} \ . 
\end{align}
Comparing this with the gravity answer \eqref{2ptGrav}, we can determine $F(E_1,E_2)$
\begin{equation}
\label{eq:Fab}
	F(E_1,E_2) = e^{S_0} \left(  \frac{\Gamma(\Delta \pm i \sqrt{E_1} \pm i \sqrt{E_2})}{ \Gamma(2 \Delta)}\right)^{-1} \ .
\end{equation}
We express this match between the matrix model and gravity computations of the 2-point function pictorially
\begin{align}
\label{eq:twopointcomparison}
\raisebox{-.3in}{\includegraphics[scale=.2]{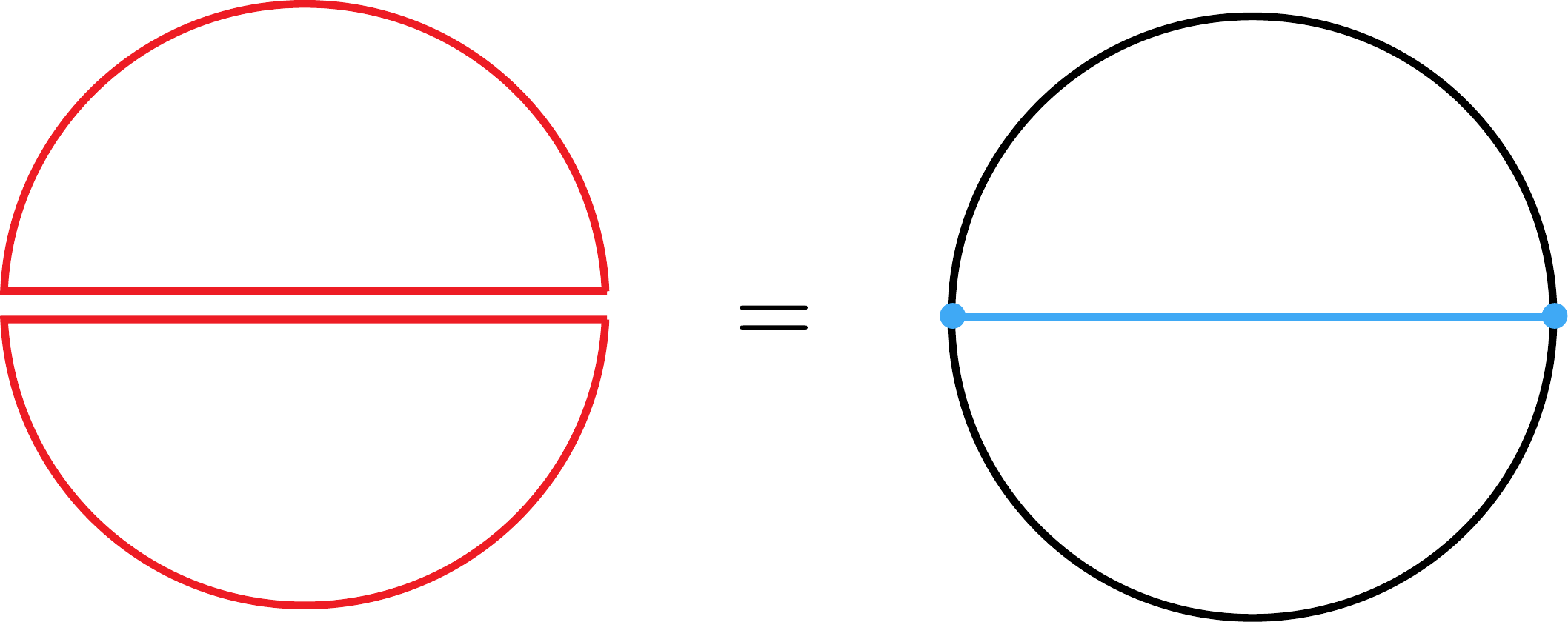}} \ ,
\end{align}
where the LHS represents the sum over t'Hooft diagrams computing the 2-point function in the matrix model. The red double-line is the propagator of the $\calo$ matrix. Each single red line represents an energy, and the two closed red loops should be filled in like a disk (that is, with $\calo$ bubble diagrams and $H$ double-lines in all possible ways). Filling in a loop like a disk corresponds to integrating over the corresponding energy with the disk density of states $e^{S_0} \rho_0(E)$. The RHS is the Feynman diagram in JT gravity coupled to a free scalar \eqref{2ptGrav}.

\bigskip

We can similarly compute the 4-point function. In \eqref{4.10} to \eqref{4ptToy} only, we imagine that the spectrum of $H$ has been fixed to some instance, and $\braket{\dots}$ refers to the expectation value in the ensemble defined by the $\calo$ matrix integral. 
\begin{align}
\label{4.10}
    &\la  \tr  \prod_{j=1}^4 e^{-\beta_j H} \calo \ra
    = \sum_{a_1,\dots ,a_4}  e^{-\sum_{j=1}^4\beta_j E_{a_j}} 
    \la \calo_{a_1 a_2} \calo_{a_2 a_3} \calo_{a_3 a_4} \calo_{a_4 a_1} \ra
    \\
        =& \sum_{a_1,\dots ,a_4} e^{-\sum_{j=1}^4\beta_j E_{a_j}}  
        \big( 
        \la \calo_{a_1 a_2} \calo_{a_2 a_3} \ra \la \calo_{a_3 a_4} \calo_{a_4 a_1} \ra
        +
         \la \calo_{a_1 a_2} \calo_{a_4 a_1} \ra \la  \calo_{a_2 a_3} \calo_{a_3 a_4}   \ra \\
         & \hspace{2in} 
        +\la \calo_{a_1 a_2}  \calo_{a_3 a_4} \ra \la \calo_{a_2 a_3} \calo_{a_4 a_1} \ra
        \big)
            \\
        =&\sum_{a_1,\dots ,a_4} e^{-\sum_{j=1}^4\beta_j E_{a_j}}  
       F(E_{a_1} E_{a_2})^{-1}F(E_{a_3} E_{a_4})^{-1} ( \delta_{a_1a_3} + \delta_{a_2a_4}  ) \\
        & \hspace{2in} 
       + \sum_a e^{- (\beta_1 + \dots+ \beta_4) E_a} F(E_a, E_a)^{-2} \ . 
       \label{4ptToy}
\end{align}
We used Wick contractions above. The propagator in the Gaussian model is $\la \calo_{a_1a_2} \calo_{a_3a_4} \ra = \delta_{a_1a_4} \delta_{a_2a_3} F(E_{a_1}, E_{a_2})^{-1} $. The last term \eqref{4ptToy} is non-planar and we neglect it.

To get the disk four-point function in the double-scaled two-matrix model, we use \eqref{eq:Fab}, we substitute $\delta_{ab}$ with ${\delta(E_a-E_b) \over e^{S_0} \rho_0(E_a)}$, and we substitute $\sum_a$ with $\int dE_a e^{S_0} \rho_0(E_a)$. The result is that we reproduce the first two terms of the gravity answer \eqref{4pt diag}
\begin{align}
\la  \tr  \prod_{j=1}^4 e^{-\beta_j H} \calo \ra_{\disk} 
&= 
e^{S_0}
\int_0^\infty \prod_{j=1}^4 \left(dE_j ~ e^{-\beta_j E_j} \rho_0(E_j) \right)
~
\\
& 
{\Gamma(\Delta \pm i \sqrt{E_1} \pm i \sqrt{E_2}) \over \Gamma(2\Delta)}
{\Gamma(\Delta \pm i \sqrt{E_3} \pm i \sqrt{E_4}) \over \Gamma(2\Delta)}
\left( 
{\delta(E_1-E_3) \over   \rho_0(E_1)} + 
{\delta(E_2-E_4) \over   \rho_0(E_2)}
\right) \ .
\end{align}
We therefore find a match between t'Hooft diagrams in the matrix model and gravitational Feynman diagrams in JT gravity
\begin{align}
\raisebox{-.45in}{\includegraphics[scale=.4]{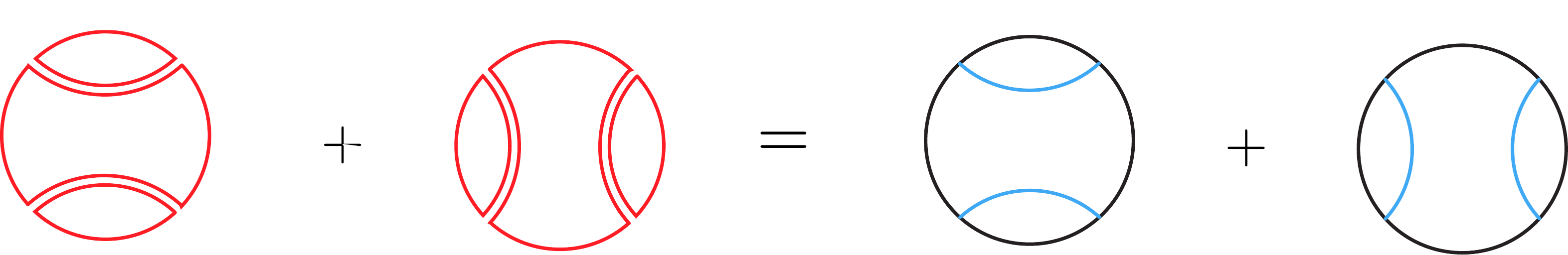}\ .} 
\label{eq:twopointfunctiongaussian}
\end{align}
Each $\calo$ propagator in the matrix model may be interpreted as a bulk line in the gravitational Feynman rules.

Note that the third term in \eqref{4pt diag} is not captured by our toy model. This is because the third Wick contraction \eqref{4ptToy} is non-planar, while in \eqref{4pt diag} the third term contributes at the same order $e^{S_0}$.

\bigskip

It should now be clear that the Gaussian in $\calo$ matrix model considered in this section captures correctly all gravitational Feynman diagrams that do not have bulk line intersections. While the Feynman diagrams with intersections (that depend on the 6j-symbols) are not reproduced. To deal with this issue we will eventually add interactions in the matrix model for $\calo$ in section \ref{sec:corrections}.

\subsection{Double-trumpet}

\label{sec:double-trumpet}

Before adding interactions, we would like to consider the double-trumpet in the Gaussian matrix model. This will facilitate a similar discussion in the interacting case later on.

\bigskip

We consider the connected correlator $\braket{ \text{Tr } e^{- \beta_L H} \, \text{Tr } e^{- \beta_R H}}_c$ in the toy matrix model. To leading order in the genus expansion, this is computed by summing `t Hooft diagrams of cylinder topology. In the SSS model, it was shown \cite{Saad:2019lba} that that this sum correctly reproduces the gravity answer, i.e. equation \eqref{eq:dttoy} without $Z_{\text{scalar}}(b)$. Therefore, we naively want to show that adding $\wt V(H)$ in \eqref{eq:29} is equivalent to inserting the 1-loop determinant $Z_{\text{scalar}}(b)$ in \eqref{eq:dttoy}. This will turn out not to be true, but it will be instructive to go through this computation.

\bigskip 

We will compute the effect of $\wt V(H)$ in \eqref{eq:29} in a perturbative expansion in the SSS matrix model. In order to do so, it is convenient to use an integral representation of $\wt V$. We use that 
\begin{align}\label{logGammab}
\log \Gamma(\Delta \pm i k_1\pm i k_2) =  \int_0^\infty db ~ b ~ {e^{-\Delta b} \over 1-e^{-b}} ~ {2\over b} \cos(b k_1) ~ {2\over b} \cos(b k_2) + \text{const} \ .
\end{align}
The additive constant here is divergent and cancels the $b\to 0 $ divergence of the integral, but it is independent of $k_1, k_2$. More precisely 
\begin{align}
\log \Gamma(z+1) = - \gamma z + \int_0^\infty {db \over b(e^b-1)} [e^{-z b} - (1-z b)] \ .
\end{align}
When we sum four such integrals to compute $\log \Gamma(\Delta \pm i k_1 \pm i k_2)$, the dependence on $k_1, k_2$ drops out everywhere except for the first term in the integral. The potential \eqref{eq:tildeV}, \eqref{eq:Fab} can now be written 
\begin{align}\label{VXrel}
-\wt V(H) = {1\over 2} \int_0^\infty db ~ b ~ {e^{- \Delta b} \over 1- e^{-b}} X(b)^2 + \text{const} \ , 
\end{align}
where we defined
\begin{align}\label{Xdef}
X(b) &= {2\over b} \left( \tr \cos (b \sqrt{H}) - \int_0^\infty dE ~ e^{S_0}\rho_0(E) \cos(b \sqrt{E}) \right)
\\
&={2\over b} \left( \tr \cos (b \sqrt{H}) - \braket{ \tr \cos (b \sqrt{H})}_{\disk}
\right) \ .
\end{align}
The reader might worry that these expressions look divergent in the double-scaling limit. The important point is that divergences in the two terms \eqref{Xdef} cancel and correlators of $X(b)$ are finite. 

\bigskip 

Now we explain correlation functions of $X(b)$ in the SSS matrix model. We call $X(b)$ a ``geodesic loop'', the name that will be justified momentarily. This observable was also considered in \cite{Goel:2020yxl}.

Consider the correlators of resolvents in the SSS matrix model \cite{Saad:2019lba} related to Weil-Petersson volumes $V_{g,n}$
\begin{align}\label{Wdef}
W(z) &= \tr {2z \over z^2 + H} = \tr \left(  {1\over z+ i \sqrt{H}} +{1\over z- i \sqrt{H}} \right) \ , \\
\braket{W(z_1) \dots W(z_n)}_{g,n} &= \int_0^\infty \prod_{j=1}^n (db_j ~ b_j e^{-b_j z_j}) V_{g,n}(b_1, \dots, b_n) \ .
\label{WVrel}
\end{align}
To get $V_{g,n}$ directly as correlators in the matrix model, we can do the inverse Laplace transform
\begin{align}
x(b) &= {1\over b} \int_{\e - i\infty}^{\e + i \infty} {dz \over 2\pi i } e^{bz} W(z)\\
& = {2\over b} \tr \cos(b\sqrt{H}) \ ,
\end{align}
where we closed the contour to the left and used \eqref{Wdef} to compute by the residue theorem. From \eqref{WVrel} we now find
\begin{align}
\braket{x(b_1) \dots x(b_n)}_{g,n} = V_{g,n}(b_1, \dots, b_n) \ .
\end{align}
The cases $g=0, n=1,2$ (disk and double-trumpet) are special. For the double-trumpet, the equation \eqref{WVrel} still holds if we assume\footnote{In this case $\braket{W(z_1) W(z_2)}_{g=0,n=2} = \int_0^\infty db_1 db_2 ~ b_1 b_2~ e^{-b_1 z_2 - b_2 z_2} ~ {\delta(b_1 - b_2)\over b_1}  = {1\over (z_1 + z_2)^2}$.} $V_{g=0,n=2}(b,b') = {1\over b} \delta(b-b')$. Therefore
\begin{align}
\braket{x(b) x(b')}_{\cyl} = {1\over b} \delta(b-b') \ .
\end{align}
The disk correlator $\braket{x(b)}_{\disk}$ diverges and we subtract it explicitly in \eqref{Xdef}
\begin{align}
X(b) = x(b) - \braket{x(b)}_{\disk} \ .
\end{align}
To properly define this difference, we imagine first computing correlators of $X(b)$ at large $N$, without the double-scaling limit. For example, in a 1-cut matrix model with finite support of the density of states $\braket{x(b)}_{\disk}$ is finite. And then we take the double-scaling limit. The result is that the correlators of $X(b)$ are finite
\begin{align}
&\braket{X(b)}_{\disk} = 0 \ , \\
&\braket{X(b)X(b')}_{\cyl} = {1\over b} \delta(b-b') \ , \\ 
&\braket{X(b_1) \dots X(b_n)}_{g,n} = V_{g,n}(b_1, \dots, b_n) \ .
\end{align}

Taking the inverse Laplace only on some of $z_j$ in \eqref{WVrel} we can compute mixed correlators of $W$ and $X$. In particular
\begin{align}
\braket{W(z) X(b)}_{\cyl} &= \int_0^\infty db' ~ b' ~ e^{-b' z} V_{g=0,n=2}(b', b)\\
&=e^{-b z} \ , \\
\braket{Z(\beta) X(b)}_\cyl &= \oint {dw \over 2\pi i }~ e^{\beta w} ~ {1\over 2\sqrt{w}} \braket{W(\sqrt{w}) X(b) }_\cyl \\
&= \oint {dw \over 2\pi i }~ e^{\beta w - b \sqrt{w}} ~ {1\over 2\sqrt{w}} \\
\label{ZXoint}
&= Z_{\text{tr}}(\beta, b) \ , \qquad Z(\beta) = \tr e^{-\beta H} \ .
\end{align}
In the 2nd computation we used $e^{-\beta H} = \oint {dw\over 2\pi i} {e^{\beta w} \over w+H} $. The contour in \eqref{ZXoint} goes around the branch cut $w \in (-\infty, 0]$ of $\sqrt{w}$.

\bigskip

\begin{figure}
	\centering
	\includegraphics[scale=.35]{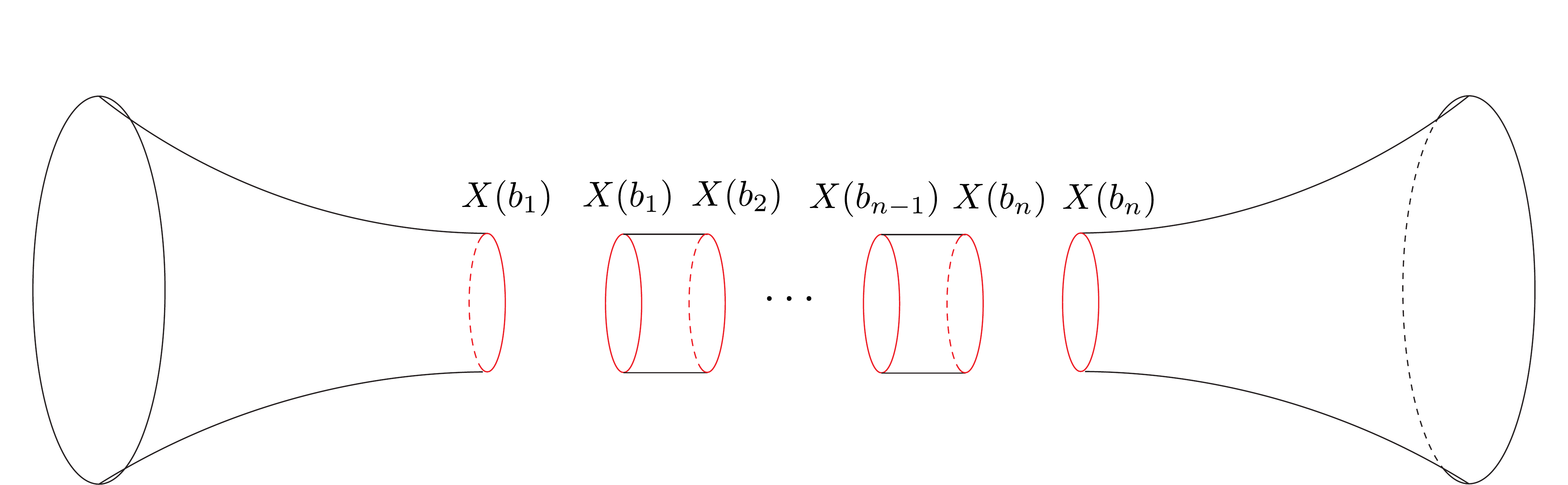}
	\caption{Two adjacent red line loops $X(b_j)$ represents the double-trace term in \eqref{VXrel}. These double-lines separate the double-trumpet into $n+1$ regions, and one should imagine filling in these regions with diagrams in the SSS model with cylinder topology. Any diagrams that contain a contractible red double-line loop are canceled because $\braket{X(b)}_\disk = 0$.}
	\label{fig:doubletrumpettwoloops}
\end{figure}

We now return to the potential \eqref{VXrel}. To leading order in the genus expansion, we compute the connected two-boundary correlator in the toy model perturbatively in $\wt V$
\begin{align}\label{eq:toydt0}
\braket{Z(\beta_L) Z(\beta_R)}_{\cyl} &=
\sum_{n=0}^\infty  {1\over n!} \braket{Z(\beta_L)~ (-\wt V)^n ~Z(\beta_R) } 
\\
& = \sum_{n=0}^\infty {1\over n!} \int_0^\infty \prod_{j=1}^n \left( db_j ~ b_j ~ {e^{-\Delta b_j} \over 1- e^{-b_j}} \right)
\braket{Z(\beta_L) ~ {X(b_1)^2 \over 2} \dots {X(b_n)^2 \over 2}~ Z(\beta_R)}
\\
&\approx \sum_{n=0}^\infty \int_0^\infty \prod_{j=1}^n \left( db_j ~ b_j ~ {e^{-\Delta b_j} \over 1- e^{-b_j}} \right)\\
&  \braket{Z(\beta_L) X(b_1)}_\cyl ~ \braket{X(b_1) X(b_2)}_\cyl \dots  \braket{X(b_{n-1}) X(b_n)}_\cyl
~\braket{X(b_n) Z(\beta_R)}_\cyl
\\
&=\int_0^\infty db~b ~ Z_{\text{tr}}(\beta_L,b) Z_{\text{tr}}(\beta_R,b) \sum_{n = 0}^\infty \left(\frac{e^{- \Delta b}}{1 - e^{-b}}\right)^n \ .
\label{eq:toydt}
\end{align}
Here, the correlator in the LHS $\braket{Z(\beta_L) Z(\beta_R)}_{\cyl} $ is in the toy model \eqref{eq:29}, while all other correlators are in the SSS model. In the third line we retained only the leading order in genus expansion. The $n$th term in the sum corresponds to a diagram that looks like the Figure \ref{fig:doubletrumpettwoloops} with $n$ red double-line loops.

Only the $n = 0,1$ terms in \eqref{eq:toydt} agree with the corresponding terms in \eqref{eq:second}. We will correct the matrix model to reproduce \eqref{eq:second} below.

\bigskip

\subsection{Corrected effective potential}

The double-trumpet in \eqref{eq:toydt} does not agree with the result in \eqref{eq:dttoy}, \eqref{eq:first} from JT minimally coupled to a scalar. However, having understood the computation \eqref{eq:toydt0} - \eqref{eq:toydt}, it is not difficult to find the potential that agrees with the gravity result. Instead of the potential $V_{SSS}(H) + \wt V(H)$ consider
\begin{equation}\label{1-loopEffPot}
     \int dH ~ \exp\left(-V_{SSS}(H)  +{1\over 2} \int_0^\infty db ~ b \left[ 1-(e^{- \Delta b},e^{-b})_\infty \right] X(b)^2 \right).
\end{equation}
A computation similar to \eqref{eq:toydt0} - \eqref{eq:toydt} shows 
\begin{align}
	\braket{Z(\beta_L) Z(\beta_R)}_{\cyl} &= \int_0^\infty db ~ b ~ Z_{\text{tr}}(\beta_L,b) Z_{\text{tr}}(\beta_R,b) \sum_{n = 0}^\infty \left[1 - (e^{- \Delta b},e^{-b})_\infty\right]^n
	\\
	&= \int_0^\infty db ~ b ~ Z_{\text{tr}}(\beta_L,b) Z_{\text{tr}}(\beta_R,b) \frac{1}{(e^{- \Delta b},e^{-b})_\infty} \ ,
\end{align}
where the Pochhammer symbol is $(e^{- \Delta b},e^{-b})_\infty = \prod_{n=0}^\infty (1-e^{-b(\Delta + n)})$. This agrees with the gravity result \eqref{eq:dttoy}, \eqref{eq:first}. In the next section, we will explain how one can incorporate the $\calo$ matrix into \eqref{1-loopEffPot}.

\section{Outline of the rest of the paper}

\label{sec:summary}

In the last section we considered a toy model that is Gaussian in $\calo$. This model is undesirable for two reasons:
\begin{enumerate}
    \item The Gaussian model fails to correctly reproduce the disk $2n$-point functions for $n \geq 2$. Instead, its disk correlators are given by sums over gravitational Feynman diagrams with no intersections of the bulk lines. While the disk correlators in JT gravity minimally coupled to a scalar also include diagrams where the bulk lines cross.
    \item The empty double-trumpet, given in \eqref{eq:toydt}, differs from \eqref{eq:dttoy}. If we insist that the Gaussian model is dual to JT gravity minimally coupled to some matter theory, then the partition function of this theory in AdS$_2$ differs from that of a scalar field.
\end{enumerate}

In this section, we summarize two strategies for finding matrix models that do not suffer from the above issues. We then comment on what these models can potentially teach us about UV divergences in wormhole amplitudes.

\subsection{A well-motivated guess for the matrix potential}

Our first strategy is to make a well-motivated guess for the matrix potential and then argue that it correctly reproduces the gravitational correlators. This was discussed in section \ref{sec:ethasamatrixmodel}. We identified an operator equation that is quadratic in $\calo$ (but has a complicated dependence on $H$) that holds in any disk correlator. In the eigenbasis of $H$, this expression is given in \eqref{eq:constraint}. It is reasonable to expect that in any instance of a matrix ensemble that reproduces the correct gravitational disk correlators, the two matrices should represent (to a good approximation) an abstract operator algebra where the operators $\calo$ and $H$ obey \eqref{eq:constraint}. Hence, we wrote in \eqref{eq:constraintsquared} a model where \eqref{eq:constraint} is manifestly obeyed in the limit $\Lambda \rightarrow \infty$. In section \ref{sec:crossingsymmetric}, we write down rules for constructing `t Hooft diagrams in this model. Because these rules involve the 6j symbol, `t Hooft diagrams greatly simplify due to the unlacing rule in \eqref{eq:6jorthog}. We use \eqref{eq:6jorthog} to show that a large class of Schwinger-Dyson equations is solved by the correct gravitational disk correlators. 

\subsection{Solving for the matrix potential given the gravitational correlators}

Our second strategy is to start with the correct gravitational disk correlators and work backwards to find the corresponding matrix potential. To do this, we generalize the gravitational correlators by introducing a parameter $\epsilon$ such that the original correlators are recovered as $\epsilon \rightarrow 1$, and the correlators of the Gaussian toy model are recovered as $\epsilon \rightarrow 0$. We think of $\epsilon$ as either a regulator or backing away from the double-scaling limit. There are many ways to define the regulated correlators for intermediate values of $\epsilon$, and in section \ref{sec:corrections} we introduce two specific regulators: the Selberg and the $q$-deformed regulators.\footnote{In section \ref{sec:regtwomatrixmodel}, we carefully define the regulators and show how they correspond to a `t Hooft-scaled matrix model. The double-scaling limit corresponds to removing the regulator.} Given a regulator, we explain in section \ref{sec:corrections} that one may algorithmically determine the matrix potential as a series expansion in $\epsilon$.\footnote{We have not fully answered the question of whether this expansion converges for $|\epsilon| < 1$. We discuss this issue more in sections \ref{sec:corrections} and \ref{sec:regtwomatrixmodel}.} After sending $\epsilon \rightarrow 1$, the matrix integral takes the form
\begin{align}
\label{eq:schematicmodel}
\mathcal Z = \int dH d\calo~ \exp\Big( -\tr \left[ V_{SSS}(H) + V_{c.t.}(H)\right] 
- e^{S_0} \sum_{ab} {F_{ab}\over 2}  \calo_{ab} \calo_{ba}  + \text{interactions} \Big) \ .
\end{align}
The interaction terms, which include higher powers of $\calo$, are designed so that the matrix integral produces the correct gravitational disk correlators. As in the toy model, the role of the counter-term $V_{c.t.}(H)$ is to ensure that the density of states of $H$ is the same as in the SSS model.\footnote{The explicit formula for $V_{c.t.}(H)$ differs from that of the toy model due to the interaction terms, which create new $\calo$ bubble diagrams whose disk contributions need to be canceled.} In analogy with one-matrix models, we are fixing the model using only the disk data.

Next, we would like to compute correlation functions with more boundaries and/or handles and compare the results with gravity. In section \ref{EmptyDt}, we study the empty double-trumpet, or the connected two-boundary correlator 
\begin{equation}
\braket{\tr e^{-\beta_L H} \tr e^{-\beta_R H}}_c
\label{eq:connectedtwoboundary}
\end{equation} at leading order in the genus expansion. Surprisingly, we find that despite the presence of the interaction terms in \eqref{eq:schematicmodel}, the empty double-trumpet agrees with the $\epsilon = 0$ answer \eqref{eq:toydt}. We could not find a simple reason for this, so the interested reader is encouraged to read section \ref{EmptyDt} for the technical details. This result does not depend on whether the $q$-deformed or Selberg regulator was used.

In section \ref{sec:twopointfunctiondoubletrumpet}, we compute the double-trumpet two-point function with one $\calo$ inserted on each AdS boundary. The result depends on whether the $q$-deformed or Selberg regulator is used. We now briefly sketch some of the ingredients that go into this computation, leaving the full computation for section \ref{sec:twopointfunctiondoubletrumpet}. Readers who are only interested in the answer may jump to section \ref{sec:constructingselbergandqdeformed}.

\subsubsection{The double-trumpet two-point function}

\label{sec:regdependenceindoubletrumpet}

Consider the matrix model computation of the disk four-point function, which is designed to agree with the gravitational answer. In analogy to \eqref{eq:twopointfunctiongaussian} in the toy model, we represent this computation as follows:
\begin{equation}
    \includegraphics[scale=0.45]{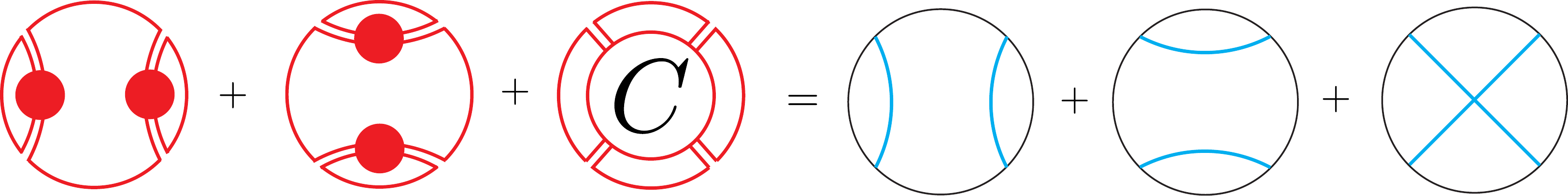},
    \label{eq:fourpointfunctioninteractingmodel}
\end{equation}
where a double-line with a red blob refers to the exact planar two-point function, and the blob labeled ``C'' refers to the sum over connected planar four-point diagrams. Each diagram on the left side corresponds to a diagram on the right side. It is also convenient to amputate the connected four-point function and define a new blob labelled ``A'' such that
\begin{equation}
    \includegraphics[scale=0.5]{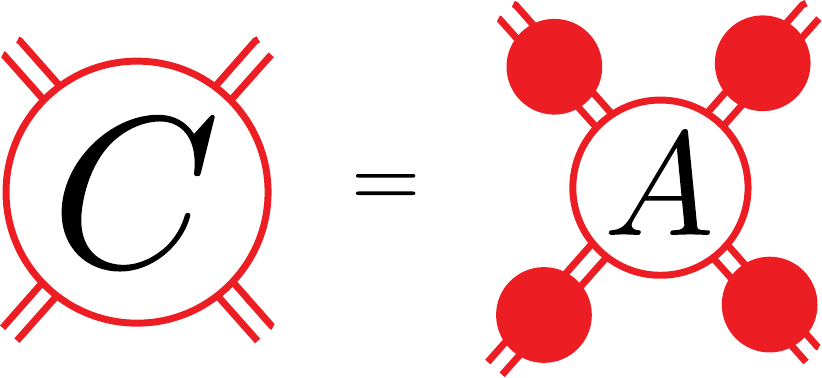} \quad.
    \label{eq:CA}
\end{equation}
The blobs labeled ``C'' and ``A'' refer to smooth functions of four energies, where each energy corresponds to one of the four single-lines shown in the above graphical representation.\footnote{It looks like there are eight single-lines emanating from the ``C'' or ``A'' blobs, but in planar diagrams they are connected to each other in pairs through the blob. For example, the simplest contribution to each of the two blobs is a tree four-point vertex.} It is convenient to label an energy $E$ with an $s$ parameter, where $E = s^2$.

We can build a class of diagrams that contribute to the double-trumpet two-point function by gluing together two opposite propagators in \eqref{eq:CA} after removing one of the red blobs. Explicitly, we have
\begin{equation}
    \includegraphics[scale=0.15]{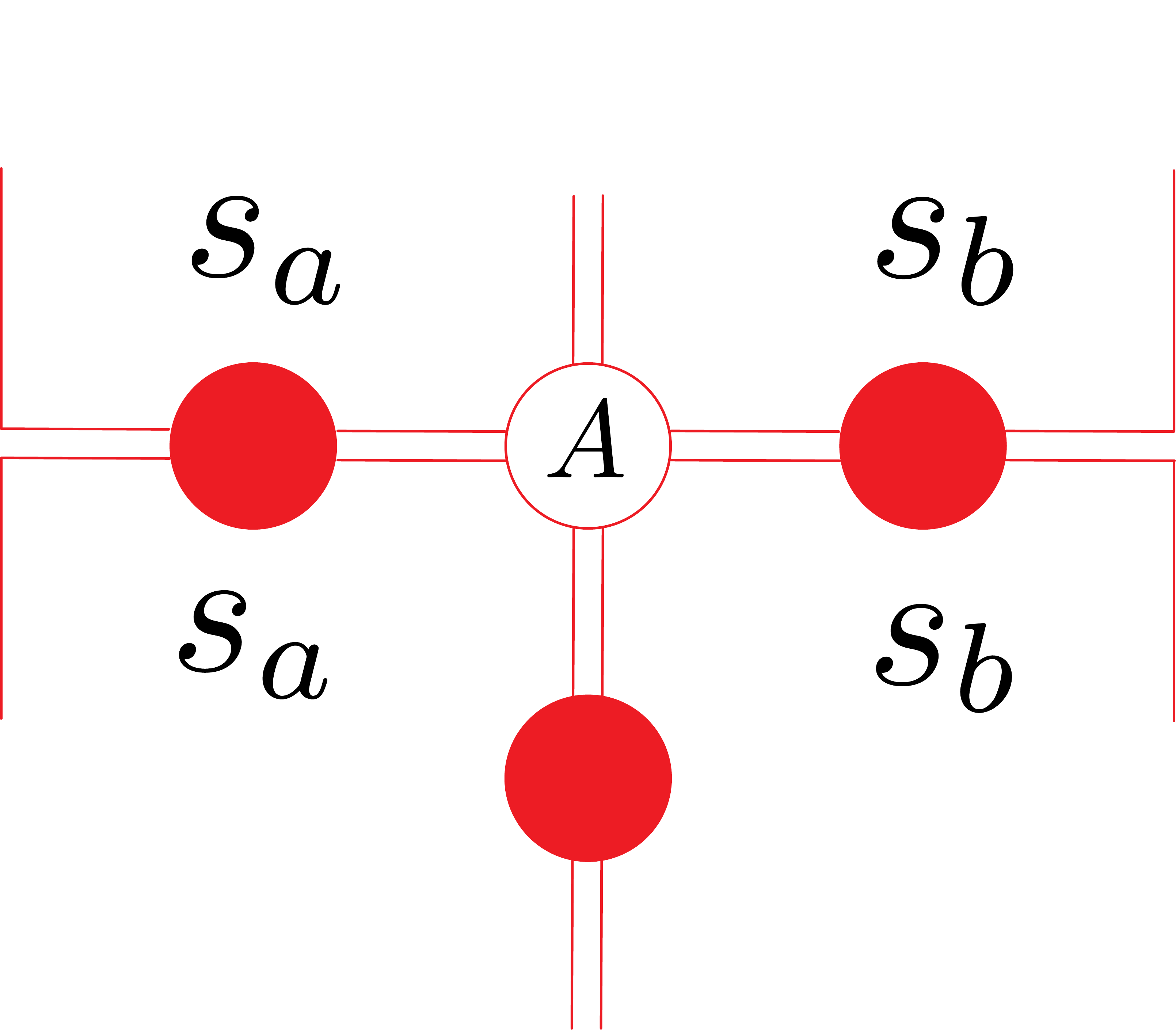}
\end{equation}
where the vertical single-lines on the far left and far right of the diagram represent the two traces in \eqref{eq:connectedtwoboundary}, and the top and bottom ends of this diagram are identified to obtain the cylinder topology. Removing the red blob is necessary to avoid overcounting diagrams. Another way to write this is as follows,
\begin{equation}
    \includegraphics[scale=0.2]{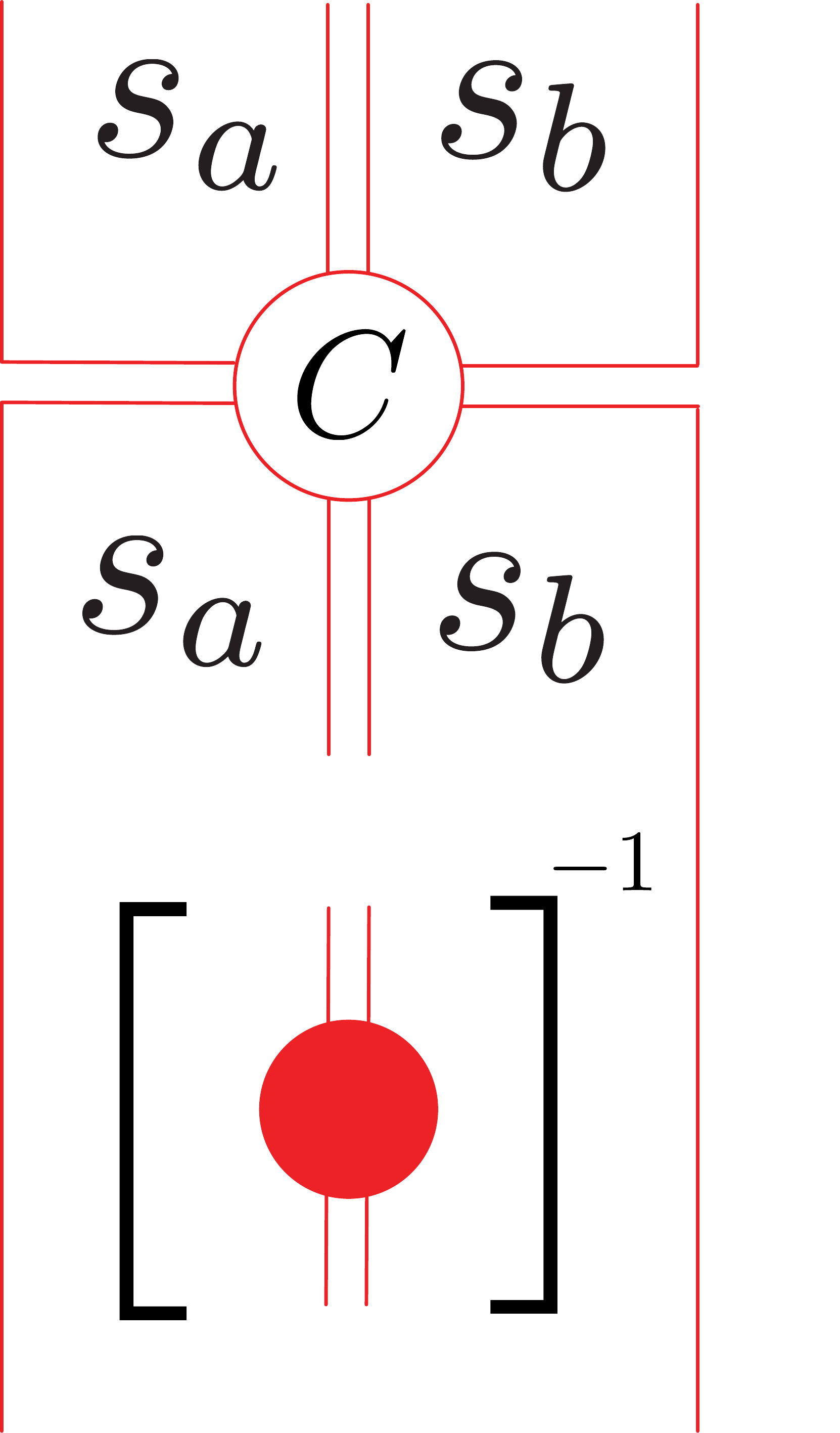}
    \label{eq:divideredblob}
\end{equation}
where raising the red blob to the $-1$ power is another way to express that a red blob should be removed from one of the external propagators of the ``C'' blob before making the identification. In practice, the red blob is a function of $s_a$ and $s_b$, and we want to divide the ``C'' blob by this function. The result is a function of the two energies $E_a = s_a^2$ and $E_b = s_b^2$ that are associated to the two AdS boundaries.

So far, this discussion did not depend on whether the Selberg or $q$-deformed regulators were used. Now, we will build another class of diagrams starting from the connected six-point function. As above, we will identify opposite propagators to construct diagrams with cylinder topology. However, to avoid overcounting diagrams, we must remove certain subdiagrams from the connected six-point function. The appropriate generalization of \eqref{eq:divideredblob} is
\begin{equation}
    \includegraphics[scale=0.14]{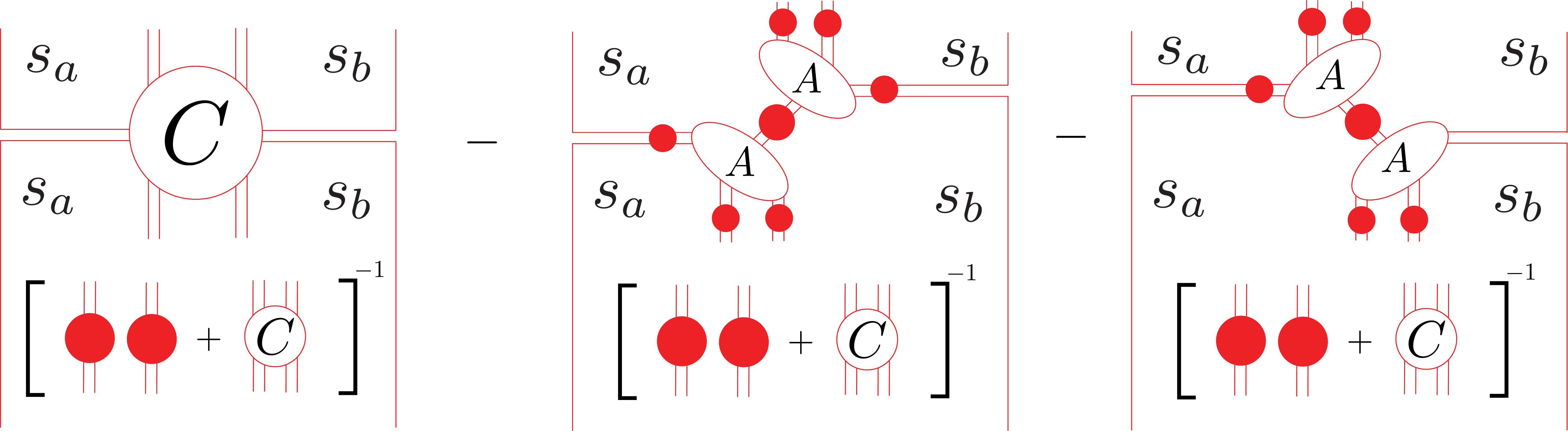}.
    \label{eq:winding2stuff}
\end{equation}
Without the $[\cdots]^{-1}$ insertions above, the process of identifying the top two double-lines with the bottom two double-lines would result in an overcounting of diagrams, as the computation would not only sum over diagrams but also sum over the locations where the diagrams can be cut in two places to obtain connected planar six-point diagrams. The two subtractions are introduced so that \eqref{eq:winding2stuff} does not count any diagrams that were already counted in \eqref{eq:divideredblob}.\footnote{A more detailed explanation of our procedure for systematically classifying diagrams is provided in section \ref{sec:doubletrumpet}.} Unlike \eqref{eq:divideredblob}, \eqref{eq:winding2stuff} contains additional explicitly-shown closed single-line loops aside from those associated with $s_a$ and $s_b$. This means that to remove the diagrams inside the brackets from the blob labeled ``C,'' we must integrate one of the energies of the ``C'' blob against the inverse of an appropriate ``two-to-two propagator'' that we will return to shortly. 

Explicit formulas for the blobs in \eqref{eq:winding2stuff} are known because by construction, the matrix model disk correlators reproduce the gravitational disk correlators (for example, we may use \eqref{eq:fourpointfunctioninteractingmodel} to obtain the explicit formula for the four-point ``C'' blob). We can thus convert \eqref{eq:winding2stuff} into an expression that may be evaluated using the gravitational Feynman rules,
\begin{equation}
    \includegraphics[scale=0.2]{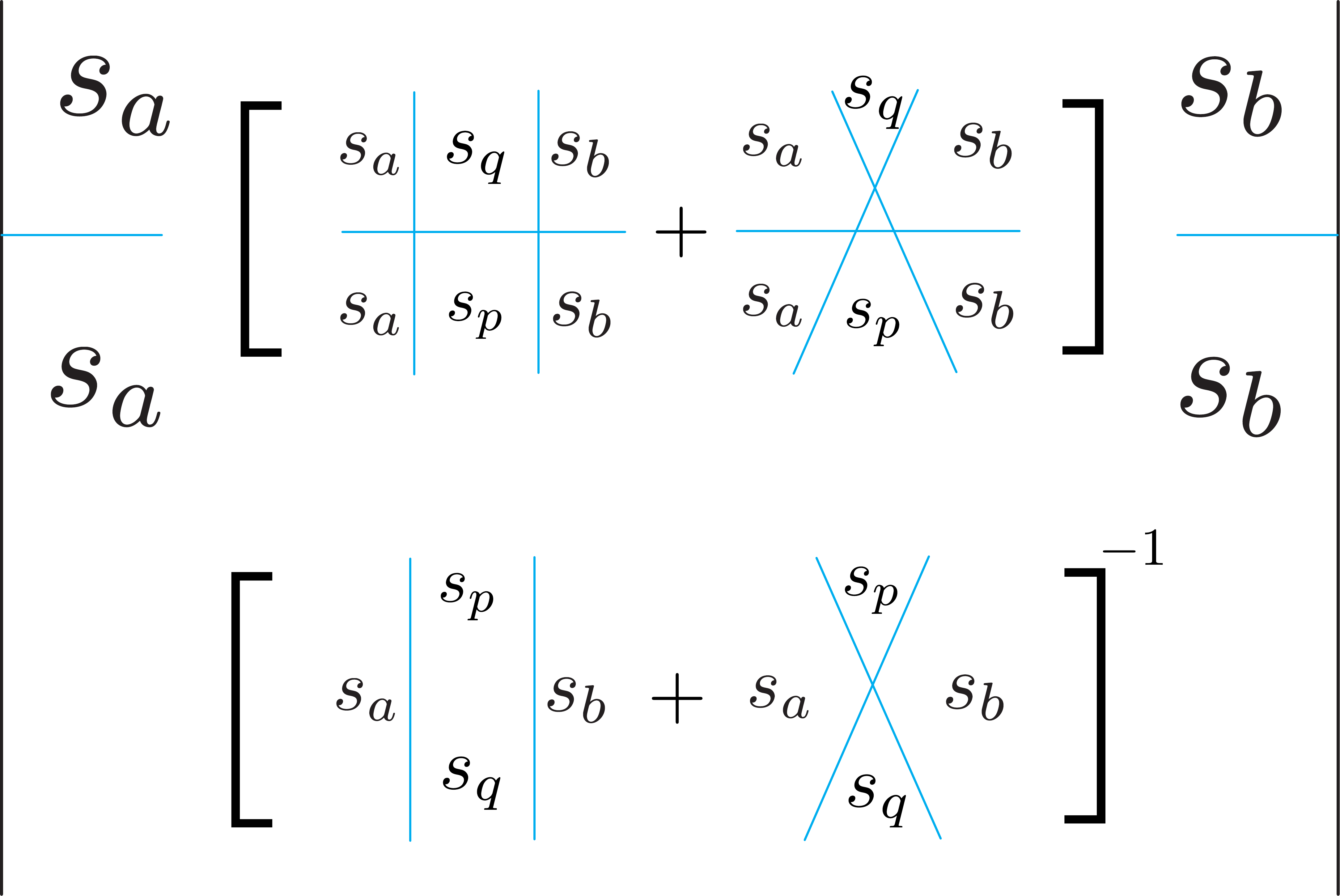}.
    \label{eq:winding2stuffgravity}
\end{equation}
The first set of brackets (without the $-1$ exponent) is meant to be expanded out into two terms, where each term is evaluated using the gravitational Feynman rules. The blue lines attached to the black boundaries correspond to insertions of $(\Gamma^\Delta_{aa})^{1/2}$ and $(\Gamma^\Delta_{bb})^{1/2}$, as per the Feynman rules. The diagrams contained in the second set of brackets (with the $-1$ exponent) become, using the Feynman rules,
\begin{equation}
    \includegraphics[scale=0.2]{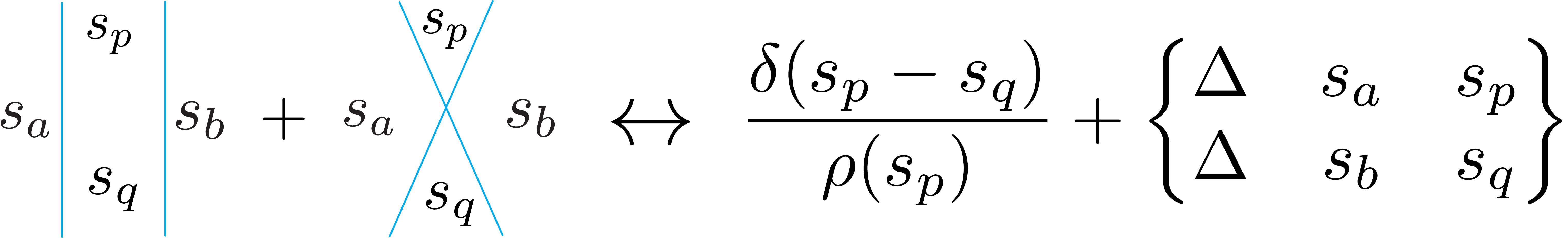}.
    \label{eq:oneplus6j}
\end{equation}
We should think of \eqref{eq:oneplus6j} as a ``two-to-two propagator'' that acts on a function of $s_q$ to produce a function of $s_p$. The first term (with the delta function, or the two parallel lines) acts as the identity operator. The two-to-two propagator is a projector because it squares to itself, thanks to the unlacing rule \eqref{eq:6jorthog} and the fact that \eqref{eq:oneplus6j} is invariant under exchanging the two top (or two bottom) endpoints of the blue lines. The $[\cdots]^{-1}$ term in \eqref{eq:winding2stuffgravity} corresponds to the inverse propagator. Having defined the two bracketed terms in \eqref{eq:winding2stuffgravity}, the parameters $s_p$ and $s_q$ should be integrated over using the density of states $\rho(s)$. The $s_p$, $s_q$ integrals are associated to the additional internal single-line loops mentioned in the previous paragraph. The two vertically separated bracketed terms are conveniently interpreted as two operators that are multiplied together, and the identification of the top and bottom ends of the diagram corresponds to taking a trace.

A subtlety arises because the two-to-two propagator is not invertible, as it is a projector. This can lead to puzzles in the evaluation of \eqref{eq:winding2stuffgravity}. For instance, the first bracketed term is rescaled by a factor of two under the action of the two-to-two propagator, as it is invariant under exchanging the top two (or bottom two) endpoints. This naively suggests that the inverse propagator in \eqref{eq:winding2stuffgravity} may be replaced by a factor of $\frac{1}{2}$,
\begin{equation}
    \includegraphics[scale=0.25]{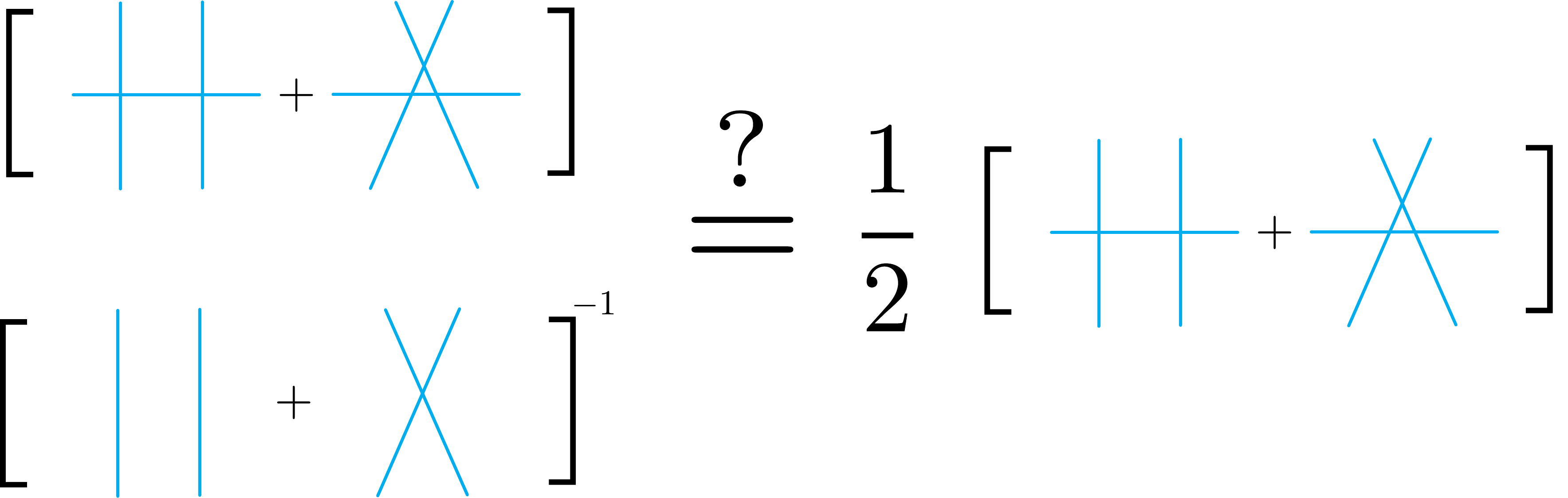}.
    \label{eq:selberganswer}
\end{equation}
Another naive guess is that the inverse propagator undoes the sum over the crossed and uncrossed vertical lines that pass through the horizontal line in the first bracketed term:
\begin{equation}
    \includegraphics[scale=0.25]{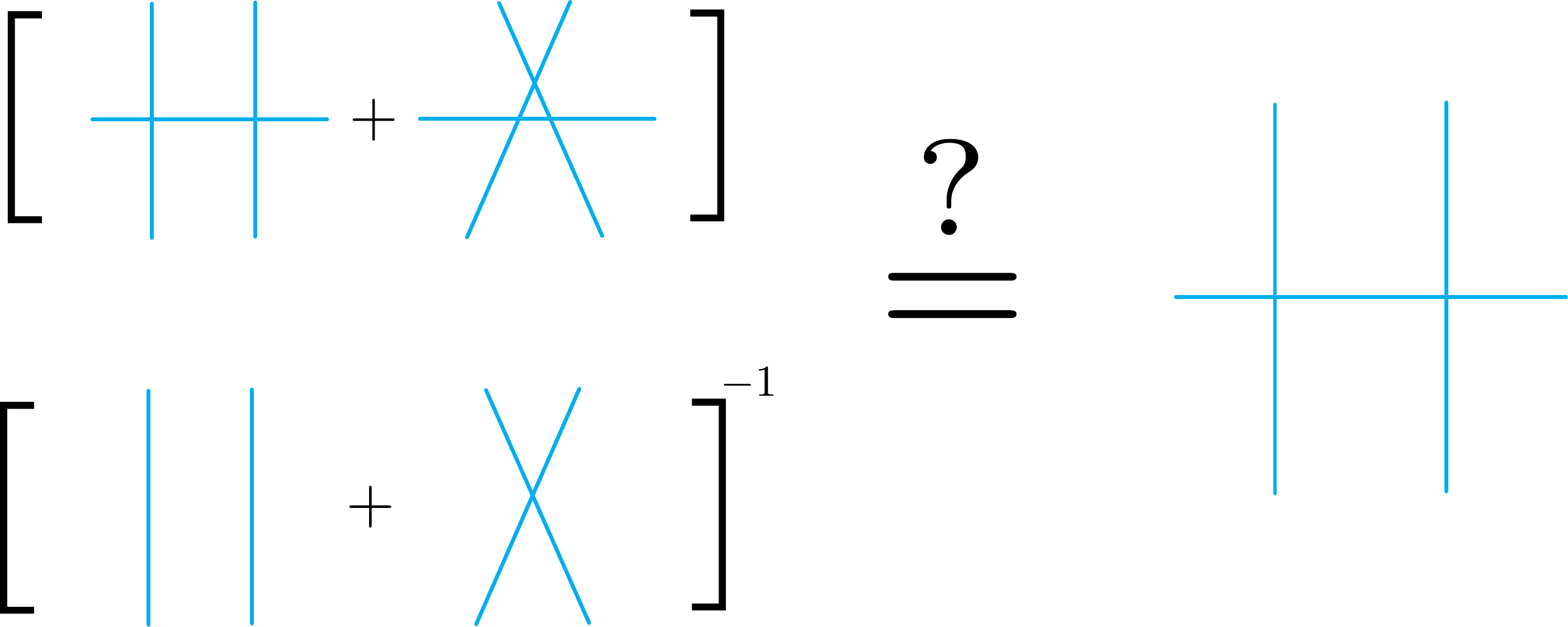}.
    \label{eq:qdeformedanswer}
\end{equation}
Using the regulators mentioned above, the two-to-two propagator becomes invertible and one can obtain definite answers. Equation \eqref{eq:selberganswer} is correct using the Selberg regulator, while \eqref{eq:qdeformedanswer} is correct using the $q$-deformed regulator. The Selberg result for \eqref{eq:winding2stuffgravity} reproduces the third term in equation \eqref{2pt-dt2}. The result for the $q$-deformed regulator is the same except the sum over dimensions $2 \Delta + 2m$ for $m \in \mathbb{Z}_{\ge 0}$ becomes instead a sum over dimensions $2 \Delta + m$ for $m \in \mathbb{Z}_{\ge 0}$.

In appendix \ref{sec:twopointdoubletrumpetappendix}, we explicitly compute one further class of diagrams. Using the Selberg regulator, the analogue of \eqref{eq:selberganswer} becomes
\begin{equation}
    \includegraphics[scale=0.3]{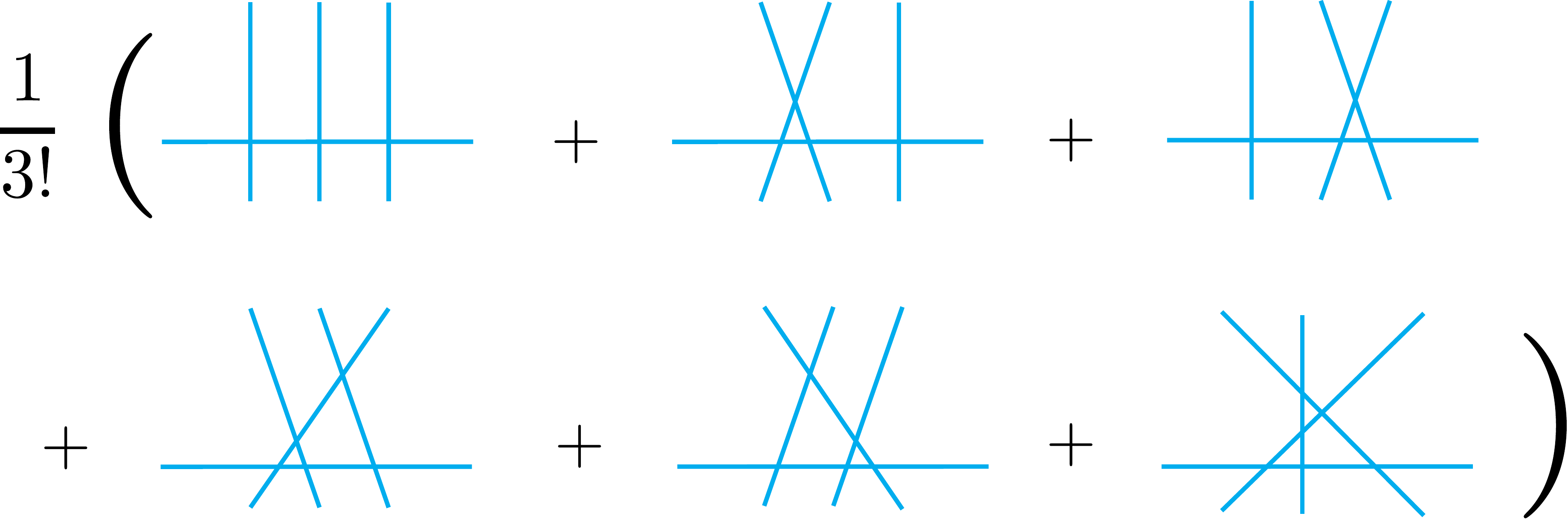}.
    \label{eq:dtwinding3}
\end{equation}
We further show in appendix \ref{sec:dtw=3} that after identifying the top and bottom energies of \eqref{eq:dtwinding3}, we obtain a contribution to the 2-point function on the double-trumpet
\begin{align}
 \sum_{n,m=0}^\infty 
\int_0^\infty ds_a ds_b ~ \rho(s_a) \rho(s_b) ~ e^{-\beta_L s_a^2 - \beta_R s_b^2}
 ~(\Gamma_{aa}^\Delta \Gamma_{bb}^\Delta )^{1/2} ~
 \left\{ 
\begin{matrix}
3\Delta + 2n + 3m & s_a & s_b \\ 
\Delta & s_b & s_a
\end{matrix}
\right\}
\ .
\end{align}
This is indeed the gravity answer if we continued the expansion \eqref{2pt-dt2} to the next term. Seeing a pattern, we conjecture that using the Selberg regulator, the class of `t Hooft diagrams where the lowest number of double-line propagators crossed by a left-right path is $n$ returns the result
\begin{align}
\includegraphics[scale=.4]{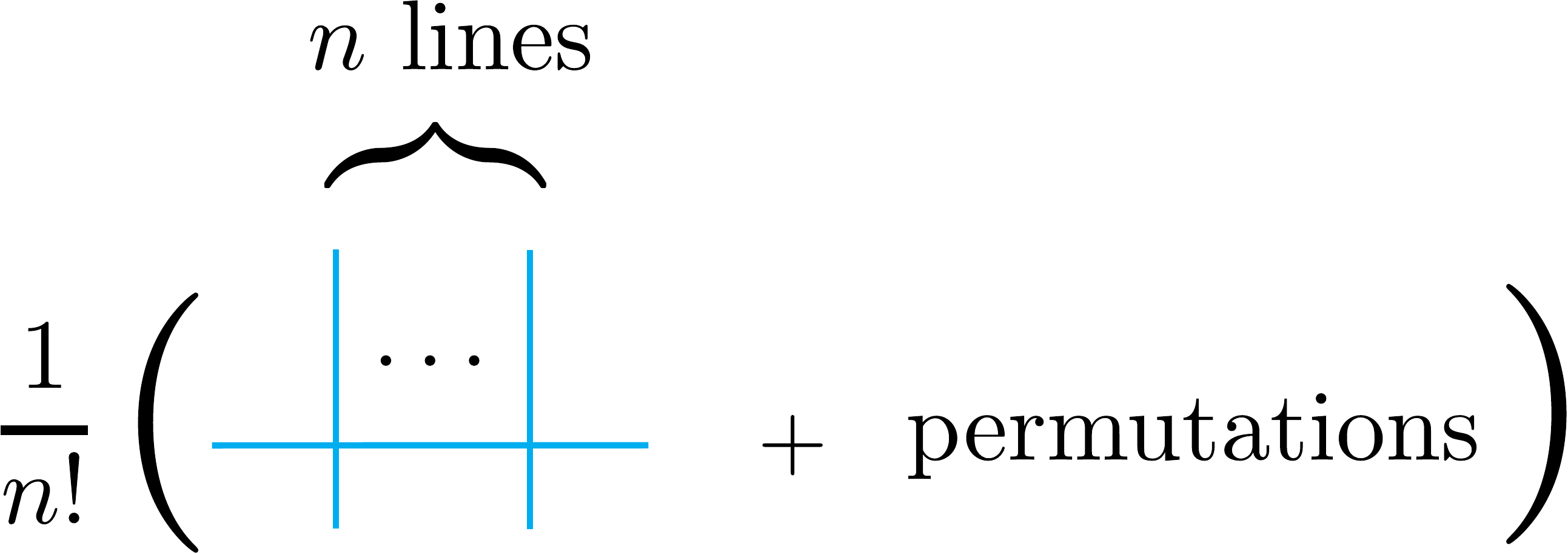},
\label{eq:dt2ptw=n}
\end{align}
and that this term reproduces the $n$th term in the sum in \eqref{2pt-dt2}. Using the $q$-deformed regulator, the pattern is that only the trivial permutation in \eqref{eq:dt2ptw=n} appears, with no prefactor. This trivial permutation may be evaluated using successive applications of the pentagon identity \eqref{Pent1}, which allows two vertical lines (crossing over a horizontal line) to be fused into a sum over single vertical lines with different dimensions, and the orthogonality relation of the Wilson polynomials (see Appendix \ref{sec:specialfunctions} for an introduction to the relevant special functions and identities). The final result after removing the regulator (or taking the JT limit) is the $n$th term in the sum in \eqref{eq:toydt}.

\subsubsection{Constructing the Selberg and $q$-deformed matrix models}

\label{sec:constructingselbergandqdeformed}

The result of the computation sketched above is that the connected correlator 
\begin{equation}
\braket{\tr \calo e^{-\beta_L H} \tr \calo e^{-\beta_R H}}_c
\label{eq:connectedtwopointdt}
\end{equation}
 at leading order in the genus expansion is equal to a sum over infinitely many 't Hooft diagrams in the matrix model \eqref{eq:schematicmodel}, which we have systematically classified. Each class of diagrams may be summed to obtain an explicit formula. The gravitational computation of the double-trumpet two-point function may be written as an infinite series as in \eqref{2pt-dt2}, where the $n$th term computes the contribution from the $n$th term in \eqref{eq:second} to the final answer.\footnote{In \eqref{2pt-dt2} we only explicitly showed the $n=0,1,2$ terms, leaving the rest in the $\cdots$.} We explicitly checked that using the Selberg regulator, the first four classes of `t Hooft diagrams reproduce the first four terms in \eqref{2pt-dt2}. We conjecture that the full set of `t Hooft diagrams reproduces all of the terms in \eqref{2pt-dt2}.

Thus, the model \eqref{eq:schematicmodel} defined using the Selberg regulator succeeds in reproducing the double-trumpet two-point function of JT gravity minimally coupled to a scalar field. However, the model does not reproduce the correct empty double-trumpet. We can amend the matrix potential in \eqref{eq:schematicmodel} to make the model compute the correct empty double-trumpet without affecting the computation of \eqref{eq:connectedtwopointdt}. The Selberg matrix model is defined by replacing \eqref{eq:schematicmodel} by
\begin{align}
\label{eq:schematicmodelselberg}
\mathcal Z = \int dH d\calo~ \exp\Big( -&\tr \left[ V_{SSS}(H) + V_{c.t.}(H)\right]
\\
+&{1\over 2} \int_0^\infty db ~ b ~ \left(\left[1 - (e^{- \Delta b},e^{-b})_\infty \right] - {e^{- \Delta b} \over 1- e^{-b}}\right) X(b)^2
\label{eq:addedterm}
\\
- &e^{S_0} \sum_{ab} {F_{ab}\over 2}  \calo_{ab} \calo_{ba}  + \text{interactions} \Big) \ , \label{eq:lastterm}
\end{align}
which is the same potential up to the addition of \eqref{eq:addedterm}.\footnote{The matrix model no longer appears to be single-trace because \eqref{eq:addedterm} is a double-trace term. However, if we rescale $\calo_{ab}$ by a function of $E_a$ and $E_b$, the transformation of the measure $d\calo$ causes a double-trace term to be added to the potential. Hence, we can change variables in the $\calo$ integral to make the potential single-trace again.} Because \eqref{eq:addedterm} does not depend on $\calo$, the disk correlators of $\calo$ are the same as in \eqref{eq:schematicmodel}. In particular, the saddle-point density of states for $H$ is unaffected due to the subtractions involved in the definition of $X(b)$, \eqref{Xdef}. If we neglected the interactions above (so that the potential is that of the toy Gaussian model plus \eqref{eq:addedterm}) and integrated out $\calo$, then from \eqref{eq:29}, \eqref{VXrel}, and \eqref{1-loopEffPot}, it is clear that the model would reproduce the correct empty double-trumpet. We pointed out above that the interactions in \eqref{eq:schematicmodel} do not change the empty double-trumpet result from that of the toy model. Hence, the Selberg model computes the correct empty double-trumpet \eqref{eq:dttoy}.

\bigskip 

We conjecture that the Selberg model correctly computes all higher genus and multiboundary correlators in JT gravity minimally coupled to a scalar. This conjecture is based on the success of the Selberg model in reproducing the scalar partition function on the double-trumpet (to the extent that we explicitly checked). We expect that the correlators on general topologies may also be computed from the disk correlators using similar methods to our double-trumpet computation. We make some comments on the pair of pants in section \ref{sec:pairofpantscomments}.

\bigskip

The double-trumpet two-point function using the $q$-deformed regulator takes a form that resembles \eqref{2pt-dt2}, except the dimensions of the states/operators that propagate around the closed geodesic differ. The partition function that counts the states/operators that propagate in \eqref{2pt-dt2} is given in \eqref{eq:second}. If we instead use the $q$-deformed regulator, we find that the relevant partition function is
\begin{equation}
    Z(b) = \sum_{n = 0}^\infty \left(\frac{e^{- \Delta b}}{1 - e^{-b}}\right)^n = \frac{1 - e^{-b}}{1 - e^{-b} - e^{-\Delta b}}.
    \label{eq:Zbpartfunc}
\end{equation}
As in the Selberg-regulated computation, we found this by computing the $n=0,1,2,3$ terms explicitly and conjecturing that the pattern continues for arbitrary $n$. This partition function agrees with the partition function that appears in \eqref{eq:toydt}. Hence, the $q$-deformed regulator seems to compute amplitudes in JT gravity minimally coupled to a scalar field up to a modification of the 1-loop determinant. For this interpretation to hold, there is no need to add additional double trace terms to \eqref{eq:schematicmodel}, so we will define the $q$-deformed matrix model to simply be \eqref{eq:schematicmodel}, where the interaction terms and counterterms are determined using the $q$-deformed regulator.

An interesting feature of the bulk dual of the $q$-deformed model is the modification of the matter 1-loop determinant in global AdS$_2$. In particular, \eqref{eq:Zbpartfunc} has a Hagedorn temperature, which is discussed more in section \ref{sec:hagedorn}. This is suggestive of the physics of strings, which would then appear in the dual of the closely related double-scaled SYK model.\footnote{It would be interesting to relate this idea to the work of \cite{Goel:2021wim} as well as \cite{Susskind:2021esx,Susskind:2022dfz,Lin:2022nss}.} We will show that the $q$-deformed model can be naturally defined away from the double-scaling limit such that its disk correlators compute the SYK correlators studied in \cite{Berkooz:2018jqr}. This connection is detailed in section \ref{sec:regtwomatrixmodel}.

\subsection{Matrix model interpretation of UV divergences in wormhole amplitudes}

The double-trumpet amplitudes in both the Selberg and $q$-deformed matrix models suffer from UV divergences. In the Selberg model, $Z_{\text{scalar}}(b)$ diverges as $e^{\frac{\pi^2}{6 b}}$ for small $b$. In the $q$-deformed model, the partition function has a Hagedorn temperature. Although these amplitudes are ill-defined, we can reproduce them from the matrix model because we can write each amplitude as an infinite sum (such as in \eqref{2pt-dt2}), and we know which `t Hooft diagrams reproduce each term in the sum. Hence, the sum over all `t Hooft diagrams with cylinder topology does not converge. This is a sign that the saddle point around which the perturbative genus expansion is defined is unstable. In section \ref{sec:generalstability}, we analyze the effective potential for $H$ (after integrating out $\calo$) and show that the empty double-trumpet is directly determined by the Hessian of the matrix potential evaluated at the saddle-point of the eigenvalue integral, similarly to how the disk is directly determined by the location of the saddle-point. In the $q$-deformed model, the Hessian has negative eigenvalues, which implies an instability. In the Selberg model, infinitely many eigenvalues become arbitrarily close to zero in the double-scaling limit, which effectively also amounts to an instability.\footnote{The matrix model interpretation of these UV divergences is different from the analysis of \cite{Gao:2021uro}, which studied UV divergences arising from dynamical End-of-the-World branes. Integrating out the branes only modified the single-trace matrix potential. To make the saddle unstable, one would need to modify the inter-eigenvalue repulsive force associated to the Vandermonde determinant.
}

Having a matrix model description of the UV divergences might allow us to understand how the gravitational theory, viewed as an effective theory, can be non-perturbatively completed. By modifying the matrix potential far away from the location of the saddle point, it should be possible to make the model non-perturbatively well-defined. Then, it would be interesting to find a stable saddle that the unstable saddle can decay to. Note that the disk correlators computed using this new hypothetical saddle could be very different from the original disk correlators. Hence, a more physical approach to understanding the model might involve introducing new bulk modes/interactions that could possibly render the original saddle stable.

Sufficiently far away from the double-scaling limit, the saddle in the $q$-deformed model becomes perturbatively stable. In section \ref{sec:regtwomatrixmodel}, we explain how the $q$-deformed model can be naturally generalized to a three-parameter model,\footnote{In section \ref{sec:regtwomatrixmodel}, we call these parameters $q_A$, $q_B$, and $\tilde{q}$.} and in this model the double-trumpet can be explicitly computed (see \eqref{eq:fixednm}). Having a bulk interpretation of this three-parameter model would allow us to understand the physical mechanism that determines whether there is a Hagedorn temperature or not. While there are infinitely many ways to define the $q$-deformed model (or any double-scaled matrix model) away from the double-scaling limit, the three-parameter model is canonically defined by its relation to the double-scaled SYK model computations in \cite{Berkooz:2018jqr}. Unfortunately, we do not have a canonical way to define the Selberg model away from the double-scaling limit, so we do not have explicit regulated formulas that could admit a nice physical interpretation.

\section{Constrained matrix ensemble}

\label{sec:crossingsymmetric}

In this section we analyze the model defined in \eqref{eq:constraintsquared} in more detail. We will show that in the double-scaling limit the disk amplitudes of JT gravity coupled to a scalar (discussed in section \ref{sec:JTgravitywithmatter}) are solutions to a large class of planar Schwinger-Dyson equations for this model. The orthogonality relation \eqref{eq:6jorthog} will play a key role. This section does not contain any prerequisite material that is necessary for understanding sections \ref{sec:corrections} and beyond.

We will first work directly in the double-scaling limit and give an argument that the JT disk correlators solve the Schwinger-Dyson equations. The argument is based on ``unlacing'' relations of the 6j-symbol. Some terms in the Schwinger-Dyson equation turn out to be divergent at high energies. These divergences arise because of the double-scaling (low energy) limit, where the right edge of the spectrum is taken to infinity and the density of states is supported on a semi-infinite interval. To properly deal with these divergences, in the second part of this section we back away from the double-scaling limit and consider a regularized version of the matrix model potential \eqref{eq:constraintsquared}. The regularized matrix model is related to a certain q-deformation of JT correlators \cite{Berkooz:2018jqr}, that will be discussed in more detail in section \ref{sec:regtwomatrixmodel}. In the q-deformed model the density of states has a finite support and high-energy divergences are regulated.

\subsection{Working directly in the double-scaling limit}

First, we rewrite the part of the potential that depends on $\calo$. We find it more convenient to work with the rescaled matrix $R_{ab}$, defined by
\begin{equation}
\label{Rdef}
\calo_{ab} := R_{ab} \left( e^{-S_0} \Gamma(\Delta \pm i \sqrt{E_a} \pm i \sqrt{E_b}) \over \Gamma(2\Delta) \right)^{1/2}  := (e^{-S_0} \Gamma_{ab}^{\Delta})^{1/2} ~ R_{ab}.
\end{equation} 
Note that the matrix elements of $R$ depend on the eigenvalues of $H$. For added simplicity we set $S_0 = 0$ in this subsection.\footnote{To restore the $e^{S_0}$ factors, one should insert $e^{-S_0}$ next to every 6j symbol, $e^{S_0}$ next to every $\rho(s)$, and $e^{-S_0}$ next to every $\delta(s_1 - s_2)$. This preserves the orthogonality relation \eqref{eq:2.16}.} We have that
\begin{align}
{\Lambda \over 2}\sum_{nac}|M_{ac}^n|^2 =& 
{\Lambda \over 8}\sum_{nac} ~ 
 \sum_b 
\left(\left\{
        \begin{array}{ccc}
	        \Delta & s_a & s_b \\
	 	    \Delta & s_c & s_n
	    \end{array}
	 \right\} 
	 - \delta_{bn}
\right) 
\left(R_{a b} R_{bc} - \delta_{ac} \right) \\
&\qquad \quad \sum_d
\left(\left\{
            \begin{array}{ccc}
	 		\Delta & s_a & s_n \\
	 		\Delta & s_c & s_d
	 	    \end{array}
	   \right\} 
	   - \delta_{dn}
\right)
\left(R_{c d} R_{da} - \delta_{ac} \right) \\ 
=& {\Lambda \over 4} \sum_{abcd} 
\left( 
\delta_{bd} - 
\left\{
        \begin{array}{ccc}
	        \Delta & s_a & s_b \\
	 	    \Delta & s_c & s_d
	    \end{array}
	 \right\} 
\right)
\left(R_{a b} R_{bc} - \delta_{ac} \right) 
\left(R_{cd} R_{da} - \delta_{ac} \right) \\
:= &
\Lambda 
\left( 
{1\over 2} \sum_{ab}g_{ab} R_{ab}R_{ba} 
+ {1\over 4} \sum_{abcd} g_{abcd}R_{ab}R_{bc}R_{cd}R_{da} 
+ \text{const} 
\right)
\label{eq:c2potential}
\end{align}
where in the second equality we have used the orthogonality relation \eqref{eq:6jorthog}.\footnote{Before we use \eqref{eq:6jorthog}, we replace the sum $\sum_d$ by an integral $\int ds_d \rho(s_d)$. This is because we work at large $N$ and consider only disk amplitudes in this section. We will make similar substitutions when we write down the diagrammatic rules of this model.} The couplings are
\begin{align}\label{g2ds}
g_{ab} &= -1 + {1\over 2} \sum_{cd} (\delta_{ac}+\delta_{bd}) 
\left\{
\begin{array}{ccc}
\Delta & s_a & s_b \\
\Delta & s_c & s_d
\end{array}\right\}  \ ,\\
g_{abcd} &= {1\over 2} \left( \delta_{ac} + \delta_{bd} \right) 
- \left\{\begin{array}{ccc}
\Delta & s_a & s_b \\
\Delta & s_c & s_d
\end{array}\right\} \ .
\label{g4ds}
\end{align}
From \eqref{eq:c2potential}, we may determine the propagator and four-point interaction vertex for the $R$ matrix that is used to compute `t Hooft diagrams. The propagator is
\begin{equation}
\label{eq:constraintsquaredpropagator}
    \includegraphics[scale=0.25]{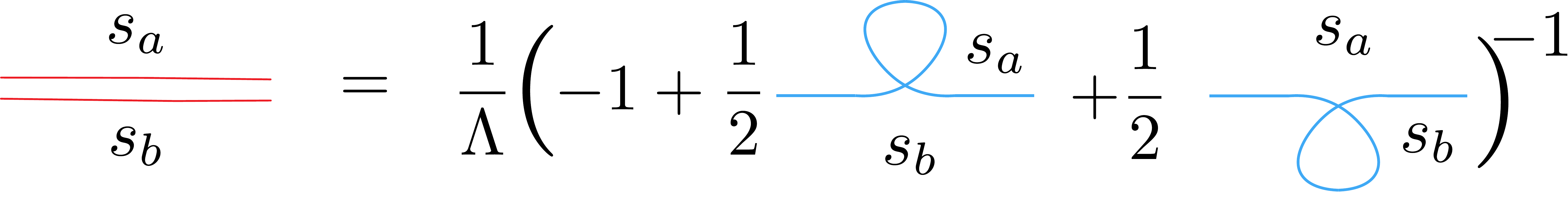}
\end{equation}
where on the left hand side, each index line in the double-line propagator corresponds to an energy (this convention was used in section \ref{sec:toymodel}). On the right hand side, the diagrams should be evaluated using the gravitational Feynman rules. That is,
\begin{equation}
    \raisebox{-.4in}{\includegraphics[scale=0.3]{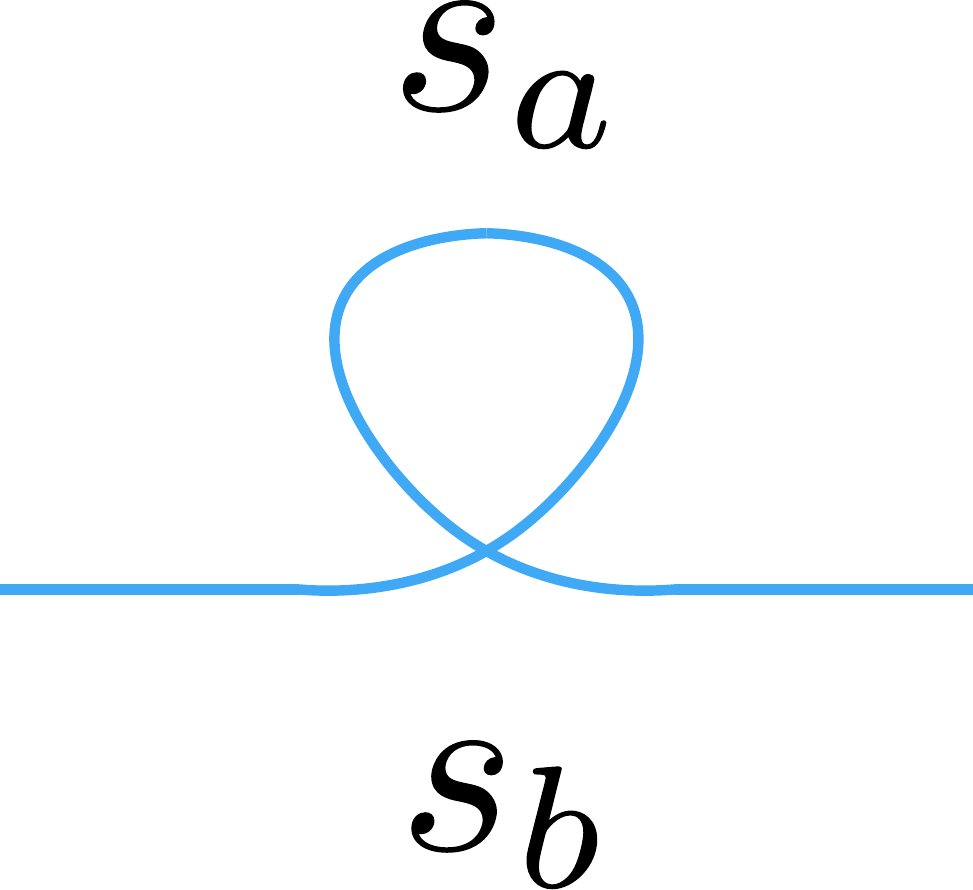}} 
    ~~~=~~~ 
    \int_0^\infty ds \, \rho(s) 
    \left\{\begin{array}{ccc}
	 		\Delta & s_a & s_b \\
	 		\Delta & s_a & s
	 	\end{array}\right\} \ .
\end{equation}
This integral actually diverges because at high energies (see Appendix \ref{sec:specialfunctions})
\begin{align}
\left\{\begin{array}{ccc}
	 		\Delta & s_a & s_b \\
	 		\Delta & s_a & s
	 	\end{array}\right\} \sim e^{-\pi s} , \qquad (s\to \infty) \ ,
\end{align}
while $\rho(s) \sim e^{2\pi s}$ at large $s$. We will see however that such divergences will formally cancel out of our final expressions. We will deal with these divergences more carefully in a regularized model in the next subsection.

The quartic interaction vertex for $R$ is given by
\begin{equation}
    \includegraphics[scale=0.2]{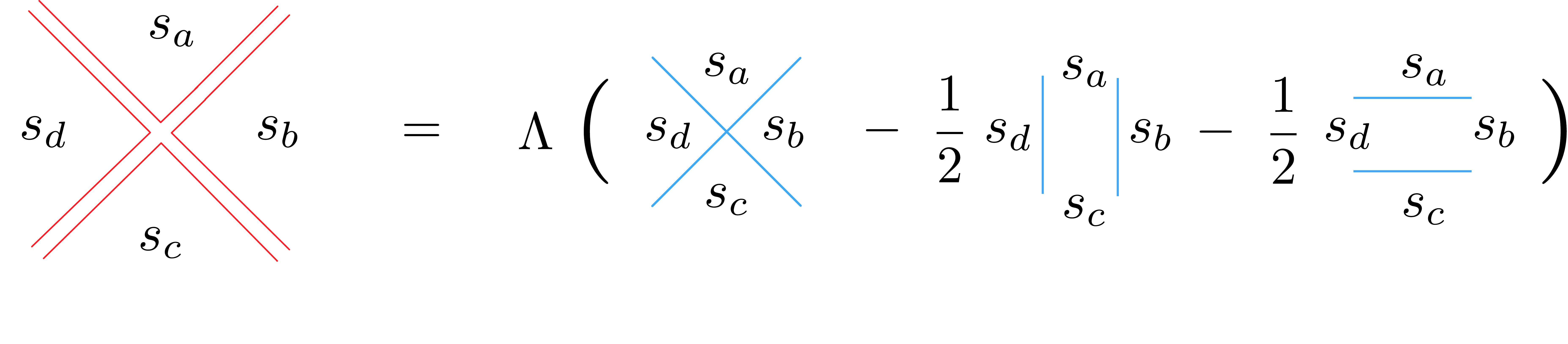}
    \label{eq:constraintsquaredvertex}
\end{equation}
where again the right hand side should be evaluated using the gravitational Feynman rules. The first term on the right hand side is the 6j symbol. The last two terms are delta functions ${\delta(s_a-s_c) \over \rho(s_a)}$ and ${\delta(s_b-s_d) \over \rho(s_b)}$ respectively.

\bigskip

From the diagrammatic rules above we may compute planar correlators of the $R$ matrix. Any closed loop corresponds to an integral over the energy with measure $ds \, \rho(s)$. Next, we introduce a Schwinger-Dyson equation that relates the couplings in the matrix potential to the correlators. The equation is represented diagrammatically as follows:
\begin{equation}
\includegraphics[scale=0.2]{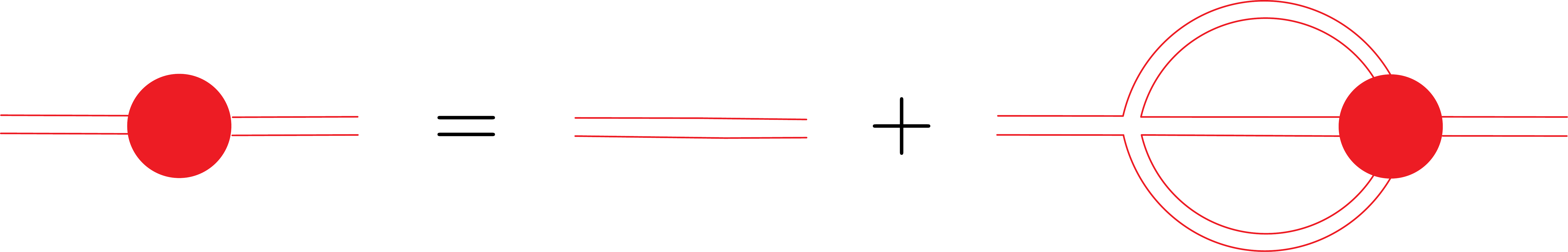}
\label{eq:firstsdeqn}
\end{equation}
The left hand side represents the sum over all planar two-point diagrams and is equal to $\braket{R_{ab}R_{ba}}_{\disk}$. The red blob on the right hand side represents the sum over all planar (connected and disconnected) four-point diagrams and is given by $\braket{R_{ab}R_{bc} R_{cd} R_{da}}_{\disk}$.\footnote{By $\braket{R_{ab}R_{bc} R_{cd} R_{da}}_{\disk}$, we are referring to the smooth function of four energies that is obtained by performing inverse Laplace transforms on the four-point function in the planar limit of the matrix model. See section \ref{sec:corrections} for more comments.} To understand the above equation, note that the leftmost double-line can either directly connect to the rightmost double-line, or it can connect to a vertex. The other three legs of this vertex and the rightmost double-line become the external lines of the four-point red blob, which accounts for all the remaining diagrams that contribute to the left hand side. The Schwinger-Dyson equation \eqref{eq:firstsdeqn} can also be derived from the identity
\begin{align}\label{SDder1}
\int dR ~{\p \over \p R_{ab} } \left( R_{ab}  e^{-V(R)} \right) 
=0 
\end{align}
and is expressed as
\begin{align}\label{SD1}
\braket{R_{ab} R_{ba}}_\disk = 
{1\over \Lambda}g_{ab}^{-1} 
- g_{ab}^{-1} \sum_{cd} g_{abcd} \braket{R_{ab} R_{bc} R_{cd} R_{da} }_\disk
 \ ,
\end{align}
where $g_{ab}, g_{abcd}$ are defined in \eqref{g2ds}, \eqref{g4ds}.

We will now show that the gravitational correlators that correspond to $\braket{R_{ab}R_{ba}}_{\disk}$ and $\braket{R_{ab}R_{bc} R_{cd} R_{da}}_{\disk}$ solve \eqref{eq:firstsdeqn} or equivalently \eqref{SD1}. Of the two terms on the right hand side of \eqref{SD1}, the first is order $\Lambda^{-1}$, while the second is order one in the large $\Lambda$ expansion. Hence, we may drop the first term. To evaluate the second term, we replace the interaction vertex by the sum over the three diagrams in \eqref{eq:constraintsquaredvertex}, and we replace the red blob by a sum over the three four-point gravitational Feynman diagrams (two uncrossed and one crossed), which correspond to the three terms in \eqref{eq:fourpointfunctioninteractingmodel}.\footnote{The red four-point blob in \eqref{eq:firstsdeqn} is by definition equal to the sum of the three terms on the left hand side of \eqref{eq:fourpointfunctioninteractingmodel}.} We then find that $-g_{abcd} \braket{R_{ab}R_{bc}R_{cd}R_{da}}_\disk$ is given by
\begin{equation}\label{g4R^4}
    \includegraphics[scale=0.3]{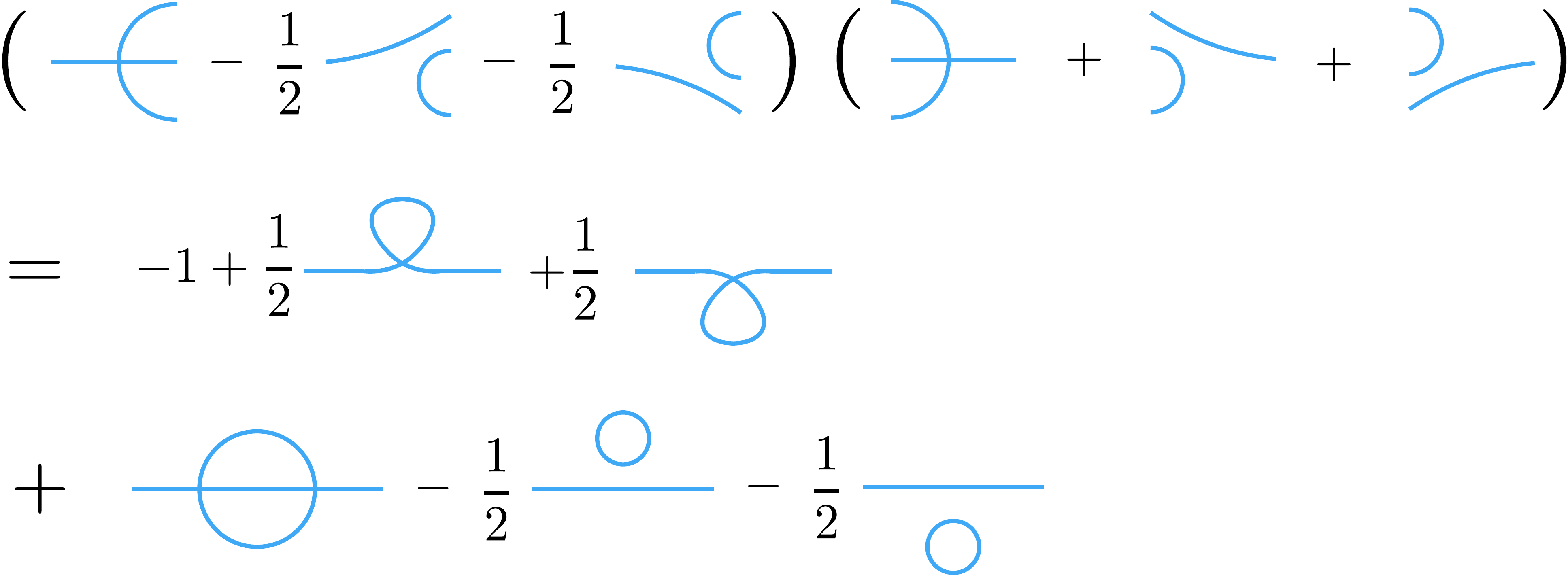}
\end{equation}
In the LHS, a product of two diagrams means that we connect the three lines in the middle, which corresponds to doing the sum $\sum_{cd}$. To get the RHS, we expanded the first line into nine terms and simplified the algebra. 

Now we would like to argue that the last three terms in \eqref{g4R^4} cancel out. We know that 6j-symbols obey ``unlacing'' rules, such as the orthogonality relation \eqref{eq:6jorthog}. So heuristically, one might expect that more unlacing relations are obeyed, such that in the first term in the last line of \eqref{g4R^4} we can move the circle away from the horizontal line. Then the last three terms in \eqref{g4R^4} would cancel out. In practice, if we compute the integral corresponding to the first term in the last line of \eqref{g4R^4}, we find a divergent result $\delta(0)$, delta function at zero argument. This can be seen from the orthogonality relation \eqref{eq:6jorthog}. The latter two terms are proportional to a square of the delta function and therefore contain $\delta(0)$ as well. For now we assume that such unlacing rules work and the last three terms cancel out. We will deal with this more carefully in the next subsection. The remaining terms in the second line of \eqref{g4R^4} are canceled by $g_{ab}^{-1}$ in \eqref{SD1}. giving one in total and Schwinger-Dyson equation becomes $\la R_{ab}R_{ba} \ra_\disk = 1$, which is the correct result.

\bigskip

The next Schwinger-Dyson equation we consider is diagrammatically represented as follows:
\begin{equation}
    \includegraphics[scale=0.12]{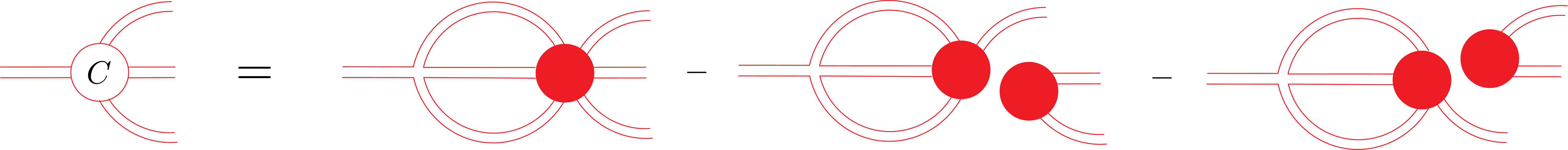}
    \label{eq:second2dequation}
\end{equation}
where the blob with six external double-lines represents the sum over all planar (connected and disconnected) six-point diagrams. The blob labeled ``C'' on the left represents the sum over all connected four-point planar diagrams. To understand this equation, note that the leftmost double-line must be attached to a vertex, or else the resulting diagram cannot be both connected and planar. The other three double-lines of the vertex together with the three double-lines on the right hand side should be connected using additional propagators and vertices in all possible ways such that the entire diagram is connected. There are some contributions to the six-point blob that result in a disconnected diagram, and so these contributions are subtracted off as shown. The equation \eqref{eq:second2dequation} can also be derived from the identity 
\begin{align}\label{SDder2}
\int dR ~ {\p \over \p R_{ad} } \left( R_{ab} R_{bc} R_{cd} e^{-V(R)} \right) = 0 \ .
\end{align}

If we plug the gravitational correlators into the expressions for the matrix model correlators, the ``C'' blob becomes a 6j symbol, and the blob with six external double-lines becomes a sum over $15$ gravitational Feynman diagrams, but $6$ of those are subtracted off. The right hand side of \eqref{eq:second2dequation} is order one in the large $\Lambda$ limit. Plugging in the diagrammatic rules \eqref{eq:constraintsquaredpropagator} and \eqref{eq:constraintsquaredvertex}, we may again verify that \eqref{eq:second2dequation} holds, assuming unlacing rules. This can again be dealt with more carefully using the regularization in the next section.

\bigskip

We can continue to write down more Schwinger-Dyson equations where the left hand side is analogous to the left hand side of \eqref{eq:second2dequation}, except with additional double-lines emanating from the right side of the ``C'' blob. This entire class of Schwinger-Dyson equations is solved by the gravitational Feynman rules. This is evidence that \eqref{eq:constraintsquared} correctly reproduces JT gravity minimally coupled to a scalar at disk level.

\subsection{Backing away from the double-scaling limit and q-deformation}

\label{sec:backingaway}

In the previous subsection we considered the matrix model directly in the double-scaling limit and gave a suggestive argument that JT correlators solve the Schwinger-Dyson equations. However, some of the expressions were not completely well-defined due to high-energy divergences. This is to be expected in the double-scaling limit. To give a more precise matrix model description, in this section we consider a particular way to back away from the double-scaling limit. The model will depend on the size of matrices $N$ and an extra parameter $q$. We will show that in the double-scaling limit $N \to \infty, q\to 1$, while keeping a particular combination (to be specified below) of $N$ and $q$ fixed, this matrix model is solved by JT correlators.

\bigskip

Let us consider a quartic matrix model of $N\times N$ matrices
\begin{align}
V(R) = N \left( {1\over 2} \sum_{ab} g_{ab} |R_{ab}|^2 + {1\over 4} \sum_{abcd} g_{abcd} R_{ab} R_{bc} R_{cd} R_{da}  \right) \ .
\end{align}
To get ourselves oriented, let us first consider the usual large $N$ limit (no double-scaling). At large $N$ we take the couplings to be order one: $g_{ab}, g_{abcd} \sim 1$, as it is common in matrix models. We can again derive the planar $N \to \infty$ Schwinger-Dyson equation from \eqref{SDder1}
\begin{align}\label{SD1reg}
{1\over N}  = g_{ab} \la |R_{ab}|^2 \ra_{\disk} + 
\sum_{cd} g_{abcd} \la R_{ab} R_{bc} R_{cd} R_{da} \ra_\disk \ . 
\end{align}
 At large $N$, the correlators scale in the standard way
 \begin{align}\label{2ptNscaling}
 &\la |R_{ab}|^2 \ra_{\disk} \sim {1\over N}  \ , \\
 &\la R_{ab} R_{bc} R_{cd} R_{da} \ra_\disk \sim {1\over N^3} \ .
 \label{4ptNscaling}
 \end{align}
For the four-point function the disconnected part scales as $\la R_{ab} R_{bc} \ra \la R_{cd} R_{da} \ra_\disk \sim {1\over N^2} \delta_{ac}$, but the extra Kronecker $\delta_{ac}$ can be thought of as effectively $1\over N$. The scaling is such, that if we sum over all indices the result is of order $N$: $\sum_{ab} \la |R_{ab}|^2 \ra_{\disk} \sim N^2 \cdot {1\over N} = N$ and $\sum_{abcd} \la R_{ab} R_{bc} R_{cd} R_{da} \ra_\disk \sim N^4 \cdot {1\over N^3} = N$. This is indeed what we expect for disk correlators. 

Using the above scaling it is easy to check that all three terms in \eqref{SD1reg} are of the same order and must be kept in the large $N$ limit
\begin{align}
{1\over N}  = 
\underbrace{g_{ab}}_{\sim 1} \ 
\underbrace{\la |R_{ab}|^2 \ra_{\disk}}_{\sim {1\over N}} + 
\underbrace{\sum_{cd}}_{\sim N^2} \
\underbrace{g_{abcd}}_{\sim 1}  \ 
\underbrace{\la R_{ab} R_{bc} R_{cd} R_{da} \ra_\disk}_{\sim {1\over N^3}} \ . 
\end{align}

Now we turn to the double-scaling limit. In this case we will see that the situation is different. In the limit that we define below, the LHS of \eqref{SD1reg} can be dropped, while the rest is solved by JT correlators.

\bigskip

To define the double-scaling limit we need to introduce a certain q-deformation of JT correlators. This will be described in much more detail in section \ref{sec:disk q-def}. Here we give a few results necessary for this section. First, instead of the Schwarzian density of states $\rho(s)$ we consider its q-deformation 
\begin{align}
\rho_q(s) = {1\over 2\pi \Gamma_q(\pm 2is)} \ , \qquad 
\int_0^{\pi/|\log q|} ds ~ e^{S_0}\rho_q(s) = e^{S_0}N_q \ , 
\end{align}
where $N_q$ is 
\begin{align}
N_q=  {1\over |\log q| (1-q)^2 (q;q)_\infty^3} \ .
\end{align}
In particular, $N_q \to \infty$ as $q\to 1$. The density of states $e^{S_0}\rho_q(s)$ is supported on a finite interval $(0, {\pi \over |\log q|})$, as in a 1-cut matrix model. It is also normalized as shown above. Therefore it is natural to identify
\begin{align}
N = e^{S_0} N_q \ .
\end{align}
The double-scaling limit that we will consider is defined by taking $N \to \infty, q\to 1$, while keeping ${N\over N_q} = e^{S_0}$ finite.

In addition, we also deform the 6j-symbol to a q-deformed one
\begin{align}
J_{abcd} \equiv
\left\{
\begin{matrix}
\Delta & s_a & s_b \\
\Delta & s_c & s_d
\end{matrix}
\right\}_q \ .
\end{align}
The definition of the q-deformed 6j-symbol is given in appendix \ref{sec:specialfunctions}. For now, we only need to know that in the limit $q\to 1$ it gives back the classical 6j-symbol appearing in JT correlators. We take the exact correlators to be
\begin{align}\label{2ptSDq}
\la |R_{ab}|^2 \ra_\disk &= {N_q \over N} = e^{-S_0} \ , \\
\la R_{ab} R_{bc} R_{cd} R_{da} \ra_\disk &= \left( N_q \over N \right)^2 (\delta_{ac} + \delta_{bd}) + \left( N_q \over N \right)^3 
\left\{
\begin{matrix}
\Delta & s_a & s_b \\
\Delta & s_c & s_d
\end{matrix}
\right\}_q
\\ 
& = e^{-2S_0} \left( \delta_{ac} + \delta_{bd} + e^{-S_0} 
\left\{
\begin{matrix}
\Delta & s_a & s_b \\
\Delta & s_c & s_d
\end{matrix}
\right\}_q
\right) \ .
\label{4ptSDq}
\end{align}
It is interesting to note that the two-point function is enhanced by a factor $N_q$ in comparison to the large $N$ scaling \eqref{2ptNscaling}. Similarly for four-point and higher correlators.

Now we would like to find the couplings $g_{ab}, g_{abcd}$ such that in the double-scaling limit the Schwinger-Dyson equation \eqref{SD1reg} is solved by correlators \eqref{2ptSDq} - \eqref{4ptSDq}. We first define a smeared delta-function
\begin{align}\label{SmearedDeltaDef}
\delta(k,k'; s_a,s_c|q) : =
\int_0^{\pi/|\log q|} ds ~ \rho_q(s) ~
\left\{
\begin{matrix}
\Delta &s_a & k \\
\Delta & s_c & s
\end{matrix}
\right\}_q
\left\{
\begin{matrix}
\Delta &s_a & k' \\
\Delta & s_c & s
\end{matrix}
\right\}_q   \ .
\end{align}
In the limit $q\to 1$ this equation reduces to the orthogonality relation of the classical 6j-symbols \eqref{eq:2.16} and
\begin{align}\label{SmearedDeltaLimit}
\lim_{q\to 1} \delta(k,k'; s_a,s_c|q) = {\delta(k-k') \over \rho(k)} \ .
\end{align}
But for $0<q<1$ the function $\delta(k,k'; s_a,s_c|q) $ is smooth and finite. It is bell-shaped and defines a smearing of the delta-function.

Now we are ready to define the couplings. We choose them to be
\begin{align}\label{g2qdef}
g_{ab} =& - e^{-S_0}{1\over 2}\sum_{cd} \left[
  \delta_{ac} \delta(s_b,s_d; s_a,s_c|q) + \delta_{bd} \delta(s_a,s_c; s_b,s_d|q)
\right] \\
&+e^{-S_0}\sum_{cd} J_{abcd}
\left(
\delta_{ac} + \delta_{bd} - e^{-S_0}{1\over 2}  \left[ \delta(s_a,s_c; s_b,s_d|q) + \delta(s_b,s_d; s_a,s_c|q) \right]
\right) \ , \\
g_{abcd} = & 
{1\over 2}  \delta(s_a,s_c; s_b,s_d|q) +{1\over 2} \delta(s_b,s_d; s_a,s_c|q) - J_{abcd} \ .
\label{g4qdef}
\end{align}
Several comments are in order. The powers of $e^{S_0}$ are chosen to be such that if we substitute $e^{S_0} = {N \over N_q}$ and take $N\to \infty$, $q$ - fixed limit, the couplings are of order $1$. On the other hand, in the double-scaling limit $N \to \infty, \ q\to 1$ with $e^{S_0} = {N\over N_q}$ - fixed we recover the couplings \eqref{g2ds}, \eqref{g4ds} where we now restored factors of $e^{S_0}$. To see the latter, the Kronecker delta and discrete sums are substituted in the double-scaling limit as
\begin{align}\label{dsRules}
\delta_{ab} \to {\delta(s_a-s_b) \over e^{S_0}\rho_q(s_a)} , \qquad \sum_a \to \int_0^{\pi/|\log q|} ds ~ e^{S_0} \rho_q(s) \ .
\end{align}

Now we can check that the Schwinger-Dyson equation is satisfied in the double-scaling limit. First, the LHS of \eqref{SD1reg} can be dropped. Some of the individual terms in the RHS are in fact divergent in double-scaling limit. Though of coure all divergences cancel, as we now show. We compute
\begin{align}
\sum_{cd}g_{abcd} \la R_{ab} R_{bc} R_{cd} R_{da} \ra_\disk 
=& 
\sum_{cd} 
\left( 
{1\over 2}  \delta(s_a,s_c; s_b,s_d|q) +{1\over 2} \delta(s_b,s_d; s_a,s_c|q) - J_{abcd}
\right) \\
&  
e^{-2S_0}\left( 
\delta_{ac} + \delta_{bd} + e^{-S_0}J_{abcd}
\right)
\\
=& -g_{ab} \la |R_{ab}|^2 \ra_\disk \\
\label{d(0)1}
&+ e^{-2S_0} {1\over 2} \sum_{cd} \left[ \delta_{ac} \delta(s_a,s_c;s_b,s_d|q) + 
\delta_{bd} \delta(s_b,s_d;s_a,s_c|q)\right] \\
\label{d(0)2}
&- e^{-3S_0} \sum_{cd} J_{abcd}^2 \\
=& - g_{ab} \la |R_{ab}|^2 \ra_\disk \ .
\end{align}
In the last equality we used that \eqref{d(0)1} and \eqref{d(0)2} cancel out. This follows from the definition of the smeared delta-function \eqref{SmearedDeltaDef}, both give a smeared delta function at zero argument. We thus showed that the Schwinger-Dyson equation \eqref{SD1reg} is satisfied in the double-scaling limit with the LHS dropped.

\bigskip

In the double-scaling (IR) limit, we showed that we can drop the LHS of \eqref{SD1reg}. This is equivalent to neglecting the bare propagator of the free theory, first term in the RHS of \eqref{eq:firstsdeqn}, \eqref{SD1}. It is interesting to note that this seems similar to the way Schwinger-Dyson equation is solved in the SYK model \cite{Sachdev_1993, Kitaev1, Kitaev2, Maldacena:2016hyu}. There in the low energy limit one neglects the bare UV part of the two-point function.

\bigskip

The next Schwinger-Dyson equation is derived from \eqref{SDder2} and takes the form 
\begin{align}
{1\over N} 
\left(
\delta_{ac} \la |R_{ab}|^2 \ra_\disk +\delta_{bd} \la |R_{cd}|^2 \ra_\disk
\right)
=g_{ad} \la R_{ab} R_{bc} R_{cd} R_{da} \ra_\disk +
\sum_{ef} g_{adef}
\la R_{ab} R_{bc} R_{cd} R_{de} R_{ef} R_{fa} \ra_\disk \ .
\end{align}
In the equation \eqref{SDder2} there is an extra term from differentiating $R_{bc}$ giving $\delta_{ab}\delta_{cd} \la R_{ab} R_{cd} \ra$. However this contians too many deltas and, after summing over all energy indices, this would give a non-planar contribution and is therefore suppressed at large $N$. The two-point coupling $g_{ab}$ can be excluded using the first Schwinger-Dyson equation \eqref{SD1reg}. After some algebra we obtain
\begin{align}
\sum_{ef} g_{adef} 
\la 
R_{ab} R_{bc} R_{cd}\left(\la |R_{ad}|^2 \ra_\disk - |R_{da}\ra \la R_{ad} |\right) R_{de} R_{ef} R_{fa} 
\ra_\disk
=
-{1\over N}
\la 
R_{ab} R_{bc} R_{cd} R_{da} 
\ra_{\disk,conn} \ .
\label{SD2reg}
\end{align}
In the LHS we have a 6-point function up to subtractions. The subtraction is the product of two 4-point functions. There are 15 chord diagrams contributing to the 6-point function. However, if two $R$'s in $R_{ab}R_{bc}R_{cd}$ are connected by a chord, then such a term is canceled by the second term in the LHS. Therefore the only remaining contributions in the LHS are chord diagrams where all three $R$'s in $R_{ab}R_{bc}R_{cd}$ are connected to one of the $R$'s in $R_{de} R_{ef} R_{fa}$. There are $3! = 6$ such chord diagrams
\begin{equation}
    \includegraphics[scale=0.12]{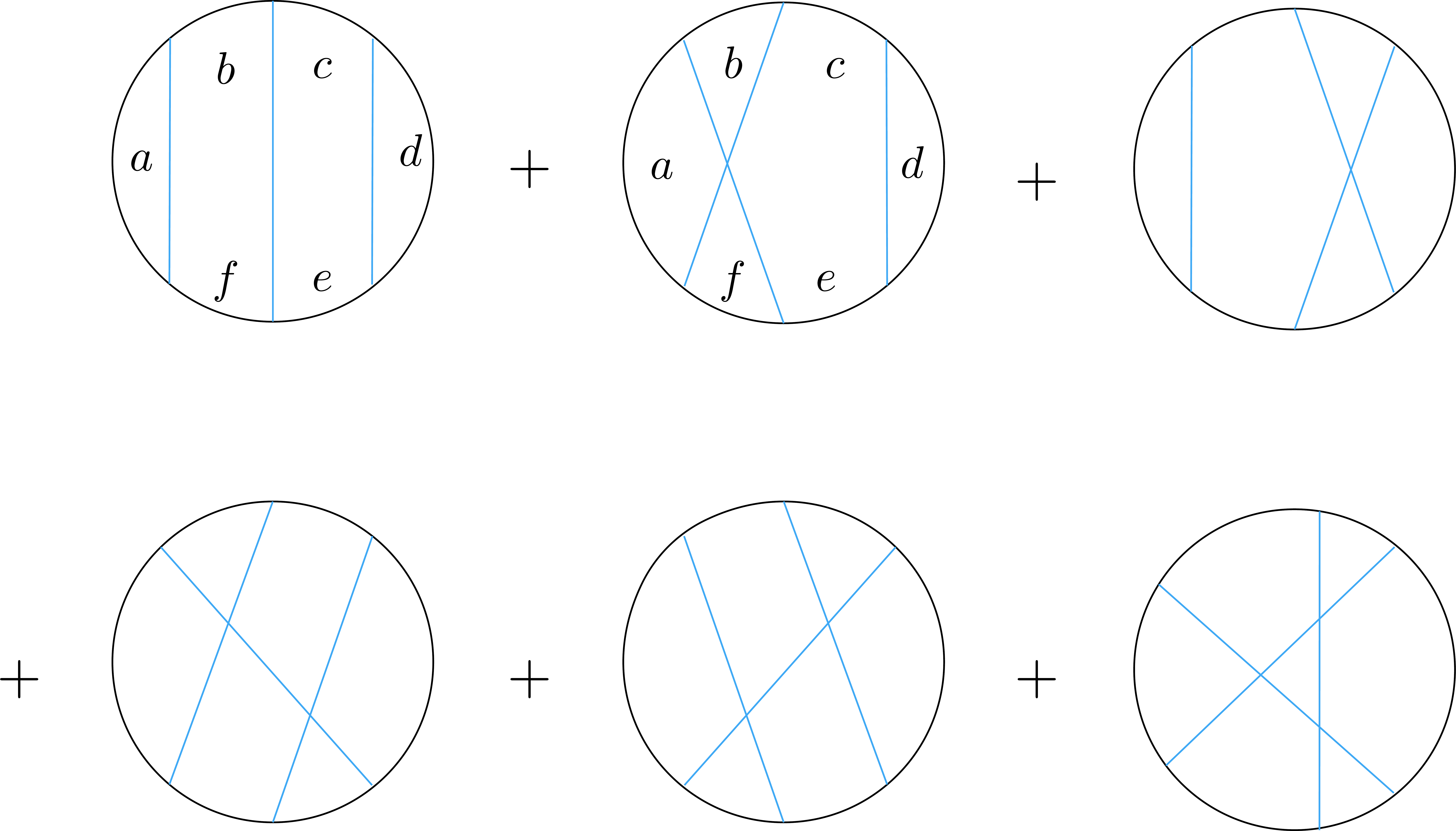} 
    \label{6ptSix}
\end{equation}
We need to insert this and the coupling \eqref{g4qdef} into the Schwinger-Dyson equation \eqref{SD2reg}. In the double-scaling limit we can drop the RHS in \eqref{SD2reg}. The computation is somewhat involved and is easier to do pictorially. We checked that in the double-scaling limit $q\to 1$ it is indeed satisfied. Crucially, one has to use the Yang-Baxter equation for the (q-deformed) 6j-symbol.

\bigskip

One can consider more general Schwinger-Dyson equations that involve higher-point correlators. They can be derived similarly to \eqref{SDder1}, \eqref{SDder2}.

\section{Matrix model potential}

\label{sec:corrections}

We now return to the discussion of the matrix model potential and argue that it is determined by the disk correlators. This is analogous to the statement that in a one-matrix model, the disk density of states determines the matrix potential and vice versa.

\bigskip

We will outline a systematic procedure for computing non-Gaussian corrections to the potential of $\calo$, such that the matrix model correctly computes disk correlators with an arbitrary number of $\calo$ insertions. In section \ref{sec:perttheory}, we explicitly compute the leading correction to the potential in a regulated model.

\bigskip

We write the matrix integral as follows:
\begin{equation}
	\label{eq:interactingmodel}
	\begin{split}
	\mathcal{Z} &= \int dH d\calo \, e^{-V(H,\calo)} 
	\\
	&= \int dH d\calo \, 
	\exp \left(- \sum_a \left[V_{SSS}(E_a) + V_{c.t.}(E_a)\right]  \right.
	\\
	&\quad \quad \left. - e^{S_0}  \sum_{a,b} \calo_{ab} \calo_{ba}  \frac{F_{ab}}{2} + e^{S_0} \sum_{a,b,c,d} \calo_{ab} \calo_{bc} \calo_{cd} \calo_{da} \frac{G^{(4)}_{abcd}}{4} + \cdots \right),
	\end{split}
\end{equation}
where as usual we have chosen to work in the eigenbasis of $H$. We define $F_{ab}$ and $G^{(2 n)}_{a_1 a_2 \cdots a_{2n}}$, $n \in \mathbb{Z}_{\ge 2}$, to be smooth real functions of the eigenvalues of $H$. That is, $F_{ab} = F(E_a,E_b)$ and $G^{(4)}_{abcd} = G^{(4)}(E_a,E_b,E_c,E_d)$. The $\cdots$ includes terms that are sixth and higher order in $\calo$ (we include even powers of $\calo$ only so that the model has a $\calo \rightarrow - \calo$ symmetry). The full set of coupling constants is specified by $V_{SSS}$ and $V_{c.t.}$ together with the functions $F_{ab}$ and $G^{( 2n)}_{a_1 a_2 \cdots a_{2n}}$ for $n \in \mathbb{Z}_{\ge 2}$. These functions are invariant under cyclic shifts of their indices (e.g. $G^{(4)}_{abcd} = G^{(4)}_{dabc}$) as well as reversals (e.g. $G^{(4)}_{abcd} = G^{(4)}_{dcba}$). As in the previous section, we are working directly in the double-scaled limit, so the number of eigenvalues is infinite. The factors of $e^{S_0}$ in the action ensure that `t Hooft diagrams have the correct factors of $e^{S_0}$ according to their topology. In Section \ref{sec:regmodeldef} we will carefully consider a regulated model where the number of eigenvalues is large but finite.

\bigskip

As in the toy model, the requirement that the matrix integral \eqref{eq:interactingmodel} correctly computes the correlator $\braket{\text{Tr } e^{- \beta H}}$ at the level of the disk determines the counter-term potential $V_{c.t.}$ in terms of the other coupling constants.  If we integrate out $\calo$, the resulting matrix potential for $H$ becomes a sum of multi-trace terms. In analogy to Figure \ref{fig:threedisks}, these multi-trace terms generate $\calo$ bubble diagrams that correct the disk partition function in the SSS model. We choose the single-trace counterterm potential to cancel all of these corrections.

\bigskip

Next, we consider the disk two-point function. Some of the `t Hooft diagrams that contribute are shown below:
\begin{align}
\raisebox{-.3in}{\includegraphics[scale=.13]{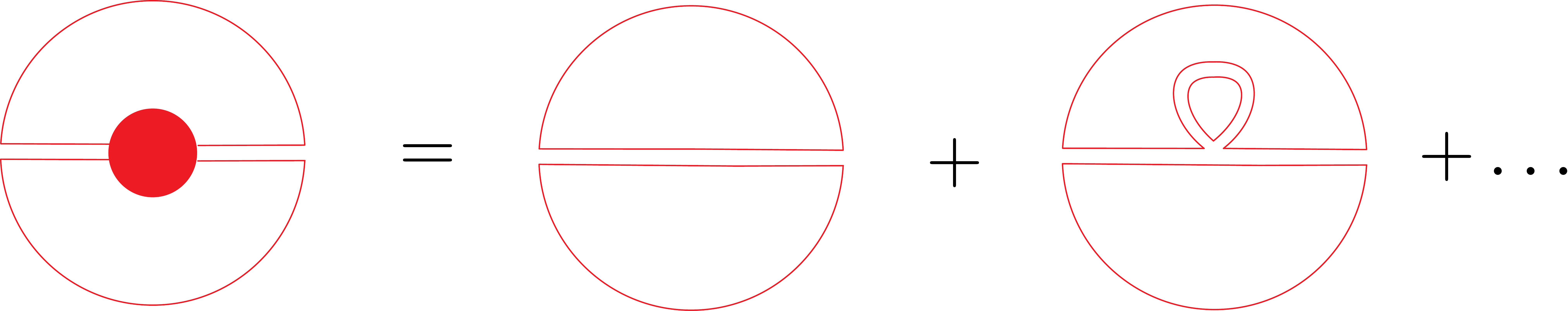}}
\label{eq:interactingtwopointthooft}
\end{align}
The red blob is defined to be the sum over all planar `t Hooft diagrams with two external double-lines. As in \eqref{eq:twopointcomparison} and \eqref{eq:twopointfunctiongaussian}, we have declined to draw any $\calo$ bubble diagrams or double-lines of the $H$ matrix. These are automatically accounted for by the rule that each single-line loop corresponds to an integral over the corresponding energy with the measure $dE ~ e^{S_0} \rho_0(E)$. Mathematically, \eqref{eq:interactingtwopointthooft} is represented by
\begin{equation}
	\braket{
	\text{Tr } e^{- \beta_1 H} \calo e^{-\beta_2 H} \calo}_{\text{disk}} =  \int ds_a ds_b ~ e^{S_0} \rho(s_a) ~ e^{S_0} \rho(s_b)~ e^{- \beta_1 s_a^2} e^{-\beta_2 s_b^2} \braket{\calo_{a b} \calo_{ba}}_{\text{disk}}, 
\end{equation}
where $\rho(s)$ was defined in \eqref{eq:rhos}\footnote{For comparsions with gravitational amplitudes, it is more convenient to work with the $s$ variables rather than the energies $E$. They are simply related by $E = s^2$.} and $\braket{\calo_{a b} \calo_{ba}}_{\text{disk}}$ is a smooth function of $s_a$ and $s_b$ that is interpreted as a microcanonical two-point function.\footnote{Note that $\braket{\calo_{a b} \calo_{b a}}_{\text{disk}}$ is the same object as (6.80) of \cite{Yang:2018gdb} up to factors of the density of states.} To be precise, we define
\begin{equation}
	\braket{\calo_{ab} \calo_{ba}}_{\text{disk}} := \frac{1}{\delta s_a e^{S_0} \rho(s_a) } \frac{1}{\delta s_b e^{S_0} \rho(s_b) }  \braket{\text{Tr } P(s_a) \calo P(s_b) \calo}_{\text{disk}},
	\label{eq:40}
\end{equation}
where $P(s_a)$ is a projection onto a microcanonical window centered around energy $E_a = s_a^2$ that has $\delta s_a e^{S_0} \rho(s_a)$ eigenvalues (to leading order in $e^{S_0}$). Taking inverse Laplace transforms after the large $N$ limit results in microcanonical-averaged correlators because the information about the fine-grained details of the spectrum is washed away at large $N$. We can only deduce the microcanonical-averaged correlators from the gravitational path integral. The requirement that the matrix model computes the correct two-point function implies that
\begin{equation}
	\braket{\calo_{ab} \calo_{ba}}_{\text{disk}} = e^{- S_0} \Gamma^{\Delta}_{ab}.
	\label{eq:9.5}
\end{equation}

\bigskip

We next consider the disk four-point function, which may be computed by summing `t Hooft diagrams with four external $\calo$ double-lines. These diagrams may be organized into connected and disconnected diagrams, as shown in \eqref{eq:fourpointfunctioninteractingmodel}, which we reproduce here:
\begin{equation}
    \includegraphics[scale=0.45]{figures/fourpointfunctioninteractingmodel.pdf}.
    \label{eq:9.7}
\end{equation}
On the right hand side we specify the gravitational four-point function that the `t Hooft diagrams on the left hand side must reproduce. We may write
\begin{equation}
	\begin{split}
		\braket{\calo_{a b} \calo_{b c} \calo_{c d} \calo_{d a}}_{\text{disk}} 
	    &= \braket{\calo_{a b} \calo_{b a}}_{\text{disk}} \braket{\calo_{b c} \calo_{c b}}_{\text{disk}} \frac{\delta(s_b - s_d)}{e^{S_0} \rho(s_b)} 
		\\
		&+
		\braket{\calo_{a b} \calo_{b a}}_{\text{disk}} \braket{\calo_{a d} \calo_{d a}}_{\text{disk}} \frac{\delta(s_a - s_c)}{e^{S_0} \rho(s_a)}
		\\
		&+ \braket{\calo_{a b} \calo_{b c} \calo_{c d} \calo_{d a} }_{\text{disk},c}.
	\end{split}
\end{equation}
A correlator with the subscript $\text{disk},c$ refers to a function of the energies that represents the sum over connected planar `t Hooft diagrams only.\footnote{Here, we mean diagrams where the external $\calo$ double-lines are connected through the bulk.} Because the two-point function has already been fixed in \eqref{eq:9.5}, the first two terms on the left side of \eqref{eq:9.7} reproduce the first two terms on the right hand side. It follows that
\begin{equation}
    \braket{\calo_{a b} \calo_{b c} \calo_{c d} \calo_{d a} }_{\text{disk},c} = e^{-3 S_0} (\Gamma^\Delta_{ab} \Gamma^\Delta_{bc} \Gamma^\Delta_{cd} \Gamma^\Delta_{da})^{1/2} \left\{
\begin{matrix}
\Delta & s_a & s_b \\
\Delta & s_c & s_d
\end{matrix}
\right\}.
\end{equation}

\bigskip

Continuing in a similar fashion, we can compare the matrix model `t Hooft diagrams with gravitational Feynman diagrams as above and derive expressions for the connected $2n$-point functions
$
	\braket{ \calo_{a_1a_2} \cdots \calo_{a_{2n} a_1}}_{\text{disk},c}.
$
It follows from the Feynman rules in section \ref{sec:JTgravitywithmatter} that the connected $2n$-point function is given by a sum over all connected gravitational $2n$-point Feynman diagrams (defined such that the bulk lines of the Feynman diagrams form a connected graph). 

\bigskip

\bigskip

Until now, we have assumed that for a suitable choice of the couplings, the matrix integral correctly computes all of the gravitational disk correlators. It follows that the sum over all connected planar $2n$-point `t Hooft diagrams is completely determined by the gravitational Feynman rules, as illustrated above for $n = 1,2$. We now show how this data can be used to systematically determine the couplings.\footnote{We will soon emphasize that there should be multiple ways to choose the couplings such that the disk correlators of the matrix model agree with the gravitational answers in the appropriate scaling limit. Here, we just specify one way.} First, we modify the gravitational Feynman rules such that each crossing of two blue lines comes with an additional factor of $\epsilon$, where $0 < \epsilon < 1$. We may determine the matrix potential order by order in $\epsilon$. In particular, we write
\begin{align}
\label{eq:vhexpansion}
V_{c.t.}(H) &= \sum_{m = 0}^\infty \epsilon^m V_{c.t.}^{(m)} \ , \\
\label{eq:fexpansion}
F &= \sum_{m = 0}^\infty \epsilon^m F^{(m)} \ , \\
\label{eq:gexpansion}
G^{(2n)} &= \sum_{m = 1}^\infty \epsilon^m G^{(2n),(m)} \ ,
\end{align}
where we have omitted the indices on $F_{ab}$ and $G^{(2n)}_{a_1 \cdots a_{2n}}$ for convenience. Note that $\braket{ \calo_{a_1a_2} \cdots \calo_{a_{2n} a_1}}_{\text{disk},c}$ is $O(\epsilon^{n - 1})$ in the $\epsilon$ expansion. That is, any connected gravitational $2n$-point Feynman diagram must have at least $n - 1$ crossings. When $\epsilon = 0$, the matrix potential must reduce to $\mathcal{Z}_{\text{toy}}$ in \eqref{eq:29}, which is why the sum in \eqref{eq:gexpansion} starts from $m = 1$.

\bigskip 

\begin{figure}
	\centering
	\includegraphics[scale=.1]{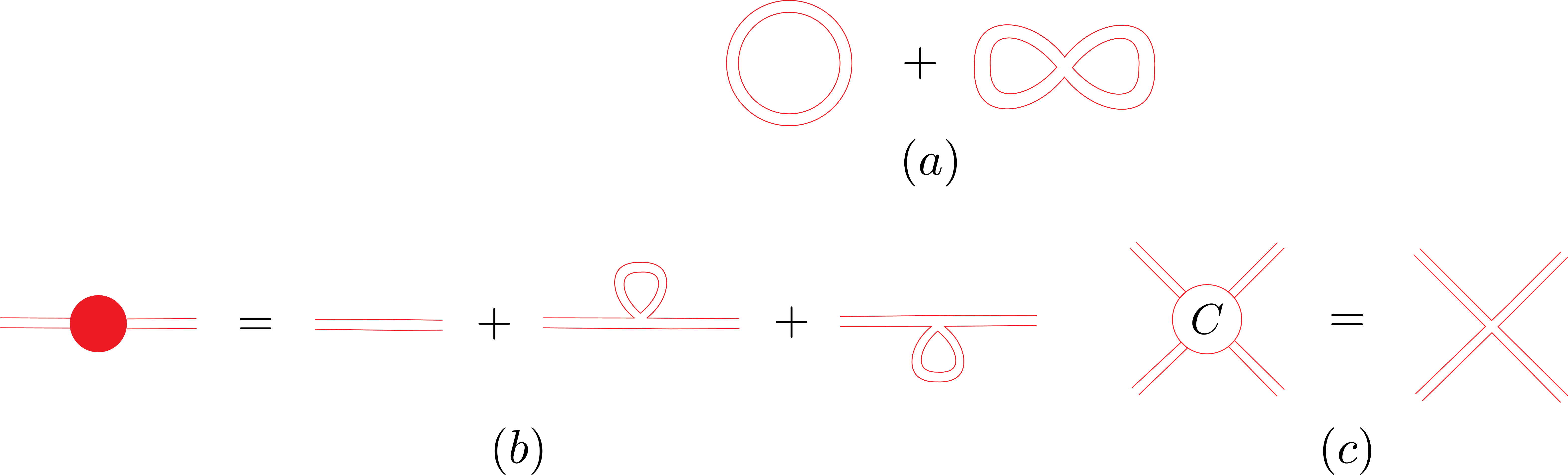}
	\caption{To compute the correct gravitational observables to order $\epsilon$, it suffices to let the quartic $\calo$ coupling be the only interaction to order $\epsilon$. 
	(a) $\calo$ bubble diagrams that contribute to the disk partition function at order $\epsilon$. As in Figure \ref{fig:threedisks}, these are cancelled by counterterms in $V_{c.t.}$. (b) The `t Hooft diagrams that contribute to the two-point function. The order $\epsilon$ correction to the quadratic $\calo$ term is chosen to ensure that these diagrams sum to the known gravitational answer for the two-point function. 
	(c) The value of the quartic $\calo$ coupling is  determined by matching the four-point tree diagram to the known gravitational answer for the connected four-point function.}
	\label{fig:orderepsilondiagrams}
\end{figure}

Working to first order in $\epsilon$, we set $G^{(2n),(1)} = 0$ for $n \ge 3$. Then, we choose $G^{(4),(1)}$ to ensure that the matrix model computes the correct connected four-point function, which is order $\epsilon$. The only contribution is from a tree diagram:
\begin{equation}
  \frac{\epsilon G^{(4),(1)}_{a_1 a_2 a_3 a_4}}{F_{a_1 a_2}^{(0)}F_{a_2 a_3}^{(0)}F_{a_3 a_4}^{(0)}F_{a_4 a_1}^{(0)}}    = \braket{ \calo_{a_1a_2} \calo_{a_2a_3} \calo_{a_3a_4}  \calo_{a_{4} a_1}}_{\text{disk},c}.
\end{equation}
After choosing $G^{(4),(1)}$, we note that the four-point vertex appears in loop diagrams that contribute to the two-point function at order $\epsilon$. We choose $F^{(1)}$ to cancel these contributions. Next, note that both the four-point vertex and the propagator contribute to $\calo$ bubble diagrams at order $\epsilon$ (and the result depends on $F^{(1)}$ and $G^{(4),(1)}$). As in Figure \ref{fig:threedisks}, we choose $V_{c.t.}^{(1)}$ to cancel the order $\epsilon$ contributions of $\calo$ bubble diagrams to the disk density of states. This defines the matrix potential to order $\epsilon$. See Figure \ref{fig:orderepsilondiagrams} for a list of the diagrams that appear in this calculation to order $\epsilon$.\footnote{Note that the loop integrations in these diagrams may diverge. In section \ref{sec:regtwomatrixmodel}, we will show that the gravitational Feynman rules can be further modified by a $q$-deformation such that the $q$-deformed density of states has compact support, which implies that the energies running in the loops are integrated over a finite range so that all the loop integrations converge. We will properly address this point in section \ref{sec:regtwomatrixmodel}. Until then, we ignore the issue of divergent loop integrals.}

\bigskip

We now outline the general procedure to determine the matrix potential to any order in $\epsilon$. The basic idea is that to any given order in $\epsilon$, there are a finite number of nonzero connected planar $2n$-point functions. Hence, we only need to adjust finitely many terms in \eqref{eq:gexpansion} to make the matrix model correctly compute the gravitational correlators to a given order in $\epsilon$. Thus, it is possible to systematically determine the matrix potential order by order in $\epsilon$.

\bigskip 

Let us explain the procedure in more detail. First, assume that the matrix potential has been chosen to order $\epsilon^{p-1}$ such that the matrix model correctly computes all of the connected correlators to order $\epsilon^{p-1}$. Assume also that the couplings obey $G^{(2 n),(m)} = 0$ for $n > 1+m$. As shown above, these assumptions are true for $p = 2$, which we take as the base case of an inductive argument. To determine the couplings at order $\epsilon^p$, we first choose $G^{(2n),(p)} = 0$ for $n > p+1$. Next, note that the only contribution from $G^{(2p + 2),(p)}$ to the connected $(2p + 2)$-point function at order $\epsilon^p$ is in a tree-level connected $(2p + 2)$-point diagram with a single $(2p + 2)$-point vertex:
\begin{equation}
  \frac{\epsilon^{p} G^{(2p+2),(p)}_{a_1 \cdots a_{2p + 2}}}{F_{a_1 a_2}^{(0)} \cdots F_{a_{2p + 2} a_1}^{(0)}}.
  \label{eq:contribution}
  \end{equation}
 We choose $G^{(2p + 2),(p)}$ such that the sum of \eqref{eq:contribution} and all of the other diagrams that contribute (which involve lower-point vertices only) yield the correct order $\epsilon^p$ result for $\braket{ \calo_{a_1a_2} \cdots  \calo_{a_{2p + 2} a_1}}_{\text{disk},c}$.  Next, we consider the connected $2p$-point function at order $\epsilon^p$. This cannot depend on $G^{(2n),(p)}$ for $n < p$ because any diagram with a $(2n)$-point vertex with $n < p$ must have at least one other vertex, and every vertex is at least order $\epsilon$. Hence, the only order $\epsilon^p$ terms in the matrix potential that contribute to the $(2p)$-point function at order $\epsilon^p$ are $G^{(2p),(p)}$ and $G^{(2p+2),(p)}$, and $G^{(2p),(p)}$ only contributes as part of a tree diagram. We choose $G^{(2p),(p)}$ such that the sum of this tree diagram and all the other diagrams yields the correct result for the $(2p)$-point function. We may continue to choose $G^{(2n),(p)}$ for successively lower values of $n$ together with $F^{(p)}$ to ensure that the matrix model correctly computes all of the connected $2n$-point functions at order $\epsilon^p$. Finally, we choose $V_{c.t.}^{(p)}$ to cancel the corrections from  $\calo$ bubble diagrams to the disk density of states. Thus, it is possible to choose the couplings to all orders in $\epsilon$ such that the matrix model agrees with the gravitational answers at disk level. After determining the matrix potential to all orders in $\epsilon$, we can take $\epsilon \rightarrow 1$ at the end of any calculation.

\bigskip

We expect that there are many ways to define a double-scaled matrix model whose disk correlators agree with those of the gravitational theory. While we have described above a specific procedure for determining a suitable set of couplings, there could be other ways to determine the couplings. For instance, in the previous paragraph we modified the gravitational Feynman rules by including an additional factor of $\epsilon$ whenever two blue lines cross. We will refer to the associated matrix model as the ``$q$-deformed matrix model'' for reasons that will become clear in the next section. Another way to modify the gravitational Feynman rules is to weight each connected $2n$-point function with a  factor of $\epsilon^{n - 1}$. The inductive argument outlined above can still be used to write the matrix potential in an $\epsilon$ expansion. The associated matrix model will be referred to as the ``Selberg matrix model.'' The $\epsilon \rightarrow 1$ limit corresponds to the double-scaling limit of the matrix model. We use the term ``regulated matrix model'' to refer to either of these models away from the double-scaling limit. The $q$-deformed and Selberg matrix models are double-scaled in different ways, but their disk correlators agree. However, as we will see later, their connected two-boundary (or double-trumpet) correlators disagree. Thus, it is important to distinguish between the two different double-scaling limits.

An important question for either the $q$-deformed or Selberg matrix models is whether the potential, written as a power series expansion in $\epsilon$, actually converges for $0 < \epsilon < 1$. We want the matrix potential to be defined for $\epsilon$ in this range, so that the $\epsilon \rightarrow 1$ limit defines a scaling limit of a well-defined matrix model. We do not have a rigorous proof that the sums in \eqref{eq:vhexpansion}, \eqref{eq:fexpansion} , and \eqref{eq:gexpansion} converge for $0 < \epsilon < 1$. However, in the next section, we will provide highly nontrivial evidence in favor of the conclusion that the sums converge for the $q$-deformed matrix model. This evidence follows from the relationship between the $q$-deformed matrix model and the double-scaled SYK model, which was studied in \cite{Berkooz:2018jqr}. We do not have an analogous argument for why the matrix potential converges in the Selberg matrix model. However, in the remainder of this paper, we will assume that the Selberg matrix model potential is well-defined for $0 < \epsilon < 1$ so that we can compute the double-trumpet in this model.

Even if the matrix potential is well-defined in the regulated models, these models may be non-perturbatively ill-defined. As an analogy, note that the potential $V(x) = x^2 - x^4$ has a local minimum at $x = 0$, which implies that the associated single-matrix model has a perturbatively stable saddle with a single-cut density of states centered on $x = 0$. However, the matrix integral itself is non-perturbatively ill-defined because the potential is unbounded below. We note here that the matrix model in \cite{Saad:2019lba} is non-perturbatively ill-defined (at least when the eigenvalue contour is $\mathbb{R}$), and the regulated matrix models we consider also appear to be non-perturbatively ill-defined. It would be interesting to find non-perturbatively well-defined models with the same genus expansion as the regulated models, in analogy to \cite{Johnson:2019eik,Johnson:2020exp,Johnson:2020heh,Johnson:2020mwi,Johnson:2021owr,Johnson:2021tnl,Johnson:2021zuo,Post:2022dfi,Altland:2022xqx} but for two-matrix models.

\section{The regulated two-matrix models}
\label{sec:regtwomatrixmodel}

In the previous section we showed that it is possible to systematically determine the coupling constants of single-trace, two-matrix models that compute the disk correlators of JT gravity minimally coupled to a scalar field. When computing `t Hooft diagrams in these models, one encounters divergent loop integrals due to the noncompact support of the disk density of states $\rho(s)$. In order for the correlators to be finite in the double-scaling limit, the coupling constants cannot be well-defined in the double-scaling limit. This is simply because counterterms are needed to cancel the loop divergences.

We introduced two different regulated matrix models whose double-scaling limits reproduce the gravitational disk correlators. We referred to these models as the $q$-deformed model and the Selberg model. However, we did not explain how the divergent loop integrals should be regulated in these models. In this section, we begin by carefully defining the $q$-deformed matrix model. We will explain how the construction of the Selberg model differs. Then, we present a nontrivial calculation in support of the conclusion that the matrix potential of the $q$-deformed model is well-defined.

The reader who is mainly interested in the `t Hooft diagram computations in section \ref{sec:doubletrumpet} can skip most of this section without loss of continuity. To understand the results in section \ref{sec:doubletrumpet} (aside from part of section \ref{sec:6.1.1} and section \ref{sec:hagedorn}), the only important point from this section is that the special functions appearing in the gravitational Feynman rules may be deformed by a parameter $q \in [0,1]$ (where $q = 1$ corresponds to the original, undeformed Feynman rules). We use these $q$-deformed special functions to regulate the aforementioned loop divergences. These special functions are defined in appendix \ref{sec:specialfunctions}.

\subsection{Summary of chord diagram combinatorics}

\label{sec:disk q-def}

\begin{figure}
	\centering
	\includegraphics[scale =.4]{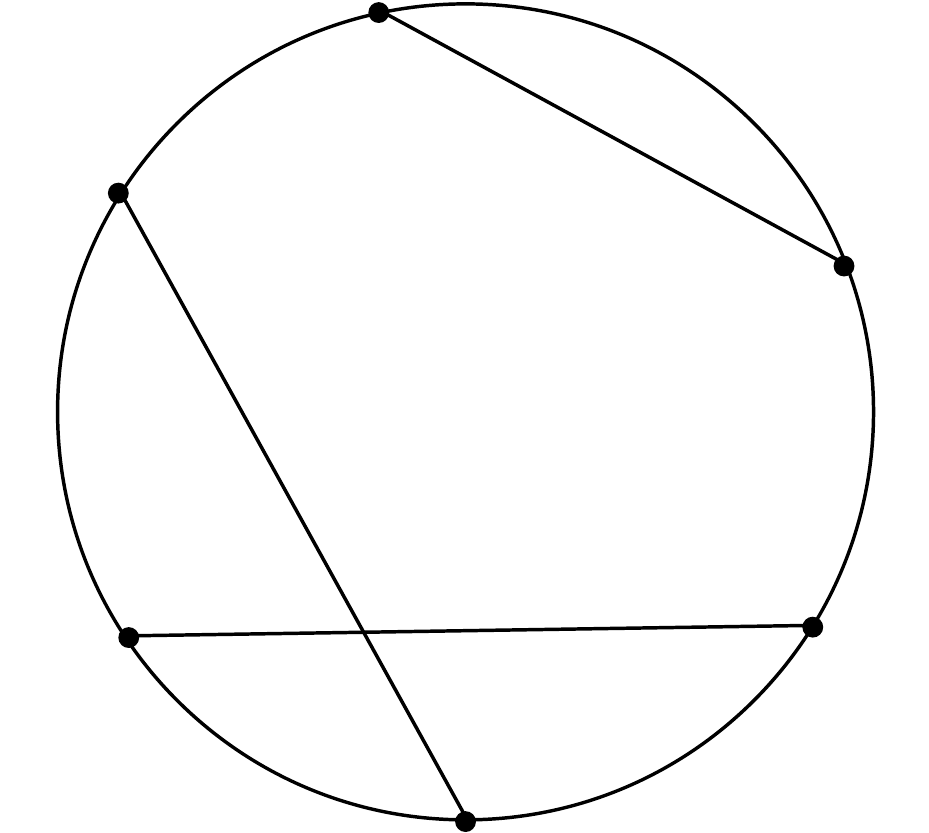}
	\caption{An example of a 6-point chord diagram. Pairs of points are connected with chords. This diagram is assigned the value $q$ because there is one crossing.}
	\label{fig:chorddiagramexample}
\end{figure}

The $q$-deformed model has close connections with the results of \cite{Berkooz:2018jqr}, which we review in this subsection. The authors of \cite{Berkooz:2018jqr} studied the SYK model in the double-scaled limit. If we write the SYK Hamiltonian as
\begin{equation}
	H_{\text{SYK}} = i^{p/2} \sum_{i \leq i_1 < \cdots < i_p \leq N} J_{i_1 i_2 \cdots i_p} \psi_{i_1}\cdots \psi_{i_p},
\end{equation}
then the double-scaled limit is a large $N$ limit where
\begin{equation}
	\lambda \equiv \frac{2 p^2}{N}
\end{equation}
is held fixed. If we define
\begin{equation}
	q = e^{- \lambda},
\end{equation}
then \cite{Berkooz:2018jqr} showed that the moments of $H_{\text{SYK}}$ only depend on $q$ and may be computed by summing chord diagrams. A $2n$-point chord diagram is defined to be a circle with $2n$ labeled points on the circumference and $n$ chords. Each chord connects two points and each point is attached to exactly one chord. An example of a chord diagram is given in Figure \ref{fig:chorddiagramexample}. Each chord diagram is assigned a value of $q^n$, where $n$ is the number of involuntary chord crossings (we always assume that $q \in [0,1)$). The authors of \cite{Berkooz:2018jqr} computed an elegant formula for the sum over all $2n$-point chord diagrams:
\begin{equation}
\text{sum over all $2n$-point chord diagrams} = 	\int_0^\pi \frac{d\theta}{ 2 \pi} (q,e^{\pm 2 i \theta};q)_\infty \left(\frac{2 \cos \theta}{\sqrt{1-q}} \right)^{2n}.
\label{eq:1.58}
\end{equation}
Note that if one replaces $2n$ in \eqref{eq:1.58} with an odd number, then \eqref{eq:1.58} vanishes.

\bigskip

\begin{figure}
	\centering
	\includegraphics[scale=.4]{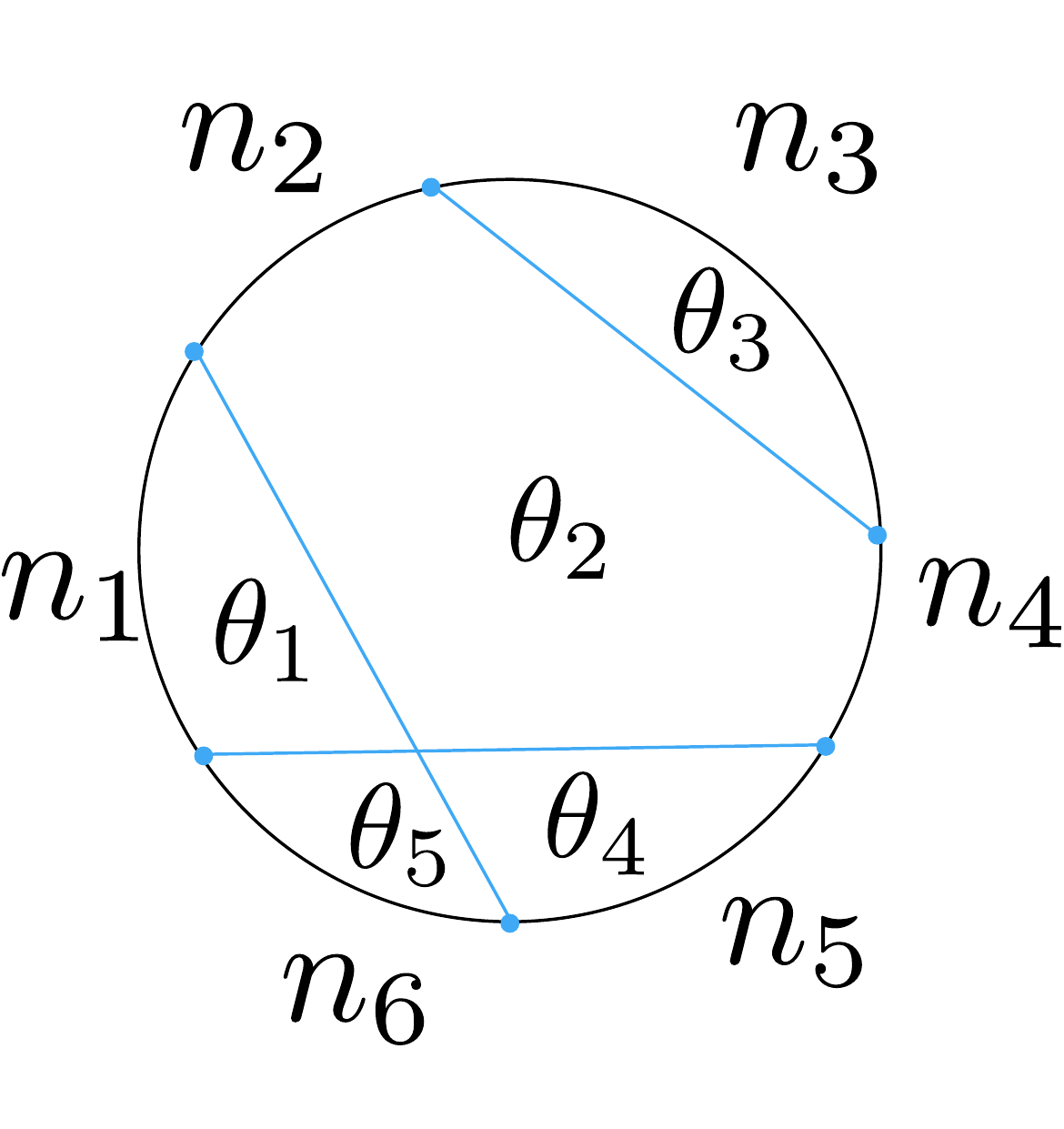}
	\caption{The B-type points of this two-species chord diagram are labeled in blue. The labels $n_1, n_2,\ldots$ indicate how many A-type points (not shown) are between two B-type points. The sum over all ways of drawing the A-type chords is equal to a finite-dimensional integral which is determined by a set of Feynman rules. The Feynman rules require that each disk-shaped region is labeled with a $\theta$ parameter.}
	\label{fig:twospecieschorddiagram}
\end{figure}

The authors of \cite{Berkooz:2018jqr} also considered chord diagrams with two chord species. A two-species chord diagram is defined to be a circle with some number of labeled points on the circumference, and each point is either of type A or type B. All of the points are attached to a chord, and each chord connects exactly two points of the same type. If there are an odd number of either type A or type B points, then it is not possible to pair up all of the points and the diagram is assigned a value of zero. Otherwise, the diagram is assigned a value of $q_A^{n_1} q_B^{n_2} {\wt q}^{n_3}$, where $n_1$ is the number of involuntary crossings of A-type chords, $n_2$ is the number of involuntary crossings of B-type chords, and $n_3$ is the number of involuntary crossings between an A-type and B-type chords. Another result of \cite{Berkooz:2018jqr} is that the sum over all two-species chord diagrams with a fixed configuration of points on the circle and a fixed configuration of B-type chords may be computed using a set of Feynman rules that mirror the rules described in section \ref{sec:disk}.\footnote{Technically, \cite{Berkooz:2018jqr} only proved this result for diagrams with up to one intersection of $B$-type chords. For general two-species chord diagrams, this result is a well-motivated conjecture.} Figure \ref{fig:twospecieschorddiagram} represents a sum over such chord diagrams.

We will often set $q_A = q, \wt q = q^{\Delta}$. The chord-diagram Feynman rules are as follows (special functions are explained in Appendix \ref{sec:specialfunctions}):

\begin{itemize}

\item For each boundary segment labeled by $n$, we include a factor of \begin{equation}
	\label{eq:59}
		\left(\frac{2 \cos \theta}{\sqrt{1 - q}}\right)^n,
	\end{equation}
where $\theta$ labels the disk-shaped region that is adjacent to the boundary segment.
\item For each B-type point (given by a blue dot in Figure \ref{fig:twospecieschorddiagram}), we include a factor of 
\begin{align} 
	\left( \frac{(q^{2\Delta};q)_\infty}{(q^{\Delta} e^{i(\pm \theta_1 \pm \theta_2)};q)_\infty} \right)^{1/2}
&= (1-q)^\Delta \left( {|\log q| \over 1-q} N_q \right)^{1/2} ~ 
\left( {\Gamma_q(\Delta \pm i s_1 \pm i s_2) \over \Gamma_q(2\Delta)} \right)^{1/2} \\
&\equiv (1-q)^\Delta \left( {|\log q| \over 1-q} N_q \right)^{1/2} ~ 
(\Gamma_{12,q}^\Delta)^{1/2}
	\ ,
	\label{eq:1.60}
\end{align}
where $\theta_j = s_j |\log q|$ are associated to the two disk-shaped regions adjacent to the blue dot and we introduced for later convenience
\begin{align}\label{eq:Nq}
N_q = {1\over |\log q| (1-q)^2 (q;q)_\infty^3 } \ .
\end{align}
We expressed the RHS of \eqref{eq:1.60} in a way that will be convenient for taking the limit $q \to 1$ below.

\item For each involuntary crossing of two B-type chords, we include a factor ($q_A \equiv q$)
\begin{equation}
	\begin{split}
	S_{\tilde{q}_1,\tilde{q}_2}^{q_A,q_B}(e^{i \theta_1},e^{i \theta_2},e^{i \theta_3},e^{i \theta_4}) \equiv	
	q_B   N_q  \left\{\begin{array}{ccc}
		\Delta_1 & s_1 & s_2 \\
		\Delta_2 & s_3 & s_4
	\end{array}\right\}_q \ , \qquad \theta_j =  s_j |\log q| \ ,
\end{split}
\label{eq:1.61}
\end{equation}
where $\wt q_1 = q^{\Delta_1}, \wt q_2 = q^{\Delta_2}$ and we need to set $\tilde{q}_1 = \tilde{q}_2 = \tilde{q} = q^{\Delta}$. Equation \eqref{eq:1.61} for $\tilde{q}_1 \neq \tilde{q}_2$ is needed if one wants to introduce yet another chord species. The parameters $\theta_1,\dots,\theta_4$ represent the four disk-shaped regions that surround the crossing.

\item After including all of the appropriate factors as specified above, we integrate over each $\theta \in (0,\pi)$ with the measure
\begin{align}
 \frac{d\theta}{2 \pi} (q;q)_\infty (e^{\pm 2 i \theta};q)_\infty 
 &=
 {1\over N_q} ~ {ds \over 2\pi \Gamma_q(\pm 2i s)} \\
 & \equiv {1\over N_q} ~ ds ~ \rho_q (s)
 \ ,
\label{eq:1.62}
\end{align}
where as usual $\theta = s |\log q|$ and $s \in (0, \pi/|\log q|)$. This measure is normalized 
\begin{align}
\int_0^\pi  \frac{d\theta}{2 \pi} (q;q)_\infty (e^{\pm 2 i \theta};q)_\infty =1 \ .
\end{align}

\end{itemize}

\bigskip

\noindent For example, the two-point correlator of $B$'s described by these rules is 
\begin{align}
&\la \tr B e^{-\beta_1 A} B e^{-\beta_2 A} \ra_{\disk, q} \\
=& 
(1-q)^{2\Delta}  {|\log q| \over 1-q} ~{1\over N_q} \int_0^{\pi\over |\log q|} \prod_{j=1}^2 \left[ ds_j ~\rho_q(s_j) \exp\left(-\beta_j  {2\cos \theta_j \over \sqrt{1-q}} \right) \right] ~ 
 {\Gamma_q(\Delta \pm i s_1 \pm i s_2) \over \Gamma_q(2\Delta)} \ .
\end{align}

\subsection{Scaling limit}
\label{sec:scalinglimit}

We now show that in the limit $q \to 1$ the $q$-deformed correlators reduce to JT correlators described in section \ref{sec:disk}. The density of states $\rho_q(s)$ and the 6j-symbol \eqref{eq:1.61} in the limit $q\to 1$ simply reduce to the corresponding JT values
\begin{align}
\rho_q(s) \rightarrow \rho(s) \ , \qquad  
\left\{\begin{array}{ccc}
		\Delta_1 & s_1 & s_2 \\
		\Delta_2 & s_3 & s_4
	\end{array}\right\}_q  
\rightarrow  
\left\{\begin{array}{ccc}
		\Delta_1 & s_1 & s_2 \\
		\Delta_2 & s_3 & s_4
	\end{array}\right\} \ , \qquad (q \to 1) \ .
\end{align}
The counting of the factors $N_q$ is as follows. Consider a chord diagram with $2n$ operators $B$. One can show that factors of $N_q$ from all crossings \eqref{eq:1.61} and densities of states \eqref{eq:1.62} combine to give $N_q^{-(n+1)}$. Further, each operator $B$ contributes $N_q^{1/2}$ \eqref{eq:1.60}, giving in total $N_q^n$ from $2n$ operators. Combining these factors we have for each chord diagram $N_q^{-(n+1)} N_q^n = {1\over N_q}$. To absorb the remaining factors in \eqref{eq:1.60} we define an operator $\calo$
\begin{align}
B =  (1-q)^\Delta \calo \ .
\end{align}
The operator $\calo$ will reduce to the same $\calo$ considered in JT gravity in section \eqref{sec:disk}.

Now let's discuss the spectrum of $A$. Near the right edge in the limit $q\to 1$ it is 
\begin{align}
{2\cos(\theta) \over \sqrt{1-q} } &= {2\cos(s |\log q|) \over \sqrt{1-q} } \\
& \approx {2\over \sqrt{1-q}} \left( 1 - {(1-q)^2\over 2} s^2 + \dots \right) \ , \qquad (q\to 1) \ .
\end{align}
To zoom into the right edge of the spectrum we define an operator $H$
\begin{align}\label{eq:HA}
A = {2\over \sqrt{1-q}} \left( 1 - {(1-q)^2\over 2} H\right) 
\end{align}
such that in the limit $q\to 1$ the spectrum of $H$ is $s^2$ with $s \in (0,\infty)$. This is the energy in JT limit. Combining everything together we find
\begin{align}
\lim_{q \to 1 \atop q_B \to 1} N \left\la \tr  \calo e^{-\beta_1 H} \dots \calo e^{-\beta_{2n}H}  \right\ra_{\disk,q} = \left\la \tr  \calo e^{-\beta_1 H} \dots \calo e^{-\beta_{2n}H}  \right\ra_\disk \ ,
\end{align}
where we also defined $N = e^{S_0} N_q$.\footnote{Not to be confused with $N$ in SYK.} Here, the RHS is defined by the Feynman rules in JT gravity from section \eqref{sec:disk}. Later, in the matrix model $N$ will be the size of matrices.

 The parameter $q_B$ is equivalent to $\epsilon$ in section \ref{sec:corrections}. The parameter $q$ was not introduced in section \ref{sec:corrections} but is needed to carefully define the matrix model below. In the remainder of this paper, we refer to the limit $q\to 1, q_B \to 1$ as the ``JT limit'' of the $q$-deformed model.

\subsection{Two two-matrix models regulating JT correlators}

\label{sec:regmodeldef}

As we explained earlier, we expect JT gravity coupled to a free scalar to be dual to a double-scaled two-matrix model. To properly define the matrix model, we would like to describe it away from the double-scaled limit. We will consider two different ways to move away from the double-scaled limit: ``$q$-deformed matrix model'' and ``Selberg matrix model''.\footnote{The name ``Selberg matrix model'' is due to the fact that this model will reproduce the contribution of the 1-loop determinant on the double-trumpet, which in turn is computed by Selberg trace formula.} The reason for considering two different regulators is the following. In the double-scaling limit both models will agree with the disk JT correlators. However, on the double-trumpet in the double-scaling limit they give rise to different results. The Selberg matrix model will agree with JT correlators on the double-trumpet, while the $q$-deformed matrix model will lead to a different answer. 

We define the regulated two-matrix models by first specifying their regulated disk correlation functions. The matrix potential is then obtained by applying the algorithm in section \ref{sec:corrections} which produces a potential that computes a given set of disk correlators.

\subsubsection*{$q$-deformed matrix model}

We define the ``$q$-deformed matrix model'' such that its disk correlation functions are those of the double-scaled SYK model, which are described by the Feynman rules in section \ref{sec:disk q-def}. It depends on the parameters $q_A \equiv q$, $q_B$, and $\tilde{q} \equiv q^\Delta$. The potential may be found using the algorithm of section \ref{sec:corrections}.

In particular, the $q$-deformed model is a large $N$, `t Hooft-scaled two-matrix model with a single-trace potential, and we refer to the two Hermitian matrices as $A$ and $B$. A single-trace, planar correlator of an arbitrary string of matrices, such as
\begin{equation}
    \lim_{N \rightarrow \infty} \frac{1}{N}\braket{\text{Tr } AABBABABA},
    \label{eq:exampleobservable}
\end{equation}
is computed by placing the same string of A- and B-type points on the circle of a two-species chord diagram and summing over all A- and B-type chord configurations.

\subsubsection*{Selberg matrix model}

The Selberg regulator differs from the $q$-deformed regulator in two ways. First, we set $q_B = 1$. Second, we introduce a weight $\e^{n-1}$ for each connected (in energy basis\footnote{That is before integrating over parameters $s_j$.}) $2n$-point correlator. In other words, the blue chords of a Feynman diagram form a graph. For each connected subgraph with $2n$ external points, we include an additional factor of $\epsilon^{n - 1}$. For example, the connected 4-point function using the Selberg regulator has a factor $\e$.

\bigskip 

The $q$-deformed and Selberg regulators return the same results for the two and four-point functions if we identify $q_B$ with $\e$. But the $2n$-point correlators with $n \geq 3$ are different between the two regulators. The $q$-deformed regulator gives weights $q_B$ according to the number of crossings in the Feynman diagram, while the Selberg model gives weights $\e$ according to the number of external operators in each connected subgraph.

\bigskip 

 With these regulated gravitational Feynman rules, we may use the systematic procedure described toward the end of section \ref{sec:corrections} to construct a matrix potential to all orders in $\epsilon$. The result accounts for \eqref{eq:schematicmodelselberg} and \eqref{eq:lastterm}. The double-scaling limit of the Selberg model corresponds to taking $q \rightarrow 1$ and then $\epsilon \rightarrow 1$ (the order is important, as we will see in section \ref{sec:doubletrumpet}). The matrix potential of the Selberg model also includes a double-trace term \eqref{eq:addedterm} that is added by hand; as explained around \eqref{eq:addedterm}, its presence is required for the Selberg model to compute the correct empty double-trumpet.
 
 In the remainder of this section, we focus only on the $q$-deformed model. We will revisit the Selberg model when we discuss the double-trumpet.

\subsection{A single-matrix model warmup}
\label{sec:singlematrixwarmup}

Before we explain why the $q$-deformed two-matrix model should exist, we will first consider the easier problem of finding a one-matrix model whose single-trace, planar correlators are computed by sums over single-species chord diagrams. This is possible because the density of states \eqref{eq:1.62} is supported on a finite interval $s \in (0, {\pi \over |\log q|})$, just like in a single-cut matrix model.

We let $M$ refer to a $N \times N$ Hermitian matrix, and the matrix integral is given by
\begin{equation}
	\mathcal{Z}_q = \int dM \,  e^{-N \text{Tr } V_q(M)}.
\end{equation}
Our goal is to choose $V_q(M)$ such that the correlators of $M$ obey
\begin{equation}
\lim_{N \rightarrow \infty} \frac{1}{N} \braket{\text{Tr } M^k} = \int_0^\pi \frac{d\theta}{ 2 \pi} (q,e^{\pm 2 i \theta};q)_\infty \left(\frac{2 \cos \theta}{\sqrt{1-q}} \right)^k, \quad k \in \mathbb{Z}_{\ge 0}. \label{eq:1.73}
\end{equation}
Let $\rho_{q,0}(E)$ be the normalized tree-level eigenvalue distribution of $M$, which obeys the saddle-point equation \cite{Saad:2019lba}
\begin{equation} 
	\int_{a_-}^{a_+} d\lambda \, \frac{\rho_{q,0}(\lambda)}{E - \lambda} = \frac{V_q^\prime(E)}{2}, \quad E \in (a_-,a_+),
	\label{eq:saddlepoint}
\end{equation}
where the integral is a principal value integral, and the endpoints of the single-cut eigenvalue spectrum are
\begin{equation}
	a_\pm = \pm \frac{2}{\sqrt{1 - q}}. \label{eq:apm}
\end{equation}
Next, we define the rescaled quantities $\tilde{\rho}_{q,0}(x)$ and $\tilde{V}_q(x)$ such that
\begin{equation}
	d \lambda \, \rho_{q,0}(\lambda) = dx \, \tilde{\rho}_{q,0}(x), \quad \tilde{V}_q(x) = V_q(\lambda), \quad \quad \lambda = \frac{2 x}{\sqrt{1-q}}. 
\label{eq:rescaledef}
\end{equation}
Then, \eqref{eq:saddlepoint} becomes
\begin{equation} 
	\int_{-1}^{1} dx \, \frac{\tilde{\rho}_{q,0}(x)}{y-x} = \frac{\tilde{V}_q^\prime(y)}{2}.
	\label{eq:rescaledsaddle}
\end{equation}
Next, note that
\begin{equation}
	(q,e^{\pm 2 i \theta};q)_\infty = \sum_{n \in \mathbb{Z}} (-1)^n q^{\frac{n^2}{2}}\left[q^{-\frac{n}{2}} + q^{\frac{n}{2}}\right] T_{2n}(\cos \theta),
\end{equation}
where $T_n(\cos \theta) \equiv \cos n \theta$ is a Chebyshev polynomial of the first kind. Then, \eqref{eq:1.73} implies that
\begin{equation}
\label{eq:5.26}
	\tilde{\rho}_{q,0}(x) = \frac{1}{ 2 \pi} \frac{1}{\sqrt{1 - x^2}} \left[\sum_{n \in \mathbb{Z}} (-1)^n q^{\frac{n^2}{2}}\left[q^{-\frac{n}{2}} + q^{\frac{n}{2}}\right] T_{2n}(x)\right] .
\end{equation}
We will need to use the mathematical fact that
\begin{equation}
	\int_{-1}^{1} \frac{dx}{\sqrt{1-x^2}} \, \frac{T_{2n}(x)}{y-x} = - \pi U_{2 n - 1}(y), \quad n \ge 0, \quad y \in (-1,1),
\end{equation}
where $U_n(\cos \theta) \equiv \frac{\sin((n + 1)\theta)}{\sin \theta}$ is the Chebyshev polynomial of the second kind and the integral is a principal value integral. We can then use \eqref{eq:rescaledsaddle} to directly compute $\tilde{V}_q^\prime(y)$, and the result is
\begin{equation}
	\frac{\tilde{V}_q^\prime(y)}{2} = -  \sum_{n = 1}^\infty (-1)^n q^{\frac{n^2}{2}}\left[q^{-\frac{n}{2}} + q^{\frac{n}{2}}\right]  U_{2n-1}(y).
\end{equation}
Integrating the above with respect to $y$, we obtain
\begin{equation}
	\label{eq:vtilde}
	\tilde{V}_q(y) = -  \sum_{n = 1}^\infty (-1)^n q^{\frac{n^2}{2}}\frac{\left[q^{-\frac{n}{2}} + q^{\frac{n}{2}}\right]}{ n}  T_{2n}(y).
\end{equation}
 The expression for $V_q(\lambda)$ follows from \eqref{eq:vtilde} and \eqref{eq:rescaledef}:
 \begin{equation}
 V_q\left( \lambda\right) = -  \sum_{n = 1}^\infty (-1)^n q^{\frac{n^2}{2}}\frac{\left[q^{-\frac{n}{2}} + q^{\frac{n}{2}}\right]}{ n}  T_{2n}(\frac{\sqrt{1-q}}{2} \lambda). \label{eq:vq}
 \end{equation}
  We will refer to \eqref{eq:vq} in the next subsection. This single-matrix model is a regulated version of the SSS model \cite{Saad:2019lba}.
  
  \label{sec:5.4}
 
\subsection{Solvable limits of the $q$-deformed model}

\label{sec:5.5}
 
In section \ref{sec:scalinglimit}, we showed that in a certain limit, the $q$-deformed two-matrix model computes the disk correlators of JT gravity minimally coupled to a scalar field. In this subsection, we consider other limits of the $q$-deformed model, in which the model is solvable. Although we cannot write a closed-form expression for the entire matrix potential of the model, we can write the potential in these solvable limits. In the next subsection, we provide nontrivial evidence that the model also exists away from these solvable limits.

We remind the reader that the circle of a two-species chord diagram represents a trace in the matrix model, and each A- or B-type point on the circle corresponds to an insertion of the matrix $A$ or $B$ into the trace. In the $q$-deformed two-matrix model, the expectation value of this trace at planar order (normalized by $N$) is given by a sum over all the chord configurations that connect the boundary points. Two intersecting A-type chords come with a factor of $q_A$, two intersecting B-type chords come with a factor of $q_B$, and an intersection of an A-type and B-type chord comes with a factor of $\tilde{q}$. We let $\mathcal{Z}_{q_A,q_B,\tilde{q}}$ denote the matrix integral of the regulated two-matrix model.

The first solvable limit that we consider is $\tilde{q} = 0$. In this limit, an A-type chord cannot intersect a B-type chord. In the matrix model, this means that every connected planar `t Hooft diagram contains either $A$ double-lines or $B$ double-lines, but not both. Thus, there can be no interaction terms between $A$ and $B$ in the potential. The matrix integral must take the form
\begin{equation}
	\mathcal{Z}_{q_A,q_B,\tilde{q} = 0} = \int dA \, dB  \, e^{-N \left(\text{Tr } V_{q_A} (A) + V_{q_B} (B) \right) },
	\label{eq:1.83}
\end{equation}
where $V_{q}$ was defined in \eqref{eq:vq}.

The next solvable limit of interest is $q_B = 0$. In this case, two B-type chords may not intersect. Given that a B-type chord may be interpreted as a bulk line in the JT limit and that the toy matrix model $\mathcal{Z}_{\text{toy}}$ does not allow two bulk lines to cross, it is clear that the $q$-deformed model with $q_B = 0$ is an appropriate deformation of $\mathcal{Z}_{\text{toy}}$. In particular, the potential of the $q$-deformed model at $q_B = 0$ must be quadratic in $B$. The matrix integral for $q_B = 0$ is
\begin{equation}
	\mathcal{Z}_{q_A,q_B = 0,\tilde{q}} = \int d A d B \, \exp\left(- N \text{Tr } V_{q_A}(A) - N \left[\sum_{a,b} B_{ab} B_{ba} \frac{F^{q_A,\tilde{q}}_{ab}}{2}\right] - N \text{Tr } V^{q_A,\tilde{q}}_{c.t.}(A) \right),
	\label{eq:1.85}
\end{equation}
where $V_{q_A}$ is defined in \eqref{eq:vq}, and
\begin{equation}
	F^{q,\tilde{q}}_{ab} \equiv F^{q,\tilde{q}}(\lambda_a,\lambda_b) \equiv \tilde{F}^{q,\tilde{q}}(\frac{\sqrt{1-q}}{2} \lambda_a,\frac{\sqrt{1-q}}{2} \lambda_b), \quad \tilde{F}^{q,\tilde{q}}(\cos \theta_1,\cos \theta_2) \equiv   \frac{(\tilde{q} e^{i (\pm \theta_1 \pm \theta_2)};q)_\infty}{(\tilde{q}^2;q)_\infty},
	\label{eq:1.86}
\end{equation}
where $\lambda_a$ refers to an eigenvalue of $A$, and
\begin{equation}
	V_{c.t.}^{q,\tilde{q}}(E) = - \int_{a_-}^{a_+} d\lambda \,  \rho_{q,0}(\lambda) \log F^{q,\tilde{q}}(\lambda,E),
	\label{eq:1.87}
\end{equation}
where $a_{\pm}$ is defined in \eqref{eq:apm}. The smooth function $F^{q,\tilde{q}}(\lambda_a,\lambda_b)$ is chosen so that the two-point function of $B$ agrees with the chord-diagram Feynman rule \eqref{eq:1.60}. As in section \ref{sec:toymodel}, the counterterm potential is chosen so that the tree-level eigenvalue distribution of $A$ is $\rho_{q,0}$. To evaluate \eqref{eq:1.87}, note that
\begin{equation}
	-   \log (\tilde{q} e^{\pm i \theta} ; q)_\infty = 2 \sum_{n = 1}^\infty \frac{ \cos n \theta}{(1 - q^n)n} \tilde{q}^n
\end{equation}
and thus
\begin{equation}
	-   \log (\tilde{q} e^{\pm i \theta_1 \pm i \theta_2} ; q)_\infty = 4 \sum_{n = 1}^\infty \frac{ \cos n \theta_1 \, \cos n \theta_2}{(1 - q^n)n} \tilde{q}^n.
\label{eq:89}
\end{equation}
It then follows from \eqref{eq:1.87} that up to an unimportant additive constant,
\begin{align}
	V_{c.t.}^{q,\tilde{q}}(E) &= \tilde{V}_{c.t.}^{q,\tilde{q}}(\frac{\sqrt{1-q}}{2} E)    
	\label{eq:5.37}
	\\
		\tilde{V}_{c.t.}^{q,\tilde{q}}(x) &=       \sum_{n = 1}^\infty (-1)^n q^{\frac{n^2}{2}} \frac{q^{-\frac{n}{2}} + q^{\frac{n}{2}}}{(1 - q^{2n})n}   T_{2n}(x) \tilde{q}^{2n}.
\end{align}
And thus,
\begin{equation}
	V_q(E) + V^{q,\tilde{q}}_{c.t.}(E) =  \sum_{n = 1}^\infty (-1)^n q^{\frac{n^2}{2}}  T_{2n}(\frac{\sqrt{1 - q}}{2} E) \frac{q^{-\frac{n}{2}} + q^{\frac{n}{2}}}{ n}\left(-1 + \frac{1}{(1 - q^{2n})}    \tilde{q}^{2n} \right).
\end{equation}

The last solvable limit we consider is $q_A = 0$. This limit is the same as the previous limit up to exchanging the labels $A$ and $B$. This is because the matrix potential of the $q$-deformed two-matrix model must have a symmetry that exchanges the $A$ and $B$ labels everywhere (including the $A$, $B$ matrices themselves as well as the $q_A$, $q_B$ parameters). This symmetry follows from the simple fact that the two types of chords in a two-species chord diagram are treated on an equal footing. Of course, the JT limit defined in section \ref{sec:scalinglimit} breaks this symmetry.

\subsection{Perturbation theory around a solvable limit}

\label{sec:perttheory}

In the previous subsection, we described three solvable limits of the $q$-deformed matrix model, corresponding to the cases where either $q_A$, $q_B$, or $\tilde{q}$ are set to zero. To test our conjecture that the $q$-deformed model exists, one can compute perturbative corrections to the potential around each of these limits and check that they are mutually consistent. In this subsection, we show that the $O(q_B)$ correction to the $q_B = 0$ potential in \eqref{eq:1.85} is consistent with the other two solvable limits. In particular, we will first compute this correction when $\tilde{q} = 0$, and the result will agree with the $O(q_B)$ term of the potential in \eqref{eq:1.83}. Then, we will compute the $O(1)$ and $O(q_A)$ parts of this correction in the small-$q_A$ expansion (for arbitrary $\tilde{q}$) and find that the terms in the potential we obtain are symmetric under relabeling $A$ and $B$. 

Our strategy to compute corrections to the $q_B = 0$ potential is the same as in section \ref{sec:corrections}, where we explained how one could systematically compute corrections to the matrix potential in $\mathcal{Z}_{\text{toy}}$ order by order in $\epsilon$ (see Figure \ref{fig:orderepsilondiagrams} for a description of this procedure to first order). Here, $q_B$ plays the role of $\epsilon$. We write the matrix integral of the regulated model as follows:
\begin{align}
	\label{eq:leadingcorrection}
	\begin{split} \mathcal{Z}_{q_A,q_B,\tilde{q}} &= \int dA dB \, \exp\biggl(- V_{q_A,q_B,\tilde{q}} \biggr),
		\\
 V_{q_A,q_B,\tilde{q}}		=   &N \text{Tr } \left( V_{q_A}(A) + V_{c.t.}^{q_A,\tilde{q}}(A) + q_B \Delta V^{q_A,\tilde{q}}(A) \right) 
\\
+&N \sum_{a,b} B_{ab} B_{ba} \frac{(F^{q_A,\tilde{q}}_{ab} + q_B \Delta F^{q_A,\tilde{q}}_{ab})}{2}
\\
- &N q_B \sum_{a,b,c,d} B_{ab} B_{bc} B_{cd} B_{da} \frac{G^{q_A,\tilde{q}}_{abcd}}{4} + O(q_B^2)
	\end{split}
\end{align}
We allow for a generic $O(q_B)$ correction to the terms that are already present in \eqref{eq:1.85}, and we also include a  new term that is quartic in $B$. This quartic term makes the planar connected four-point function of $B$ nontrivial, which allows B-type chords to intersect each other. In order for the single-trace, planar correlators of this matrix model to agree to order $q_B$ with the corresponding sums over two-species chord diagrams, we must set 
\begin{align}
	\label{eq:1.92}
	\begin{split}
	&G_{abcd}^{q,\tilde{q}} = \tilde{G}^{q,\tilde{q}}\left(\frac{\sqrt{1-q}}{2} \lambda_a,\frac{\sqrt{1-q}}{2} \lambda_b,\frac{\sqrt{1-q}}{2} \lambda_c,\frac{\sqrt{1-q}}{2} \lambda_d\right),
	\\
	&\tilde{G}^{q,\tilde{q}}(\cos(\theta_1),\cos(\theta_2),\cos(\theta_3),\cos(\theta_4)) =
	\\
& \sqrt{\tilde{F}^{q,\tilde{q}}(\cos \theta_1,\cos \theta_2) \tilde{F}^{q,\tilde{q}}(\cos \theta_2,\cos \theta_3) \tilde{F}^{q,\tilde{q}}(\cos \theta_3,\cos \theta_4) \tilde{F}^{q,\tilde{q}}(\cos \theta_4,\cos \theta_1)} 		S_{\tilde{q},\tilde{q}}^{q,1}(e^{i \theta_1},e^{i \theta_2},e^{i \theta_3},e^{i \theta_4}).
	\end{split}
\end{align}
This is because in any planar `t Hooft diagram with a quartic $B$ vertex, the propagators can be set to the uncorrected propagator $(N F_{ab}^{q_A,\tilde{q}})^{-1}$ (since the $O(q_B)$ correction to the propagator becomes an $O(q_B^2)$ correction to the entire diagram, which we are not interested in). Thus the quartic interaction in \eqref{eq:leadingcorrection} is entirely determined by the chord-diagram Feynman rule for two crossing B-type chords, \eqref{eq:1.61}.

To determine the $O(q_B)$ correction to the two-point coupling $\Delta F_{ab}^{q,\tilde{q}}$, we choose $\Delta F_{ab}^{q_A,\tilde{q}}$ to cancel the corrections to $\braket{B_{a b} B_{b a}}_{\text{disk}}$ from loop diagrams involving the quartic coupling, such that
\begin{equation}
	\braket{B_{a b} B_{b a}}_{\text{disk}} = (N F_{ab}^{q_A,\tilde{q}})^{-1}.
\end{equation}
We obtain
\begin{equation}
\Delta F^{q_A,\tilde{q}}_{ab} =   \int d\lambda_c  \rho_{q_A,0}(\lambda_c) 
\frac{ G^{q_A,\tilde{q}}_{abcb}}{F^{q_A,\tilde{q}}_{bc}} + \int  d\lambda_c  \rho_{q_A,0}(\lambda_c)
\frac{ G^{q_A,\tilde{q}}_{abac}}{F^{q_A,\tilde{q}}_{ac}} .
\label{eq:deltaF}
\end{equation}

To determine the correction to the $B$-independent part of the potential $\Delta V^{q_A,\tilde{q}}$, we should integrate out $B$ in \eqref{eq:leadingcorrection} to leading order in $q_B$. The result is
\begin{align}
	\label{eq:intoutB}
	\begin{split} \mathcal{Z}_{q_A,q_B,\tilde{q}} = \int dA dB \, \exp\biggl( - &N \text{Tr } \left( V_{q_A}(A) + V_{c.t.}^{q_A,\tilde{q}}(A) + q_B \Delta V^{q_A,\tilde{q}}(A) \right) 
		\\
		- \frac{1}{2} \sum_{a,b} \log F^{q_A,\tilde{q}}_{ab}
		- &\frac{q_B}{2} \sum_{a,b}  \frac{\Delta F^{q_A,\tilde{q}}_{ab}}{F^{q_A,\tilde{q}}_{ab}}
		+ N^{-1} \frac{q_B}{2} \sum_{a,b,c}   \frac{G^{q_A,\tilde{q}}_{abac}}{F^{q_A,\tilde{q}}_{ab} F^{q_A,\tilde{q}}_{ac}} 
		\\
		+ &N^{-1}  \frac{q_B}{4} \sum_{a}      \frac{G^{q_A,\tilde{q}}_{aaaa}}{F^{q_A,\tilde{q}}_{aa} F^{q_A,\tilde{q}}_{aa}}  
		+ O(q_B^2) \biggr)
		. 
	\end{split}
\end{align}
We must choose $\Delta V^{q_{A},\tilde{q}}$ to cancel the contributions from the other $O(q_B)$ terms to the saddle-point equation for the tree-level eigenvalue distribution of $A$. Thus,
\begin{align}
	\begin{split}
		\Delta V^{q_{A},\tilde{q}}(\lambda_a) &=   	
		-  \int d\lambda_b \, \rho_{q_A,0}(\lambda_b)  \frac{\Delta F^{q_A,\tilde{q}}_{ab}}{F^{q_A,\tilde{q}}_{ab}}
		\\
		&+  \frac{1}{2} \int d\lambda_b \rho_{q_A,0}(\lambda_b) d\lambda_c \rho_{q_A,0}(\lambda_c)   \frac{G^{q_A,\tilde{q}}_{abac}}{F^{q_A,\tilde{q}}_{ab} F^{q_A,\tilde{q}}_{ac}}
		\\
		&+   \int d\lambda_b \rho_{q_A,0}(\lambda_b) d\lambda_c \rho_{q_A,0}(\lambda_c)   \frac{G^{q_A,\tilde{q}}_{babc}}{F^{q_A,\tilde{q}}_{ba} F^{q_A,\tilde{q}}_{bc}},
	\end{split}
\end{align}
and if we use \eqref{eq:deltaF}, this becomes
\begin{equation}
		\Delta V^{q_{A},\tilde{q}}(\lambda_a) = -  \frac{1}{2} \int d\lambda_b \rho_{q_A,0}(\lambda_b) d\lambda_c \rho_{q_A,0}(\lambda_c)   \frac{G^{q_A,\tilde{q}}_{abac}}{F^{q_A,\tilde{q}}_{ab} F^{q_A,\tilde{q}}_{ac}}
		.
		\label{eq:deltaV}
\end{equation}

We now consider the $\tilde{q} = 0$ limit. From \eqref{eq:deltaF}, \eqref{eq:1.92}, \eqref{eq:1.86}, \eqref{eq:1.61}, and \eqref{eq:8w7}, we find that 
\begin{equation}
	G^{q_A,0}_{abcd} = 1, \quad \Delta F_{ab}^{q_A,0} = 2, \quad \Delta V^{q_A,0}(A) = - \frac{1}{2},
\end{equation}
and in particular we may ignore $\Delta V^{q_A,0}(A)$ because it is a constant. Then, 	\eqref{eq:leadingcorrection} becomes
\begin{align}
	\begin{split} \mathcal{Z}_{q_A,q_B,0} = \int dA dB \, \exp\biggl( - N \text{Tr }  \left[  V_{q_A}(A)     
		+ B^2 \left(\frac{1}{2} + q_B\right)
		-  \frac{q_B}{4} B^4  + O(q_B^2) \right] \biggr)
		. 
	\end{split}
\end{align}
Using \eqref{eq:vq}, we see that this agrees with  \eqref{eq:1.83}.

We now consider the computation of \eqref{eq:1.92}, \eqref{eq:deltaF}, and \eqref{eq:deltaV} to order $q_A$ for arbitrary $\tilde{q}$. Because \eqref{eq:1.92} depends on the $q$-deformed 6j symbol, we use \eqref{eq:diagrelation} to write the $q$-deformed 6j symbol as an infinite sum. To a finite order in $q_A$, the sum truncates to a finite number of terms, and collecting the $O(1)$ and $O(q_A)$ terms is straightforward (but tedious). After determining $G^{q_A,\tilde{q}}_{abcd}$, the integrals in \eqref{eq:deltaF} and \eqref{eq:deltaV} are straightforward to evaluate to $O(q_A)$. We now present an explicit formula for the potential in \eqref{eq:leadingcorrection} to $O(q_A)$:

\begin{align}
\label{eq:bigpotentialformula}
	\begin{split}
		 &\frac{1}{N} V_{q_A,q_B,\tilde{q}}	
		 \\	=   &\text{Tr }  \left[\frac{(1 - \tilde{q}^2)}{2} A^2 +  \frac{(1- \tilde{q}^2)}{2}  B^2 + \frac{\tilde{q}^2}{1 - \tilde{q}^2} B^2 A^2 - \frac{\tilde{q}(1 + \tilde{q}^2)}{2(1 - \tilde{q}^2)}  A B A B  \right]
		 \\
		 + q_A &\text{Tr } \biggl[ (1 - \tilde{q}^4)  A^2 + \frac{(\tilde{q}^4 - 1)}{4} A^4 
		 +   \frac{\qt^2(1 - \qt^2)}{2}  B^2 -  \qt^2  A^2 B^2 
		 \\
		 &\quad +  \qt^3  A B A B - \frac{\qt^3}{1 - \qt^2} A^3 B A B  + \frac{\qt^2 (1 + \qt^2)}{2(1 - \qt^2)}  A^2 B A^2 B \biggr]
		 \\
+ q_B &\text{Tr } \biggl[ \frac{\qt^2(1 - \qt^2)}{2}  A^2 +  (1 - \qt^4)  B^2   -  \qt^2  A^2 B^2 +  \qt^3  ABAB
\\
 &\quad +  \frac{(\qt^4 - 1)}{4} B^4 - \frac{\qt^3}{(1 - \qt^2)} A B A B^3 + \frac{ \qt^2 (1 + \qt^2)}{2(1 - \qt^2)} A B^2 A B^2   \biggr]
		 \\
+q_B q_A &\text{Tr } \biggl[ \qt^4(1 - \qt^2)  A^2 - \frac{\qt^4(1 - \qt^2)}{2}  A^4
+  \qt^4(1 - \qt^2) B^2 - \frac{ \qt^4(2 - 5 \qt^2  + \qt^4)}{1 - \qt^2} B^2 A^2 
\\
&\quad +  \qt^4 B^2 A^4 - \frac{2 \qt^7}{1 - \qt^2} A B A B  -  \qt^5 A^3 B A B   - \frac{\qt^4(1 - \qt^2)}{2}  B^4 
\\ 
&\quad + \qt^4 A^2 B^4 - \qt^5 A B A B^3 - \frac{\qt^4}{2(1 - \qt^2)^2} A^2 B^2 A^2 B^2 + \frac{\qt^5}{(1-\qt^2)^2} A^2 B A B A B^2
\\
&\quad - \frac{\qt^6}{(1-\qt^2)^2} A^2 B A B^2 A B + \frac{\qt^5}{(1 - \qt^2)^2} A^2 B^2 A B A B + \frac{\qt^4 (\qt^4 - 2 \qt^2 - 1)}{4(1 - \qt^2)^2} (AB)^4
 \biggr] + \cdots.
	\end{split}
\end{align}
By inspection, the terms in \eqref{eq:bigpotentialformula} are symmetric under $A \leftrightarrow B$. At the level of this perturbative calculation, this result is highly nontrivial, and would be a coincidence if the disk observables of the $q$-deformed model did not have a chord diagram interpretation.

The above calculation is the first step of the systematic procedure described in section \ref{sec:corrections} for determining the couplings of the $q$-deformed model. We expect that if we continue to determine the couplings to higher order in $q_B$, we will continue to see the symmetry that interchanges $A$ and $B$. This is strong evidence in favor of the conclusion that the matrix potential of the $q$-deformed model is well-defined. If \eqref{eq:bigpotentialformula} did not have the $A \leftrightarrow B$ symmetry, we would have to conclude that the power series representation of the matrix potential\footnote{We are referring to \eqref{eq:vhexpansion}, \eqref{eq:fexpansion}, and \eqref{eq:gexpansion}, where $q_B$ plays the role of $\epsilon$.} is merely a formal power series that does not converge (if it did converge, it would converge to a potential without the $A \leftrightarrow B$ symmetry, which contradicts the fact that the chord diagram Feynman rules treat $A$ and $B$ on equal footing). Although we lack a rigorous proof, we believe that the potential of the $q$-deformed model is well-defined. We will learn more about the $q$-deformed model in the next section, where we compute the double-trumpet. In particular, we will compute the empty double-trumpet as well as the double-trumpet with one $\calo$ inserted on each boundary. From either of these calculations, we can read off the partition function of the bulk matter theory on the double-trumpet. The fact that these two separate calculations return the same result for the matter partition function provides further nontrivial evidence that the $q$-deformed model exists.

We do not have analogous arguments for why the Selberg model exists because we have no {\it a priori} reasons to expect a symmetry in the Selberg model. Furthermore, in the Selberg model, we view the empty double-trumpet as part of the data that defines the model, as opposed to something that is nontrivially determined by the defining data.

\section{The double-trumpet}

\label{sec:doubletrumpet}

Previously, we argued that there exist single-trace two-matrix models which, in the double-scaling limit, correctly compute the disk amplitudes of JT gravity minimally coupled to a scalar field. Although we explained in section \ref{sec:corrections} that there is a systematic way to determine all of the terms in the matrix potential, actually computing these terms in practice quickly becomes tedious. The purpose of this section is to demonstrate that one does not need to know the detailed form of the matrix potential to compute the double-trumpet in a single-trace, two-matrix model. It is sufficient to know the (regulated) disk correlators. In both the $q$-deformed and Selberg models, we will compute contributions to the double-trumpet that can be directly compared against our calculations in section \ref{sec:doubletrumpetreview}. We will consider the empty double-trumpet as well as the double-trumpet with an $\calo$ operator inserted on each boundary. For simplicity, we set $S_0 = 0$ in this section.

Let us briefly summarize the technical results of this section. In the $q$-deformed model, our result for the empty double trumpet in the JT limit is in agreement with \eqref{eq:toydt}, which we reproduce here:
\begin{align}
	\braket{ \text{Tr } e^{- \beta_L H} \, \text{Tr } e^{- \beta_R H}}_{\cyl} &= \int_0^\infty db ~b \, Z_{\text{tr}}(\beta_L,b) Z_{\text{tr}}(\beta_R,b) Z(b)
	\label{eq:6.1}
	\\
	\label{eq:6.2}
	Z(b) &= \sum_{n = 0}^\infty \left(\frac{e^{- \Delta b}}{1 - e^{-b}}\right)^n = \frac{1 - e^{-b}}{1 - e^{-b} - e^{- \Delta b}}.
\end{align}
We may interpret $Z(b)$ as a partition function of some matter theory with inverse temperature $b$. We found this result by classifying all of the `t Hooft diagrams that contribute to the empty double-trumpet, and we explicitly computed the first few classes. The first class of diagrams agrees with the $\epsilon = 0$ result \eqref{eq:toydt}. The next few classes of diagrams, which represent finite $\epsilon$ corrections, actually make a vanishing contribution to the empty double-trumpet. We conjecture that the sum over all the `t Hooft diagrams agrees with the $\epsilon = 0$ result, so that the matter partition function $Z(b)$ computed using the $q$-deformed regulator is the same as in the toy model of section \ref{sec:toymodel} that is Gaussian in $\calo$. Our computation of the double-trumpet with one $\calo$ inserted on each boundary provides more evidence in support of this result. As shown in Appendix \ref{2pt-dt}, the gravitational computation of this leads to an integral over the closed geodesic length $b$. The integrand is a product of the integral over the ``boundary wiggles'' at fixed $b$ (which includes an $e^{- \Delta \ell}$ weighting factor, where $\ell$ is the renormalized geodesic length between the two $\calo$ insertions) and a partition function $Z(b)$. In the $q$-deformed model, we classify all of the `t Hooft diagrams, and we compute analytic formulas for the sums over the first four classes of diagrams. We express the results in a way that allows us to read off what $Z(b)$ is. We find that the first four classes of diagrams reproduce the $n = 0,1,2,3$ terms in the sum in \eqref{eq:6.2}. We conjecture that the remaining classes of diagrams that we do not explicitly compute continue to match the remaining terms in the sum. Thus, both the empty double-trumpet and the two-point function return the same result \eqref{eq:6.2} for the matter partition function. We comment on the Hagedorn behavior of $Z(b)$ in section \ref{sec:hagedorn}.

Using the Selberg regulator, the empty double-trumpet is the same as in the $q$-deformed model. Thus, as explained in section \ref{sec:summary}, we add a double-trace term \eqref{eq:addedterm} to the matrix potential of the Selberg model to ensure that the empty double-trumpet becomes \eqref{eq:6.1} except with $Z(b)$ replaced by $Z_{\text{scalar}}(b)$, which was defined in \eqref{eq:first}. Our freedom to add this term arises from an ambiguity in defining the integration measure for the matrix integral. We have identified two reasons for this ambiguity. First, by rescaling the matrix elements $\calo_{ab}$ by a function of $E_a$ and $E_b$ and transforming the measure $d \calo$, we can add an arbitrary double-trace term to the potential that only depends on $H$. In other words, if we let $d \calo$ in \eqref{eq:theansatz} refer to the standard flat measure for the rescaled matrix elements, then the measure written in terms of the original matrix elements includes a Jacobian factor that adds a double-trace term to the matrix potential that only depends on $H$. Unlike the $q$-deformed model, we have no way to canonically specify the measure for $\calo$.\footnote{If we wish to preserve the $A \leftrightarrow B$ symmetry in the $q$-deformed model, we should not redefine the $A$ matrix in a $B$-dependent way. As explained in section \ref{sec:regtwomatrixmodel}, in the double-scaling limit $A$ becomes $H$ and $B$ becomes $\calo$ up to trivial rescalings.} Second, if we view this model as an effective matrix model, then we should consider additional matrices other than $\calo$ that can couple to $H$. Upon integrating these other matrices out, we obtain the matrix potential of the two-matrix model, but perhaps with a modified measure for $H$. Note that the single-trace potential was obtained from a disk-level matching calculation (analogous to how one determines the effective Lagrangian of an EFT from matching). This matching calculation cannot rule out the presence of a double-trace term like \eqref{eq:addedterm} that does not change the disk answers. This double-trace term can be determined by matching to the gravitational double-trumpet.

Hence, in the Selberg model, the empty double-trumpet is viewed as part of the data that defines the model, while in the $q$-deformed model, the empty double-trumpet is a nontrivial prediction of the model. In the Selberg model, we compute the double-trumpet two-point function using similar techniques as in the $q$-deformed model, and we find that the first four classes of `t Hooft diagrams reproduce the $n=0,1,2,3$ terms in the sum in \eqref{eq:second}. We then conjecture that the sum over all `t Hooft diagrams in the Selberg model reproduces, in the double-scaling limit, the double-trumpet two-point function in JT gravity minimally coupled to a massive scalar field.

\subsection{Empty double-trumpet}

\label{EmptyDt}

In this section, we study the double-trumpet with no $\calo$ insertions by explicitly analyzing `t Hooft diagrams. We illustrate our strategy by working directly in the double-scaling limit (as in sections \ref{sec:toymodel} and \ref{sec:corrections}), although we will eventually write down explicit regulated expressions.

\begin{figure}
	\centering
	\includegraphics[scale=.5]{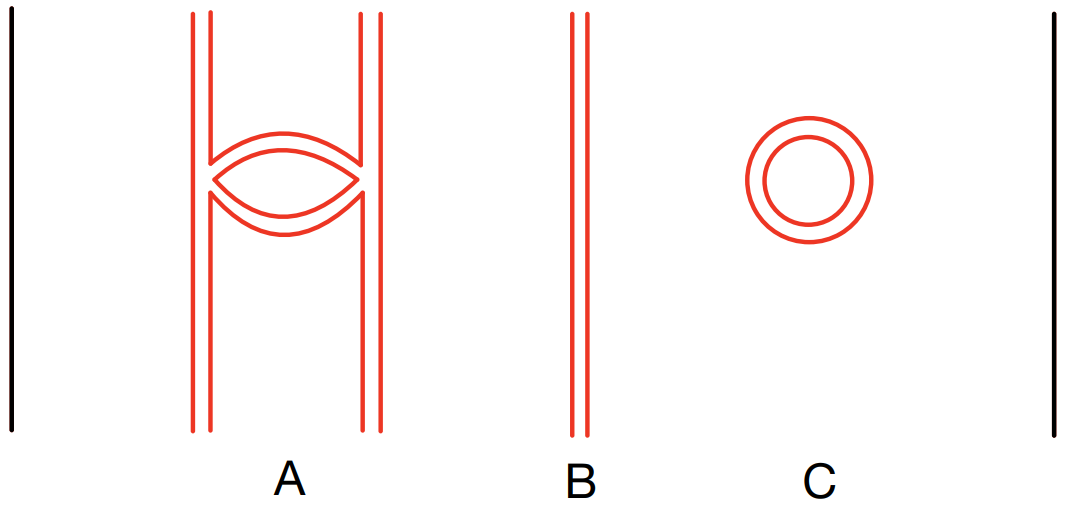}
	\caption{A `t Hooft diagram that contributes to the connected double-trace correlator \eqref{eq:connectedtwopoint} at leading order in the genus expansion. The top and bottom ends of the diagram are identified to obtain the cylinder topology. Each of the two traces in \eqref{eq:connectedtwopoint} is represented by a black line. Each red double-line corresponds to the $\calo$ matrix. There can be many additional black double-lines which we have not explicitly drawn (these correspond to the $H$ matrix). Ignoring the $H$ double-lines, a general diagram can be separated into connected diagrams. Each of the three connected diagrams above is labelled by a letter. Diagrams A and B wrap the double-trumpet. Diagram C is contractible to a point. The contractible diagrams (namely, the $\calo$ bubble diagrams) will be cancelled by counterterms (just as in Figure \ref{fig:threedisks}), so we will no longer draw them in subsequent figures. Note that diagrams B and C both arise from the same double-trace term that appears in the action for $H$ after integrating out $\calo$ (in the toy model, this term is the second term in \eqref{eq:1.23}).
	}
	\label{fig:doubletrumpetthooft}
\end{figure}

We are interested in the connected double-trace correlator
\begin{equation}
    \braket{ \text{Tr } e^{- \beta_L H} \, \text{Tr } e^{- \beta_R H} }_c \equiv \braket{ \text{Tr } e^{- \beta_L H} \, \text{Tr } e^{- \beta_R H} } - \braket{ \text{Tr } e^{- \beta_L H} } \braket{ \text{Tr } e^{- \beta_R H} }.
    \label{eq:connectedtwopoint}
\end{equation}
Let $\braket{ \text{Tr } e^{- \beta_L H} \, \text{Tr } e^{- \beta_R H} }_{\text{cyl}}$ denote the leading contribution to \eqref{eq:connectedtwopoint} in the genus expansion. An example of a `t Hooft diagram that contributes to $\braket{ \text{Tr } e^{- \beta_L H} \, \text{Tr } e^{- \beta_R H} }_{\text{cyl}}$ is given in Figure \ref{fig:doubletrumpetthooft}. To compute the sum over all such diagrams, it is convenient to first consider the sum over all the connected $\calo$ diagrams that wrap the double-trumpet (such as diagrams A and B in Figure \ref{fig:doubletrumpetthooft}). Each diagram contains two non-contractible index loops and hence may be thought of as a function of the two energies running in these loops.\footnote{The contractible index loops are also associated with energies, but these energies should be integrated over with the disk density of states for $H$. That is, these loops should be filled in like a disk.} Let $\cald(s_a,s_b)$ denote a smooth function that represents the sum over all connected $\calo$ diagrams that wrap the double-trumpet. For convenience, we will often write $\cald_{a b}$ in place of $\cald(s_a,s_b)$.\footnote{As always, $s$ is related to energy by $s^2 = E$. The two noncontractible index loops of a diagram are averaged over microcanonical windows centered on energies $s_a^2$ and $s_b^2$ respectively to obtain the contribution to $\cald_{a b}$. In the toy model of section \ref{sec:toymodel}, the only contribution to $\cald_{a b}$ comes from diagram B in Figure \ref{fig:doubletrumpetthooft} and is given by $\log \Gamma(\Delta \pm i s_a \pm i s_b)$. Note that $\cald_{a b}$ may be defined to include or not include a symmetry factor of $2$ associated with the left-right symmetry. Our conventions for $\cald_{a b}$ are clear from \eqref{eq:105}.} A general `t Hooft diagram (such as Figure \ref{fig:doubletrumpetthooft}) is obtained by connecting together the diagrams that contribute to $\cald_{a b}$ with $H$ double-lines and contractible $\calo$ bubble diagrams. Because the contractible $\calo$ bubble diagrams are canceled by counterterms, we can use the double-trumpet of SSS to connect the diagrams that appear in $\cald_{a b}$. Explicitly, we have that
\begin{align}
\begin{split}
&\braket{ \text{Tr } e^{- \beta_L H} \, \text{Tr } e^{- \beta_R H} }_{\cyl} \\
&= \int_0^\infty dE_L dE_R e^{- \beta_L E_L} e^{- \beta_R E_R} \biggl[ \braket{\rho_H(E_L) \rho_H(E_R)}_{\cyl,SSS} 
\\
&+ \int_0^\infty dE_1 dE_2
\braket{\rho_H(E_L) \rho_H(E_1)}_{\cyl,SSS}
    \cald(\sqrt{E_1}, \sqrt{E_2})
\braket{\rho_H(E_2) \rho_H(E_R)}_{\cyl,SSS}
    \\
&+ \int_0^\infty dE_1 dE_2 dE_3 dE_4
\braket{\rho_H(E_L) \rho_H(E_1)}_{\cyl,SSS}
    \cald(\sqrt{E_1}, \sqrt{E_2})
    \\
&\quad \times
\braket{\rho_H(E_2) \rho_H(E_3)}_{\cyl,SSS}
    \cald(\sqrt{E_3}, \sqrt{E_4})
\braket{\rho_H(E_4) \rho_H(E_R)}_{\cyl,SSS}
+ \cdots \biggr]
,
\end{split}
\label{eq:105}
\end{align}
where $\rho_H$ was defined in \eqref{eq:rhoe}, and $\braket{\rho_H(E_L) \rho_H(E_R)}_{\cyl,SSS}$ denotes the leading order connected correlator in the SSS model and may be computed from the inverse Laplace transform of the SSS double-trumpet. Using the expression for $\braket{\rho_H(E_L) \rho_H(E_R)}_{\cyl,SSS}$ from \cite{Saad:2019lba}, equation \eqref{eq:105} is equal to
\begin{align}\label{dtDMM}
\begin{split}
&\braket{ \text{Tr } e^{- \beta_L H} \, \text{Tr } e^{- \beta_R H} }_c \\
&= \int_0^\infty b_L db_L b_R db_R Z_{\text{tr}}(\beta_L,b_L) Z_{\text{tr}}(\beta_R,b_R) \biggl[  \frac{1}{b_R}\delta(b_R - b_L)   +  
    \tilde{\cald}(b_L,b_R)
    \\
&+ \int_0^\infty b_1 db_1 
    \tilde{\cald}(b_L, b_1) 
    \tilde{\cald}(b_1, b_R)
+ \cdots \biggr]
,
\end{split}
\end{align}
where
\begin{equation}
    \tilde{\cald}(b_1,b_2) \equiv \int_0^\infty ds_1 ds_2 \frac{\cos(s_1 b_1)}{\pi} \cald(s_1,s_2) \frac{\cos(s_2 b_2)}{\pi},
\end{equation}
and $Z_{\text{tr}}(\beta,b)$ was defined in \eqref{eq:ztr}. Note that adding an energy-independent constant to $\cald$ does not affect the connected double-trace correlator because
\begin{equation}
    \int_0^\infty dE_1 \braket{\rho_H(E_1) \rho_H(E_2)}_{\cyl,SSS} = 0.
\end{equation}
The integral of the density of states is equal to the number of eigenvalues, which does not fluctuate in the ensemble.

By deriving \eqref{eq:105}, we have reduced the task of computing the empty double-trumpet to the task of computing $\cald_{a b}$, which is the sum over connected $\calo$ diagrams that wrap the double-trumpet. Our next step is to classify these diagrams by studying paths from the left to the right side of a diagram which are only allowed to cross over $\calo$ double-lines (and not vertices). Let $\cald^{(n)}_{a b}$ denote the contribution to $\cald_{a b}$ from all the diagrams obeying the condition that the lowest number of $\calo$ double-lines that are crossed by a path from the left to the right side is $n$. For example, diagram B in Figure \ref{fig:doubletrumpetthooft} belongs to $\cald^{(1)}_{a b}$ because every path from the left to the right side crosses exactly one $\calo$ double-line, while diagram A belongs to $\cald^{(2)}_{a b}$ because one can draw left-to-right paths that cross over two $\calo$ double-lines but not one. 
It follows that
\begin{equation}
    \cald_{a b} = \sum_{n = 1}^\infty \cald^{(n)}_{a b}.
    \label{eq:dterms}
\end{equation}
In the next subsection, we compute $\cald^{(1)}_{a b}$, $\cald^{(2)}_{a b}$, and $\cald^{(3)}_{a b}$ using the $q$-deformed and Selberg regulators and then claim that there is a pattern that generalizes to arbitrary $n$.


As mentioned above, by redefining the measure for $\calo$, we can obtain different answers for the empty double-trumpet. In particular, adding the double-trace term
\begin{equation}
    -\frac{1}{2}\sum_{a,b} \log f(E_a,E_b)
    \label{eq:logf}
\end{equation}
to the matrix potential is equivalent to adding $\log f(E_1,E_2)$ to $\cald(\sqrt{E_1},\sqrt{E_2})$. To add \eqref{eq:logf} to the matrix potential, we may define $\tilde{\calo}_{ab}$ by
\begin{equation}
    \tilde{\calo}_{ab} := \left[f(E_a,E_b)\right]^{-1/2} \calo_{ab}
\end{equation}
and then define the measure to be $d \tilde{\calo}$, or the standard flat measure for the $\tilde{\calo}$ matrix elements. In this way, we can set $\cald(\sqrt{E_1},\sqrt{E_2})$ to whatever we want. In the Selberg model, we fix the ambiguity in the definition of the measure for $\calo$ by demanding that the empty double-trumpet agrees with \eqref{eq:dttoy}. On the other hand, in the $q$-deformed model, we explicitly chose the standard measure for the $A$ and $B$ matrices. We should not modify the measure for $B$ because the $q$-deformed model is supposed to treat the $A$ and $B$ matrices symmetrically.

\subsubsection{Computing $\cald^{(1)}$}

To compute $\cald^{(1)}_{a b}$, we consider diagrams that allow for left-right paths that cross over one $\calo$ double-line. We enumerate all of the possible diagrams in Figure \ref{fig:d1diagrams}.

\begin{figure}
	\centering
	\includegraphics[scale=.3]{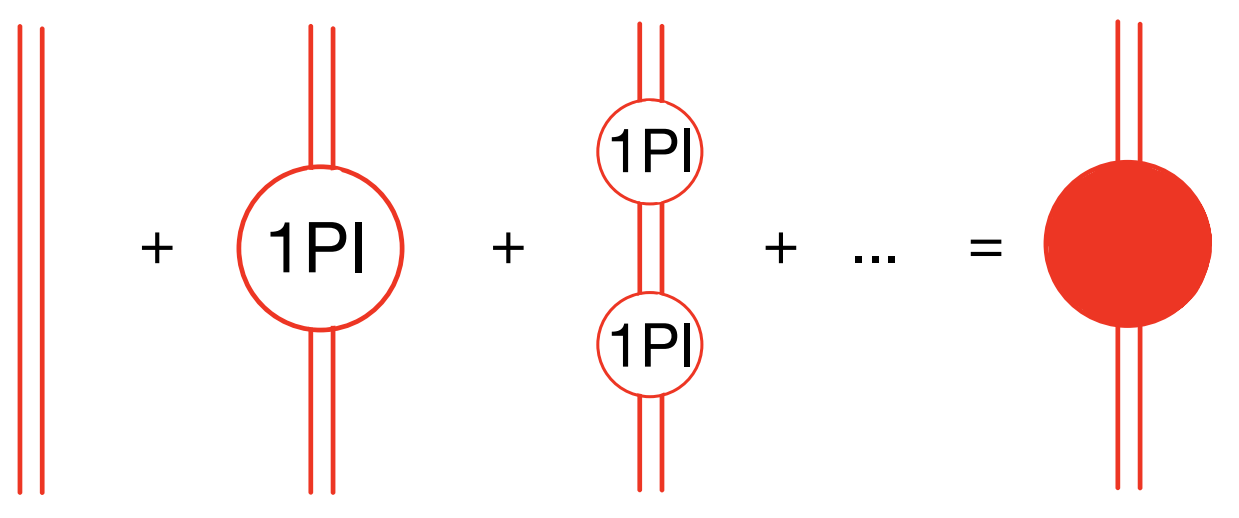}
	\caption{The 1PI two-point function is related to the two-point function as shown. The red blob corresponds to $\braket{\calo_{a b} \calo_{b a}}_{\text{disk}}.$ The top and bottom ends of the diagrams are {\it not} identified. We define $\calx_{a b}$ to be the function of two energies that represents the 1PI blob above.}
	\label{fig:twopoint1PI}
\end{figure}

\begin{figure}
	\centering
	\includegraphics[scale=.3]{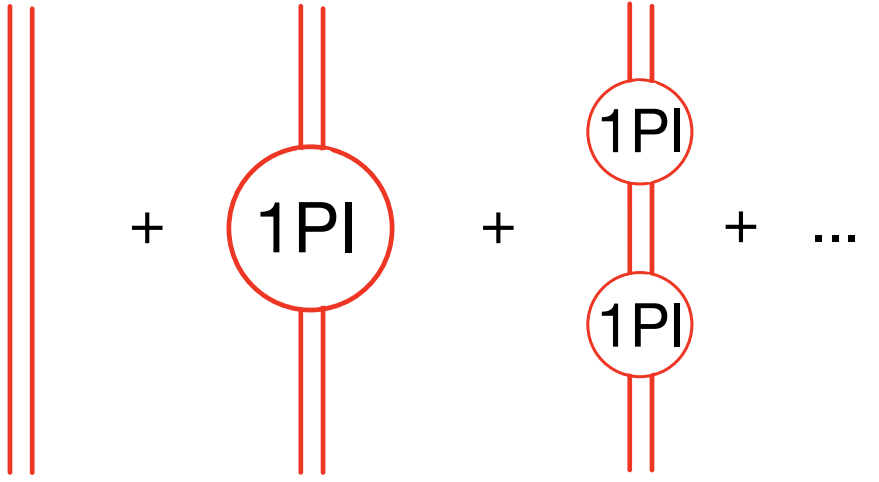}
	\caption{The diagrams that contribute to $\cald^{(1)}_{a b}$ consist of 1PI two-point functions threaded together. The top and bottom ends of the diagrams are identified.}
	\label{fig:d1diagrams}
\end{figure}

Let $\calx_{a b}$ refer to the 1PI two-point function (see Figure \ref{fig:twopoint1PI}). Referring to the matrix integral in \eqref{eq:interactingmodel}, we find that $\calx_{a b}$ obeys
\begin{equation}
    \braket{\calo_{a b} \calo_{b a}}_{\text{disk}} = \frac{F_{a b}^{-1}}{1 - F_{a b}^{-1} \calx_{a b}},
\end{equation}
which implies that
\begin{align}
\label{eq:111}
    \begin{split}
    \cald^{(1)}_{a b} &= -  \log F_{a b} + \sum_{n = 1}^\infty \frac{\left(F^{-1}_{a b} \calx_{a b} \right)^n}{n} = -  \log F_{a b} - \log \left(1 - F^{-1}_{a b} \calx_{a b} \right),
    \\
    &=  \log \braket{\calo_{a b} \calo_{b a}}_{\text{disk}} \\
    &= \log {\Gamma(\Delta \pm i s_a \pm i s_b) \over \Gamma(2\Delta)} \ .
\end{split}
\end{align}
The $\frac{1}{n}$ factor is a symmetry factor arising from the $\mathbb{Z}_n$ symmetry of the diagrams in Figure \ref{fig:d1diagrams}. Using \eqref{logGammab} we then find that 
\begin{align}
\widetilde{\cald}^{(1)}(b ,b') = {1\over b} \delta(b - b') ~ {e^{-\Delta b} \over 1- e^{-b}} \ .
\end{align}
This contribution to the double-trumpet \eqref{dtDMM} precisely equals the double-trumpet result \eqref{eq:toydt} in the toy model of section \ref{sec:toymodel}. Changing the normalization of $\calo$ changes $\cald^{(1)}$ by an additive constant that does not affect the double-trumpet, as explained above.

\label{sec:6.1.1}

\bigskip

Before continuing on to $\cald^{(2)}$, we will compute the contribution from $\cald^{(1)}$ to the double-trumpet using \eqref{eq:105} in the $q$-deformed model. To be precise, we will compute
\begin{equation}
    \braket{\text{Tr } e^{- \beta_L A} \, \text{Tr } e^{- \beta_R A}}_{\cyl}.
\end{equation}
In Figures \ref{fig:twopoint1PI} and \ref{fig:d1diagrams}, the red double-lines now correspond to the $B$ matrix, and we will let the $E$ dummy variable in \eqref{eq:105} refer to an eigenvalue of the $A$ matrix. Using the chord-diagram Feynman rules beginning from \eqref{eq:59}, we have that
\begin{equation}
    \braket{ \text{Tr } B A^{n_1} B A^{n_2} }_{\text{disk}} = 
    N_{q_A} \int_0^\pi \prod_{j = 1}^2 \left[\frac{d\theta_j}{2 \pi} (q_A;q_A)_\infty (e^{\pm 2 i \theta_j};q_A)_\infty\
    \left( \frac{2 \cos \theta_j}{\sqrt{1 - q_A}}\right)^{n_j}
    \right]
    \frac{(\tilde{q}^2;q_A)_\infty}{(\tilde{q} e^{i (\pm \theta_1 \pm \theta_2)};q_A)_\infty},
\end{equation}
where $N_{q_A}$, defined in \eqref{eq:Nq}, is the number of eigenvalues of each matrix. This implies that
\begin{align}
\label{eq:regodisk}
\begin{split}
 \frac{1}{\delta E_1 N_{q_A} \rho_{q_A,0}(E_1)} \frac{1}{\delta E_2 N_{q_A} \rho_{q_A,0}(E_2)}   \braket{ \text{Tr } B P(E_1) B P(E_2) }_{\text{disk}} &= N_{q_A}^{-1} \frac{(\tilde{q}^2;q_A)_\infty}{(\tilde{q} e^{i (\pm \theta_1 \pm \theta_2)};q_A)_\infty},
 \\
E_j &= \frac{2 \cos \theta_j}{\sqrt{1 - q_A}}, \quad j = 1,2,
\end{split}
\end{align}
where $\rho_{q,0}$ was defined in \eqref{eq:1.73} and \eqref{eq:saddlepoint}. Equation \eqref{eq:regodisk} is defined analogously to \eqref{eq:40}, so $P(E_1)$ is a projector onto a microcanonical window of width $\delta E_1$ centered around $E_1$, with $\delta E_1 N_{q_A} \rho_{q_A,0}(E_1)$ eigenvalues to leading order in $N_{q_A}$. Equation \eqref{eq:regodisk} is the analogue of $\braket{\calo_{a b} \calo_{b a}}_{\text{disk}}$ in the $q$-deformed model, so $\cald^{(1)}$ immediately follows from \eqref{eq:111}.

Next, we need to compute
\begin{equation}
    \braket{\rho_A (E_1) \rho_A (E_2)}_{\cyl,\text{1-matrix}}
    \label{eq:rhoa1m}
\end{equation}
in the $q$-deformed model, where $\rho_A$ is the density of states of $A$. Equation \eqref{eq:rhoa1m} refers to the cylinder in the single-matrix model of section \ref{sec:singlematrixwarmup}. That is, \eqref{eq:rhoa1m} is the analogue of $\braket{\rho_H(E_1) \rho_H(E_2)}_{\cyl,SSS}$ in the $q$-deformed model. The density-density correlator in the $q$-deformed 1-cut matrix model takes the universal form  \cite{Ambjorn:1990ji, Brezin:1993qg}
\begin{equation}
    \braket{\rho_A(E_1) \rho_A(E_2)}_{\cyl,\text{1-matrix}} = \frac{1}{(2 \pi)^2} \frac{1}{(E_1 - E_2)^2} \frac{2\left(\frac{E_1 E_2}{a^2} - 1\right)}{\sqrt{\left( 1 - \left( \frac{E_1}{a}\right)^2\right) \left( 1 - \left( \frac{E_2}{a}\right)^2\right)}},
\end{equation}
where
\begin{equation}
    a = \frac{2}{\sqrt{1 - q_A}}.
\end{equation}
Next, note that the Chebyshev polynomials of the first kind have the following generating function\footnote{This may be derived from the results of \cite{Schmidlin}.} for $|t| < 1$
\begin{align}
    \begin{split}
&\sum_{n = 1}^\infty n T_n(x) T_n(y) t^n =
\\
\small
 &\frac{\left(x^2+y^2\right) \left(4 t^3 \left(t^2+1\right) x y-4 t^2 \left(t^4+1\right)\right)+\left(t^6+7
   t^4+7 t^2+1\right) t x y+2 \left(1-t^2\right)^2 t^2-16 t^4 x^2 y^2}{\left(4 t^2 \left(x^2+y^2\right)-4 \left(t^2+1\right) t x
   y+\left(1-t^2\right)^2\right)^2}.
    \end{split}
\end{align}
Taking $t \rightarrow 1$, we formally obtain that
\begin{align}
\sum_{n = 1}^\infty n \ 2T_n(x) \ 2T_n(y)  = \frac{2(x y - 1)}{ (x - y)^2},
\end{align}
which implies that
\begin{equation}
     \braket{\rho_A(E_1) \rho_A(E_2)}_{\cyl,\text{1-matrix}} = \frac{1}{(2\pi a)^2 } \sum_{n = 1}^\infty \frac{n ~ 2T_n\left(\frac{E_1}{a}\right) \ 2T_n\left(\frac{E_2}{a}\right)}{\sqrt{\left( 1 - \left( \frac{E_1}{a}\right)^2\right) \left( 1 - \left( \frac{E_2}{a}\right)^2\right)}}.
     \label{eq:reg1}
\end{equation}
Meanwhile, from \eqref{eq:regodisk}, \eqref{eq:111}, and \eqref{eq:89}, we have
\begin{equation}
    \cald^{(1)}(\sqrt{E_1},\sqrt{E_2}) =  \sum_{n = 1}^\infty \frac{\tilde{q}^n }{n(1 - q_A^n)}
   \  2T_n\left(\frac{E_1}{a}\right) \ 2 T_n\left(\frac{E_2}{a}\right)
    . \label{eq:reg2}
\end{equation}
Noting that the Chebyshev polynomials are orthogonal,
\begin{equation}
    \int_{-1}^1 \frac{dx}{2\pi \sqrt{1-x^2}} 2T_n(x) \  2T_m(x) = \delta_{n m}, \quad n,m \ge 1, 
\end{equation}
we may plug \eqref{eq:reg1} and \eqref{eq:reg2} into \eqref{eq:105} to obtain
\begin{equation}
\braket{ \text{Tr } e^{- \beta_L A} \, \text{Tr } e^{- \beta_R A} }_{\cyl} \supset \sum_{n = 1}^\infty n \ \frac{1 - q_A^n}{1 - q_A^n - \tilde{q}^n} \int_{-a}^a \  \prod_{i \in \{L,R\}} \left[  \frac{d E_i}{2\pi a} e^{- \beta_i E_i}  \frac{ 2T_n\left(\frac{E_i}{a}\right) }{\sqrt{ 1 - \left( \frac{E_i}{a}\right)^2 }}\right].
\label{eq:123}
\end{equation}
The $\supset$ symbol indicates that \eqref{eq:123} only contains the contribution from $\cald^{(1)}$. However, below we will see that neither $\cald^{(2)}$ nor $\cald^{(3)}$ make any nontrivial contributions to $\cald$, and we will conclude that in general, only $\cald^{(1)}$ contributes nontrivially to $\cald$ in the $q$-deformed model. Hence, we may replace the $\supset$ by an $=$ above. Note that \eqref{eq:105} only converges when\footnote{Given that $q_A$ and $q$ are in the interval $(0,1)$, it is enough to write \eqref{eq:condition} for $n = 1$ only.}
\begin{equation}
\label{eq:condition}
    q_A^n + \tilde{q}^n < 1, \quad \forall n \ge 1.
\end{equation}
The integrals over $E_i$ in \eqref{eq:123} can be computed
\begin{align}
\int_{-1}^1 {dx \over 2\pi \sqrt{1-x^2}} ~ e^{-\beta x} ~ 2 T_n(x) = (-1)^n I_n(\beta) \ .
\end{align}
So we get
\begin{equation}\label{ZZqdefMM}
\braket{ \text{Tr } e^{- \beta_L A} \, \text{Tr } e^{- \beta_R A} }_{\cyl} = 
\sum_{n = 1}^\infty n ~ \frac{1 - q_A^n}{1 - q_A^n - \tilde{q}^n} ~
I_n(\beta_L a) I_n(\beta_R a) \ .
\end{equation}
Without the factor $\frac{1 - q_A^n}{1 - q_A^n - \tilde{q}^n}$ this is the equation (50) from \cite{Brezin:2007aa}\footnote{See also eq. (2.34) in \cite{Blommaert:2021gha}}. Our formula \eqref{ZZqdefMM} is a generalization of that to the 2-matrix model.

In section \ref{sec:hagedorn}, we will interpret the matrix model in the regime where \eqref{eq:condition} is violated. A simpler correlator is
\begin{equation}
    \braket{ \text{Tr } 2T_n\left(\frac{A}{a}\right)  \, \text{Tr } 2T_m\left(\frac{A}{a}\right) }_{\cyl} = n \frac{1 - q_A^n}{1 - q_A^n - \tilde{q}^n} \delta_{nm}.
\label{eq:fixednm}
\end{equation}
It would be interesting to reproduce this result for the double-trumpet in q-deformed theory from the transfer matrix approach in the double-scaled SYK \cite{Berkooz:2018jqr}, \cite{Lin:2022rbf}. This computation can be thought of as the bulk dual of our matrix model result above. One can construct the bulk Hilbert space by slicing open the chord diagrams \cite{Lin:2022rbf} and use it to compute the double-trumpet.

We can also take the JT limit of \eqref{eq:fixednm}. We set $q_A = q$, $\tilde{q} = q^\Delta$, and we use \eqref{eq:HA} to relate $H$ and $A$. Then define $b_1$ and $b_2$ via
\begin{equation}
    q_A^n = e^{- b_1}, \quad q_A^{m} = e^{- b_2}.
\end{equation}
Taking $q \rightarrow 1$ for fixed $b_1,b_2$, we obtain
\begin{equation}
    \braket{ \text{Tr } {2\over b_1} \cos \left( b_1 \sqrt{H}\right) \, \text{Tr } {2\over b_2}\cos \left( b_2 \sqrt{H}\right) }_{\cyl} = \frac{1 - e^{-b_1}}{1 - e^{-b_1} - e^{- \Delta b_1}} \ {1\over b_1} \delta(b_1 - b_2),
    \label{eq:6.27}
\end{equation}
which is in agreement with \eqref{eq:toydt}.

\subsubsection{Computing $\cald^{(2)}$}

\begin{figure}
	\centering
	\includegraphics[scale=.15]{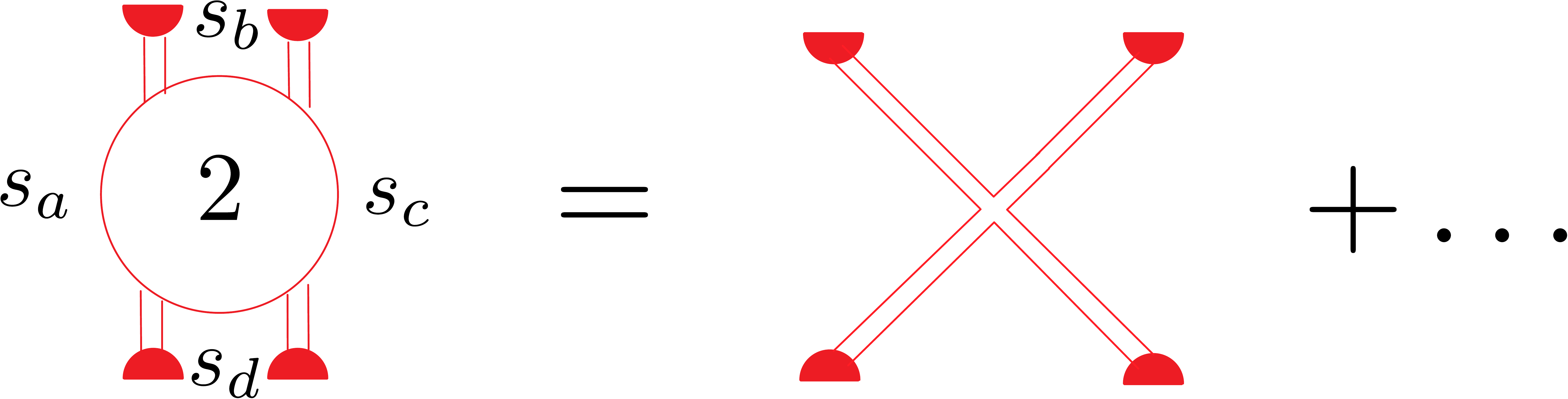}
	\caption{
 The blob labeled ``2'' is the sum over two-particle irreducible diagrams, i.e. all connected planar four-point diagrams such that any left-right path through the blob crosses more than two double-lines. The simplest contribution to this blob comes from a tree four-point diagram (in this case, there are no left-right paths that can go through the blob, because we only let left-right paths cross over double-lines, and all of the double-lines are external. So it is vacuously true that all left-right paths through the blob cross over more than two double-lines). The blob labeled ``2'' is a function of the four energies shown above. We define $\calx^{(2)}_{a c}(s_b,s_d)$ to be the product of this function and  $(\braket{\calo_{a b} \calo_{ba}}_{\disk} \braket{\calo_{b c} \calo_{c b}}_{\disk} \braket{\calo_{c d} \calo_{d c}}_{\disk} \braket{\calo_{a d} \calo_{da }}_{\disk})^{1/2}$, which is represented by the four red semicircles. Including these factors ensures that $\calx^{(2)}_{a c}(s_b,s_d)$ does not depend on the normalization of $\calo$.}
	\label{fig:b2}
\end{figure}

To compute $\cald^{(2)}$, we consider diagrams that allow for left-right paths that cross over two $\calo$ double-lines, but we exclude those diagrams that were already counted in $\cald^{(1)}$. It is convenient to think of each diagram as an operator that acts on functions of a single energy. Let $\calx^{(2)}_{a b}(s_c,s_d)$ refer to a smooth function of $s_a$, $s_b$, $s_c$, and $s_d$ that is defined in Figure \ref{fig:b2}. We may think of $\calx^{(2)}_{a b}(s_c,s_d)$ as an operator (which depends on the parameters $s_a$ and $s_b$) that acts on a function of $s_d$ to produce a function of $s_c$.

\begin{figure}
	\centering
	\includegraphics[scale=.1]{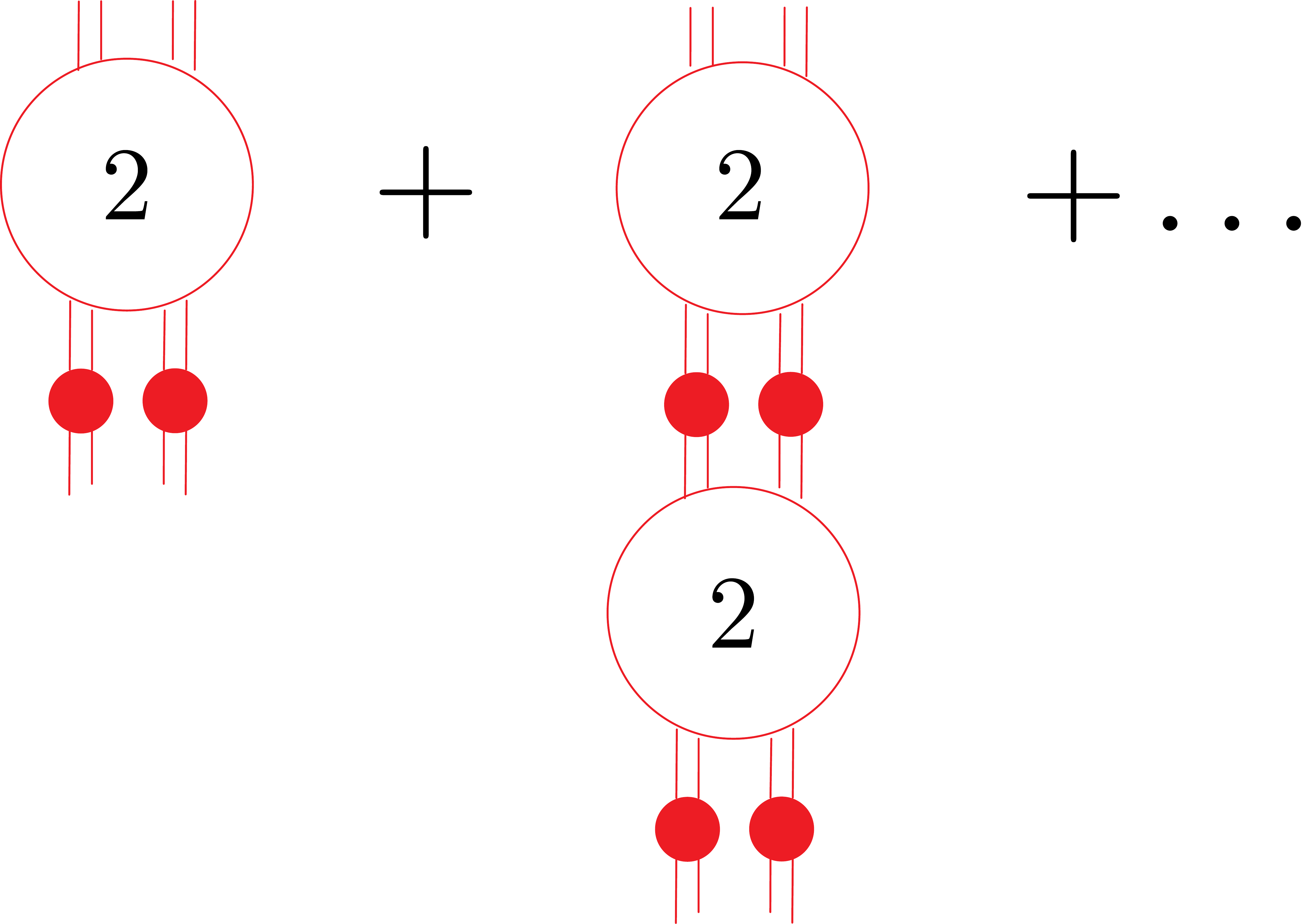}
	\caption{The diagrams that contribute to $\cald^{(2)}$. The top and bottom ends of these diagrams are identified. The red blob represents the exact two-point function $\braket{\calo_{a b} \calo_{b a}}_{\disk}$.}
	\label{fig:dtdiagrams}
\end{figure}

We enumerate all of the diagrams in $\cald^{(2)}$ in Figure \ref{fig:dtdiagrams}, where we explicitly show the locations where a left-right path may be drawn to cross over two double-lines. The remaining parts of the diagrams are encapsulated in $\calx^{(2)}_{a b}(s_c,s_d)$.

The sum over the diagrams in Figure \ref{fig:dtdiagrams} is given by
\begin{align}
\label{eq:128}
    \begin{split}
    \cald^{(2)}_{a b} &= \int_0^\infty ds \, \rho(s) \, \calx^{(2)}_{a b}(s,s)  + \frac{1}{2} \int_0^\infty \, ds_1 ds_2 \, \rho(s_1) \rho(s_2) \calx^{(2)}_{a b}(s_1,s_2) \calx^{(2)}_{a b}(s_2,s_1) 
    \\
    &+ \frac{1}{3} \int_0^\infty \, ds_1 ds_2  ds_3 \, \rho(s_1) \rho(s_2) \rho(s_3) \calx^{(2)}_{a b}(s_1,s_2) \calx^{(2)}_{a b}(s_2,s_3) \calx^{(2)}_{a b}(s_3,s_1) + \cdots,
    \end{split}
\end{align}
where the integrals refer to the closed index loops that are explicitly depicted in Figure \ref{fig:dtdiagrams}. The $n$th term in the sum has a symmetry factor of $\frac{1}{n}$ due to a $\mathbb{Z}_n$ symmetry.

\begin{figure}[!ht]
	\centering
	\includegraphics[scale=.4]{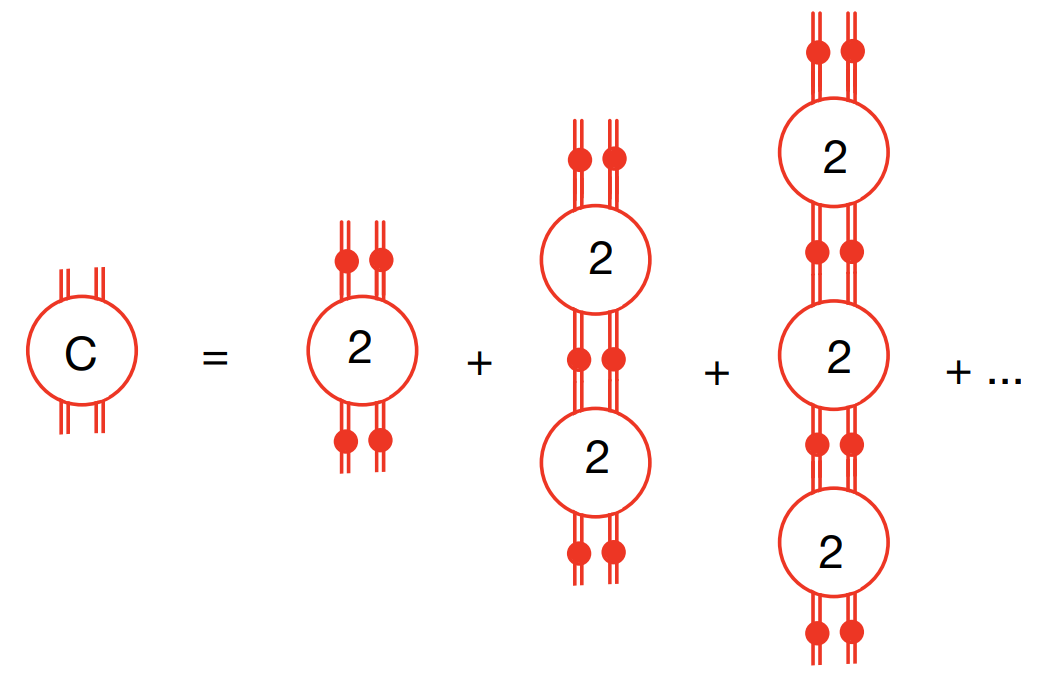}
	\caption{
	A geometric series involving $\calx^{(2)}_{a b}(s_c,s_d)$ computes the connected planar four-point function, which is represented by a blob labeled ``C.'' The top and bottom ends of these diagrams are {\it not} identified.}
	\label{fig:b2def}
\end{figure}

The function $\calx^{(2)}_{a b}(s_c,s_d)$ may be determined from the disk four-point function. See Figure \ref{fig:b2def}. To determine $\calx^{(2)}_{a c}(s_b,s_d)$, we write
\begin{align}
    \begin{split}
	&\frac{\braket{\calo_{a b} \calo_{b c} \calo_{c d} \calo_{d a} }_{\text{disk},c}}{\sqrt{\braket{\calo_{a b} \calo_{b a}}_{\disk} \braket{\calo_{b c} \calo_{c b}}_{\disk} \braket{\calo_{c d} \calo_{d c}}_{\disk} \braket{\calo_{d a} \calo_{a d}}_{\disk}}} 
	\\
	&= \calx^{(2)}_{a c}(s_b,s_d) + \int_0^\infty ds \, \rho(s)\,  \calx^{(2)}_{a c}(s_b,s)
	\calx^{(2)}_{a c}(s,s_d) + \cdots,
    \\
	&= \calx^{(2)}_{a c} + \calx^{(2)}_{a c} \calx^{(2)}_{a c} + \cdots,
	\\
	&= \calx^{(2)}_{a c} \left[1- \calx^{(2)}_{a c}\right]^{-1},
    \end{split}
    \label{eq:129}
\end{align}
where in the third line above we have suppressed some of the $s$ parameters and the integration to simplify the notation,\footnote{We will continue to suppress these $s$ parameters as long as doing so does not cause confusion.} and in the fourth line we summed the geometric series (which is depicted in Figure \ref{fig:b2def}). We have
\begin{align}
\label{eq:130}
    \begin{split}
        &\frac{\braket{\calo_{a b} \calo_{b c} \calo_{c d} \calo_{d a} }_{\text{disk},c}}{\sqrt{\braket{\calo_{a b} \calo_{b a}}_{\disk} \braket{\calo_{b c} \calo_{c b}}_{\disk} \braket{\calo_{c d} \calo_{d c}}_{\disk} \braket{\calo_{d a} \calo_{a d}}_{\disk}}} 
        =  \left\{\begin{array}{ccc}
		\Delta & s_b & s_c \\
		\Delta & s_d & s_a
	\end{array}\right\} \equiv \cals_{a c}(s_b,s_d),
    \end{split}
\end{align}
from which it follows that
\begin{equation}
	\label{eq:131}
    \calx^{(2)}_{a c} = \left[1 + \cals_{a c}\right]^{-1} \cals_{a c},
\end{equation}
where we have formally inverted the geometric series in \eqref{eq:129}. Note however that $1 + \cals_{a c}$, viewed as an operator acting on functions of a single $s$ variable, is not invertible because
\begin{equation}
\left[1 + \cals_{a c}\right](s_b,s_d) = \frac{\delta(s_b - s_d)}{\rho(s_b)} +  \left\{\begin{array}{ccc}
		\Delta & s_a & s_b \\
		\Delta & s_c & s_d
	\end{array}\right\} = 2\sum_{n \in 2 \mathbb{Z}_{\ge 0}.} P_n^{\Delta,\Delta}(s_b;s_a,s_c) P_n^{\Delta,\Delta}(s_d;s_a,s_c),
	\label{eq:132}
\end{equation}
where we have used \eqref{6jdiag}, \eqref{IdRes}. Because the sum runs over even $n$ only, it follows that any $P_n^{\Delta,\Delta}(s_d;s_a,s_c)$ function with $n$ odd is annihilated by $1 + \cals_{a c}$. The $q$-deformed and Selberg models regulate these expressions. We define $\cals^{q}_{a c}$ as in \eqref{eq:130} except with the $q$-deformed 6j symbol. 
Using either of the two regulators, \eqref{eq:132} becomes
\begin{align}
    \begin{split}
&\left[1 + \epsilon ~ \cals^{q}_{a c}\right](s_b,s_d) = \frac{\delta(s_b - s_d)}{\rho_q(s_b)} + \epsilon \left\{\begin{array}{ccc}
		\Delta & s_a & s_b \\
		\Delta & s_c & s_d
	\end{array}\right\}_q 
	\\
	&= \sum_{n \in  \mathbb{Z}_{\ge 0}.} P^{\Delta,\Delta}_n(s_b;s_c,s_a|q) P^{\Delta,\Delta}_{n}(s_d;s_c,s_a|q) \left( 1 + \epsilon (-1)^n q^{\frac{n(n-1)}{2}} q^{2 \Delta n} \right),
	\label{eq:133}
    \end{split}
\end{align}
which reduces to \eqref{eq:132} as $q \rightarrow 1$ and $\epsilon \rightarrow 1$. In the $q$-deformed model, we have set $q_A = q$ and $q_B = \epsilon$. From \eqref{eq:133} we see that $1 + \epsilon~\cals^{q}_{a c}$ is invertible. A simpler way to write \eqref{eq:128} is
\begin{equation}
        \cald^{(2)}_{a b} = \text{Tr } \sum_{n = 1}^\infty \frac{1}{n} \left[\calx^{(2)}_{a b}\right]^n.
\end{equation}
Using \eqref{eq:131}, we finally have
\begin{equation}
\cald^{(2)}_{a b} =     - \text{Tr } \log \left( 1 - \calx^{(2)}_{a b} \right) = - \text{Tr } \log \left( 1 - \frac{\epsilon~ \cals^{q}_{a b}}{1 + \epsilon~\cals^{q}_{a b}}  \right) =  \text{Tr } \log \left(  1 + \epsilon~\cals^{q}_{a b}  \right).
    \label{eq:134}
\end{equation}
It follows that $\cald^{(2)}_{a b}$ only depends on the eigenvalues of $1 + \epsilon~\cals^{q}_{a b}$, but from \eqref{eq:133} the eigenvalues do not depend on $s_a$ or $s_b$. Hence, using the $q$-deformed or Selberg regulators, we find that $\cald^{(2)}_{a b}$ is a constant that does not contribute to the empty double-trumpet.

The computation of $\cald^{(3)}_{ab}$ is presented in appendix \ref{sec:computed3}.

\subsection{Two-point function on the double-trumpet}

\label{sec:twopointfunctiondoubletrumpet}

In this section, we consider the double-trumpet two point function where a single $\calo$ operator is inserted on each boundary:
\begin{equation}
\braket{\text{Tr } e^{- \beta_L H} \calo \, \,  \text{Tr } e^{- \beta_R H} \calo}_{\cyl},
\label{eq:twopointdt}
\end{equation}
where as before $\cyl$ refers to the leading contribution in the genus expansion. As in the previous section, we can systematically classify and compute the `t Hooft diagrams that contribute. We obtain results for both the $q$-deformed and Selberg models and explain for the $q$-deformed model why these results are consistent with our results for the empty double-trumpet.

\begin{figure}
	\centering
	\includegraphics[scale=.3]{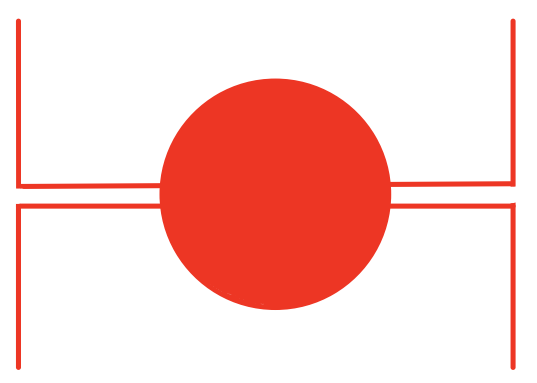}
	\caption{The simplest class of `t Hooft diagrams that contribute to the two-point function on the double-trumpet consists of diagrams that admit a path from the left to the right boundary that does not cross over any $\calo$ double-lines. The red blob represents a sum over connected planar two-point diagrams. The top and bottom ends of this diagram are identified.}
	\label{fig:dttwopoint0}
\end{figure}

\bigskip 

The simplest class of diagrams that contribute to \eqref{eq:twopointdt} is depicted in Figure \ref{fig:dttwopoint0}. In either the $q$-deformed model or the Selberg model, their total contribution in the JT limit is
\begin{equation}
    \int_0^\infty  ds_a \rho(s_a) \,  e^{-(\beta_L + \beta_R)s_a^2} \,  \braket{\calo_{a a}\calo_{a a}}_{\disk,c} =     \int_0^\infty  ds_a \rho(s_a) \, e^{-(\beta_L + \beta_R)s_a^2} \,  \Gamma^{\Delta}_{aa},
\end{equation}
which reproduces the first term in \eqref{2pt-dt2}.

\begin{figure}
	\centering
	\includegraphics[scale=.4]{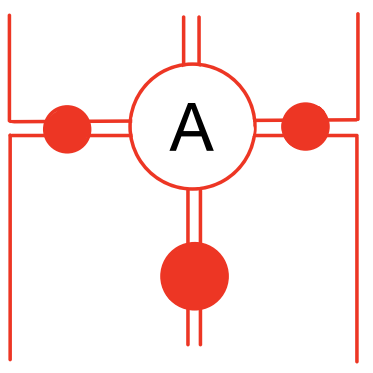}
	\caption{We enumerate diagrams where a left-right path can cross a single $\calo$ double-line. The top and bottom ends of these diagrams are identified. As in Figure \ref{fig:AFblob}, a blob with an ``A'' represents a sum over planar connected amputated diagrams.}
	\label{fig:dttwopoint1}
\end{figure}

\bigskip 

Next, we consider diagrams such that the minimum number of $\calo$ double-lines that are traversed by a left-right path is one. These diagrams are depicted in Figure \ref{fig:dttwopoint1}. In either the Selberg or $q$-deformed models, they become in the JT limit
\begin{align}
\begin{split}
    &\int ds_a ds_b\, e^{- \beta_L s_a^2} e^{- \beta_R s_b^2} \rho(s_L) \rho(s_R) \, \frac{\braket{\calo_{a a} \calo_{a b} \calo_{b b} \calo_{b a}}_{\disk,c}}{\braket{\calo_{a b} \calo_{b a}}_{\disk,c}}
    \\
    =& \int ds_a ds_b\, e^{- \beta_L s_a^2} e^{- \beta_R s_b^2} \rho(s_L) \rho(s_R) \, 
    (\Gamma_{aa}^\Delta \Gamma_{bb}^\Delta)^{1/2}
    \left\{\begin{array}{ccc}
		\Delta & s_a & s_b \\
		\Delta & s_b & s_a
	\end{array}\right\},
\end{split}
\end{align}
which reproduces the second term in \eqref{2pt-dt2}. 

\begin{figure}
	\centering
	\includegraphics[scale=.4]{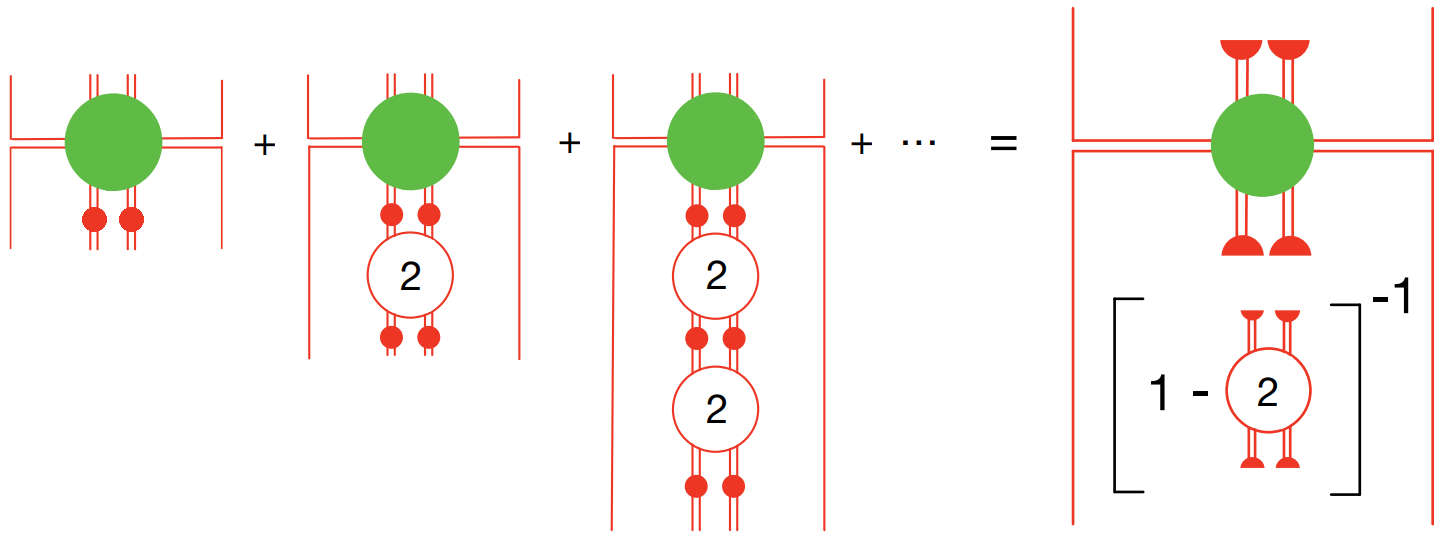}
	\caption{In these diagrams, the minimum number of $\calo$ double-lines that are traversed by a left-right path is two. The blob labelled ``2'' refers to the same object as in Figure \ref{fig:b2}. The top and bottom ends of these diagrams are identified. The green blob with six external double-lines is defined in Figure \ref{fig:dttwopoint2_greenblob}. On the right side, we represent the sum as a geometric series, where multiplication is vertical.}
	\label{fig:dttwopoint2}
\end{figure}

\begin{figure}
	\centering
	\includegraphics[scale=.3]{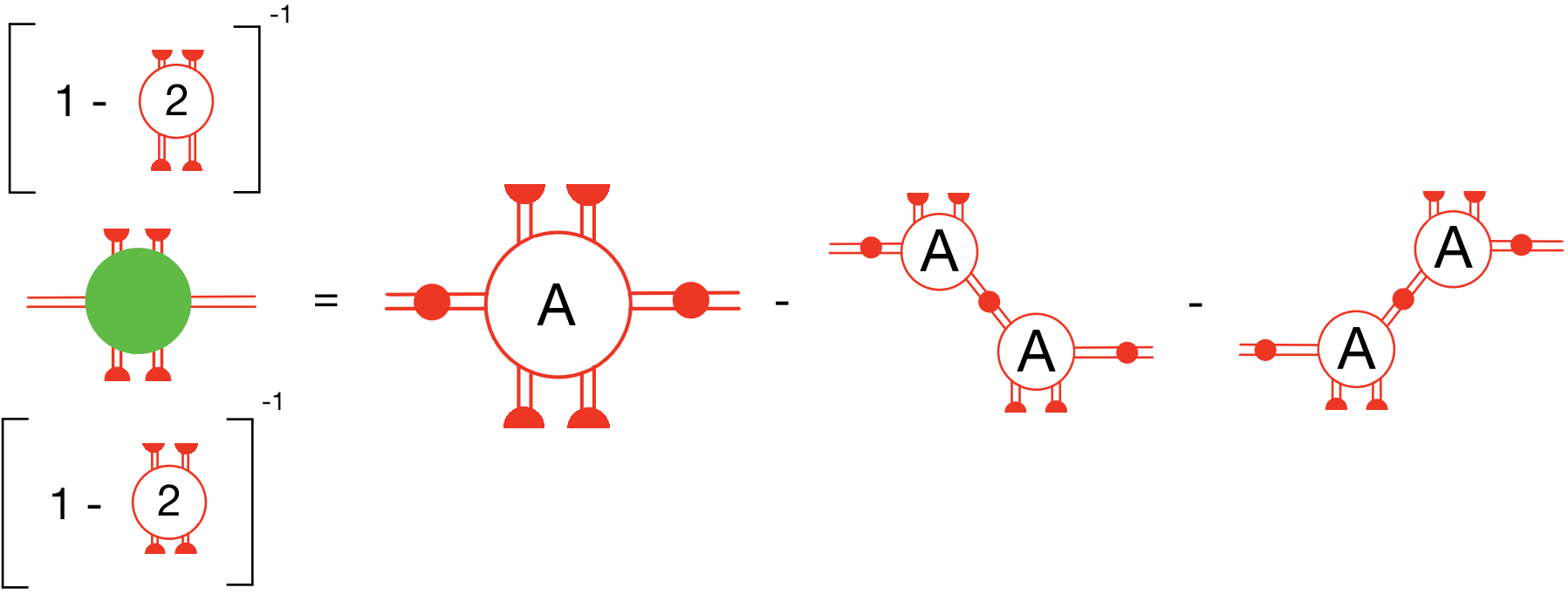}
	\caption{This equation is used to solve for the green blob with six external double-lines. A blob labelled ``A'' represents a sum over connected amputated planar diagrams (for the blob with four external double-lines, this is equivalent to summing over 1PI diagrams). The green blob should consist of connected diagrams only, or else the diagrams in Figure \ref{fig:dttwopoint2} would permit left-right paths that cross fewer than two double-lines. Furthermore, any diagrams that contribute to the ``2'' blob should be amputated off the bottom two and top two legs of this green blob, because in Figure \ref{fig:dttwopoint2} we want to make all of the appearances of the ``2'' blob explicit. That is, there should not be any ``2'' blobs hidden in the green blob. If it were not for the two subtractions on the right hand side, then the green blob would be defined to be the sum of connected six-point diagrams with ``2'' blobs amputated off the top two and bottom two legs. However, we must also exclude from this green blob certain connected diagrams that would permit left-right paths that cross over one double-line (this explains the subtractions on the right hand side). The top and bottom ends of these diagrams are not identified.}
	\label{fig:dttwopoint2_greenblob}
\end{figure}

\bigskip

Next, we consider diagrams such that the minimum number of $\calo$ double-lines that are traversed by a left-right path is two. These are depicted in Figure \ref{fig:dttwopoint2}. To evaluate these diagrams, we should first solve for the green blob in Figure \ref{fig:dttwopoint2_greenblob}. The sum of the diagrams in Figure \ref{fig:dttwopoint2} differs between the $q$-deformed and Selberg models. We remind the reader that for both the Selberg and $q$-deformed regulators, the gravitational Feynman rules are replaced by their $q$-deformed counterparts. Furthermore, in the $q$-deformed model, each 6j symbol is accompanied by a power of $\epsilon$.\footnote{In section \ref{sec:regmodeldef}, we referred to $\epsilon$ as $q_B$. Here, we will return to a notation that is consistent with section \ref{sec:corrections}.} In contrast, in the Selberg model, each connected $2n$-point function is weighted by $\epsilon^{n-1}$.

\pagebreak

In the $q$-deformed model, the sum of the diagrams on the right hand side of Figure \ref{fig:dttwopoint2_greenblob} is 
\begin{equation}
    \includegraphics[scale=0.2]{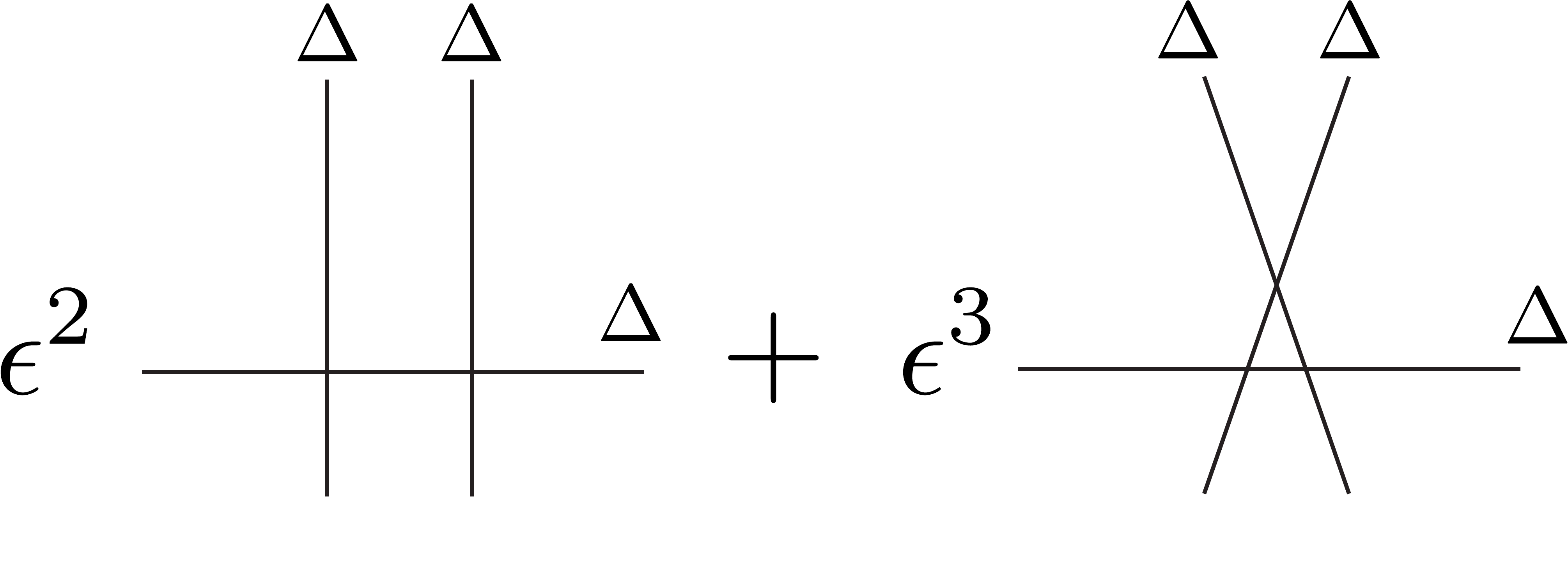}
    \label{eq:189}
\end{equation}
We choose to make the factors of $\epsilon$ explicit in our graphical representations of products of 6j symbols. Furthermore, Figure \ref{fig:b2def} implies that
\begin{equation}
    \includegraphics[scale=0.1]{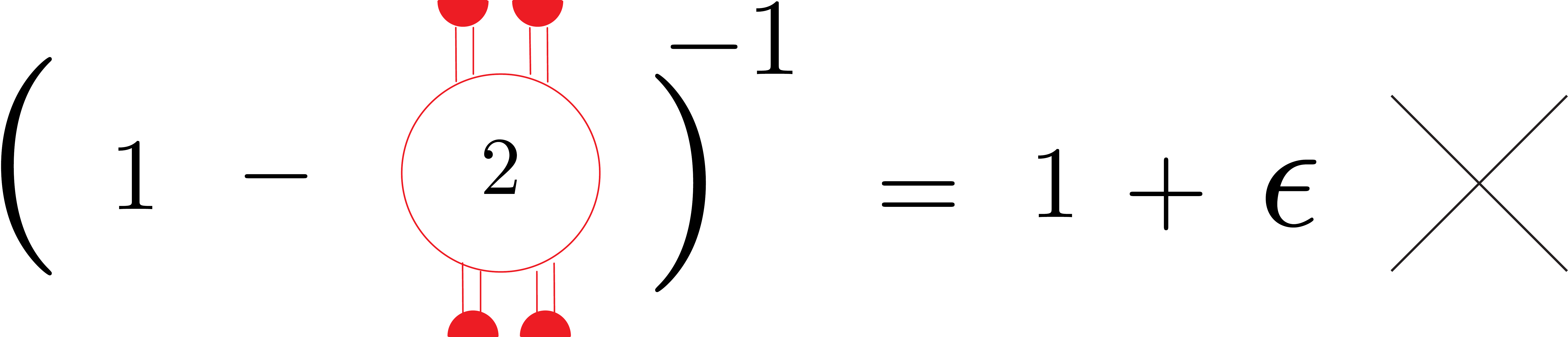}.
    \label{eq:190}
\end{equation}
Note that if we multiply the left hand side of Figure \ref{fig:dttwopoint2_greenblob} by the inverse of \eqref{eq:190} and then identify the top and bottom ends of the diagram, we obtain the left hand side of Figure \ref{fig:dttwopoint2}. Thus, we need to multiply \eqref{eq:189} by the inverse of \eqref{eq:190}. In the $q$-deformed model, the sum of the diagrams in Figure \ref{fig:dttwopoint2} then becomes 
\begin{equation}
    \includegraphics[scale=0.2]{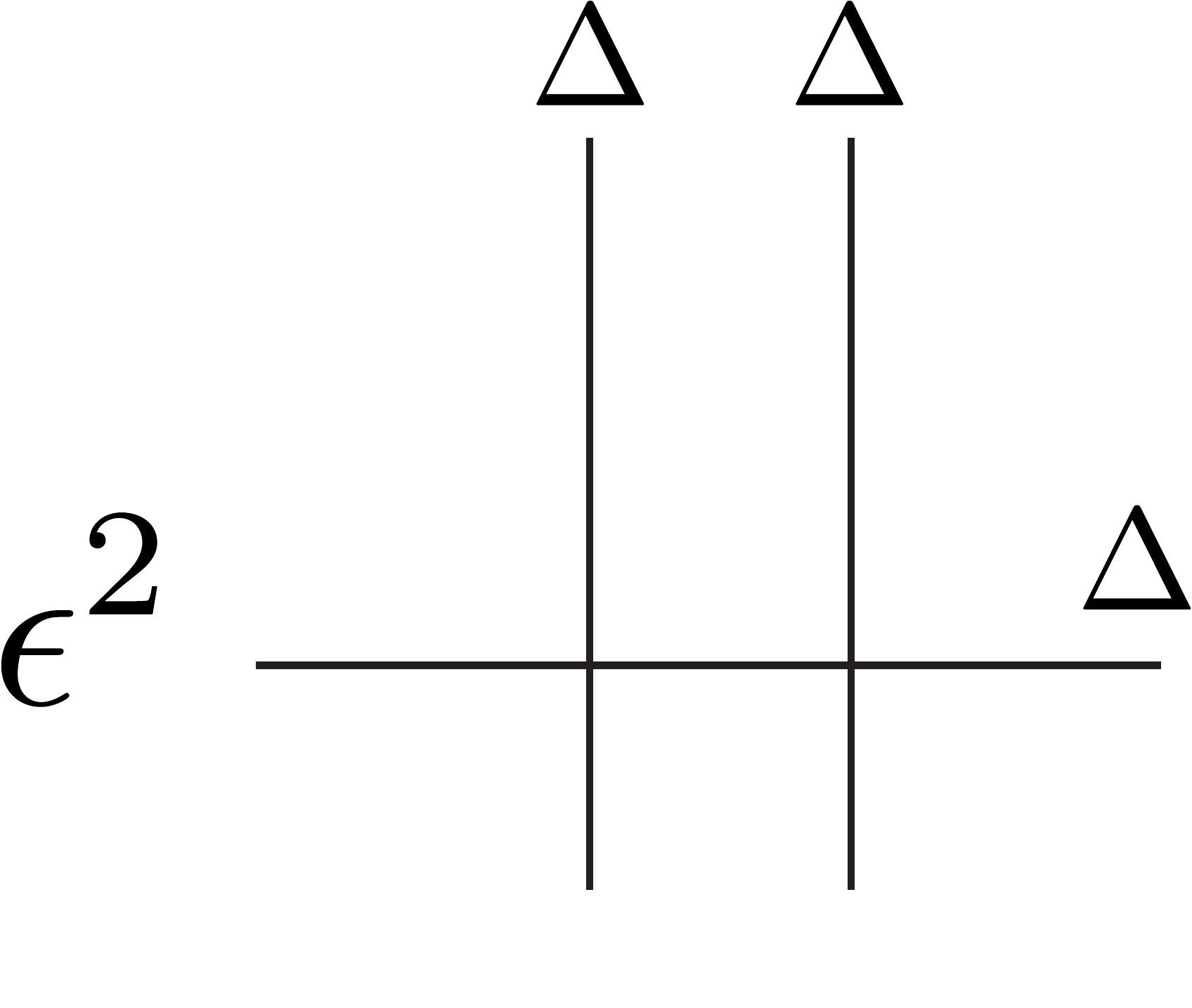}
    \label{eq:dtdiagramtwopoint}
\end{equation}
where the top and bottom ends of this diagram are identified.
Equation \eqref{eq:dtdiagramtwopoint} is a graphical representation of the following expression:
\begin{equation}
    \int ds_a \rho_q(s_a) \, ds_b \rho_q(s_b) \, ds_c \rho_q(s_c) \, e^{-\beta_L s_a^2 - \beta_R s_c^2}
    (\Gamma^\Delta_{aa,q} \Gamma^\Delta_{cc,q})^{1/2}
    \epsilon^2 \left\{\begin{array}{ccc}
		\Delta & s_a & s_b \\
		\Delta & s_b & s_a
	\end{array}\right\}_q
	\left\{\begin{array}{ccc}
		\Delta & s_b & s_c \\
		\Delta & s_c & s_b
	\end{array}\right\}_q.
\end{equation}
Using the pentagon identity in \eqref{eq:pentagon}, this becomes
\begin{equation}
    \int ds_a \rho_q(s_a)  \, ds_c \rho_q(s_c) \, e^{-\beta_L s_a^2 - \beta_R s_c^2}
    ~(\Gamma^\Delta_{aa,q} \Gamma^\Delta_{cc,q})^{1/2}
   ~ \epsilon^2 \sum_{m = 0}^\infty q^{\Delta x}  \left\{\begin{array}{ccc}
		2\Delta+m & s_a & s_c \\
		 \Delta  & s_c & s_a
	\end{array}\right\}_q \ ,
	\label{eq:afterpentagon}
\end{equation}
which gives in the JT limit
\begin{equation}
    \int ds_a \rho(s_a)  \, ds_c \rho(s_c) \, e^{-\beta_L s_a^2 - \beta_R s_c^2}
    ~(\Gamma^\Delta_{aa} \Gamma^\Delta_{cc})^{1/2}
     \sum_{m = 0}^\infty   \left\{\begin{array}{ccc}
		2\Delta + m & s_a & s_c \\
		\Delta  & s_c & s_a
	\end{array}\right\}.
\end{equation}
From \eqref{eq:afterpentagon}, it is clear that the $\epsilon \rightarrow 1$ and $q \rightarrow 1$ limits commute in the $q$-deformed model.

\bigskip 

We now consider the sum of the diagrams in Figure \ref{fig:dttwopoint2} using the Selberg regulator. The explicit expression for the amputated blob with six external double-lines in Figure \ref{fig:dttwopoint2_greenblob} is given by summing over the four connected chord diagrams with six external lines. However, two of these chord diagrams are canceled by the subtractions on the right hand side of Figure \ref{fig:dttwopoint2_greenblob}. The remaining two chord diagrams are both weighted by $\epsilon^2$, which is in contrast to \eqref{eq:189}, where one diagram is weighted by $\epsilon^3$ and the other by $\epsilon^2$. We again want to multiply the sum of these two chord diagrams by the inverse of \eqref{eq:190}. Using \eqref{eq:diagrelation} and \eqref{eq:pentagon}, we have that
\begin{equation}
    \includegraphics[scale=0.15]{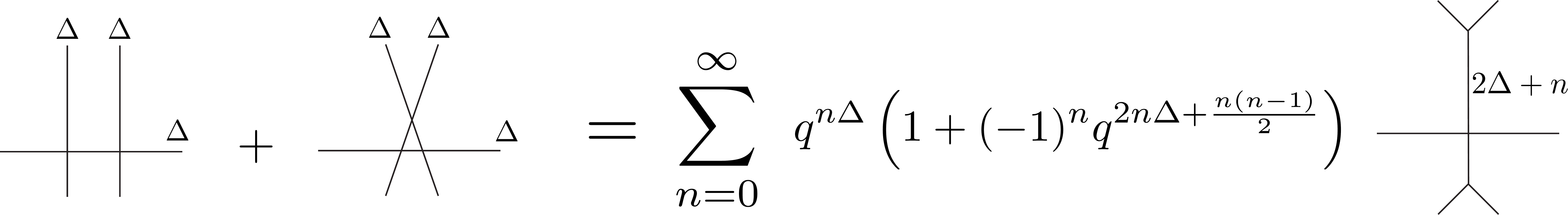}. \label{eq:diagram_sumsymmetrize6j}
\end{equation}
Next, we multiply \eqref{eq:diagram_sumsymmetrize6j} by the inverse of \eqref{eq:190} and identify the top and bottom ends. The sum over the diagrams in Figure \ref{fig:dttwopoint2} finally becomes
\begin{equation}
    \e^2 \sum_{n=0}^\infty 
    q^{n\Delta} \frac{1 + (-1)^n q^{2n\Delta + {n(n-1) \over 2}}}{1 + \e (-1)^n q^{2n\Delta + {n(n-1) \over 2}}}
    ~~~\times ~~~\raisebox{-.4in}{\includegraphics[scale=0.2]{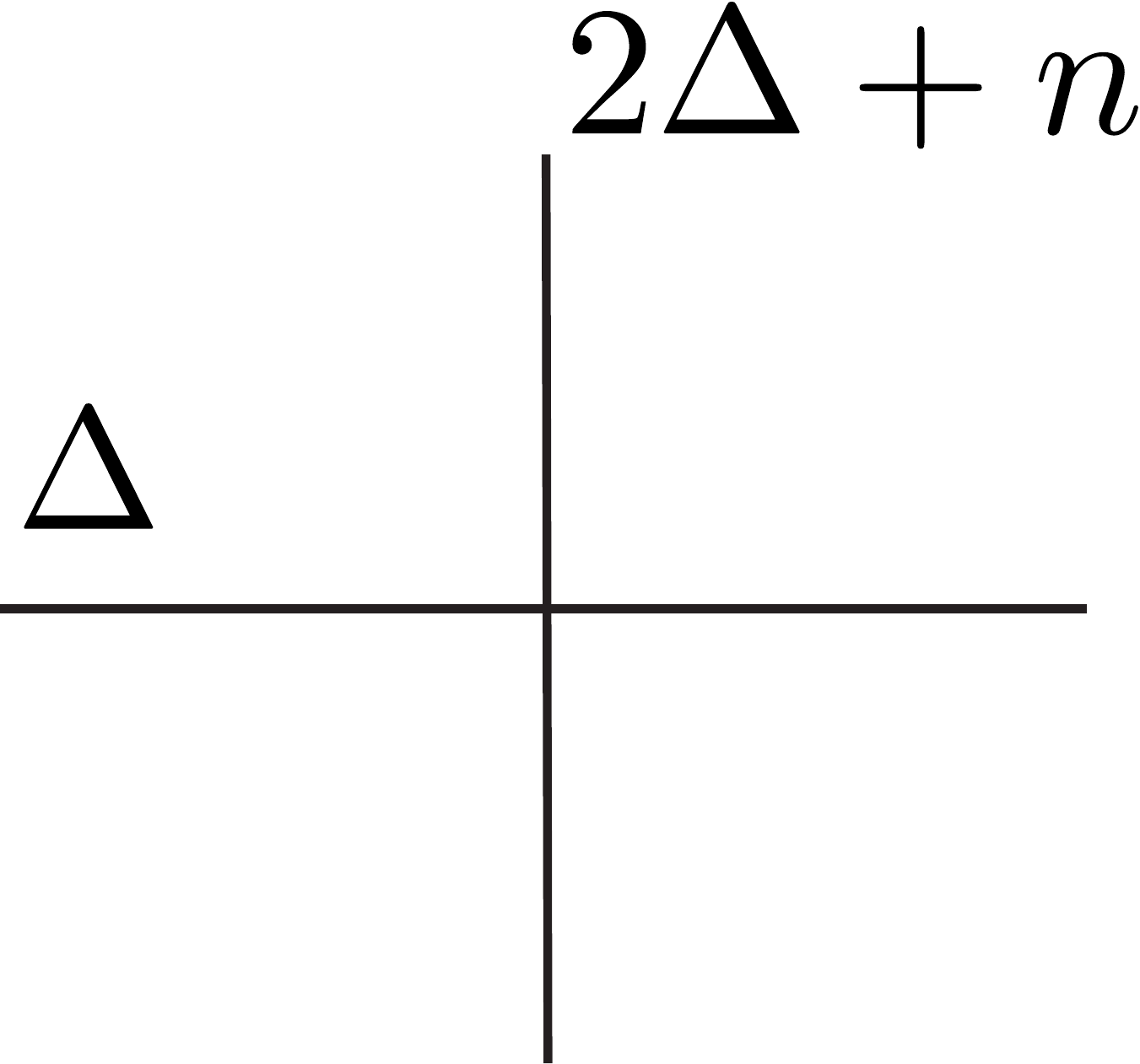}} \ .
\end{equation}
The JT limit corresponds to taking $\epsilon \rightarrow 1$ and $q \rightarrow 1$. For the Selberg model, the $q \rightarrow 1$ limit must be taken first, or else the answer would agree with the $q$-deformed model. The result is that only the even $n$ terms in the above sum contribute. In the JT limit, the result becomes
\begin{equation}
    \int ds_a \rho(s_a)  \, ds_c \rho(s_c) \, e^{-\beta_L s_a^2 - \beta_R s_c^2}
    (\Gamma^\Delta_{aa}  \Gamma^\Delta_{cc})^{1/2}
     \sum_{n = 0}^\infty    \left\{\begin{array}{ccc}
		2\Delta + 2n & s_a & s_c \\
		 \Delta  & s_c & s_a
	\end{array}\right\},
\end{equation}
which reproduces the third term in \eqref{2pt-dt2}. 

The next class of `t Hooft diagrams, in which the lowest number of $\calo$ double-lines crossed by a left-right path is three, is computed in appendix \ref{sec:twopointdoubletrumpetappendix}.

\subsection{Comments on the pair of pants}

\label{sec:pairofpantscomments}

Throughout this section, we have explained how double-trumpet correlators in the matrix model may be directly computed from the disk correlators. We expect that our computational techniques generalize from the double-trumpet to topologies with more handles and boundaries. In this subsection we sketch an approach to computing the empty pair of pants.

\begin{figure}
    \centering
    \includegraphics{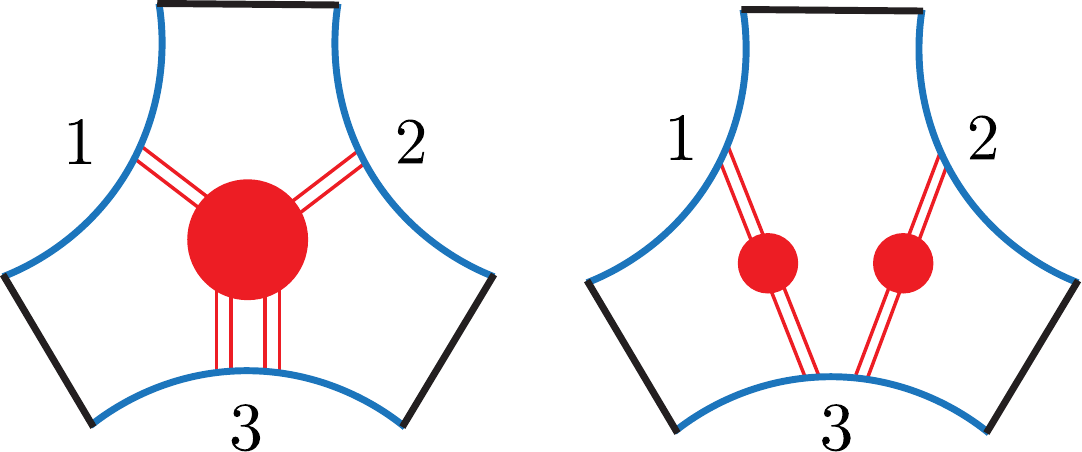}
    \caption{A representation of the pair of pants as two hexagons. The blue edges are identified in pairs according to the labels 1, 2, and 3. The black edges refer to the boundaries that correspond to the single trace insertions in the matrix integral. By drawing planar graphs on the hexagons and gluing them across the edges, we may construct graphs with the pair of pants topology. Each red blob stands for a sum of planar diagrams with the appropriate number of external double-lines.}
    \label{fig:pairofpantsmatrixmodel}
\end{figure}

It helps to represent nontrivial topologies using the plane. For instance, we represented the double-trumpet as a square with one pair of opposite edges identified. For the pair of pants, we may use two hexagons with an appropriate set of edge identifications. See Figure \ref{fig:pairofpantsmatrixmodel}. One should consider only matrix model diagrams which are connected at the level of the $\calo$ double-lines (ignoring trivial $\calo$ bubble diagrams), and non-trivially wrap all three holes.\footnote{A hole is nontrivially wrapped if every path that begins on the hole and ends on any other hole crosses one or more $\calo$ double-lines.} Otherwise, the diagram (or a disconnected subdiagram) will have already been included in a double trumpet leg. One may enumerate all the possible graphs as follows. First, one should specify the number (greater than or equal to one) of double-lines passing through each of the three blue edges in Figure \ref{fig:pairofpantsmatrixmodel}. Then, on each hexagon, one should draw all possible planar graphs with the specified number of external lines on each edge. We have drawn some examples in Figure \ref{fig:pairofpantsmatrixmodel}. Of course, this procedure will overcount the graphs. For a better count, one should use the appropriate analogue of the inverse ``two-to-two propagator'' mentioned in section \ref{sec:summary} to strip off the appropriate subdiagrams from the external lines before gluing the two hexagons together, so that one is gluing correctly amputated disk diagrams. Note that in the Selberg model, the competition between the vanishing of the total disk amplitudes between non-symmetrized states and the divergence of the inverse propagator results in only symmetrized states appearing in each glued edge, by identical reasoning to the double trumpet 2-point function calculation. Moreover, in the $q$-deformed model, we expect that the symmetrizers again cancel and a Hagedorn spectrum of states propagates across the blue edge. 

We leave a careful study of this counting problem to future work. This matrix model analysis should lead to expressions involving integrals of products of 6j symbols. We expect that further calculations like the one in section \ref{sec:PairOfPants}, which covers the simplest irreducible diagram in the pair of pants, can demonstrate the correspondence between certain classes of `t Hooft diagrams and certain geodesics on the pair of pants. Because the pair of pants has infinitely many different closed geodesics while the double-trumpet has only one closed geodesic, we expect the study of the pair of pants (as well as geometries with handles) to be more involved than the analysis presented in this paper.

\subsection{UV divergences in the double-trumpet}

\label{sec:uvdivergences}

In this section we comment on the matrix model interpretation of the problematic UV behavior in the matter partition function on the double-trumpet. We first comment on the Hagedorn temperature in the $q$-deformed model and then argue more generally that whenever the double-trumpet is undefined, the matrix model saddle is unstable.

\subsubsection{Hagedorn temperature in the $q$-deformed model}

\label{sec:hagedorn}

An interesting outcome of our analysis is that the empty double-trumpet in the $q$-deformed model does not depend on the parameter $\epsilon$ that was introduced by the regularization scheme. We explicitly found that $\cald^{(2)}$ and $\cald^{(3)}$ do not contribute to the empty double-trumpet, and we conjectured that $\cald^{(n)}$ for $n > 3$ also do not contribute. The contribution from $\cald^{(1)}$ agrees with the $\epsilon = 0$ model, which in the JT limit is given by \eqref{eq:toydt}. Equation \eqref{eq:toydt} may be interpreted as the double-trumpet path integral of a matter theory minimally coupled to JT gravity. From \eqref{eq:toydt}, we find that the matter partition function is
\begin{equation}
    Z(b) = \frac{1 - e^{-b}}{1 - e^{-b} - e^{- \Delta b}}.
    \label{eq:6.94}
\end{equation}
The matter theory associated to this partition function has many more states than the theory of a free scalar quantized in global AdS$_2$. In particular, the states counted in $Z_{\text{scalar}}(b)$ span a Fock space that is defined by acting on the vacuum with creation operators $a^\dagger_n$ for $n \ge 0$. The operator $a^\dagger_n$ raises the energy by $\Delta + n$, and all of the creation operators commute. The states that contribute to $Z(b)$ may be enumerated in the same way, except without the $[a^\dagger_n,a^\dagger_m] = 0$ condition. That is, different orderings of the creation operators correspond to different states. Put differently, we may say that $Z(b)$ counts all the words that may be constructed using the letters $\calo$ and $\partial$ such that the rightmost letter is $\calo$. The energy of a word is given by $\Delta n_\calo + n_\partial$, where $n_\calo$ and $n_\partial$ are the numbers of $\calo$ and $\partial$ letters.

For sufficiently small (but non-zero) $b$, the denominator of \eqref{eq:6.94} goes to zero, which indicates that there is a Hagedorn temperature. When $b$ is below the inverse Hagedorn temperature, the sum over the `t Hooft diagrams that contribute to \eqref{eq:6.27} does not converge. Away from the JT limit, we see from \eqref{eq:fixednm} that a non-convergent sum is still possible when $q_A + \tilde{q} > 1$. In the remainder of this section, we show that the saddle-point we are doing perturbation theory around is perturbatively unstable in this regime. Because the empty double-trumpet does not depend on $\epsilon$, we will for simplicity first analyze the $\epsilon = 0$ model for $q_A < 1$, which was introduced in equation \eqref{eq:1.85}. After integrating out $B$, the matrix integral becomes
\begin{equation}
    \mathcal{Z}_{q_A,q_B = 0,\tilde{q}} = \int d^N \lambda \exp\left(\sum_{a < b} \log(\lambda_a - \lambda_b)^2 - N \sum_a \left[ V_{q_A}(\lambda_a) + V_{c.t.}^{q_A,\tilde{q}}(\lambda_a) \right] - \frac{1}{2} \sum_{a,b} \log F^{q_A,\tilde{q}}(\lambda_a,\lambda_b) \right),
\end{equation}
where we have written the matrix as an integral over the eigenvalues of $A$. By making the change of variables
\begin{equation}
    \lambda = \frac{2 x}{\sqrt{1-q_A}}
\end{equation}
for each eigenvalue $\lambda$, the matrix integral becomes
\begin{equation}
    \mathcal{Z}_{q_A,q_B = 0,\tilde{q}} \propto \int d^N x \exp\left(\sum_{a < b} \log(x_a - x_b)^2 - N \sum_a \left[ \tilde{V}_{q_A}(x_a) + \tilde{V}_{c.t.}^{q_A,\tilde{q}}(x_a) \right] - \frac{1}{2} \sum_{a,b} \log \tilde{F}^{q_A,\tilde{q}}(x_a,x_b) \right),
\end{equation}
where $\tilde{V}_{q_A}$ was defined in \eqref{eq:rescaledef}, $\tilde{V}_{c.t.}^{q_A,\tilde{q}}$ was defined in \eqref{eq:5.37}, and $ \tilde{F}^{q_A,\tilde{q}}$ was defined in \eqref{eq:1.86}. The density of states $\tilde{\rho}(x)$ defines a saddle of the matrix integral when it extremizes the total potential energy, which is 
\begin{equation}
    V_{\text{total}} = N \int dx \tilde{\rho}(x) \left[ \tilde{V}_{q_A}(x) + \tilde{V}_{c.t.}^{q_A,\tilde{q}}(x) \right] + \frac{1}{2}\int dx_a \tilde{\rho}(x_a) dx_b \tilde{\rho}(x_b) \left[\log \tilde{F}^{q_A,\tilde{q}}(x_a,x_b) - \log(x_a - x_b)^2 \right].
\end{equation}
The function $\tilde{F}^{q_A,\tilde{q}}$ modifies the Coulomb repulsive force between different eigenvalues. Our saddle-point of interest is $\tilde{\rho}(x) = N\tilde{\rho}_{q_A,0}(x)$, which was defined in \eqref{eq:5.26}. This saddle is stable when the Hessian of $V_{\text{total}}$ evaluated for $\tilde{\rho} = N \tilde{\rho}_{q_A,0}$ is positive-definite. If we vary a single eigenvalue at position $x_0$ by $\delta x_0$, the density of states changes by
\begin{equation}
    \tilde{\rho}(x) \rightarrow \tilde{\rho}(x) - \delta x_0 \, \delta^\prime(x - x_0).
\end{equation}
More generally, if we vary all of the eigenvalues according to the rule $\delta x = f(x)$ for some smooth function $f$, then the change in the density of states is
\begin{equation}
    \delta \tilde{\rho}(x) = -\frac{d}{dx}\left[ \tilde{\rho}(x) f(x)\right].
    \label{eq:9.100}
\end{equation}
Note that the support of $\delta\tilde{\rho}(x)$ is $x \in [-1,1]$, which is the same as the support of $\tilde{\rho}_{q_A,0}(x)$. Hence, a small deformation away from the saddle point corresponds to varying the density of states by a function $\delta \tilde{\rho}(x)$ that has support for $x \in [-1,1]$ and integrates to zero. We may expand $\delta\tilde{\rho}(x)$ in a basis of Chebyshev polynomials:
\begin{equation}
    \delta\tilde{\rho}(x) = \sum_{n=1}^\infty \frac{\delta c_n}{\sqrt{1-x^2}} T_n(x),
    \label{eq:6.101}
\end{equation}
where the $n = 0$ term was omitted to ensure that
\begin{equation}
    \int_{-1}^1 dx \, \delta\tilde{\rho}(x) = 0.
\end{equation}
Note that the integral of each term in \eqref{eq:6.101} is proportional to $\sqrt{x+1}$ (resp. $\sqrt{1-x}$) near the $x = -1$ (resp. $x = 1$) endpoint. This is consistent with \eqref{eq:9.100} and the behavior of $\tilde{\rho}(x)$ near the endpoints. If we expand $V_{\text{total}}$ around the saddle to second order in the deformation and use \eqref{eq:6.101}, we obtain
\begin{align}
    V_{\text{total}} &=  \frac{1}{2}\int dx_a \delta\tilde{\rho}(x_a) dx_b \delta \tilde{\rho}(x_b) \left[\log \tilde{F}^{q_A,\tilde{q}}(x_a,x_b) - \log(x_a - x_b)^2 \right].
    \\
     &=   
    \frac{\pi^2}{2} \sum_{n = 1}^\infty \delta c_n^2 
     \left[-  \frac{\tilde{q}^n}{(1-q^n)n}  + \frac{1}{n}  \right],
     \label{eq:6.104}
\end{align}
where we have used \eqref{eq:89} as well as
\begin{equation}
    	\int_{-1}^{1} \frac{dx}{\sqrt{1-x^2}} \, T_{n}(x) \log(y - x)^2 =  - \frac{2\pi}{n}  T_{n}(y), \quad n \ge 1, \quad y \in (-1,1).
\end{equation}
The Hessian is positive-definite when each term in the sum in \eqref{eq:6.104} is positive. Thus, when $q + \tilde{q} > 1$, the saddle becomes unstable. This stability criterion exactly coincides with \eqref{eq:condition}.

\subsubsection{General stability analysis}

\label{sec:generalstability}

For the general $q$-deformed model with nonzero $\epsilon$, the analysis above does not change very much. In fact, for a general `t Hooft-scaled single-trace two-matrix model, one can show that the Hessian directly determines the double-trumpet. As a reminder, the two matrices in the `t Hooft-scaled model are $A$ and $B$, and the scale factor $a$ is chosen such that the saddle-point spectrum of $A/a$ has support on $[-1,1]$. After integrating out $B$, the total potential energy of the eigenvalues of $A$ takes the form
\begin{equation}
    V_{\text{total}} = \sum_{k = 1}^\infty N^{2-k} \int \left[\prod_{i = 1}^k dx_i \tilde{\rho}(x_i)\right] \tilde{V}^{(k)}(x_1,\ldots,x_k)
    \label{eq:vtotalgeneral}
\end{equation}
where $\tilde{\rho}(x)$ is the density of states of the rescaled matrix $A/a$ and $x \in (-1,1)$. Because this is a `t Hooft-scaled model, $N$ does not appear in $\tilde{V}^{(k)}$. If we denote the saddle for $\tilde{\rho}$ by $N \tilde{\rho}_0$ and write $\tilde{\rho} = N \tilde{\rho}_0 + \delta \tilde{\rho}$, then \eqref{eq:vtotalgeneral} takes the form
\begin{equation}
    V_{\text{total}} =   \sum_{k = 2}^\infty N^{2-k} \int \left[\prod_{i = 1}^k dx_i \delta \tilde{\rho}(x_i) \right]  \mathcal{V}^{(k)}(x_1,\ldots,x_k),
\end{equation}
where $\delta \tilde{\rho}$ does not appear in the expression for $\mathcal{V}^{(k)}$. The $k = 1$ term does not appear because we are expanding around a saddle of $V_{\text{total}}$. In the large $N$ limit the only nonvanishing term in the sum is for $k = 2$, because $\delta \tilde{\rho}$ is order one in the large $N$ expansion. Using \eqref{eq:6.101} again, we finally have that
\begin{equation}
    V_{\text{total}} = \frac{1}{2} \sum_{n,m = 1}^\infty \delta c_n V^{(H)}_{nm} \delta c_m  ,
\end{equation}
where
\begin{equation}
V^{(H)}_{nm} =   2 \int dx_1 dx_2 \frac{T_n(x_1)}{\sqrt{1-x_1^2}}\mathcal{V}^{(2)}(x_1,x_2) \frac{T_n(x_2)}{\sqrt{1-x_2^2}}
\end{equation}
is the Hessian. For general $N$, the range of the $\delta c_n$ coefficients is restricted by the condition that the total density of states should be nonnegative. However, in the large $N$ limit, the $\delta c_n$ coefficients are valued on the entire real line because the perturbation $\delta \tilde{\rho}$ is always subleading compared to the saddle $N \tilde{\rho}_0$. Thus, at large $N$, the matrix integral becomes Gaussian in the $\delta c_n$ coefficients, which reflects the fact that the multiboundary correlators are dominated by disks and cylinders.\footnote{In particular, in a one-matrix model, \eqref{eq:vtotalgeneral} only contains terms for $k = 1,2$, and the $k=2$ term is fixed by the Vandermonde determinant, while the $k=1$ term is fixed by the matrix potential. It is known that the cylinder amplitude is universal once the endpoints of the spectrum have been fixed. This result is reproduced by a Gaussian measure for the $\delta c_n$ coefficients.}

From \eqref{eq:6.101} we have that
\begin{equation}
    \int_{-1}^1 dx \, \delta\tilde{\rho}(x)  T_n(x) = \frac{\pi}{2} \delta c_n , \quad n \ge 1,
\end{equation}
and it follows that at large $N$,
\begin{align}
    &\la
    \tr T_n \left(\frac{A}{a}\right) 
    \tr T_m \left(\frac{A}{a}\right)
    \ra = \int dx_1 dx_2 \la \tilde{\rho}(x_1) \tilde{\rho}(x_2) \ra T_n(x_1) T_m(x_2) 
    \\ &= \left(\int_{-1}^1 dx_1  N\tilde{\rho}_0(x_1)  T_n(x_1)\right) \left(\int_{-1}^1 dx_2 N\tilde{\rho}_0(x_2) T_m(x_2)\right)  + \frac{\pi^2}{4} \braket{\delta c_n \delta c_m}
    .
\end{align}
Extracting the connected part of the above, we have that
\begin{equation}
        \la
    \tr T_n \left(\frac{A}{a}\right) 
    \tr T_m \left(\frac{A}{a}\right)
    \ra_{\cyl} = \frac{\pi^2}{4} \braket{\delta c_n \delta c_m} =     \frac{\pi^2}{4} \left[V^{(H)}\right]^{-1}_{nm}.
\end{equation}
We have thus shown that the double-trumpet is well-defined precisely when the matrix model saddle is stable. Thus, the conclusions in section \ref{sec:hagedorn} hold for any $\epsilon$. Similarly, although we do not have an explicit expression for the regulated double-trumpet in the Selberg model, we know that in the double-scaled limit the double-trumpet correlators look like
\begin{equation}
    \braket{ \text{Tr } \cos \left( b_1 \sqrt{H}\right) \, \text{Tr } \cos \left( b_2 \sqrt{H}\right) }_{\cyl} = \frac{b_1}{4} Z_{\text{scalar}}(b_1) \delta(b_1 - b_2),
\end{equation}
which means that in the double-scaling limit, the spectrum of the Hessian extends down to zero, leading to an instability.\footnote{In the Selberg model, the regulated disk density of states is the same as in the $q$-deformed model, so we can still use \eqref{eq:HA} to relate $H$ and $A$ and set $a = \frac{2}{\sqrt{1-q}}$. In the double-scaling (or $q \rightarrow 1$) limit, we set $e^{-b} = q^n$ with fixed $b$ such that $T_n\left(\frac{A}{a}\right)$ becomes $\cos(b \sqrt{H})$ (see the discussion leading to \eqref{eq:6.27}).}

\section{Discussion}

\subsection{Summary of results}

In this work, we described various methods for studying two-matrix models dual to JT gravity minimally coupled to matter. Our aim was to demonstrate how the gravitational path integral may be used to determine the ETH ensemble that arises from coarse-graining the CFT data involving heavy operators at large but finite $N$.  

In one approach, we identified an operator equation obeyed by two operators, $\calo$ and $H$, that is analogous to the statement in 1D CFT that the only primary operators (aside from the identity) that appear in the $\calo \calo$ OPE have dimensions $2 \Delta$ plus a even non-negative integer. Together with conformal invariance and associativity of the OPE, this condition entirely fixes the $n$-point $\calo$ functions to be those of a generalized free field (GFF).\footnote{See appendix \ref{sec:gffproof}.} We constructed a two-matrix ensemble by imposing this operator equation as a constraint  on the matrices that represent the operators. This matrix model should compute the correct holographic $n$-point functions because we are representing the operators using matrices, and matrix multiplication is associative. This result is further supported by explicit checks of Schwinger-Dyson equations.

In another approach, we described an algorithm for determining a matrix potential that reproduces all of the gravitational disk correlators. To organize the calculation, we introduced a fictitious parameter $\epsilon$ such that gravitational Feynman diagrams with more crossings are weighted by more factors of $\epsilon$. The matrix potential can be written in an $\epsilon$ expansion. Different schemes for weighting gravitational diagrams with powers of $\epsilon$ correspond to different ways to take the double-scaling limit. We discussed two specific schemes: the ``Selberg regulator'' and the ``$q$-deformed regulator.'' By construction, the disk correlators of the matrix model do not depend on which regulator is used.

We showed that in any single-trace two-matrix model, the cylinder (as well as the cylinder two-point function) can be determined directly from the disk amplitudes without explicit knowledge of the potential. Our strategy was to systematically classify `t Hooft diagrams with cylinder topology. For the matrix models of interest in this paper, our procedure for determining the cylinder only returns a definite answer if the method for taking the double-scaling limit is known. We obtained formulas using both the $q$-deformed and Selberg regulators. To the extent that we checked, the Selberg model reproduces the cylinder amplitudes of JT gravity minimally coupled to a scalar field.\footnote{Due to an ambiguity in the measure of the matrix $\calo$, the empty double-trumpet in the Selberg model is treated as part of the data that defines the model, rather than a nontrivial prediction of the model. The double-trumpet two-point function is a nontrivial prediction. For the $q$-deformed model, both the  empty double-trumpet and the two-point function are nontrivial predictions.} The bulk dual of the $q$-deformed model is not known, but the cylinder amplitude indicates that the partition function of the unknown matter that propagates on the double-trumpet has a Hagedorn temperature.

After removing the regulators, the cylinder amplitudes of the matrix models (or equivalently, the double-trumpet amplitudes of their bulk duals) are formally undefined due to the UV divergence associated to the shrinking cycle in the off-shell gravitational path integral. Hence, the matrix models do not have a well-defined genus expansion in the JT limit.\footnote{Note that the regulated models can have a well-defined genus expansion.} We explained that this is because the saddle-point of the matrix integral is perturbatively unstable. It would be interesting to find a new saddle that the unstable saddle can decay to. We leave this task for future work.

\subsection{Generalization to higher dimensions}

The ETH ensembles considered in this work are analogous to the ETH ensembles that one might consider in higher dimensional AdS/CFT\footnote{We would like to thank A. Belin, J. de Boer and P. Nayak for important discussions on the issues described in this section.}. In a general holographic CFT, if we know the correlation functions, OPE coefficients, and scaling dimensions of the light operators only (including $1/N$ corrections), then we can construct an ensemble by averaging over all of the CFT data involving the heavy operators that is consistent with this information \cite{Cotler:2020ugk,Belin:2020hea,Chandra:2022bqq}.\footnote{This ensemble has the property that observables involving states that are far below the black hole threshold are not affected by ensemble averaging \cite{Schlenker:2022dyo}.} For example, the light-light-heavy structure constants will be constrained by the four-point function of light operators because heavy operators may run in the intermediate channel. These structure constants are constrained by the requirement that the four-point function of light operators is computed correctly.

Correlators with external heavy operators also lead to crossing-symmetry constraints on the structure constants. Each structure constant can be thought of as a random tensor whose rank is the number of heavy operators. The crossing equation is quadratic in the structure constants. We can impose crossing symmetry as a constraint by squaring the crossing equation and adding it to the potential with a large coefficient $\Lambda$. This means that the potential is at most quartic in the tensors (note that in the CFT ensembles proposed by \cite{Belin:2020hea,Chandra:2022bqq}, non-Gaussianities also play an important role \cite{Belin:2021ryy,Anous:2021caj}). The ensemble will also include a set of variables that represent the scaling dimensions of the heavy operators. These are analogous to the eigenvalues of $H$ in our matrix model. The potential will depend on these eigenvalues in a complicated way through their appearance in the crossing equations. The number of eigenvalues must be scaled to infinity as $\Lambda$ is sent to infinity; the precise way to do this might be similar to the scaling procedure discussed in section \ref{sec:backingaway}.

To reproduce the correlation functions of light operators at finite temperature, only tensors up to rank 2 given by light-heavy-heavy structure constants will play a significant role. The resulting matrix models will be very similar to those constructed in this paper, except that vector spaces on which they act will be graded by spin, and there will be a matrix for every light operator, including multi-twist operators. In holographic theories, it may be possible to express the ensemble only in terms of the primitive operators, dual to bulk fields, by integrating out the multi-twist matrices. 

Previous work in AdS$_3$/CFT$_2$ \cite{Belin:2020hea,Chandra:2022bqq} has shown that imposing crossing symmetry for four-point functions of only heavy operators is not necessary for the ensemble to have a gravity dual. However, if we constrain the structure constants to ensure that crossing symmetry is obeyed for {\it all} correlators, then the ensemble is an average over all the CFTs that are consistent with the initially given information on the light operator data. The given light operator data may not correspond to any solution of the crossing equations, in which case we may say that the light operator data belongs to the swampland.

Suppose we use the light operator data from a theory in the landscape to construct an ensemble by squaring the crossing equations, as described above. Suppose also that we impose crossing symmetry for all correlators, so that the solutions to the constraints are actual CFTs. Suppose for simplicity that there is a unique solution to the crossing equations given the light data. One might think that computing ensemble-averaged observables in the $\Lambda \rightarrow \infty$ limit is no easier than solving the entire theory, which is extremely difficult. Nonetheless, it still may be possible to study interesting  features of the ensemble for large $\Lambda$. To be more precise, let $\Lambda_h$ refer to the $\Lambda$ parameter that multiplies the square of the crossing equation for four external heavy operators.\footnote{In general, there may be multiple $\Lambda$ parameters associated to the multiple constraints we impose.} This parameter should be viewed as a control over how much coarse-graining is being performed.\footnote{$\Lambda_h$ is analogous to the $\gamma$ parameter in (2.48) of \cite{Blommaert:2021etf}. The authors of \cite{Blommaert:2021etf} were able to give their coarse-grained/ensemble-averaged theory a gravity description involving interacting end-of-the-world branes.} For $\Lambda_h = \infty$ and fixed $N$ (or fixed central charge), the spectrum of heavy operators will be a sum over delta functions, corresponding to the solution of the crossing equations. For $\Lambda_h$ large but finite, these delta functions will be smeared by various amounts, resulting in a coarse-grained spectrum. It is unlikely that the distribution of CFT data {\it uniformly} converges to a sum of delta functions in the $\Lambda_h \rightarrow \infty$ limit. For large but finite $\Lambda_h$, we expect that the spectrum of scaling dimensions will still look smooth (as opposed to being a sum of slightly smeared delta functions) for scaling dimensions $\gtrsim \Lambda_h$. Another way to obtain a smooth spectrum is to take the $N \rightarrow \infty$ limit in tandem with the $\Lambda_h \rightarrow \infty$ limit. Because $\Lambda_h$ is interpreted as a coarse-graining parameter, it is reasonable to expect that the smooth parts of the spectrum can be reproduced by a semiclassical bulk dual. Perhaps there are features in the smooth spectrum for scaling dimensions on the order of $\Lambda_h$ that can distinguish whether the light CFT data is in the landscape or swampland. It would also be interesting to study this ensemble using the Schwinger-Dyson techniques of section \ref{sec:crossingsymmetric}.

Note that in the ETH ensemble described in section \ref{sec:crossingsymmetric}, there are light operators but no light states, and there are heavy states but no heavy operators. The heavy states are the eigenvalues of $H$, while $\calo$ is the light operator. Our constraint is analogous to the statement in 1D CFT that the spectrum of primary operators in the $\calo \calo$ OPE should be that of the generalized free field. Given our checks of the Schwinger-Dyson equations in section \ref{sec:crossingsymmetric}, we believe that this constraint is sufficient to construct an ensemble that reproduces all of the holographic disk correlators. There is no analogue of crossing symmetry of heavy external operators in our model. While we have not explicitly studied the space of solutions to our constraint, we know that solutions exist. For example, the SYK model in the appropriate scaling regime furnishes a solution to the constraint. Our constraint holds as an operator equation in any ensemble-averaged theory whose correlators reproduce the disk correlators of JT gravity minimally coupled to a scalar field. Our ensemble is maximally ignorant in the sense that all theories that obey the constraint are averaged over.

\subsection{Other ETH ensembles}

A point we want to emphasize is that it is important to have a principle that determines the ETH ensemble beyond the requirement that it reproduces gravitational path integral calculations. Without such a principle, we would not be able to learn anything new from the ETH ensemble. One such principle is that the ensemble should be the maximally ignorant ensemble that agrees with the gravitational path integral calculations that we know how to do. In practice, one should fix an ansatz for the ensemble and tune the parameters to match the gravitational path integral results. Different ETH ans\"{a}tze correspond to different schemes for coarse-graining the CFT data. Our ETH ansatz was simply a two-matrix model with a single-trace potential. It would be interesting to compare our notion of coarse-graining to other notions of coarse-graining that exist in the literature \cite{Chandra:2022fwi,Engelhardt:2018kcs}. One might also try to connect the ETH matrix model that would arise from such coarse-graining procedures to the Goldstone effective theories of chaos described in \cite{Altland:2020ccq,Altland:2021rqn}.

Another ETH ensemble was considered in \cite{Pollack:2020gfa, Pappalardi:2022aaz, Blommaert:2020seb}. In \cite{Blommaert:2020seb}, the principle that determines the ensemble was called the assumption of local typicality. This is the statement that simple observables are unaffected by conjugating the simple operators with block-diagonal random unitary matrices that act separately in each microcanonical window. From this principle, one can construct an ansatz for multipoint thermal correlators of simple operators. To leading order in $e^{S}$, this ensemble is the same as our two-matrix ensemble. For example, the ETH diagrams in Figure 2 of \cite{Pappalardi:2022aaz} are the same as the `t Hooft diagrams on the left side of \ref{eq:fourpointfunctioninteractingmodel} after accounting for the fact that the one-point correlators of $\calo$ vanish in our model. To leading order in $e^{S}$, the diagrams in \cite{Pollack:2020gfa} are isomorphic to the planar `t Hooft diagrams in this paper after specializing to the case of having a single simple operator. However, the ensemble constructed from local typicality differs from our matrix ensemble at subleading order in $e^{S}$. Using the assumption of local typicality, \cite{Blommaert:2020seb} was able to match certain wormhole contributions to multipoint amplitudes, but without the contribution from the matter determinant. Unlike the principle of local typicality, our ensemble includes an average over the eigenvalues of $\calo$. This averaging appears to be important for reproducing the matter determinant, which is an important aspect of bulk locality. Furthermore, \cite{Lin:2022rzw,Lin:2022zxd} studied the matrix elements of simple operators in extremal (or zero-energy) microstates of certain supersymmetric black holes and found that they are approximately described by a Gaussian random matrix. Hence, including the $\calo$ operator in the matrix ensemble is natural and well-motivated.

\subsection{Future directions}

There are various directions we wish to explore, some of which have already been mentioned. We are interested in studying the fate of the double-trumpet UV divergences in our models by finding a stable saddle that the unstable saddle can decay to. This computation might be easiest in the Gaussian toy model of section \ref{sec:toymodel}, or its $q$-deformed counterpart in \eqref{eq:1.85}. In the $q$-deformed model, we can tune the parameters such that only one eigenvalue of the Hessian has the wrong sign. In this case, there is only one unstable direction.

Another interesting idea is to find a modular-invariant spectrum of primary operators in a CFT$_2$ using a constrained ensemble constructed analogously to the one that we considered in sections \ref{sec:ethasamatrixmodel} and \ref{sec:crossingsymmetric}.\footnote{We thank Tom Hartman for suggesting this idea to us.} The scaling dimensions of the primaries in a given spin sector can be represented as eigenvalues in an eigenvalue integral, and the modular crossing equation can be squared and added to the potential with a large coefficient. This question is interesting to study because the Maloney-Witten-Keller \cite{Maloney:2007ud,Keller:2014xba} partition function, which is a sum over SL(2,$\mathbb{Z}$) images of the vacuum Virasoro character, yields a modular-invariant spectrum that exhibits negativity. By representing the scaling dimensions as eigenvalues in an eigenvalue integral, it is impossible for the spectrum to become negative. The Schwinger-Dyson equations of the constrained model may be useful for finding a modular-invariant spectrum free of negativity.

It would also be interesting to use the $q$-deformed model of section \ref{sec:regtwomatrixmodel} to study the bulk reconstruction of JT gravity with matter in the same manner as \cite{Lin:2022rbf}. In particular, \cite{Lin:2022rbf} noted that in the double-scaled SYK model, one may write an ensemble-averaged correlator as an overlap between a bra and a ket state. These states live in a bulk Hilbert space, and their quantum numbers count the number of chords that pass through a spatial slice. These chord numbers encode the length of the slice as well as the positions of matter particles along the slice. Because the disk correlators of the $q$-deformed model are also computed by summing over chord diagrams, one can reconstruct the same bulk Hilbert space from the $q$-deformed model. In the planar limit, the disk correlators (such as \eqref{eq:exampleobservable}) exactly agree with the double-scaled SYK model. However, the subleading corrections to the correlators differ between the two models.\footnote{As mentioned previously, finding the subleading corrections in the $q$-deformed model in the JT limit is challenging due to UV divergences. However, away from the JT limit, we found a parameter regime \eqref{eq:condition} where the double-trumpet was finite. Without a better nonperturbative understanding of the $q$-deformed model, one should compare the $q$-deformed model and the double-scaled SYK model in a parameter regime where the matrix model saddle is stable.} It would be interesting to study how bulk reconstruction differs between these two models when subleading corrections are considered. Note that due to these corrections, the bulk to boundary map is no longer an exact isometry. This is an important feature of realistic models of AdS/CFT \cite{Akers:2022qdl}.

Finally, we want to better understand the relationship between the $q$-deformed and Selberg models of section \ref{sec:regtwomatrixmodel} and the constrained ensemble of section \ref{sec:ethasamatrixmodel}. The former models appear to be perturbatively unstable (the most explicit check is performed in section \ref{sec:hagedorn}) while the constrained model is non-perturbatively well-defined, since the potential in \eqref{eq:constraintsquared} is bounded from below.\footnote{We expect that $V(H)$ can be chosen to be bounded from below. The role of $V(H)$ is to ensure that in the double-scaling limit, the density of states of $H$ becomes, in the saddle-point approximation, the disk density of states of JT gravity. Away from the double-scaling limit, the density of states of $H$ will have compact support. This means that as the double-scaling limit is taken, $V(x)$ can be designed to approach $+ \infty$ for $x \rightarrow \pm \infty$.} We have not been able to explicitly show that the Selberg and $q$-deformed models resemble \eqref{eq:constraintsquared} in the double-scaling/JT limit, although we have not ruled out this possibility. If we cannot find a stable saddle in the Selberg and $q$-deformed models, then we can instead use the constrained ensemble \eqref{eq:constraintsquared} to understand how the UV divergences in the bulk can be regulated by non-perturbative effects. We leave more explicit studies of the constrained model to future work.


\paragraph{Acknowledgements}

We would like to thank Nick Agia, Alex Belin, Noam Chai, Jan de Boer, Anatoly Dymarsky, Lorenz Eberhardt, Akash Goel, Tom Hartman, Clifford Johnson, Zohar Komargodski, Henry Lin, Juan Maldacena, Dalimil Maz\'{a}\v{c}, Vladimir Narovlansky, Pranjal Nayak, Joaquin Turiaci, and Herman Verlinde for stimulating discussions. This work was performed in part at Aspen Center for Physics, which is supported by National Science Foundation grant PHY-1607611. This work has been partially supported by the DOE through the grant DE-SC0007870, the SNF through Project Grants 200020 182513, as well as the NCCR 51NF40-141869 The Mathematics of Physics (SwissMAP). The work of BM was supported by a grant from the Simons Foundation (651444, BM) and NSF grant PHY-2014071.


\appendix

 \section{Special functions}
 
 \label{sec:specialfunctions}
 
In this appendix, we introduce the special functions used in the main text. For the properties of Wilson polynomials and Wilson functions we follow \cite{Groenevelt2003, Groenvelt2005}.

\subsection{Wilson polynomials}
 Wilson polynomials $W_n(x)=W_n(x;a,b,c,d)$ are $n$th order polynomials of $x^2$. They depend on four complex parameters $a=a_1, b=a_2, c=a_3, d=a_4$, such that non-real parameters appear in complex conjugate pairs. They are defined by
\begin{align}\label{WilsonPolynomial}
W_n(x;a,b,c,d) = (a+b)_n (a+c)_n (a+d)_n 
~\tensor[_4]{F}{_3} \left(  {-n, n + a + b+ c +d-1, a+ix, a-ix \atop a+b, a+c, a+d} ; 1 \right) \ .
\end{align}
They are symmetric in $a,b,c,d$ and obey an orthogonality relation
\begin{align}
\int_0^\infty {dx \over 2\pi} ~M(x) W_n(x) W_m(x) &= r_n \delta_{nm} \ , \qquad
r_n = n! {(n-1 +\sum a_j)_n \over \Gamma(2n+\sum a_j) } \prod_{i<j} \Gamma(a_i + a_j + n)   \ .
\end{align}
The measure is 
\begin{align}
M(x) \equiv M(x;a,b,c,d) = {\prod_{j=1}^4 \Gamma(a_j \pm i x) \over \Gamma(\pm 2i x)} \ .
\end{align}
For a general choice of $a_j \in {\mathbb C}$ there is also a discrete part of the measure \cite{Groenevelt2003}. However, it is absent if $\re a_j>0$, the case relevant for JT gravity.

We also define rescaled Wilson polynomials\footnote{Strictly speaking, $P_n$ are not polynomials, but we abuse the language slightly and still call them Wilson polynomials.} $P_n(x) \equiv P_n(x;a,b,c,d)$ that are orthonormal with respect to the Schwarzian measure
\begin{align}
&P_n(x) = {1\over \sqrt{r_n}} \prod_{j=1}^4 \Gamma(a_j \pm i x)^{1/2} ~ W_n(x) \ , \\
&\int_0^\infty {dx \over 2\pi \Gamma(\pm 2i x)} ~ P_n(x) P_m(x) = \delta_{nm} \ .
\end{align}

\subsection*{Choice of parameters relevant for JT gravity}
The choice of parameters relevant for JT gravity is
\begin{align}\label{abcdJT}
a=d^* =  \Delta_1+i k_1, \quad b= c^* =\Delta_2 +i k_2 \ .
\end{align}
We therefore define
\begin{align}\label{Pndef}
P_n^{\Delta_1, \Delta_2}(x; k_1,k_2) := P_n(x;\Delta_1 \pm i k_1, \Delta_2 \pm i k_2) \ ,
\end{align}
where by $\Delta \pm i k $ we mean that $P_n$ depends on both parameters and the order is not important due to symmetry in $a,b,c,d$. Pictorially, we represent the Wilson polynomial as
\begin{align}
P_n^{\Delta_1, \Delta_2}(x; k_1,k_2)  
~=~ 
\raisebox{-.5in}{ \includegraphics[scale=.35]{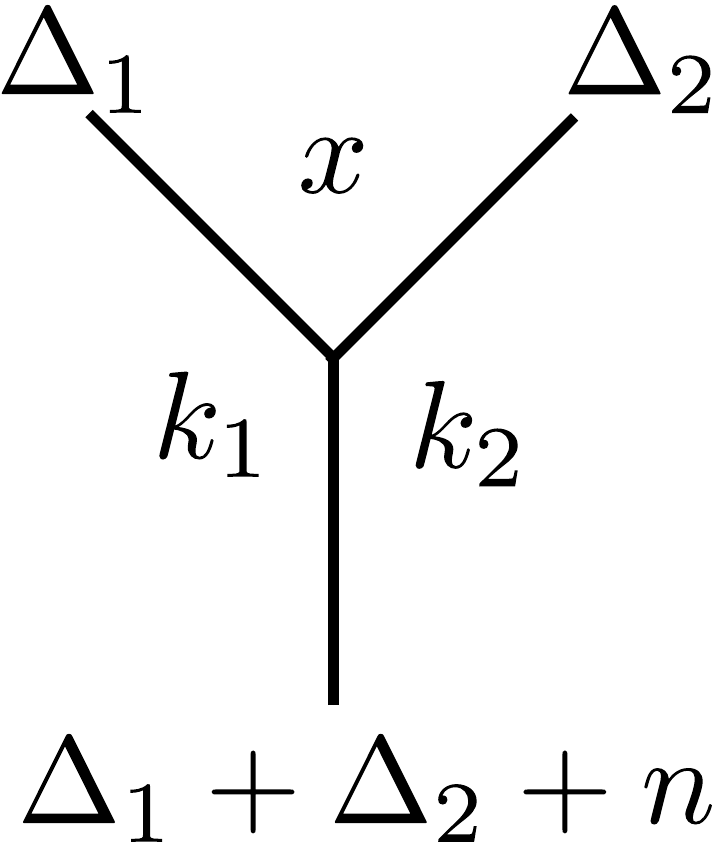} } \ .
\label{eq:A8}
\end{align}
We often denote the bottom line by $n$ instead of $\Delta_1 + \Delta_2 + n$ for brevity.

\subsection{Wilson function}

The Wilson function is a certain analytic continuation of the Wilson polynomial in its degree. It is defined by \cite{Groenevelt2003}
\begin{align}
{\cal W}_\lambda(x; a,b,c,d)  \equiv & \phi_\lambda(x; a,b,c,1-d) \\
= &
{1\over \Gamma(a+b)\Gamma(a+c)\Gamma(a+d) }
{\Gamma(g+a + i \lambda) \over \Gamma(g-a-i\lambda)  \Gamma(g+i\lambda \pm i x)} \\
& W(g+a-1 + i\lambda; a+ix,a-ix, \wt a+i\lambda, \wt b + i\lambda, \wt c + i \lambda) \ ,
\end{align}
where a ``very-will poised'' hypergeometric series is
\begin{align} 
W(a;b,c,d,e,f) \equiv & \, _7F_6 \left(\begin{array}{ccccccc}
		a, & 1 + \frac{a}{2}, & b, & c, & d, & e, & f \\
		\frac{a}{2}, & 1+a-b, & 1+a-c, & 1+a-d, & 1+a-e, & 1+a-f & 
	\end{array} \, ; \, 1
	\right)  \ .
\end{align}
The ``dual'' parameters are defined by
\begin{align}
g &= {1\over 2}\sum_{i=1}^4 a_i = {1\over 2}\sum_{i=1}^4 \wt a_i \ , \\ 
\wt a_1 &= g-a_4 \ , \qquad \wt a_4 = g-a_1  \ , \\ 
\wt a_2 &=  g-a_3  \ , \qquad \wt a_3 = g-a_2  \ .
\end{align}
Here and below we set $a_1 = a, a_2 = b, a_3 = c, a_4 = d$. We will use $a_1,a_2, a_3, a_4$ and $a,b,c,d$ interchangeably.

Two more representations that follow from some non-trivial identities of hypergeometric functions are (see equation (3.3) and Proposition 4.4 in \cite{Groenevelt2003})
\begin{align}\label{WilsonFunction}
&{\cal W}_\lambda(x; a,b,c,d) \nn  \\
&={\Gamma(d - a) \over  \Gamma(a+b)\Gamma(a+c)\Gamma(d \pm i x)\Gamma(\wt d \pm i \lambda)}
~\tensor[_4]{F}{_3} \left(  {a + ix , a- i x, \wt a + i \lambda , \wt a - i \lambda \atop a+b, a+c, a-d+1} ; 1 \right)
+ (a \leftrightarrow d) \ , \\
&{\cal W}_\lambda(x,a_1, a_2, a_3, a_4) = \nn \\
&{\Gamma(-2i\lambda) \over \Gamma( g + i \lambda \pm i x) \prod_{i=1}^4 \Gamma(\wt a_i - i \lambda)}
~\tensor[_4]{F}{_3} \left(  {\wt a_1 + i\lambda , \wt a_2+ i \lambda, \wt a_3+ i \lambda ,\wt a_4+ i \lambda \atop  g + i \lambda + i x, g + i \lambda - i x, 1+2i\lambda} ; 1 \right) + (\lambda \to -\lambda) \ , 
\label{Wrep2}
\end{align}
The representations \eqref{WilsonFunction}, \eqref{Wrep2} are useful for writing the Wilson function as contour integrals, as we will discuss below. The equation \eqref{Wrep2} is also useful for deriving asymptotics of the Wilson function (and related 6j-symbol) at large $x$.

\bigskip

The Wilson function has the following properties:

\bigskip

$\bullet$ ${\cal W}_\lambda(x;a,b,c,d)$ is symmetric in $a,b,c,d$. 

$\bullet$ It satisfies a duality ${\cal W}_\lambda(x;a,b,c,d) = {\cal W}_x(\lambda; \wt a, \wt b, \wt c, \wt d)$.

\bigskip 

For $\lambda = \pm i (\wt a +n)$ the second term in \eqref{WilsonFunction} vanishes because of $\Gamma(\wt a + i \lambda)$ in the denominator, while the first term reduces to the Wilson polynomial up to gamma-function factors.

\subsection{$\mathfrak{sl}(2, {\mathbb R})$ 6j-symbol}
The 6j-symbol used in the main text is defined by 
\begin{align}
\left\{ 
\begin{matrix}
\Delta_1 & k_1 & x \\ 
\Delta_2 & k_2 & \lambda
\end{matrix}
\right\}
& = 
(\gamma^{\Delta_1}_{1x} \gamma^{\Delta_1}_{2\lambda} \gamma^{\Delta_2}_{1\lambda} \gamma^{\Delta_2}_{2x} )^{1/2} 
~ {\cal W}_\lambda(x ; \Delta_1\pm i k_1, \Delta_2 \pm i k_2) \\
& = \prod_{i=1}^4( \Gamma(a_i \pm i x)\Gamma(\wt a_i \pm i \lambda) )^{1/2} ~ 
{\cal W}_\lambda(x;a,b,c,d) \ ,
\label{6j}
\end{align}
where in the first line by $\Delta \pm i k$ we again mean that ${\cal W}_\lambda$ depends on both and the order is not important due to symmetry in $a,b,c,d$. The parameters are chosen as in \eqref{abcdJT}. We also defined 
\begin{align}
\gamma^{\Delta}_{12} = \Gamma(\Delta \pm i k_1 \pm i k_2) \ .
\end{align}
The dual parameters exchange $\Delta_1$ and $\Delta_2$ 
\begin{align}\label{abcdJT2}
a =  d^*=  \Delta_1 + i k_1 \ , \quad b = c^* = \Delta_2 + i k_2 \ , \\
\wt a = {\wt d}^* = \Delta_2 + i k_1, \qquad \wt b = {\wt c}^* = \Delta_1 +i k_2 \ . 
\label{abcdJT3}
\end{align}
Pictorially, we represent the 6j-symbol as 
\begin{equation}
\left\{ 
\begin{matrix}
\Delta_1 & k_1 & x \\ 
\Delta_2 & k_2 & \lambda
\end{matrix}
\right\}   
~~~=~~~ 
\raisebox{-.35in}{ \includegraphics[scale=.35]{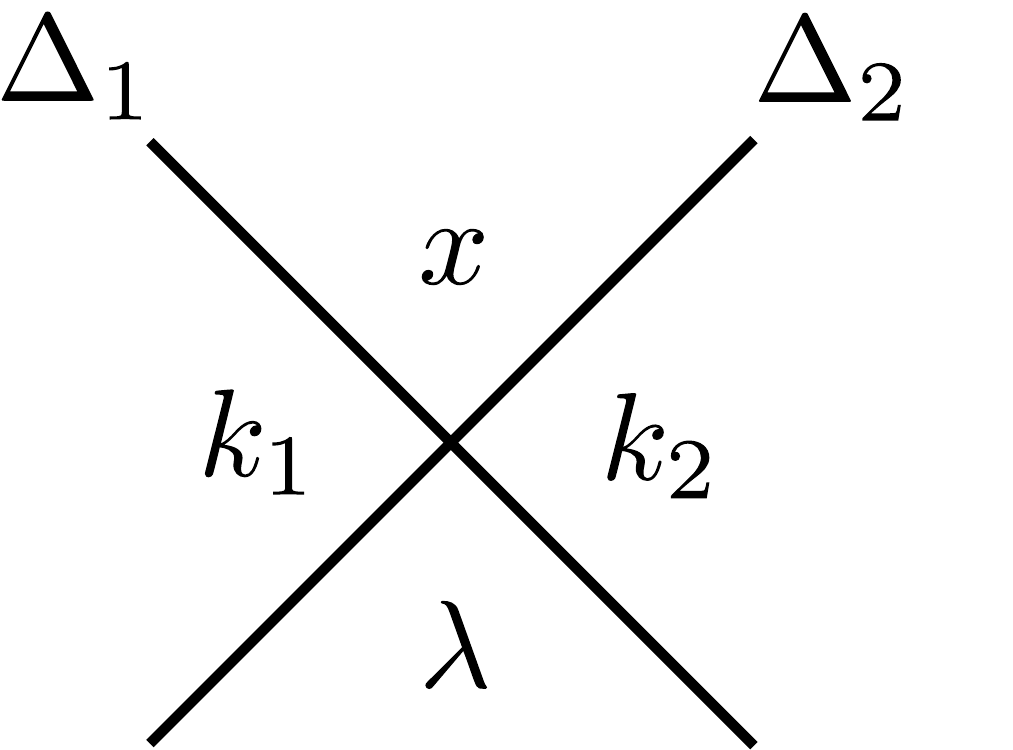} }
\end{equation}
This is a 6j symbol associated to the triple tensor product $\pi^+ \otimes \pi^P \otimes \pi^+$, where $\pi^+$ is a positive discrete series representation and $\pi^P$ is a principal series representation of $\mathfrak{sl}(2,\mathbb{R})$. See \cite{Groenvelt2005} for details.

Using the representation \eqref{Wrep2} we can derive the asymptotic at large $x$. In this limit the hypergeometric function drops out and we have
\begin{align}
\left\{ 
\begin{matrix}
\Delta & k_1 & x \\ 
\Delta & k_2 & \lambda
\end{matrix}
\right\}    \approx 
{2\pi \over x} e^{-\pi x} x^{2i \lambda} \Gamma(2i\lambda) \prod_{j=1}^4 \left( \Gamma(\wt a_j-i\lambda) \over \Gamma(\wt a_j + i\lambda ) \right)^{1/2} + (\lambda \to -\lambda) \ , \qquad (x \to \infty) \ .
\end{align}

\subsection{Properties of the 6j-symbol}

\label{sec:properties6jsymbol}

The Wilson function transform (of type I) is defined in \cite{Groenevelt2003}
\begin{align}
({\cal F} f)(\lambda) = \int_0^\infty {dx \over 2\pi} M(x) {\cal W}_\lambda(x) f(x) \ .
\end{align}
The Wilson function transform of the Wilson polynomial is the Wilson polynomial with dual parameters up to a sign
\begin{align}
({\cal F} W_n)(\lambda) &= (-1)^n  \wt W_n(\lambda) \ ,
\label{WilsonTransW}
\end{align}
where $\wt W_n(x;a,b,c,d) := W_n(x; \wt a, \wt b, \wt c, \wt d)$. This can be written in terms of the 6j-symbol and rescaled polynomials $P_n$
\begin{align}\label{WtransW}
\int_0^\infty {dx \over 2\pi \Gamma(\pm 2i x)} ~ \left\{ 
\begin{matrix}
\Delta_1 & k_1 & x \\ 
\Delta_2 & k_2 & \lambda
\end{matrix}
\right\}
P_n^{\Delta_1, \Delta_2}(x) = 
(-1)^n P_n^{\Delta_2, \Delta_1}(\lambda) \ .
\end{align}
Note that in the RHS $\Delta_1, \Delta_2$ are exchanged, which is how duality acts on the JT parameters \eqref{abcdJT3}. Since $P_n$ are a full set of orthonormal functions, \eqref{WtransW} implies
\begin{align}\label{6jdiag}
\left\{ 
\begin{matrix}
\Delta_1 & k_1 & x \\ 
\Delta_2 & k_2 & \lambda
\end{matrix}
\right\}
= \sum_{n=0}^\infty (-1)^n  P_n^{\Delta_1, \Delta_2}(x) P_n^{\Delta_2, \Delta_1}(\lambda) \ .
\end{align}
As pointed out in \cite{Groenevelt2003}, this is not actually convergent. We assume that it can be thought of as an identity of distributions and converges after integrating with a test function. We represent \eqref{6jdiag} pictorially 
\begin{align}\label{6jdiagFig}
\raisebox{-.35in}{ \includegraphics[scale=.35]{figures/6j2.pdf} }
=
\sum_{n=0}^\infty ~(-1)^n ~~
\raisebox{-.75in}{ \includegraphics[scale=.35]{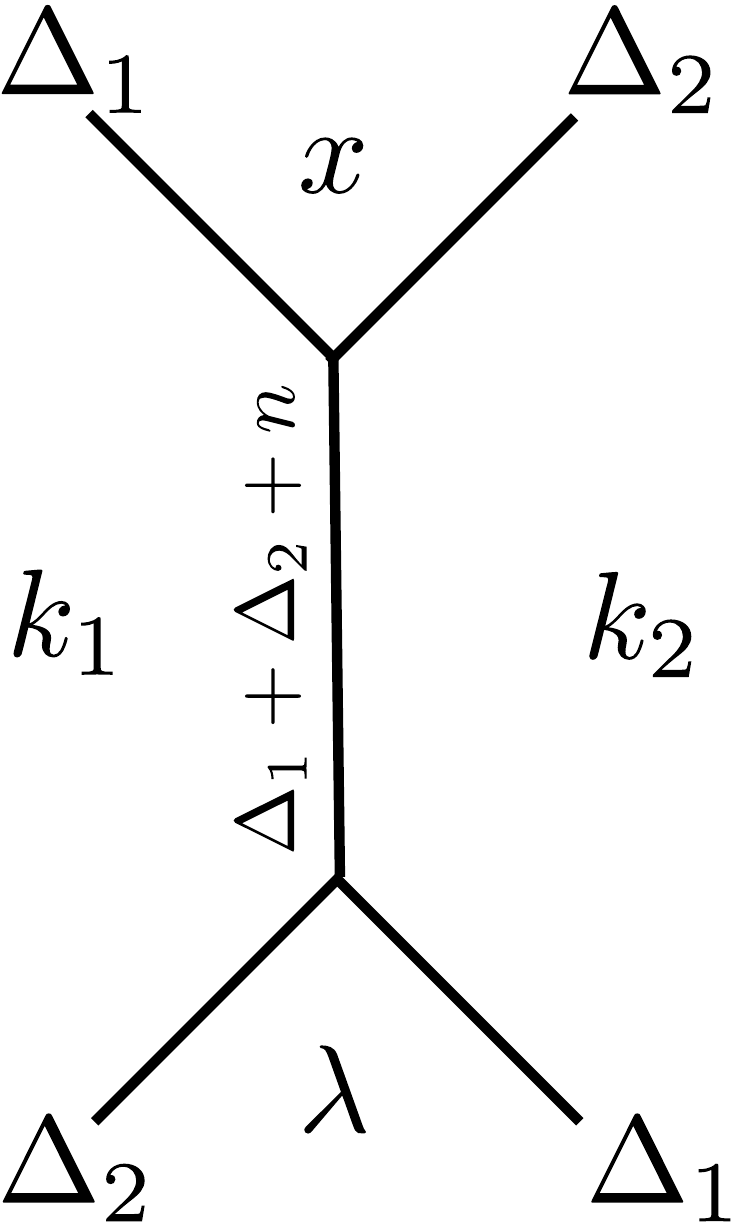} } \ .
\end{align}
An analogous equation can be written for the resolution of identity
\begin{align}\label{IdRes}
2\pi \Gamma(\pm 2i x) \delta(x-\lambda) &= \sum_{n=0}^{\infty} P_n^{\Delta_1, \Delta_2}(x) P_n^{\Delta_1, \Delta_2}(\lambda) \ , \\
\raisebox{-.35in}{\includegraphics[scale=.35]{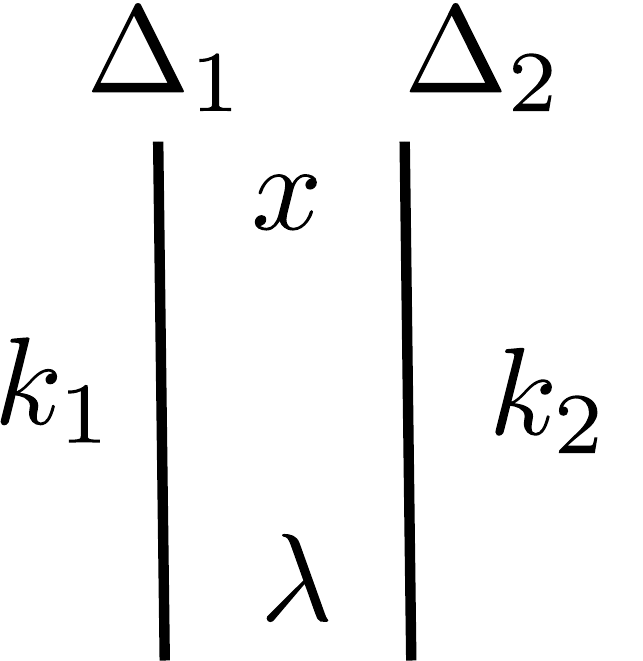}} 
~~ &=  ~~
\sum_{n=0}^\infty ~~~~~
\raisebox{-.85in}{\includegraphics[scale=.35]{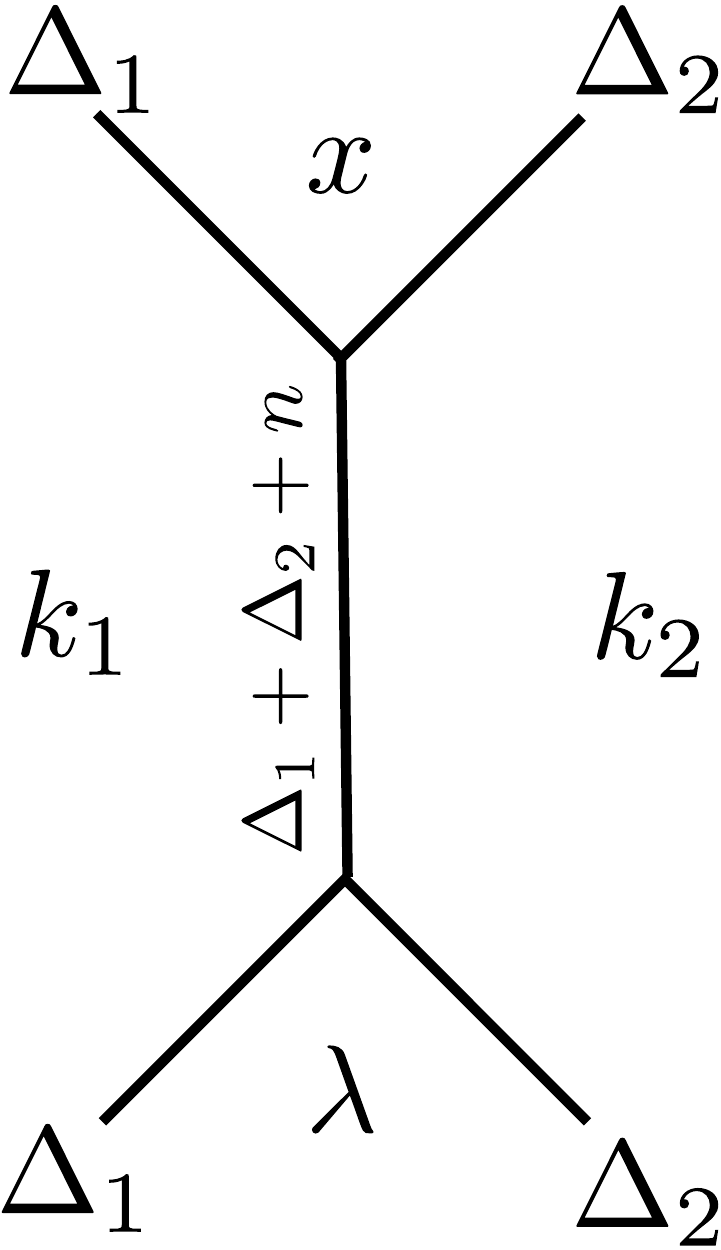} }  \ .
\label{IdResPic}
\end{align}
Here, note that $\Delta_1, \Delta_2$ are interchanged in the bottom part of the diagram in comparison to \eqref{6jdiagFig}. Furthermore, note that each term appearing in the sum on the right hand side of \eqref{6jdiagFig} or \eqref{IdResPic} may be interpreted as a conformal block (dressed with the Schwarzian mode) that corresponds to the exchange of a primary with dimension $\Delta_1 + \Delta_2 + n$ in a four-point function. For the simple case where $\Delta_1 = \Delta_2$, we verify this statement in appendix \ref{sec:confblocks}.

The ``pentagon identity'' can be derived by analytically continuing the formula at the top of p.33 in \cite{Groenvelt2005}
\begin{align}\label{Pent1}
\raisebox{-.35in}{\includegraphics[scale=.35]{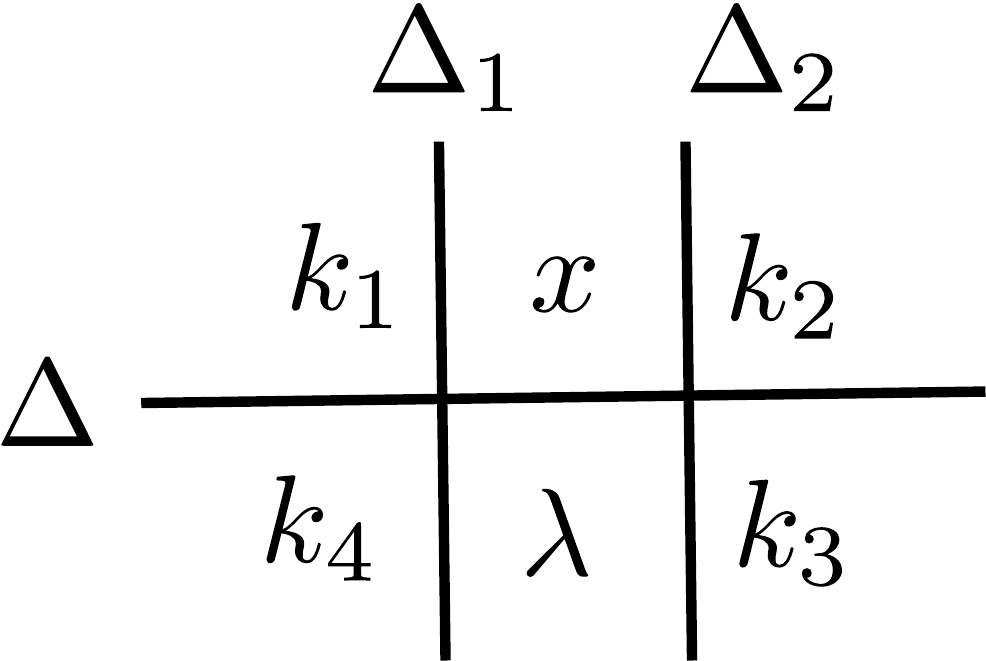}}
~~~=~~~
\sum_{n=0}^\infty  ~~~
\raisebox{-.8in}{\includegraphics[scale=.35]{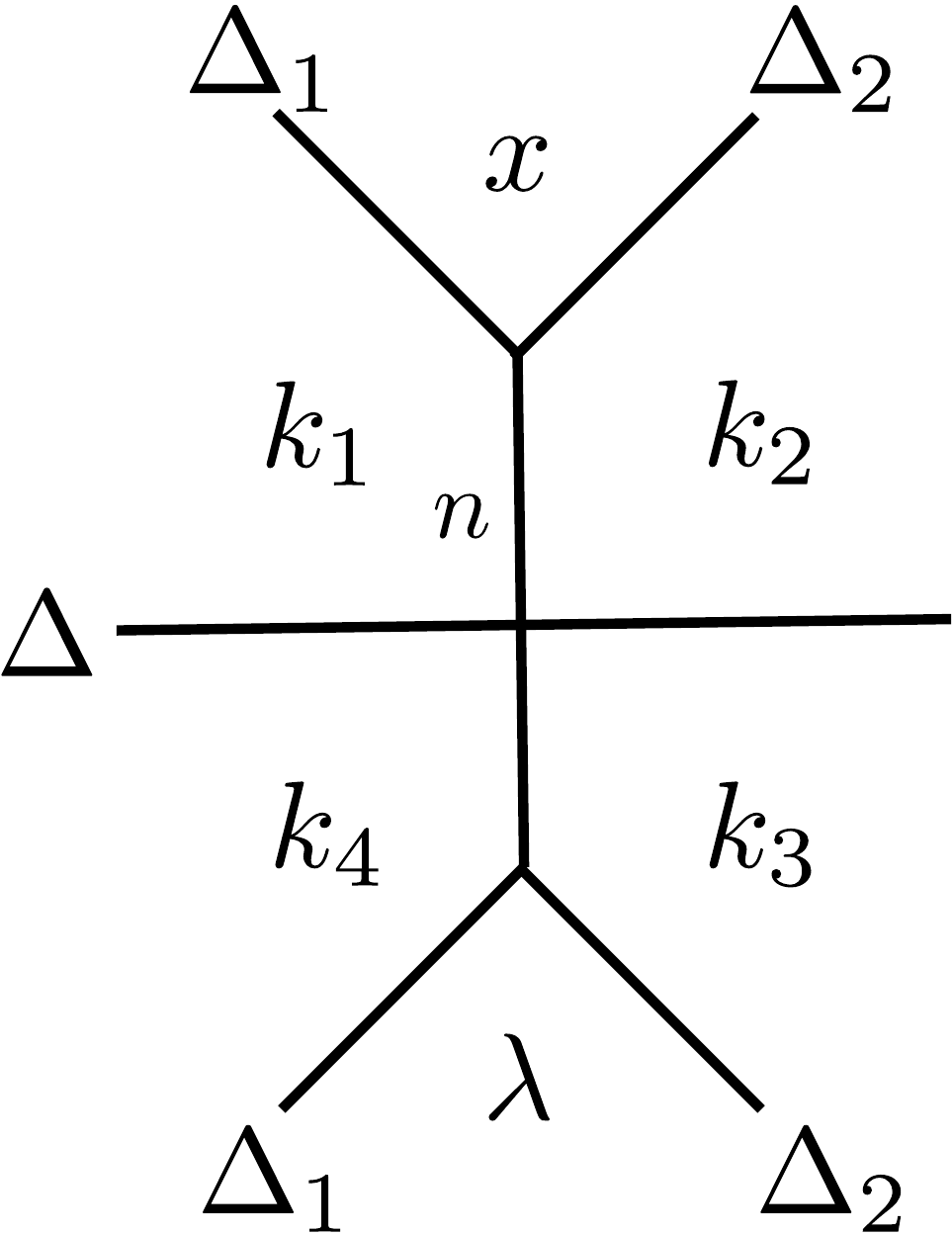}} \ .
\end{align}
This is essentially the identity resolution \eqref{IdResPic} with an additonal horizontal line $\Delta$. But written out explicitly it contains a product of two 6j-symbols in the LHS and one 6j-symbols and two Wilson polynomials $P_n$ in the RHS. One can integrate this with the Wilson polynomial $P_n$ (pictorially, attach it e.g. in the bottom), use orthogonality of Wilson polynomials in the RHS, and obtain another useful relation (Theorem 7.5 in \cite{Groenvelt2005})
\begin{align}\label{Pent2}
\raisebox{-.5in}{\includegraphics[scale=.35]{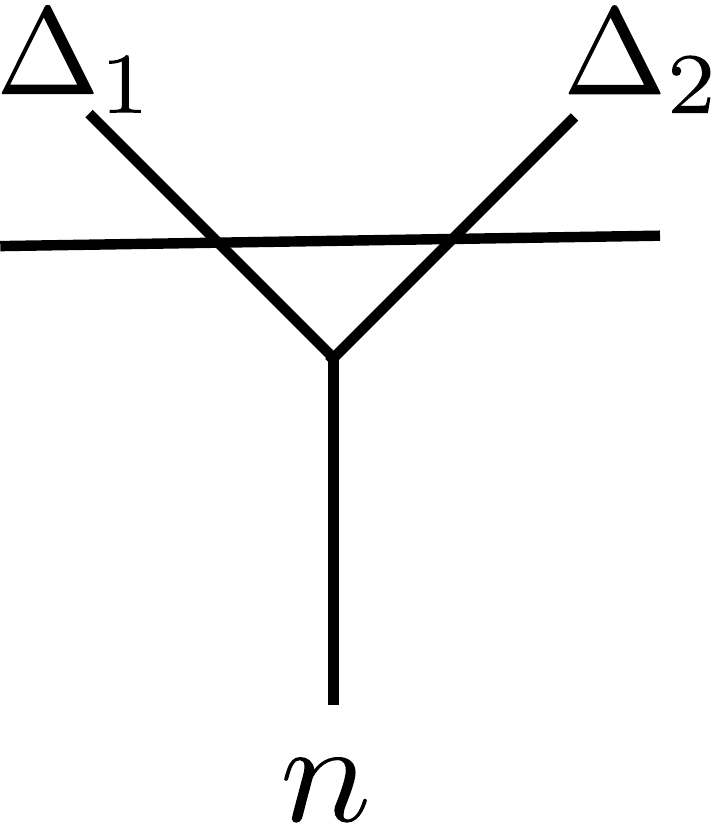} }
~~~=~~~
\raisebox{-.5in}{\includegraphics[scale=.35]{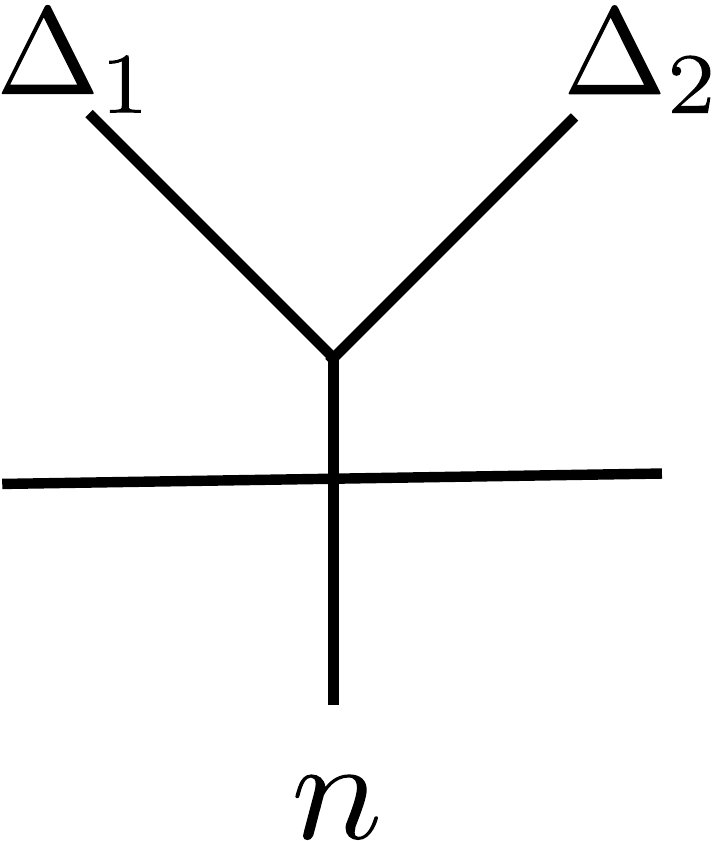} } \ ,
\end{align}
where we omitted the energy labels of regions. They are the same on both sides, except for the loop in the LHS which represents an integral with the Schwarzian measure $\int_0^\infty{dx \over 2\pi \Gamma(\pm 2i x)}$. 

The Yang-Baxter equation \eqref{eq:YB} can be derived using \eqref{6jdiagFig}, \eqref{Pent2}
\begin{align}
\raisebox{-.3in}{\includegraphics[scale=.2]{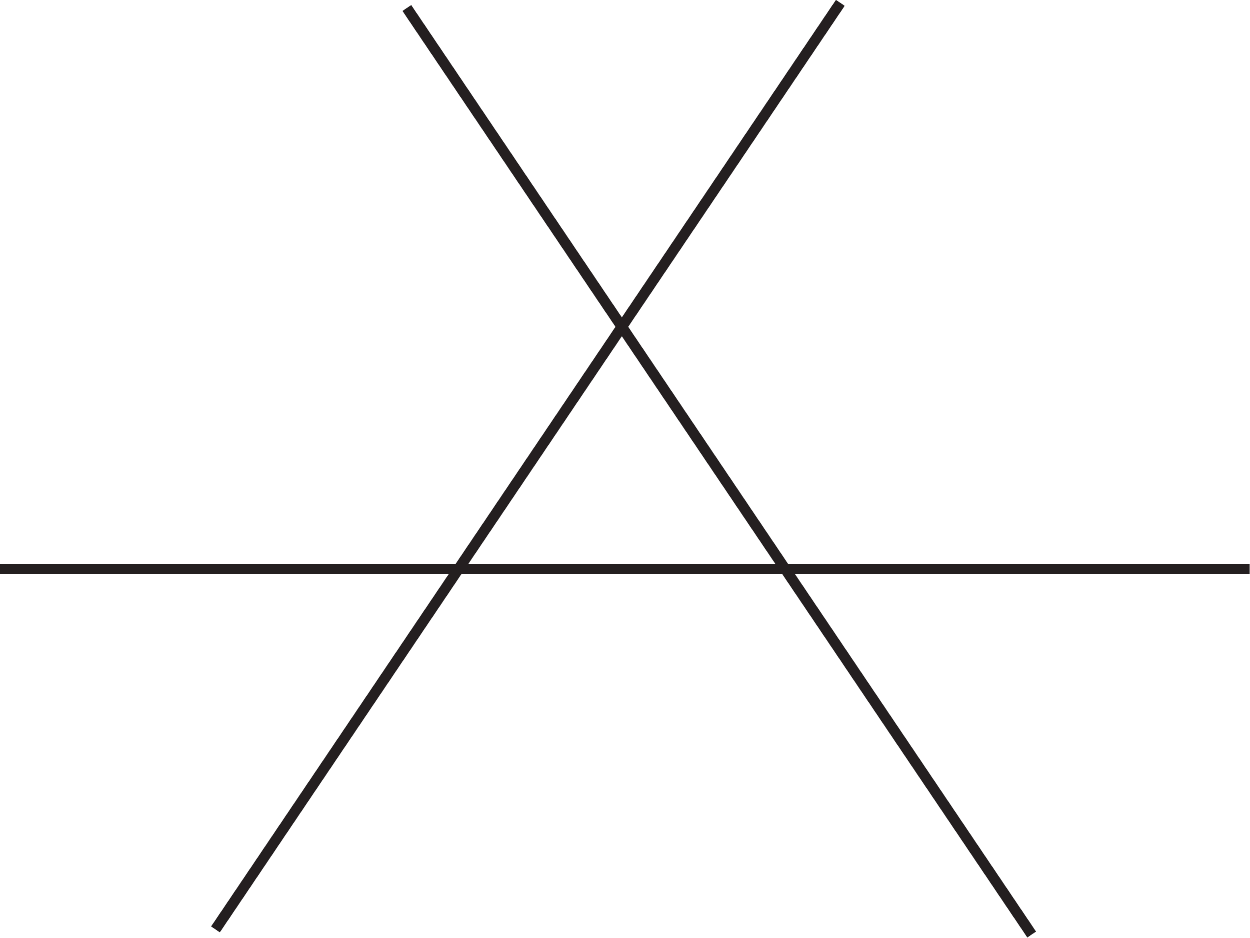}}
~~&=~~
\sum_{n=0}^\infty (-1)^n ~~
\raisebox{-.3in}{\includegraphics[scale=.2]{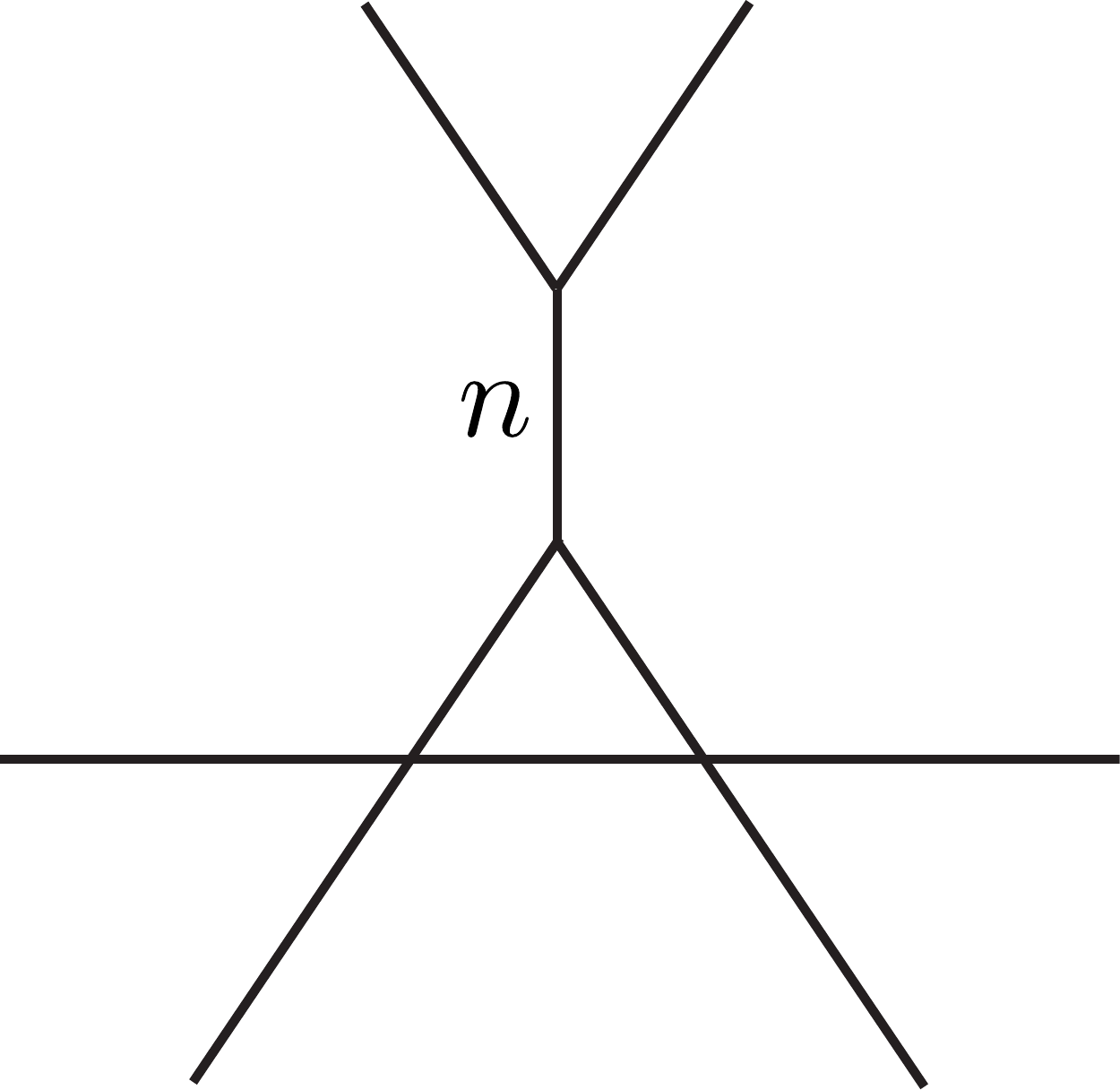}}
~~=~~\sum_{n=0}^\infty (-1)^n ~~
\raisebox{-.3in}{\includegraphics[scale=.2]{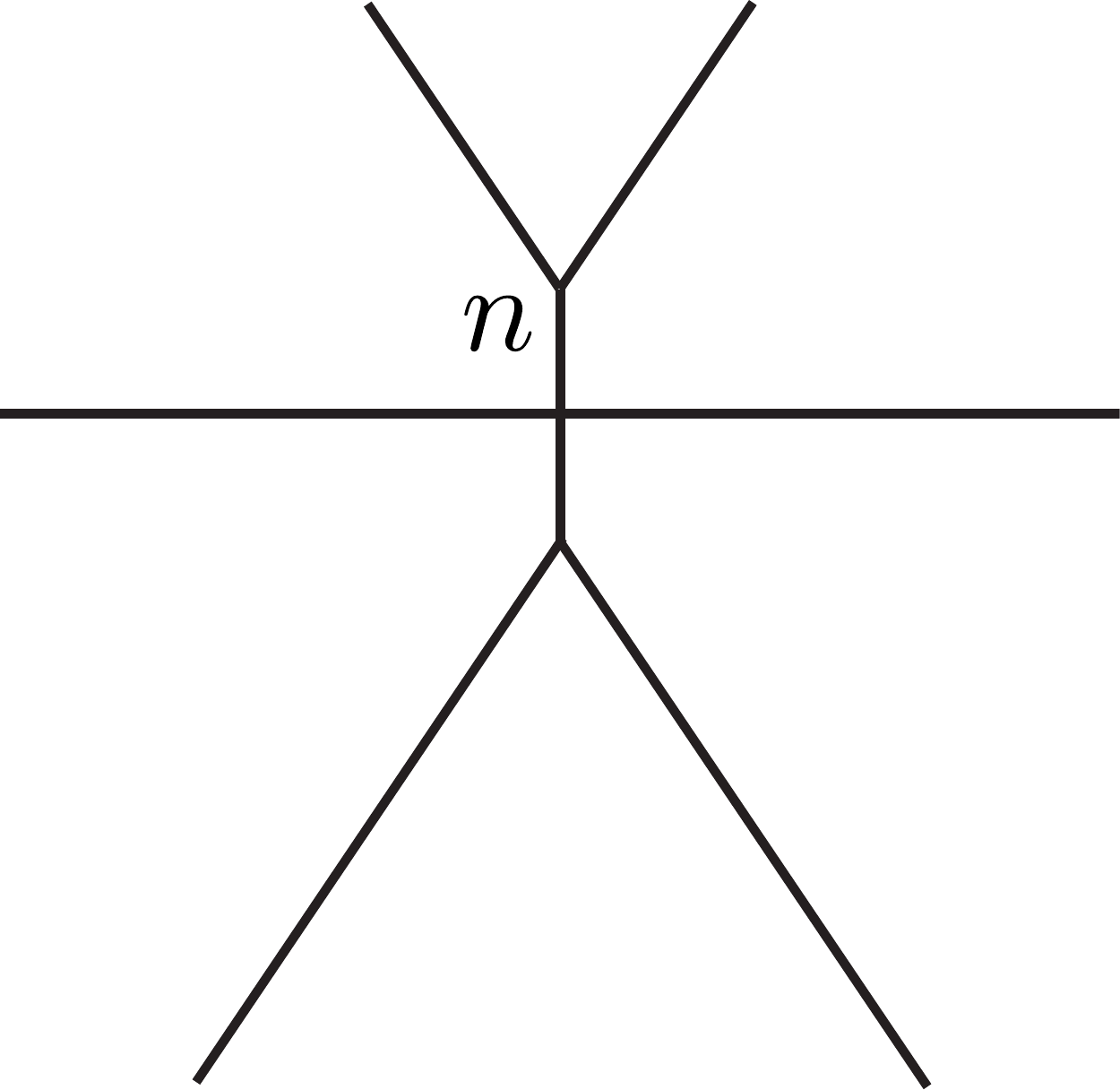}}\\
&=~~\sum_{n=0}^\infty (-1)^n ~~
\raisebox{-.3in}{\includegraphics[scale=.2]{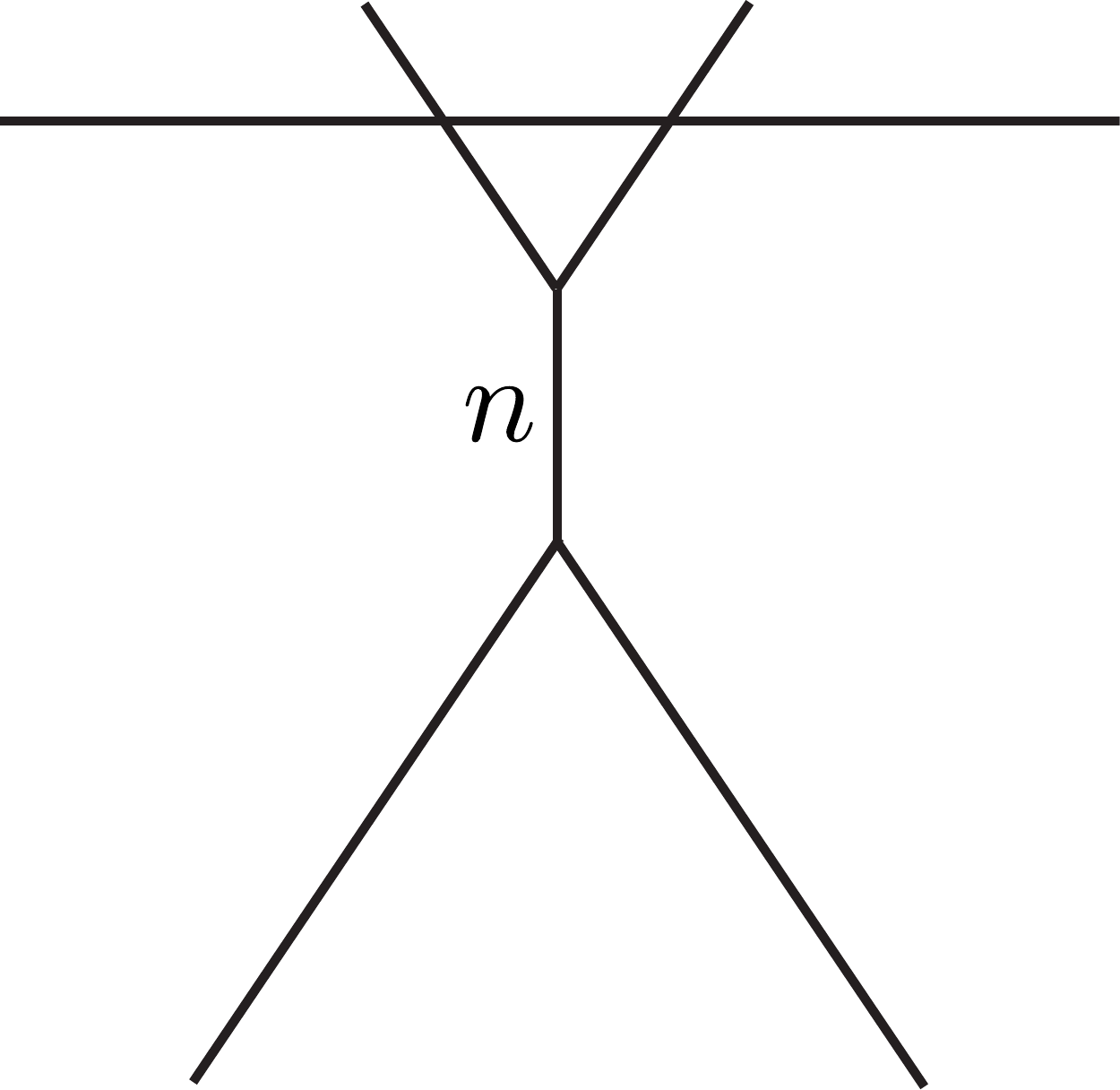}}
~~=~~
\raisebox{-.3in}{\includegraphics[scale=.2]{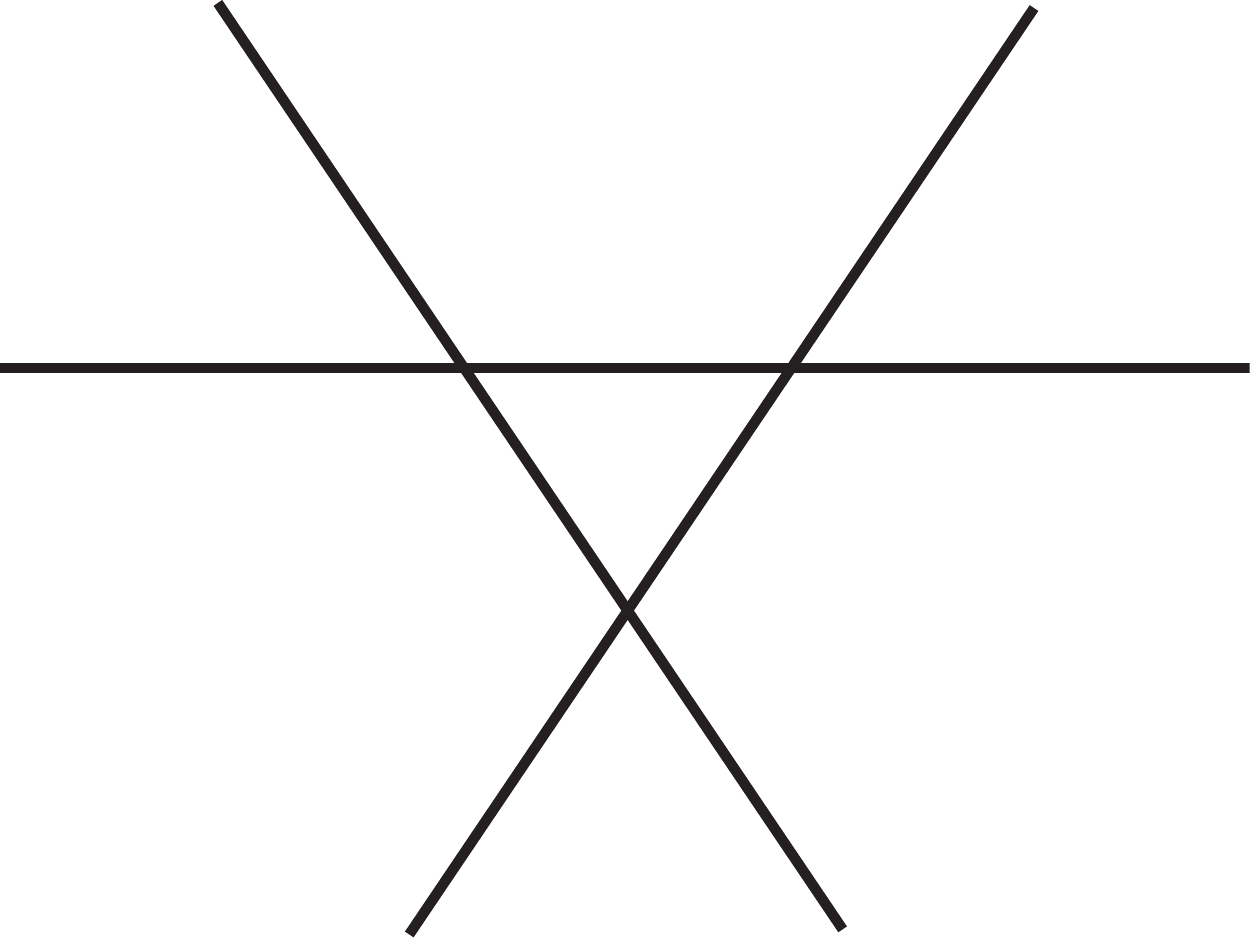}} \ .
\end{align}

\subsection{Contour integral representation}
\label{sec:6ju-int}
A useful representation of the 6j-symbol was discussed in appendix B of \cite{Mertens:2017mtv}
\begin{align}\label{6juint1}
\left\{ 
\begin{matrix} 
\Delta_1 & k_1 & k_2 \\  
\Delta_2 & k_3 & k_4 
\end{matrix} 
\right\} = 
e^{i\phi/2} \int_{ -i\infty}^{  i \infty} {du \over 2\pi i} ~ f(u)
\ ,
\end{align}
where
\begin{align}
f(u) &= \Gamma(u - {i\over 2} k_{1+3} \pm i k_2) 
\Gamma(u + {i\over 2} k_{1+3} \pm i k_4) 
{\Gamma(\Delta_1-u  - {i\over 2} k_{1-3})
\Gamma(\Delta_2-u  + {i\over 2} k_{1-3})
\over 
\Gamma(\Delta_1+u  + {i\over 2} k_{1-3})
\Gamma(\Delta_2+u  - {i\over 2} k_{1-3})} \ , \\
e^{i\phi} &= 
{
\Gamma(\Delta_1 + i k_1 \pm i k_2)
\Gamma(\Delta_1 - i k_3 \pm i k_4)
\Gamma(\Delta_2 - i k_1 \pm i k_4)
\Gamma(\Delta_2 + i k_3 \pm i k_2)
\over
\Gamma(\Delta_1 - i k_1 \pm i k_2)
\Gamma(\Delta_1 + i k_3 \pm i k_4)
\Gamma(\Delta_2 + i k_1 \pm i k_4)
\Gamma(\Delta_2 - i k_3 \pm i k_2)
}
\ .
\end{align}
We also used a short notation $k_{1+3} = k_1 + k_3, ~ k_{1-3} = k_1 - k_3$. This integral is related to the equation (B.28) in \cite{Mertens:2017mtv} by a shift of $u$. The integrand contains semi-infinite sequences of poles. The contour is such that poles of $\Gamma(u+\dots)$ are to the left of the contour and poles of $\Gamma(-u+\dots)$ are to the right. Closing the contour to the right and picking up the poles one obtains a sum of two $\tensor[_4]{F}{_3}$'s related to the representation \eqref{WilsonFunction} of the Wilson function.

The integral simplifies significantly in a special case $\Delta_1 = \Delta_2 = \Delta, k_3 = -k_1$. In this case the phase factor is absent $e^{i\phi} = 1$ and (after relabeling)
\begin{align}\label{6j124u}
\left\{ 
\begin{matrix} 
\Delta & k & k_1 \\  
\Delta & k & k_2 
\end{matrix} 
\right\} 
=
\int_{\e -i\infty}^{\e + i \infty} {du \over 2\pi i} ~ 
\Gamma(u \pm i k_1) \Gamma(u \pm i k_2)
{
\Gamma(\Delta - u \pm i k)
\over 
\Gamma(\Delta + u \pm i k)
} \ .
\end{align}
We also used that the 6j-symbol is an even function of $k_j$. For the computation of the geodesic on the pair of pants in section \ref{sec:PairOfPants}, it is convenient to take the Fourier transform of \eqref{6j124u} in $k, k_1, k_2$. We note that
\begin{align}
\Gamma(u \pm i k ) &= \Gamma(2u) \int_0^\infty db ~ {2\cos(bk) \over \left( 2\cosh {b\over 2} \right)^{2u}} \\
{\Gamma(x+i k) \over \Gamma(y+ik) } &= {1\over \Gamma(y-x)} \int_0^\infty db ~ e^{-i bk} e^{-{b\over 2} (x+y-1)} \left(2\sinh {b\over 2}\right)^{y-x-1} \ .
\end{align}
It is somewhat easier to derive the inverse Fourier of these formulas. Then one has integrals of gamma-functions that can be computed by deforming the contour and picking up residues at poles of the gamma-functions. Applying these expressions to $\Gamma(u \pm i k_1) , \Gamma(u \pm i k_2) , {\Gamma(\Delta - u + i k)
\over 
\Gamma(\Delta + u + i k)
}, 
{
\Gamma(\Delta - u - i k)
\over 
\Gamma(\Delta + u - i k)
}$ we find from \eqref{6j124u}
\begin{align}
\left\{
\begin{matrix}
\Delta & k & k_1 \\ 
\Delta & k & k_2 
\end{matrix}
\right\}
=
\int_0^\infty db_1 db_2 db db' ~ & \cos(b_1k_1) \cos(b_2k_2) e^{ik(b-b')} \\
 &  {e^{-\left( \Delta - {1\over 2} \right) (b+b')} \over \sinh{b\over 2} \sinh {b'\over 2} }
 \int_{-i\infty}^{i\infty} {du\over 2\pi i } \left( \sinh{b\over 2} \sinh {b'\over 2} \over \cosh {b_1\over 2} \cosh {b_2 \over 2} \right)^{2u}  \nn \\
  = {1\over 2}\int_0^\infty db_1 db_2 db db' ~  &\cos(b_1k_1) \cos(b_2k_2) e^{ik(b-b')} e^{-\left( \Delta - {1\over 2} \right) (b+b')} \\ 
& \delta\left( \sinh{b\over 2} \sinh {b'\over 2} - \cosh {b_1\over 2} \cosh {b_2 \over 2}  \right) \ .
\end{align}
The $u$-integral resulted in a delta function. Now we change variables to $b_{\pm} = b \pm b'$ and integrate out $b_+$ using the delta function. We have 
\begin{align}
&\int_0^\infty db db'  = {1\over 2} \int_{-\infty}^\infty db_- \int_{|b_-|}^\infty db_+ \ , \\
& \delta\left( \sinh{b\over 2} \sinh {b'\over 2} - \cosh {b_1\over 2} \cosh {b_2 \over 2}  \right) = {4\over \sinh{B\over 2}}\delta(b_+ - B) \ , 
\end{align}
where $B$ is defined by 
\begin{align}
\cosh{B \over 2}  = \cosh{b_- \over 2} + 2 \cosh{b_1 \over 2} \cosh{b_2 \over 2} \ .
\end{align}
Note that this implies $B > |b_-|$, assuming $B>0$. So $B$ is in the integration range $b_+ \in (|b_-| , \infty)$. Therefore integrating out $b_+$ leads to 
\begin{align}\label{6jb-int}
\left\{
\begin{matrix}
\Delta & k & k_1 \\ \Delta & k & k_2 
\end{matrix}
\right\}
= 
2\int_0^\infty db_1 db_2 db_- ~ \cos(b_1 k_1) \cos (b_2 k_2) \cos(b_- k) ~ { e^{-(\Delta-{1\over 2}) B} \over \sinh{B \over 2}} \ .
\end{align}

\bigskip

Finally, we give yet another integral representation of the 6j-symbol that looks a bit simpler than \eqref{6juint1}, though we do not use it in the main text. It is
\begin{align}\label{6juint2}
\left\{
\begin{matrix}
\Delta_1 & k_1 & x \\ 
\Delta_2 & k_2 & \lambda 
\end{matrix}
\right\}
=
\left( \gamma_{2x}^{\Delta_2} \gamma_{2\lambda}^{\Delta_1}
\over 
 \gamma_{1x}^{\Delta_1} \gamma_{1\lambda}^{\Delta_2}
 \right)^{1/2}
 \int_{-i\infty}^{i\infty} {du \over 2\pi i }~
{ \Gamma(\Delta_1 \pm ix +u) \Gamma(\Delta_2  \pm i \lambda +u) \over \Gamma(\Delta_1 + \Delta_2 \pm i k_2 +u) } ~\Gamma( \pm i k_1 -u)  \ .
\end{align}
Closing the contour to the right or, equivalently, using formula (B.18) in \cite{Mertens:2017mtv}, we obtain the sum of two $\tensor[_4]{F}{_3}$'s that is related to the representation of the Wilson function \eqref{Wrep2}. The integrand doesn't obey all symmetries of the 6j-symbol, so applying symmetries of the 6j-symbol we can get different integral representations.

\subsection{$q$-deformation}

We now discuss the $q$-deformed version of Wilson polynomials and functions with $0< q < 1$. First, the q-Pochhammer symbol is defined by 
\begin{align}
(A;q)_k \equiv \prod_{n = 0}^{k-1}(1 - A q^n) \ , \qquad
(A_1,\ldots,A_p;q)_k \equiv \prod_{i = 1}^p (A_i;q)_k \ .
\end{align}
Many of the $q$-deformed functions that we duscuss below can be reduced back to the undeformed case in the limit $q\to 1$ using 
\begin{align}\label{qto1}
{1-q^a \over 1-q} \to a \ , \qquad 
{(q^a;q)_n \over (1-q)^n} \to (a)_n = \prod_{j=0}^{n-1} (a+j) \ .
\end{align}
The q-hypergeometric series (often called ``basic hypergeometric'') is defined by\footnote{For the more general case $\tensor[_r]{\phi}{_p}$ see e.g. \cite{GasperRahman}}
\begin{align}
\tensor[_{r+1}]{\phi}{_r}
\left( 
{A_1 , \dots , A_{r+1}
\atop 
B_1, \dots , B_r}
; q,x
\right)
= \sum_{n=0}^\infty { (A_1 , \dots, A_{r+1} ; q)_n \over (B_1, \dots, B_r; q)_n } {x^n \over (q;q)_n} \ .
\end{align}
Using \eqref{qto1}, in the limit $q\to 1$ the basic hypergeometric series reduces to the usual hypergeometric function
\begin{align}
    \tensor[_{r+1}]{\phi}{_r}
\left( 
{q^{a_1} , \dots , q^{a_{r+1}}
\atop 
q^{b_1}, \dots , q^{b_r}}
; q,x
\right) ~\to ~
\tensor[_{r+1}]{F}{_r}
\left( 
{a_1 , \dots , a_{r+1}
\atop 
b_1, \dots , b_r}
; x
\right) \ .
\end{align}
Another useful function is the q-Gamma function
\begin{align}
\Gamma_q(x) = (1-q)^{1-x} {(q;q)_\infty \over (q^x;q)_\infty } \ .
\end{align}
In the limit $q\to 1$ this becomes the gamma function $\Gamma(x)$.

\subsection{Askey-Wilson polynomial}

The Askey-Wilson polynomial is defined by 
\begin{align}
p_n(\cos \theta; A,B,C,D | q) = A^{-n} (AB,AC,AD;q)_n
~ \tensor[_4]{\phi}{_3}
 \left( 
 {q^{-n}, q^{n-1}ABCD, A e^{i \theta }, A e^{-i \theta} 
 \atop 
 AB,AC,AD}; q,q
 \right) \ .
\end{align}
They are degree $n$ polynomials in $\cos \theta$. They are also symmetric in $A,B,C,D$. Setting $A = q^a, B=q^b, C=q^c, D=q^d$ and taking $q\to 1$ we recover the Wilson polynomial \eqref{WilsonPolynomial} up to an overall factor 
\begin{align}
p_n({q^{ix} + q^{-ix} \over 2}; q^a,q^b,q^c,q^d | q) = (1-q)^{3n} W_n(x;a,b,c,d) + \dots \ , \qquad q\to 1 \ .
\end{align}
Here, we also introduced a variable $x$ by 
\begin{align}
e^{i \theta} = q^{-i x} \ , \qquad \theta = x \log {1\over q} \ .
\end{align}
We will sometimes denote the parameters by $A_1 = A, A_2 = B, A_3 = C, A_4 = D$. Askey-Wilson polynomials satisfy an orthogonality relation. Assuming $|A_j| < 1$ (otherwise there might be an extra contribution from a discrete set of points) we have
\begin{align}
  &\int_0^\pi {d\theta \over 2\pi }~ M_q(\theta)  p_n(\cos \theta) p_m(\cos \theta) = h_n \delta_{nm} \ , \\ 
  &h_n = {1 \over 1-ABCD q^{2n-1}} 
  {(ABCD q^{n-1}; q)_\infty \over (q^{n+1};q)_\infty } ~ \prod_{i<j} {1\over (A_i A_j q^n;q)_\infty } \ ,
\end{align}
where the integration measure is
\begin{align}
M_q(\theta) &= { (e^{\pm 2i \theta};q)_\infty \over \prod_{j=1}^4  (A_j e^{\pm i\theta}; q)_\infty } \\
& = { (1-q)^{2\sum_{j=1}^4 a_j-6} \over (q;q)_\infty^6 } ~~
{ \prod_{j=1}^4 \Gamma_q(a_j \pm i x) \over \Gamma_q(\pm 2i x) } \ .
\end{align}
As before, $\theta = x \log{1\over q}$ and $A_j = q^{a_j}$.

We also define orthonormal rescaled Askey-Wilson polyonomials
\begin{align}
P_n(x; a_1,a_2,a_3,a_4|q) =|\log  q|^{1/2}
(1-q)(q;q)_\infty \left(h_n \prod_{j=1}^4 (q^{a_j\pm i x} ;q)_\infty \right)^{-1/2}
p_n({q^{ix} + q^{-ix} \over 2}; q^{a_1},q^{a_2},q^{a_3},q^{a_4} | q) \ .
\end{align}
They are orthonormal with respect to the $q$-deformed Schwarzian measure
\begin{align}\label{eq:askeywilsonpolyorth}
\int_0^{\pi/|\log q|} dx~ \rho_q(x) ~ P_n(x) P_m(x) = \delta_{nm} \ ,
\end{align}
where 
\begin{align}
    \rho_q(x) = {1\over 2\pi \Gamma_q(\pm 2ix)} \ .
\end{align}

As in the undeformed case, the parameters relevant in JT gravity in the main text are chosen to be
\begin{align}\label{abcdJTq}
a = d^*=\Delta_1 +ik_1, \qquad b = c^* = \Delta_2 + i k_2 
\end{align}
and we define
\begin{align}
P_n^{\Delta_1,\Delta_2}(x;k_1,k_2|q) := 
P_n(x; \Delta_1 \pm i k_1, \Delta_2 \pm i k_2|q) \ . 
\end{align}

\subsection{Askey-Wilson function}

The Askey-Wilson function is defined by
\begin{align}\label{qdefWilson1}
{\cal W}_{\Lambda} (X;A,B,C,D|q) &\equiv \phi_\Lambda(X;A,B,C,q/D|q) \\
&{ ( AB,AC,AD, GA^{-1}\Lambda^{-1}, G\Lambda X^{\pm 1} ;q)_\infty
\over 
(GA\Lambda ;q)_\infty} \\
&\tensor[_8]{W}{_7}
(q^{-1}GA \Lambda  ; A X, AX^{-1}, \wt A \Lambda, \wt B \Lambda, \wt C \Lambda; q , \wt D \Lambda^{-1}) \ ,
\end{align}
where the very-well-poised $\, _8 \phi_7$ series is defined by
\begin{align} \label{eq:8w7} 
_8W_7(A;~A_1,\dots,A_5;~q,z) 
= 
\tensor[_8]{\phi}{_7} 
\left(
{A, q A^{1/2}, -q A^{1/2}, A_1, \dots, A_5 
\atop 
A^{1/2},-A^{1/2},q A/A_1, \dots, q A/A_5}
;q,z
\right) \ .
\end{align}
The ``dual'' parameters are defined as
\begin{align}
&G = (ABCD)^{1/2} = (\wt A \wt B \wt C \wt D)^{1/2} , \\
&\wt A = G/D, \qquad \wt D = G/A \ , \\
&\wt B = G/C , \qquad \wt C = G/B \ .
\end{align}
Another representation that generalizes \eqref{WilsonFunction} is \cite{Groenvelt2005}\footnote{See formula (8.15) in \cite{Groenvelt2005}. Our definition differs from that paper by $D \to q/D$, such that in our convention the Askey-Wilson function is symmetric in $A,B,C,D$.}
\begin{align}\label{qdefWilson2}
&{\cal W}_{\Lambda} (X;A,B,C,D|q) \\
=&
{(AB,AC,DX^{\pm 1}, \wt D \Lambda^{\pm 1};q)_\infty \over ({D/A};q)_\infty}
\tensor[_4]{\phi}{_3}
\left( 
{AX ,AX^{-1}, \wt A \Lambda, \wt A \Lambda^{-1}
\atop 
AB,AC,qA/D}
;q,q
\right) + (A \leftrightarrow D)  \ ,
\end{align}
One can also write down a representation that generalizes \eqref{Wrep2}, see \cite{Koelink}. The Askey-Wilson function ${\cal W}_\Lambda$ is symmetric in $A,B,C,D$ and satisfies ``duality''
\begin{align}
{\cal W}_{\Lambda} (X;A,B,C,D|q)
=
{\cal W}_{X} (\Lambda;\wt A,\wt B,\wt C,\wt D|q) \ .
\end{align}
It is also symmetric in $X, \Lambda$ (as clear from \eqref{qdefWilson2})
\begin{align}
    {\cal W}_{\Lambda^{\pm 1}} (X^{\pm 1};A,B,C,D|q) = 
    {\cal W}_{\Lambda} (X;A,B,C,D|q) 
\end{align}
for any choice of signs. To recover the Wilson function \eqref{WilsonFunction}, we set $A_j = q^{a_j}, X = q^{-ix}, \Lambda = q^{-i\lambda}$ and take $q \to 1$
\begin{align}
{\cal W}_{q^{-i\lambda}} (q^{-i x};q^a,q^b,q^c,q^d|q) = 
(1-q)^{5-2\sum_j a_j} (q;q)_\infty^5 ~~
{\cal W}_{\lambda} (x;a,b,c,d) + \dots 
\end{align}

\subsection{$q$-deformed 6j-symbol}

We define the $q$-deformed 6j-symbol by 
\begin{align}
\left\{
\begin{array}{ccc}
		\Delta_1 & k_1 & x \\
		\Delta_2 & k_2 & \lambda
\end{array}
\right\}_q
&=
|\log q | (1-q)^{-6 + 2\sum_j a_j} (q;q)_\infty^{-5}~~
\prod_{j=1}^4 (\Gamma_q(a_j \pm i x) \Gamma_q(\wt a_j \pm i \lambda))^{1/2}
~~{\cal W}_{q^{-i\lambda}} (q^{-ix};q^a,q^b,q^c,q^d|q) \\
&=|\log q|(1-q)^2 (q;q)_\infty^3 ~~ 
\prod_{j=1}^4
\left[
(q^{a_j \pm i x};q)_\infty (q^{\wt a_j \pm i \lambda};q)_\infty \right]^{-1/2}
~~{\cal W}_{q^{-i\lambda}} (q^{-ix};q^a,q^b,q^c,q^d|q) \ ,
\end{align}
where the parameters $a_j$ are chosen as in \eqref{abcdJTq}. In the limit $q\to 1$ we recover the undeformed 6j-symbol
\begin{align}
\left\{
    \begin{array}{ccc}
		\Delta_1 & k_1 & x \\
		\Delta_2 & k_2 & \lambda
\end{array}
\right\}_q ~~ \to ~~ 
\left\{
\begin{array}{ccc}
		\Delta_1 & k_1 & x \\
		\Delta_2 & k_2 & \lambda
\end{array}
\right\} \ .
\end{align}
The definition of the 6j-symbol looks unpleasant, but it satisfies many beautiful identities. For example, a $q$-deformed analog of \eqref{6jdiag} is \cite{Stokman}
\begin{equation}   \left\{\begin{array}{ccc}
			\Delta_1 & k_1 & x \\
			\Delta_2 & k_2 & \lambda
		\end{array}
		\right\}_q     
		= \sum_{n = 0}^\infty (-1)^n 
		~q^{n (\Delta_1 + \Delta_2) + \frac{n(n-1)}{2}} ~P^{\Delta_1,\Delta_2}_n(x;k_1,k_2|q) P^{\Delta_2,\Delta_1}_n(\lambda; k_1,k_2|q). 
	\label{eq:diagrelation}
\end{equation}

\subsubsection*{Orthogonality relation}

The Askey-Wilson function defines the Askey-Wilson transform \cite{Koelink} and satisfies an orthogonality relation that we express in terms of the 6j-symbols (though we don't use it in the main text)\footnote{These formulas are adopted from section 8.3 in \cite{Groenvelt2005}. We set the parameter $t$ in that paper to be $t = q^{B+1/2}$.}
\begin{align}\label{6jorthqB}
 \int_0^{\pi/|\log q|} dx ~ \rho_q(x) ~ 
{\Gamma_q\left( {1\over 2} \pm B \pm i x \right) 
\over 
\sqrt{C_q}
}~
&\left\{ 
\begin{matrix}
\Delta_1 & k_1 & x \\
\Delta_2 & k_2 & \lambda
\end{matrix}
\right\}_q
\left\{ 
\begin{matrix}
\Delta_1 & k_1 & x \\
\Delta_2 & k_2 & \lambda'
\end{matrix}
\right\}_q
+ \text{discrete}
\\
=& {\delta(\lambda - \lambda') \over \rho_q(\lambda)} ~ 
{ \sqrt{\wt C_q} \over \Gamma_q\left( {1\over 2} \pm \wt B \pm i \lambda \right) } \ ,
\end{align}
where
\begin{align}
C_q = \Gamma_q\left( {1\over 2} \pm (B + \Delta_1) \pm i k_1 \right)
\Gamma_q\left( {1\over 2} \pm (B + \Delta_2) \pm i k_2 \right) \ .
\end{align}
The dual parameters $\wt B, \wt C_q$ are defined by
\begin{align}
\wt B = - B - \Delta_1 - \Delta_2 \ , \qquad \wt C_q = C_q|_{B \to \wt B} \ .
\end{align}
The discrete part is
\begin{align}
\text{discrete}=&
{\Gamma_q\left( \Delta_1 \pm (B+{1\over 2} )  \pm i k_1 \right)
\Gamma_q\left( \Delta_2 \pm (B+{1\over 2} )  \pm i k_2 \right)
\over 
\sqrt{C}
}
\\
&\sum_{k > B+1/2 }
\left\{ 
\begin{matrix}
\Delta_1 & k_1 & x_k \\
\Delta_2 & k_2 & \lambda
\end{matrix}
\right\}_q
\left\{ 
\begin{matrix}
\Delta_1 & k_1 & x_k \\
\Delta_2 & k_2 & \lambda'
\end{matrix}
\right\}_q
\Gamma_q(\Delta_1 \pm i k_1 \pm i x_k)^{-1}\Gamma_q(\Delta_2 \pm i k_2 \pm i x_k)^{-1}
\\
&{(q^{-1/2-B+\Delta_1 \pm i k_1},q^{-1/2-B+\Delta_2 \pm i k_2} ;q)_k 
\over 
(q^{1/2-B-\Delta_1 \pm i k_1},q^{1/2-B-\Delta_2 \pm i k_2} ;q)_k }
~{1- q^{2k-2B-1} \over |\log q|} ~
q^{-4k(\Delta_1 + \Delta_2 + B) + k^2 -k } \ ,
\end{align}
where $x_k = i(B+1/2 - k)$.

\subsubsection*{Physical interpretation of $B$}

The parameter $B$ can be thought of as the magnetic field. Indeed, the full density of states in the limit $q \to 1$ is
\begin{align}
\lim_{q\to 1} \rho_q (x) ~ \Gamma_q\left( {1\over 2} \pm B \pm ix \right) &= \rho (x) ~ \Gamma\left( {1\over 2} \pm B \pm ix \right) \\
&=  {2x \sinh (2\pi x) \over \cos (2\pi B) + \cosh (2\pi x) } \ .
\end{align}
This is the density of states of a non-relativistic particle moving on a hyperbolic plane in the magnetic field $B$ \cite{Yang:2018gdb, Kitaev:2018wpr}.

To recover the formulas in the Schwarzian limit we take $B \to i \infty$. In this limt
\begin{align}
{\Gamma\left( {1\over 2} \pm B \pm ix \right)  \over \sqrt{C}}
\approx 
{\sqrt{\wt C} \over \Gamma\left( {1\over 2} \pm \wt B \pm ix \right) } \approx  e^{-\pi i (\Delta_1 + \Delta_2)} \ , 
\end{align}
where $C = \lim_{q\to 1} C_q$. And the orthogonality relation \eqref{6jorthqB} becomes \eqref{eq:2.16}.

\subsubsection*{Contour integral representation}

A generalization of the integral representation \eqref{6juint2} is
\begin{align}\label{q6juint2}
&\left\{
\begin{matrix}
\Delta_1 & k_1 & x \\ 
\Delta_2 & k_2 & \lambda 
\end{matrix}
\right\}_q
=
{|\log{q}| \over 1-q}
\left( \gamma_{2x,q}^{\Delta_2} \gamma_{2\lambda,q}^{\Delta_1}
\over 
 \gamma_{1x,q}^{\Delta_1} \gamma_{1\lambda,q}^{\Delta_2}
 \right)^{1/2}
 {\Gamma_q(1+2ik_1)\Gamma_q(-2ik_1) \over \Gamma(1+2ik_1)\Gamma(-2ik_1) } \\
& \int_{-i\infty}^{i\infty} {du \over 2\pi i }~q^{u- ik_1}~
 {\Gamma_q(\Delta_1  \pm ix +u) \Gamma_q(\Delta_2 \pm i \lambda +u) 
 \over \Gamma_q(\Delta_1 + \Delta_2 \pm i k_2 +u)
 }
 { \Gamma( \pm ik_1 +u+1) \over  \Gamma_q( \pm ik_1 +u+1)} ~ 
 \Gamma(\pm i k_1 -u) \ ,
\end{align}
where 
\begin{align}
\gamma_{12,q}^\Delta = \Gamma_q(\Delta \pm i k_1 \pm ik_2) \ .
\end{align}
This representation can be derived from equation (B.23) in \cite{Berkooz:2018jqr}.


\section{4-point function on the disk}

\label{4ptCoeffCheck}

In this section we check that the relative coefficients between the 3 terms in \eqref{4pt} are indeed as written. The first two terms are computed by the same path integral up to relabeling of $\beta_j$ and must clearly enter with the same coefficient. Let's check the relative coefficient between the 2nd and 3rd terms. We note that in the limit $\beta_2 \to 0$ they must be equal 
\begin{align}\label{4ptcheck}
\includegraphics[scale=.3]{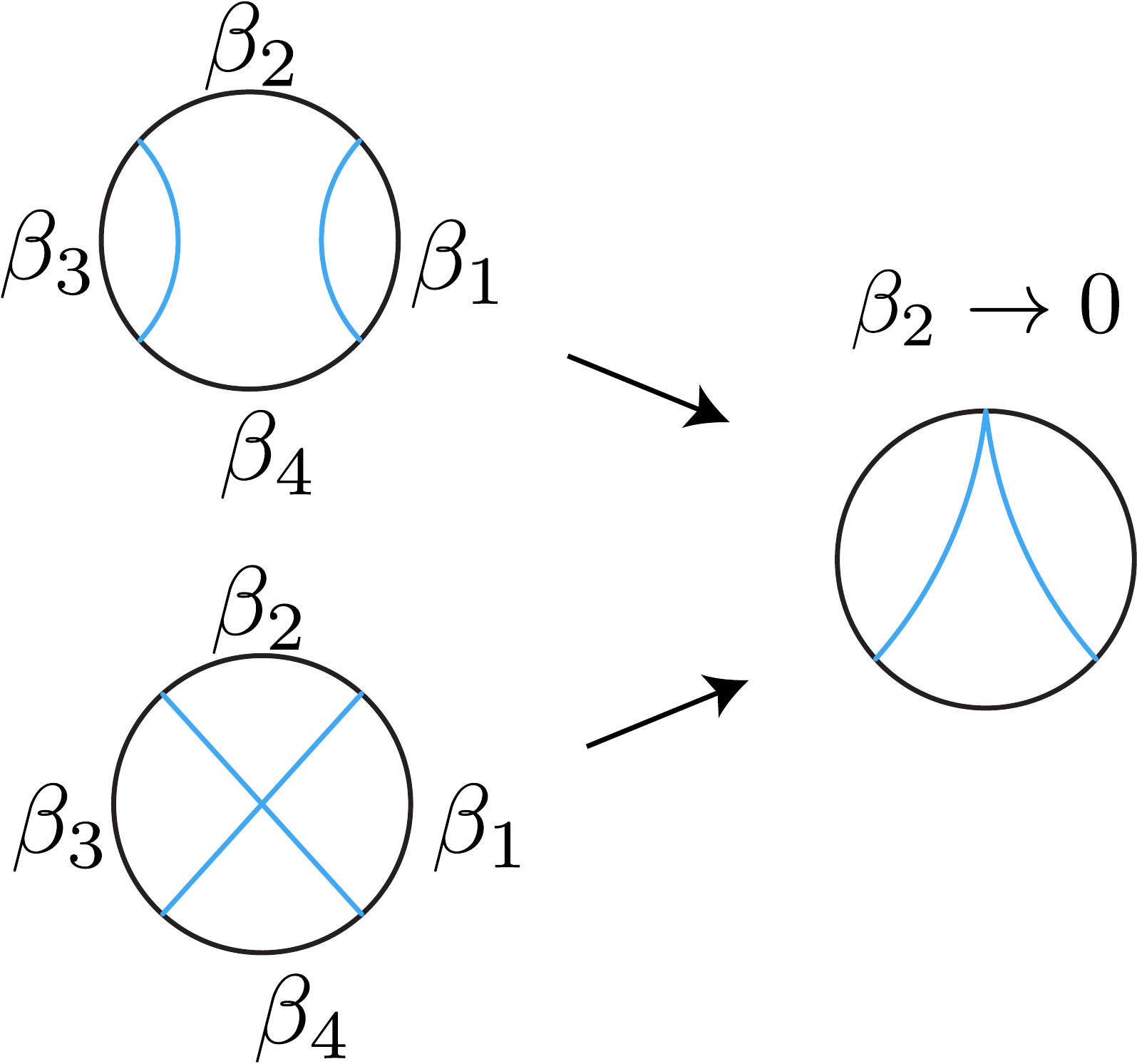}
\end{align}
Setting $\beta_2 = 0$, we can integrate over $s_2$ in the corresponding amplitudes and \eqref{4ptcheck} holds thanks to the identities
\begin{align}
\int_0^\infty ds_2 ~ \rho(s_2) ~ (\Gamma_{12}^\Delta\Gamma_{23}^\Delta)^{1/2} ~ {\delta(s_2 - s_4) \over \rho(s_2)} = 
\int_0^\infty ds_2 ~ \rho(s_2) ~ (\Gamma_{12}^\Delta\Gamma_{23}^\Delta)^{1/2} ~
\left\{
\begin{matrix}
\Delta & s_1 & s_2 \\
\Delta & s_3 & s_4
\end{matrix}
\right\} = 
(\Gamma_{14}^\Delta\Gamma_{43}^\Delta)^{1/2} \ .
\end{align}
The 2nd identity is the $n=0$ case of \eqref{WtransW}.


\section{Semi-classical limit of four-point correlator}

\subsection{Wilson polynomials and conformal blocks}

\label{sec:confblocks}

In this section we show that in the semiclassical and zero temperature limit $\beta \gg \tau_i$
\begin{align}
&{1\over Z(\beta)}\int_0^\infty  \prod_{j=1}^4 \left( ds_j ~ \rho(s_j) e^{-\beta_j s_j^2} \right) 
(\Gamma_{12}^\Delta \Gamma_{23}^\Delta \Gamma_{34}^\Delta \Gamma_{41}^\Delta)^{1/2}
~P_n^{\Delta\Delta}(s_2;s_1, s_3)
P_n^{\Delta\Delta}(s_4;s_1, s_3)\\
\approx 
&{(2\Delta)_n^2 \over n! (4\Delta +n -1)_n} 
~ (\beta_2 \beta_4)^{-2\Delta} g_{2\Delta +n} (z) \ ,
\end{align}
where $z = {\tau_{12} \tau_{34} \over \tau_{13} \tau_{24} }$. The RHS is the conformal block of the double-trace operator $2\Delta +n$. We also assume 
\begin{align}
\beta_1 = \beta - \tau_{14}  \ , \qquad 
\beta_2  = \tau_{12} \ , \qquad 
\beta_3 = \tau_{23} \ , \qquad 
\beta_4 = \tau_{34} \ ; \qquad 
\beta > \tau_1 > \tau_2 > \tau_3 > \tau_4 > 0
\end{align}
Using the definition of Wilson polynomials in section \ref{sec:specialfunctions} we have
\begin{align}
&{1\over Z(\beta)}{(2\Delta)_n^2 \over n! (4\Delta +n -1)_n} 
~ {\Gamma(4\Delta + 2n) \over \Gamma(2\Delta )^2 \Gamma(2\Delta + n)^2}
\int_0^\infty \prod_{j=1}^4 \left( ds_j ~ \rho(s_j) e^{-\beta_j s_j^2} \right) 
~
\gamma_{12}^\Delta 
\gamma_{23}^\Delta
\gamma_{34}^\Delta
\gamma_{41}^\Delta
\\
&{(2\Delta +is_1 \pm i s_3)_n^2 \over \Gamma(2\Delta +n \pm i s_1 \pm i s_3)}
\tensor[_4]{F}{_3}\left( 
{-n , 4\Delta +n -1, \Delta +is_1 \pm i s_2
\atop
2\Delta, 2\Delta +is_1 \pm is_3
}
;1
\right)
\tensor[_4]{F}{_3}\left( 
{-n , 4\Delta +n -1, \Delta +is_1 \pm i s_4
\atop
2\Delta, 2\Delta +is_1 \pm is_3
}
;1
\right) \ ,
\end{align}
where $\gamma_{nm}^\Delta = \Gamma(\Delta \pm i s_n \pm i s_m )$. To take the semiclassical limit we change variables 
\begin{align}
s_j^2 = s_1^2 + \omega_j \ , \qquad j =2,3,4
\end{align}
and take the limit $s_1^2 \gg \omega_j$. More carefully, we should first restore the factors of gravitational coupling $\gamma = {\bar \phi \over 8 \pi G_N}$ that we previously set to $2\gamma = 1$. This is easily done by dimensional analysis. The coupling $\gamma$ has dimensions of length, $k^2$ has dimensions of energy, and $\beta$ has dimensions of length. We will not do this explicitly, but instead think of it as the high energy limit $s_1^2 \gg \omega_j$.

Using that
\begin{align}
\Gamma(a\pm is) \approx  2\pi |s|^{2a -1} e^{-\pi |s|} \ , \qquad |s| \to \infty \ ,
\end{align}
we find
\begin{align}
&{1\over Z(\beta)}{(2\Delta)_n^2 \over n! (4\Delta +n -1)_n} 
~ {\Gamma(4\Delta + 2n) \over \Gamma(2\Delta )^2 \Gamma(2\Delta + n)^2}
 \int_0^\infty ds_1 ~ \rho(s_1) e^{-\beta s_1^2}  ~ (2s_1)^{4\Delta - 3}
\\
&\int_{-\infty}^\infty {d\omega_2 d\omega_3 d\omega_4 \over (2\pi)^3} 
e^{-\beta_2 \omega_2 - \beta_3 \omega_3 -\beta_4 \omega_4} ~
e^{\pi {\omega_3 \over 2s_1}}
~ 
\Gamma\left(\Delta \pm i  {\omega_2 \over 2s_1} \right)
\Gamma\left(\Delta \pm i  {\omega_4 \over 2s_1} \right)
\Gamma\left(\Delta \pm i  {\omega_{32} \over 2s_1} \right)
\Gamma\left(\Delta \pm i  {\omega_{34} \over 2s_1} \right) 
\\
& { (-1)^n\left( 2\Delta - i {\omega_3 \over 2s_1} \right)_n^2
\over 
\Gamma\left( 2\Delta +n \pm i {\omega_3 \over 2s_1} \right)}
~
\tensor[_3]{F}{_2}\left( 
{-n , 4\Delta +n -1, \Delta - i {\omega_2 \over 2s_1}
\atop
2\Delta, 2\Delta - i {\omega_3 \over 2s_1}
}
;1
\right)
\tensor[_3]{F}{_2}\left( 
{-n , 4\Delta +n -1, \Delta - i {\omega_4 \over 2s_1}
\atop
2\Delta, 2\Delta - i {\omega_3 \over 2s_1}
}
;1
\right)
\end{align}
where $\omega_{ij} = \omega_i - \omega_j$. The integral over $s_1$ is computed by a saddle approximation. The value of the saddle is determined from $\rho(s_1) e^{-\beta s_1^2} \sim e^{2\pi s_1 - \beta s_1^2}$ and is given by $s_1 = {\pi \over \beta}$. Then $\int_0^\infty ds_1 ~ \rho(s_1) e^{-\beta s_1^2} = Z(\beta)$ is the partition function, while in the rest of the integral we simply set $s_1 = {\pi \over \beta}$
\begin{align}
&{(2\Delta)_n^2 \over n! (4\Delta +n -1)_n} 
~ {\Gamma(4\Delta + 2n) \over \Gamma(2\Delta )^2 \Gamma(2\Delta + n)^2}
 ~ \left( 2\pi \over \beta \right)^{4\Delta - 3}
\\
&\int_{-\infty}^\infty {d\omega_2 d\omega_3 d\omega_4 \over (2\pi)^3} 
e^{-\beta_2 \omega_2 - \beta_3 \omega_3 -\beta_4 \omega_4} ~
e^{ {\beta \over 2}\omega_3}
~ 
\Gamma\left(\Delta \pm i  {\beta \over 2\pi} \omega_2 \right)
\Gamma\left(\Delta \pm i  {\beta \over 2\pi}\omega_4  \right)
\Gamma\left(\Delta \pm i  {\beta \over 2\pi} \omega_{32} \right)
\Gamma\left(\Delta \pm i  {\beta \over 2\pi} \omega_{34} \right) 
\\
& { (-1)^n \left( 2\Delta - i {\beta \over 2\pi} \omega_3 \right)_n^2
\over 
\Gamma\left( 2\Delta +n \pm i {\beta \over 2\pi} \omega_3 \right)}
~
\tensor[_3]{F}{_2}\left( 
{-n , 4\Delta +n -1, \Delta - i {\beta \over 2\pi} \omega_2
\atop
2\Delta, 2\Delta - i {\beta \over 2\pi} \omega_3
}
;1
\right)
\tensor[_3]{F}{_2}\left( 
{-n , 4\Delta +n -1, \Delta - i {\beta \over 2\pi} \omega_4
\atop
2\Delta, 2\Delta - i {\beta \over 2\pi} \omega_3
}
;1
\right)
\end{align}
Now we take the zero temperature limit $\beta \to \infty$. Again using the asymptotics of gamma functions at large argument, we find
\begin{align}
&{(2\Delta)_n^2 \over n! (4\Delta +n -1)_n} 
~ {\Gamma(4\Delta + 2n) \over \Gamma(2\Delta )^2 \Gamma(2\Delta + n)^2}
\\
&\int_{-\infty}^\infty {d\omega_2 d\omega_3 d\omega_4} 
e^{-\beta_2 \omega_2 - \beta_3 \omega_3 -\beta_4 \omega_4} ~
e^{ {\beta \over 2}( \omega_3 +|\omega_3| - |\omega_2| - |\omega_4| - |\omega_{32}| - |\omega_{34}| )}
\\
& 
|\omega_3|^{-4\Delta + 1}|\omega_2 \omega_4 \omega_{32} \omega_{34}|^{2\Delta - 1} 
\tensor[_2]{F}{_1}\left( 
{-n , 4\Delta +n -1
\atop
2\Delta
}
; {\omega_2 \over \omega_3}
\right)
\tensor[_2]{F}{_1}\left( 
{-n , 4\Delta +n -1
\atop
2\Delta
}
; {\omega_4 \over \omega_3}
\right)
\end{align}
One can show that at large $\beta$
\begin{align}
e^{ {\beta \over 2}( \omega_3 +|\omega_3| - |\omega_2| - |\omega_4| - |\omega_{32}| - |\omega_{34}| )}
\approx 
\begin{cases}
1 , \qquad \text{if} \quad 0< \omega_2 < \omega_3, ~ 0< \omega_4 < \omega_3 \\
0 , \qquad \text{otherwise}
\end{cases} \ .
\end{align}
Therefore we find
\begin{align}
  &{(2\Delta)_n^2 \over n! (4\Delta +n -1)_n} 
~ {\Gamma(4\Delta + 2n) \over \Gamma(2\Delta )^2 \Gamma(2\Delta + n)^2}
\\
&\int_0^\infty d\omega_3 ~ e^{-\beta_3 \omega_3}  \omega_3^{-4\Delta +1}\\
&\int_0^{\omega_3} d\omega_2 ~ e^{-\beta_2 \omega_2} (\omega_2  \omega_{32})^{2\Delta - 1} 
F(-n, 4\Delta +n-1, 2\Delta, \omega_2/\omega_3) \\
&\int_0^{\omega_3} d\omega_4 ~ e^{ -\beta_4 \omega_4}
( \omega_4 \omega_{34})^{2\Delta - 1} 
F(-n, 4\Delta +n-1, 2\Delta, \omega_4/\omega_3) \ .
\end{align}
After rescaling $\omega_2 = x \omega_3, \omega_4 = y \omega_3$ and using that 
\begin{align}
\int_0^1 dx ~ e^{-a x} [x(1-x)]^{2\Delta - 1} F(-n, 4\Delta +n-1, 2\Delta, x) = \sqrt{\pi} e^{-a/2} a^{-2\Delta + {1\over 2}} \Gamma(2\Delta) I_{2\Delta + n -{1\over 2}} (a/2)
\end{align}
we have
\begin{align}
 &{(2\Delta)_n^2 \over n! (4\Delta +n -1)_n} 
~ {\pi\Gamma(4\Delta + 2n) \over  \Gamma(2\Delta + n)^2}
\\
& (\beta_2 \beta_4)^{-2\Delta + {1\over 2}} \int_0^\infty d\omega_3 ~ e^{-\left( \beta_3 + {\beta_2 + \beta_4 \over 2} \right)\omega_3}
I_{2\Delta + n -{1\over 2}} (\beta_2 \omega_3/2)
I_{2\Delta + n -{1\over 2}} (\beta_4 \omega_3/2) \\
=&{(2\Delta)_n^2 \over n! (4\Delta +n -1)_n} 
~ (\beta_2 \beta_4)^{-2\Delta} g_{2\Delta +n} (z) , 
\end{align}
where $z = {\beta_2 \beta_4 \over (\beta_2 + \beta_3)(\beta_3 + \beta_4)} = {\tau_{12} \tau_{34} \over \tau_{13} \tau_{24}}$ and $g_h(z) = z^h F(h,h,2h,z)$. We checked this integral numerically. This is indeed the expected answer.

\subsection{Semi-classical limit of the 6j-symbol}

It is also interesting to take the semi-classical limit of the 6j-symbol and check that after an appropriate integration over energies it gives the crossed GFF Wick contraction. 

We consider
\begin{align}\label{6jSClimitdef}
\left\{
\begin{matrix}
\Delta & k_1 & k_2 \\
\Delta & k_3 & k_4
\end{matrix}
\right\} = (\gamma_{12}\gamma_{23}\gamma_{34}\gamma_{41})^{1/2} 
{\cal W}_{k_4}(k_2;\Delta \pm i k_1, \Delta \pm i k_3) \ .
\end{align}
We set 
\begin{align}
k_j^2 = k_1^2 + \omega_j \ , \qquad k_j \approx k_1 + {\omega_j \over 2k_1} \equiv k_1 + \nu_j \ , \qquad j=2,3,4
\end{align}
and take $k_1^2 \gg \omega_j$. It is convenient to use the expression for the Wilson function \eqref{WilsonFunction} giving
\begin{align}
{\cal W}_{k_4}(k_2;\Delta \pm i k_1, \Delta \pm i k_3) = 
&{\Gamma(-2ik_1) \over 
\Gamma(2\Delta +ik_1 \pm i k_3)
\Gamma(\Delta -ik_1 \pm i k_2)
\Gamma(\Delta -ik_1 \pm i k_4)
}
\\
&\tensor[_4]{F}{_3}
\left(
{\Delta + ik_1 \pm ik_2 , \Delta +ik_1 \pm i k_4 \atop
2\Delta + ik_1 \pm i k_3 , 2ik_1 + 1}
;1
\right)
+ \text{c.c.}
\end{align}
At large $k_1$ the hypergeometric function simplifies significantly ($k_{ij} = k_i -k_j$)
\begin{align}
\tensor[_4]{F}{_3}
\left(
{\Delta + ik_1 \pm ik_2 , \Delta +ik_1 \pm i k_4 \atop
2\Delta + ik_1 \pm i k_3 , 2ik_1 + 1}
;1
\right)
&\approx 
\tensor[_2]{F}{_1}
\left(
{\Delta + ik_{12} , \Delta +ik_{14} \atop
2\Delta + ik_{13} }
;1
\right) \\
&=
{\Gamma(2\Delta +i k_{13}) \Gamma(i k_{2+4-1-3})
\over 
\Gamma(\Delta +ik_{23}) \Gamma(\Delta +i k_{43})} \ ,
\end{align}
where $k_{i+j} = k_i + k_j$ etc. Also using that 
\begin{align}
\Gamma(\Delta +i x) \approx 
\sqrt{2\pi} \ |x|^{\Delta-{1\over 2}} e^{-{\pi \over 2} |x|} \ e^{i \sgn(x) {\pi \over 2} (\Delta - {1\over 2}) + i x \log {|x| \over e}} \ , \qquad x \in {\mathbb R} \ , \qquad |x| \gg 1 \ ,
\end{align}
we find 
\begin{align}
{\cal W}_{k_4}(k_2;\Delta \pm i k_1, \Delta \pm i k_3) 
\approx 
&{e^{2\pi k_1}\over 2\pi}  (2k_1)^{1-4\Delta} e^{{\pi \over 2}(\nu_2 +\nu_3 + \nu_4)} \\ 
&\left(
{(2k_1)^{i\nu_{2+4-3}}\Gamma(i\nu_{2+4-3}) \over \Gamma(\Delta +i \nu_2) \Gamma(\Delta - i \nu_{34})
 \Gamma(\Delta +i \nu_4) \Gamma(\Delta - i \nu_{32})
}
+ c.c
\right) \ .
\end{align}
At large $k_1$ it is highly oscillating unless $\nu_{2+4-3} \equiv \nu_2 + \nu_4 - \nu_3 =0$. Near this value we substitute $\nu_3 = \nu_2+\nu_4$ in the gamma-functions in the denominator, expand the gamma-function in the numerator and find 
\begin{align}
{\cal W}_{k_4}(k_2;\Delta \pm i k_1, \Delta \pm i k_3) 
\approx 
&{e^{2\pi k_1} (2k_1)^{1-4\Delta} e^{{\pi \over 2}(\nu_2 +\nu_3 + \nu_4)} 
\over 
2\pi \Gamma(\Delta \pm i \nu_2) \Gamma(\Delta \pm i \nu_4)
}
\ {2\sin(\nu_{2+4-3} \log(2k_1)) \over \nu_{2+4-3}} 
\\
\approx
&{e^{2\pi k_1} (2k_1)^{1-4\Delta} e^{{\pi \over 2}(\nu_2 +\nu_3 + \nu_4)} 
\over 
 \Gamma(\Delta \pm i \nu_2) \Gamma(\Delta \pm i \nu_4)
} \
\delta(\nu_{2+4-3})\ . 
\end{align}
In the last line we used that $\lim_{t\to \infty} {2\sin(t x) \over x} = 2\pi \delta(x)$. Also taking the limit of gamma-functions in the prefactor in \eqref{6jSClimitdef} we find a rather simple expression
\begin{align}
\left\{
\begin{matrix}
\Delta & k_1 & k_2 \\
\Delta & k_3 & k_4
\end{matrix}
\right\} 
~\approx 
~&(2\pi)^2 \  {e^{-2 \pi k} \over 2k} \ \delta(k_1 + k_3 - k_2 - k_4) \\
\approx 
~&{\delta(k_1 + k_3 - k_2 - k_4) \over \rho(k)}
\ ,
\end{align}
where we restored the original variables and defined the average momentum $k = {1\over 4}(k_1 + k_2 + k_3 + k_4)$. To reiterate, we take the limit where all $k_j$ are large, but all differences $k_i - k_j$ are finite. It is straightforward to use the above formula to show that appropriate integration over energies gives the crossed GFF Wick contraction in the four-point correlator, as expected.


\section{Operator algebra of the generalized free field}

\label{sec:gffproof}

The purpose of this appendix is to show that the condition \eqref{eq:GFFconstraint}, together with associativity of the OPE and conformal symmetry, guarantees that all $n$-point correlators of $\calo$ agree with those of the bosonic generalized free field. The upshot is that the operator algebra of $\calo$ in the generalized free field theory is completely characterized by the operator equation \eqref{eq:GFFconstraint} and conformal invariance. All other operator equations must be implied by \eqref{eq:GFFconstraint} and conformal invariance.

As mentioned in the main text, \eqref{eq:GFFconstraint} implies that the OPE takes the form \eqref{eq:3.8}, which we repeat here (in a slightly modified form):
\begin{equation}
    \calo(\tau_1) \calo(\tau_2) = \frac{1}{\tau_{12}^{2 \Delta}} + \sum_{n=0}^\infty \tau_{12}^{2 n} \,  \left.[\calo \calo]_{n}\right|_{\tau=\frac{\tau_1 + \tau_2}{2}} + \text{descendants}.
    \label{eq:symmOPE}
\end{equation}
We have chosen to expand the operators about their midpoint. Written in this way, all of the powers of $\tau_{12}$ are even. In particular, one can show that the descendants only contribute even powers of $\tau_{12}$. If we compute the Lorentzian commutator $[\calo(t_1),\calo(t_2)]$, all of the terms on the right hand side of \eqref{eq:symmOPE} drop out except for the identity contribution, which appears on the right hand side of \eqref{eq:GFFconstraint}. If any other primaries with dimensions not in $2 \Delta + 2 \mathbb{Z}_{\ge 0}$ appeared in the $\calo \calo$ OPE, these primaries would contribute to the right side of \eqref{eq:GFFconstraint}, so by imposing \eqref{eq:GFFconstraint} as a constraint on the algebra we are demanding that the OPE takes the form \eqref{eq:symmOPE}.

Let $F_{n}(\tau_1,\tau_2,\cdots,\tau_{2n})$ be the $2n$-point function of $\calo$ in a theory where the OPE \eqref{eq:symmOPE} holds, and let $G_{n}(\tau_1,\tau_2,\cdots,\tau_{2n})$ be the $2n$-point function of $\calo$ in the generalized free field theory. We will inductively show that $F_n = G_n$ for all $n \in \mathbb{N}$.\footnote{We thank Dalimil Maz\'{a}\v{c} for providing this proof to us.} The base case $n = 1$ is trivial. Next, suppose that the result has been shown for $n \leq m - 1$ with $m \in \mathbb{N}$. Define
\begin{equation}
    H_m(\tau_1,\tau_2,\cdots,\tau_{2m}) := F_m(\tau_1,\tau_2,\cdots,\tau_{2m}) - G_m(\tau_1,\tau_2,\cdots,\tau_{2m}).
\end{equation}
This function should be defined via analytic continuation from the domain $\tau_1 > \tau_2 > \cdots > \tau_{2m}$. One can easily see that $H_m$ is nonsingular for $\tau_1 \rightarrow \tau_2$ because the identity contributions to the $\calo(\tau_1) \calo(\tau_2)$ OPE cancel. Hence, we can analytically continue to $\tau_1 < \tau_2$. Furthermore, $H_m$ is invariant under $\tau_1 \leftrightarrow \tau_2$ because only even powers of $\tau_{12}$ appear in \eqref{eq:symmOPE}. Hence, the $\tau_1 \rightarrow \tau_3$ limit is nonsingular because the $\tau_2 \rightarrow \tau_3$ limit is nonsingular. One can thus show that $H_m$ is entire in $\tau_1$. Using the same logic, one can show that $H_m$ is an entire function of all of the $\tau$ variables. Furthermore, $H_m$ goes to zero as $\tau_1 \rightarrow \infty$ due to conformal symmetry. Hence, $H_m$ vanishes identically.

\section{2-point function on the double-trumpet}
\label{2pt-dt}

\begin{figure}
	\centering
	\includegraphics[scale = .2]{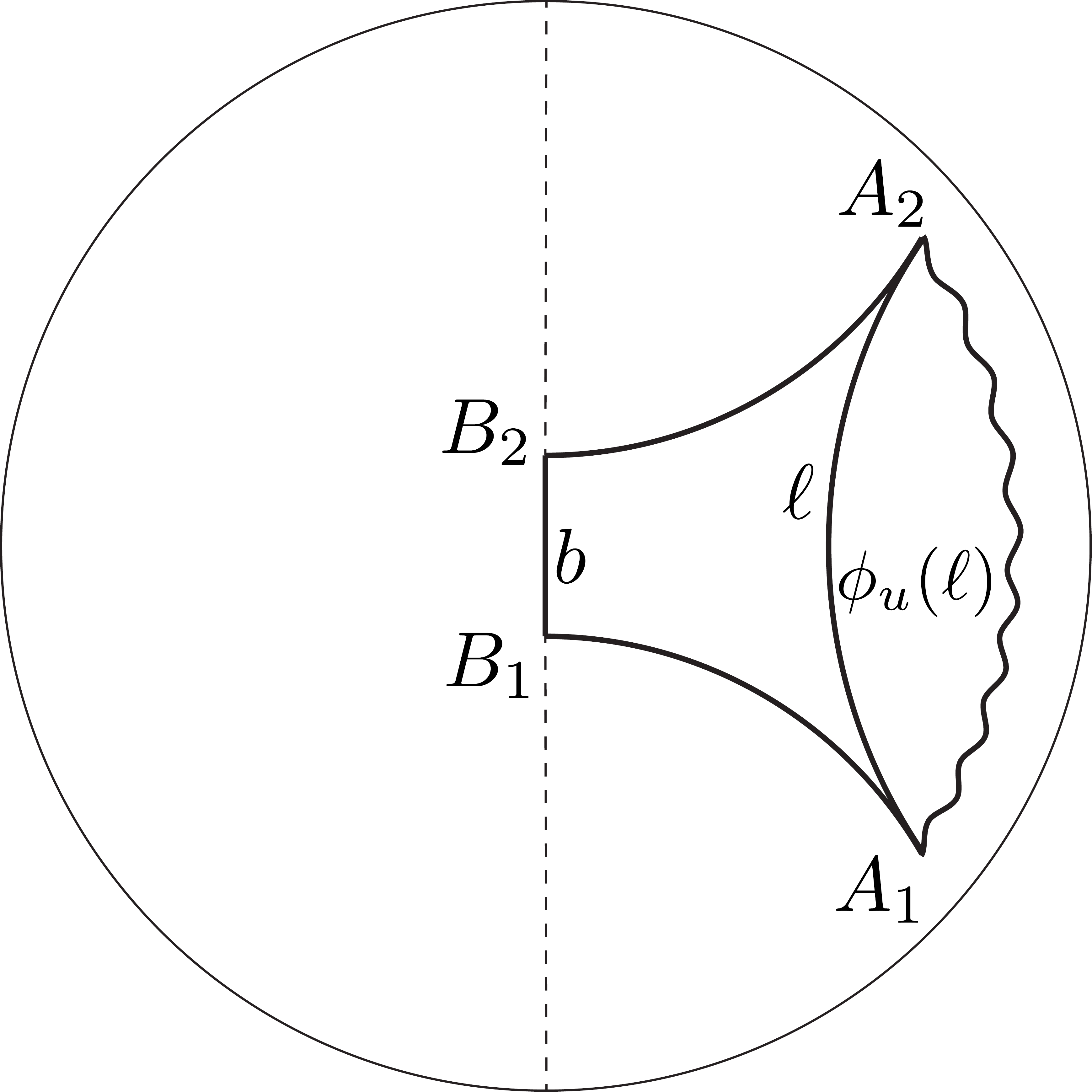}
	\caption{Geometry relevant for the computation of $g(u;A_1, A_2)$. }
	\label{fig:trumpet0}
\end{figure}

In this section we compute the 2-point correlator on the double-trumpet. We use the boundary particle formalism of \cite{Yang:2018gdb}. It is convenient to work in the coordinates $ds^2 = d\rho^2 + \cosh^2 \rho d\tau^2$. We first compute the JT path integral on a piece of the disk in the figure \ref{fig:trumpet0} bounded by four points: $A_1 = (\rho_1, \tau_1), A_2 = (\rho_2,\tau_2), B_1 = (0,\tau_1), B_2 = (0,\tau_2)$.
The wiggly line connecting $A_1,A_2$ is the boundary of physical length $u$. The embedding of this boundary segment into the hyperbolic disk is integrated over with the Schwarzian action. The other four lines connecting points $A_1,A_2, B_1,B_2$ are geodesics. We construct the path integral for $g(u;A_1, A_2)$ out of the Hartle-Hawking wave function (region bounded by the boundary and the geodesic $\ell$) and the hyperbolic quadrangle $A_1A_2B_2B_1$
\begin{align}\label{trumpet0}
g(u; A_1,A_2) = e^{\phi_b A} \phi_u(\ell) \ .
\end{align}
The factor $e^{\phi_b A}$, where $A$ is the area of the quadrangle $A_1A_2B_2B_1$, is the JT path integral on the quadrangle. An exercise in hyperbolic geometry shows that in the limit when $A_1, A_2$ are close to the boundary ($\rho_1,\rho_2 \to \infty$) this area is
\begin{align}
A \approx \pi - 2 (e^{-\rho_1} + e^{-\rho_2} )\coth
{\tau_2 - \tau_1 \over 2}   + \dots  
\end{align}
Also multiplying by $\phi_b = {\gamma\over \e} \equiv \gamma q$, rescaling the radial coordinate\footnote{This is analogous to the rescaling $z={y\over q}$ in \cite{Yang:2018gdb}. It is necessary for the particle in a magnetic field to be equivalent to the Schwarzian theory. } $e^{-\rho} = {1\over q} e^{-w}$ and setting $\gamma = 1/2$ we get a contribution 
\begin{align}
e^{\phi_b A} \approx e^{\pi q - (e^{-w_1} + e^{-w_2} )\coth {\tau_2 - \tau_1 \over 2}   } \ .
\end{align}
The first term $\pi q$ cancels out if one adds appropriate corner terms in the JT action.\footnote{Such terms are necessary for locality, see \cite{Harlow:2018tqv} for details.} In any case it doesn't depend on any of the coordinates. The second term will be important in the computation below. 

The second factor in \eqref{trumpet0} is the Hartle-Hawking wave function
\begin{align}
\phi_u(\ell) &= e^{-\ell/2} \int_0^\infty ds ~ \rho(s) e^{-u s^2} ~ 
2K_{2is}(2e^{-\ell/2}) \\
& \equiv e^{-\ell/2}\int_0^\infty ds ~ \rho(s) e^{-u s^2} ~ \widehat{\phi}_s(\ell) \ .
\end{align}
In the energy basis it obeys the orthogonality relations
\begin{align}
\int_{-\infty}^\infty d\ell ~ \widehat \phi_s(\ell) \widehat \phi_{s'}(\ell) &= {\delta(s-s') \over \rho(s)} \ , \\
\int_0^\infty ds ~ \rho(s) \widehat \phi_{s}(\ell) \widehat \phi_{s}(\ell') &= \delta(\ell - \ell') \ .
\end{align}
In \eqref{trumpet0} the (renormalized) geodesic distance $\ell$ between the points $A_1, A_2$ is $e^{\ell} = e^{w_1 + w_2} \sinh^2{\tau_2 - \tau_1 \over 2}$. Putting everything together we have
\begin{align}
g(u;A_1, A_2) = \exp\left( - (e^{-w_1} + e^{-w_2} )\coth {\tau_2 - \tau_1 \over 2}  \right)
{ e^{-{w_1 + w_2 \over 2}} \over \sinh{\tau_2 - \tau_1 \over 2}}
\int_0^\infty ds ~ \rho(s) e^{-u s^2} ~ 
2K_{2is}\left(2 e^{-{w_1 + w_2 \over 2}} \over \sinh{\tau_2 - \tau_1 \over 2} \right) \ .
\end{align}

\subsection*{Trumpet partition function}

To make sure our formulas are correct, we first compute the trumpet partition function. We set $w_1 = w_2=w, \tau_2=\tau_1 + b \equiv \tau + b$ and integrate over $w$. The correct measure of integration is\footnote{The measure which naturally arises in the boundary particle is induced by the metric $ds^2 = d\rho^2  + \cosh^2\rho d\tau^2$, which gives $\sqrt{g} d\rho d\tau= \cosh \rho d\rho d\tau  \approx q {1\over 2}e^{w} dw d\tau $. This differs from the correct measure by a factor $2\sinh{b\over 2}$. This factor is explained by the fact that the boundary particle is equivalent to the Schwarzian theory with the path integral measure $\prod_u {d\tau(u) \over \tau'(u)}$, while the correct measure on the double-trumpet is $2\sinh{b\over 2}\prod_u {d\tau(u) \over \tau'(u)}$. The factor $2\sinh{b\over 2}$ can be understood from considering the equation (2.25) in \cite{Stanford:2017thb}. We first map it to the double-trumpet coordinate by $\tan {\phi\over 2} = e^{-\tau}$. Then two solutions of (2.25) in \cite{Stanford:2017thb} become $\Psi_{\pm} = e^{\pm \tau/2}/\sqrt{\tau'}$. Under Euclidean time evolution $u\to u + 2\pi$, we have $\tau \to \tau + b$ and $\Psi_{\pm} \to e^{\pm b/2} \Psi_{\pm}$. Then we compute the trace of evolution operator around the Euclidean circle $\tr (-1)^F U = e^{b/2} - e^{-b/2} = 2\sinh{b\over 2}$, where the minus sign is from $(-1)^F$.  } 
${1\over b}d\tau dw~ e^{w} \sinh{b\over 2}$, where $1/b$ is from gauge fixing $U(1)$ symmetry. The $\tau$ integral is trivial $\int_0^b {d\tau \over b} = 1$ and the rest gives
\begin{align}
Z_{tr}(\beta,b) &= \int_{-\infty}^\infty dw ~ \left(e^{w} \sinh{b\over 2} \right) \cdot 
~e^{-2e^{-w}\coth{b\over 2}} {e^{-w} \over \sinh{b\over 2}}
\int_0^\infty ds ~ \rho(s) e^{-\beta s^2} ~ 
2K_{2is}\left(2 e^{-w} \over \sinh{b \over 2} \right) \\
&=\int_0^\infty ds ~ e^{-\beta s^2} {\cos(bs)\over \pi} \ , 
\end{align}
where we used that
\begin{align}\label{trumpetJacobi}
\int_0^\infty {dx \over x}  e^{-2x \cosh {b\over 2}} 2 K_{2is}(2x) = {1\over \rho(s)} {\cos(bs) \over \pi} \ .
\end{align}

\subsection*{2-point function without 1-loop determinant}

\begin{figure}
    \centering
    \includegraphics[scale = .2]{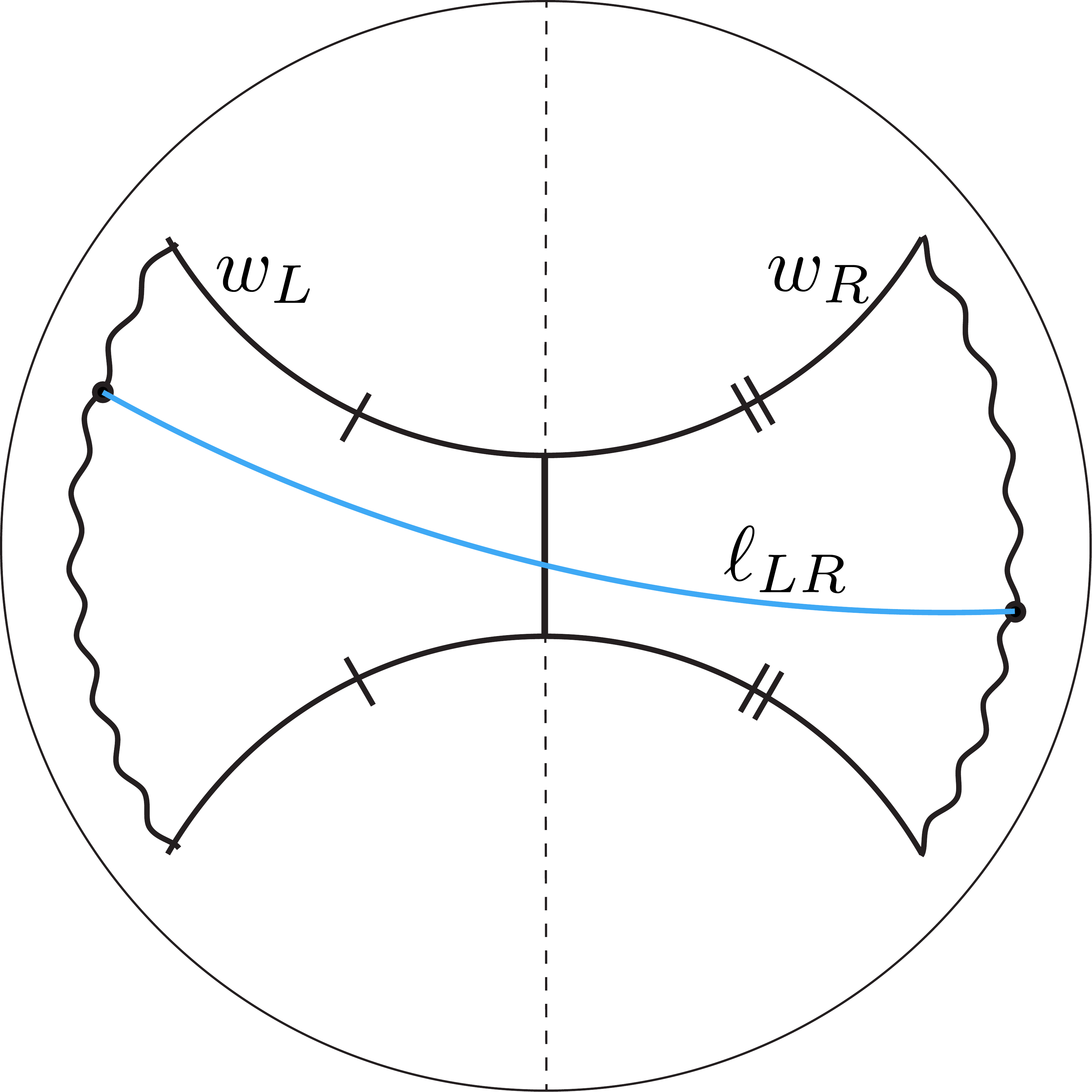}
    \caption{Geometry for the computation of the 2-point function on the double-trumpet. The blue line is the geodesic of (renormalized) length $\ell_{LR}$ connecting operators on the left and right boundaries.}
    \label{fig:2pt-dtDisk}
\end{figure}

The 2-point function on the double-trumpet without the matter determinant can be computed by putting together two factors of $g(u;A_1, A_2)$ from the previous subsection, for the left and right trumpets. We also insert $e^{-\Delta \ell_{LR}}$, where $e^{\ell_{LR}} = e^{w_L +w_R} \cosh^2{\tau_L - \tau_R \over 2}$ is the (renormalized) geodesic distance between the two operators. See figure \ref{fig:2pt-dtDisk}.
We have
\begin{align}
\la \tr e^{-\beta_L H} \calo \tr e^{-\beta_R H} \calo \ra
&\supset
\raisebox{-.3in}{\includegraphics[scale=.15]{figures/2pt-dt0.pdf}}
\\
=&\int_0^\infty db \int_0^b d\tau 
    \int_{-\infty}^{\infty} dw_L dw_R 
    \int_0^\infty ds_L ds_R ~ \rho(s_L) \rho(s_R) e^{-\beta_L s_L^2 - \beta_R s_R^2} \\
    &\times e^{-2e^{-w_L}\coth{b\over 2} } ~2 K_{2is_L}\left( 2e^{-w_L} \over \sinh{b\over 2} \right)
    ~ e^{-2e^{-w_R}\coth{b\over 2} } ~2 K_{2is_R}\left( 2e^{-w_R} \over \sinh{b\over 2} \right)\\
    &\times \sum_{n=-\infty}^{\infty} 
    \left( e^{-(w_L + w_R)} \over \cosh^2{\tau +n b \over 2} \right)^\Delta \ ,
\end{align}
where the sum over $n$ is the sum over windings of the geodesic connecting two operators. The integral over $\tau$ gives
\begin{align}
    \int_0^b d\tau   
    \sum_{n=-\infty}^\infty\left( \cosh{\tau  +n b \over 2} \right)^{-2\Delta}
    &=  \int_{-\infty}^\infty d\tau \left( \cosh{\tau   \over 2} \right)^{-2\Delta}\\
    &={2^{2\Delta} \Gamma(\Delta)^2 \over \Gamma(2\Delta)} \ .
\end{align}
To compute $w_L, w_R$ integrals we change variables to $x_L={e^{-w_L} \over \sinh{b\over 2}}$ and similarly for $w_R$. And use that 
\begin{align}
    &\int_0^\infty {dx \over x} x^\Delta e^{-2x \cosh {b\over 2}} 2 K_{2is}(2x) 
    = 2^{-2\Delta} \Gamma(\Delta \pm 2is)^{1/2} \left( \cosh{b\over 4} \right)^{-2\Delta+1} \Phi_s(b) \ , \\
    &\Phi_s(b) = {2^{2\Delta} \Gamma(\Delta) \over \Gamma(2\Delta)} \Gamma(\Delta \pm 2is)^{1/2}
    F\left({1\over 2} +2is, {1\over 2} - 2is, \Delta + {1\over 2}, - \sinh^2 {b\over 4}\right) \ .
\end{align}
This is a generalization of \eqref{trumpetJacobi}. We introduced a function $\Phi_s(b)$ which is essentially a Jacobi function that defines Jacobi transform, e.g. see \cite{Koornwinder1984}. It satisfies orthogonality relations \begin{align}
    &\int_0^\infty db \left( \sinh{b\over 4} \right)^{2\Delta}\left( \cosh{b\over 4} \right)^{2(1-\Delta)}
   \Phi_{s_L}(b) \Phi_{s_R}(b) = 
    {\delta(s_L - s_R) \over \rho(s_L)} \ , \\
    &\int_0^\infty ds ~ \rho(s)   \Phi_{s}(b) \Phi_{s}(b')
    ={\delta(b-b') \over \left( \sinh{b\over 4} \right)^{2\Delta}\left( \cosh{b\over 4} \right)^{2(1-\Delta)} } \ .
\end{align}
Using above integrals we have 
\begin{align}
   &{\Gamma(\Delta)^2 \over \Gamma(2\Delta)}
   \int_0^\infty ds_L ds_R ~ \rho(s_L) \rho(s_R) e^{-\beta_L s_L^2 - \beta_R s_R^2} 
   (\Gamma(\Delta \pm 2i s_L)\Gamma(\Delta \pm 2i s_R) )^{1/2}\\
   &\times  
   \int_0^\infty db \left( \sinh{b\over 4} \right)^{2\Delta}\left( \cosh{b\over 4} \right)^{2(1-\Delta)} 
   \Phi_{s_L}(b) \Phi_{s_R}(b) \ .
\label{2pt-dtWind0}
\end{align}
The $b$ integral here turns out to be the orthogonality relation for the Jacobi transform, so we find
\begin{align}
    \la \tr e^{-\beta_L H} \calo \tr e^{-\beta_R H} \calo \ra
    \supset
    {\Gamma(\Delta)^2 \over \Gamma(2\Delta)}\int_0^\infty ds ~\rho(s) e^{- (\beta_L + \beta_R) s^2}  \Gamma(\Delta \pm 2i s)  \ .
\end{align}

\subsection*{2-point function with 1-loop determinant}

To include the 1-loop determinant in our computation, we use the formula \eqref{eq:second} for the determinant at fixed $b$ and insert it in the computation in the previous subsection. We start with the $n=1$ term in \eqref{eq:second} corresponding to the operator $\Delta$ propagating on the closed geodesic. Instead of the integral \eqref{2pt-dtWind0} we need to compute
\begin{align}\label{strace op 6j}
    \int_0^\infty db ~ {e^{-\Delta' b} \over 1- e^{-b}} 
    \left( \sinh{b\over 4} \right)^{2\Delta}\left( \cosh{b\over 4} \right)^{2(1-\Delta)}
   \Phi_{s_L}(b) \Phi_{s_R}(b) = 
   \begin{Bmatrix}
   \Delta' & s_L & s_R \\
   \Delta & s_R & s_L
   \end{Bmatrix}
   \ .
\end{align}
We didn't find a derivation of this formula, but we checked that it holds numerically. For the $n=1$ term we need this integral with $\Delta' = \Delta$. Assuming it is correct, we have a contribution to the 2-point function
\begin{align}
 \la \tr e^{-\beta_L H} \calo \tr e^{-\beta_R H} \calo \ra
&\supset
\raisebox{-.35in}{\includegraphics[scale=.15]{figures/2pt-dt1.pdf}}\\
	=&{\Gamma(\Delta)^2  \over \Gamma(2\Delta)}\int_0^\infty ds_L ds_R ~\rho(s_L)\rho(s_R) e^{- (\beta_L s_L^2 + \beta_R s_R^2) }
	\\
   \times  &\left( 
   \Gamma(\Delta \pm 2 i s_L) \Gamma(\Delta \pm 2 i s_R)
   \right)^{1/2} 
   \begin{Bmatrix}
   \Delta & s_L & s_R \\
   \Delta & s_R & s_L
   \end{Bmatrix}
	\ .
\end{align}
For the $n=2$ term in \eqref{eq:second}, we expand it as ${e^{-2\Delta b} \over (1-e^{-b})(1-e^{-2b})} = \sum_{m=0}^\infty {e^{-(2\Delta + 2m) b} \over 1-e^{-b}}$ and use \eqref{strace op 6j} with $\Delta' = 2\Delta + 2m$.
Higher order terms $n\geq 3$ can be similarly computed as a sum over contributions of higher-trace operators.


\section{More t'Hooft diagrams on the double-trumpet}

\label{sec:dtw=3}

In this appendix, we compute t' Hooft diagrams for the empty double-trumpet and 2-point function on the double-trumpet, where a right-left path crosses 3 lines.

\subsection{Empty double-trumpet}

\label{sec:computed3}

\begin{figure}
	\centering
	\includegraphics[scale=.3]{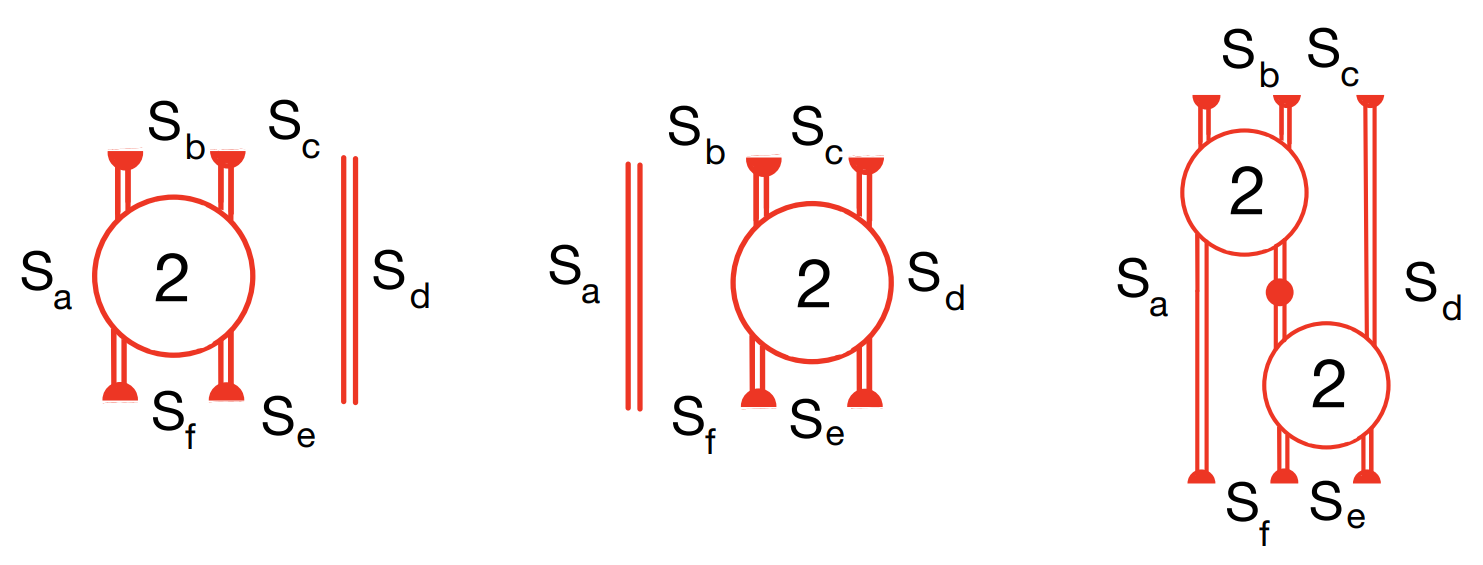}
	\caption{On the left we graphically represent $\calx^{(2)}_{1}$. In the center we graphically represent $\calx^{(2)}_2$. We may think of $\calx^{(2)}_{1}$ and $\calx^{(2)}_2$ as operators that act on a function of $s_f$ and $s_e$ to produce a function of $s_b$ and $s_c$. On the right, we represent the product $\calx^{(2)}_{1}\calx^{(2)}_{2}$.}
	\label{fig:threediagrams}
\end{figure}

\begin{figure}
	\centering
	\includegraphics[scale=.3]{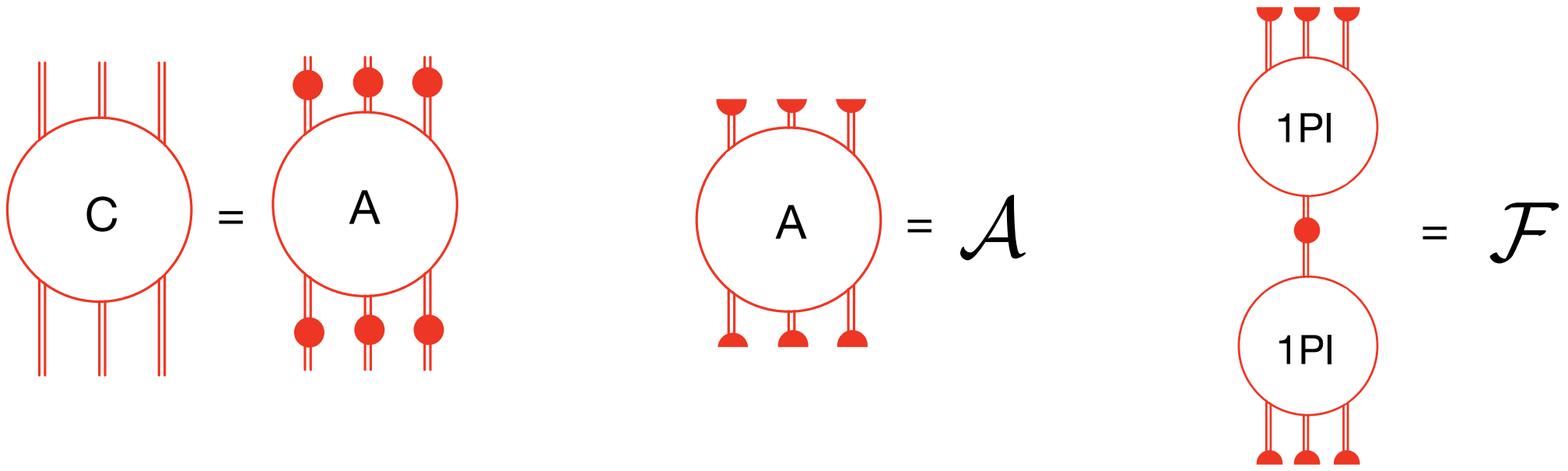}
	\caption{
	On the left we relate the sum over connected planar six-point diagrams (represented by the ``C'' blob) to the sum of amputated six-point diagrams (represented by the ``A'' blob). The operator $\cala$ is defined by multiplying the sum of amputated six-point diagrams by a factor of $\sqrt{\braket{\calo_{a b} \calo_{b a}}_{\disk}}$ for each external double-line, where $s_a$ and $s_b$ label the two energies appearing in the double-line. Each factor of $\sqrt{\braket{\calo_{a b} \calo_{b a}}_{\disk}}$ is represented by a red semicircle. The operator $\calf$ is defined in terms of the 1PI planar four-point function as shown on the right. Note that $\cala$ and $\calf$ do not depend on the normalization of $\calo$.}
	\label{fig:AFblob}
\end{figure}

\begin{figure}
	\centering
	\includegraphics[scale=.3]{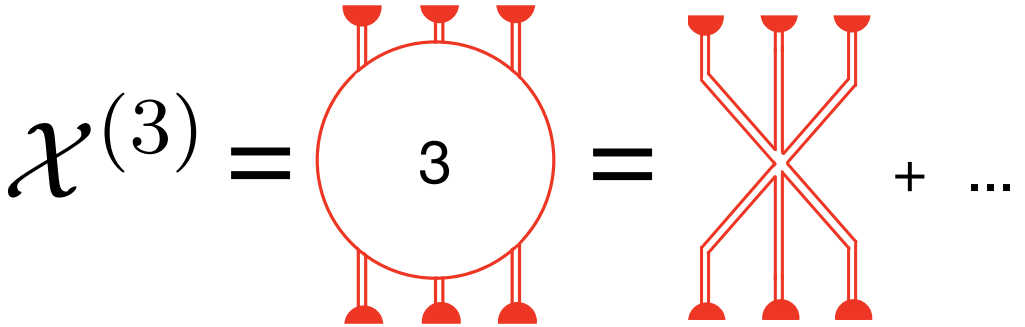}
	\caption{
	The blob with a ``3'' corresponds to the sum over all connected planar six-point diagrams such that any left-right path through the blob crosses over more than three double-lines. We define $\calx^{(3)}$ as the product of this blob with a factor of $\sqrt{\braket{\calo_{a b} \calo_{b a}}_{\disk}}$ for each external leg, where $s_a$ and $s_b$ are the two energies associated with the leg. These factors are represented by red semicircles. The simplest contribution to the blob labelled ``3'' is a tree six-point vertex (in this case, there are no left-right paths that can go through the blob, because we only let left-right paths cross over double-lines, and all of the double-lines are external. So it is vacuously true that all left-right paths through the blob cross over more than three double-lines).}
	\label{fig:blob3}
\end{figure}

\begin{figure}
	\centering
	\includegraphics[scale=.4]{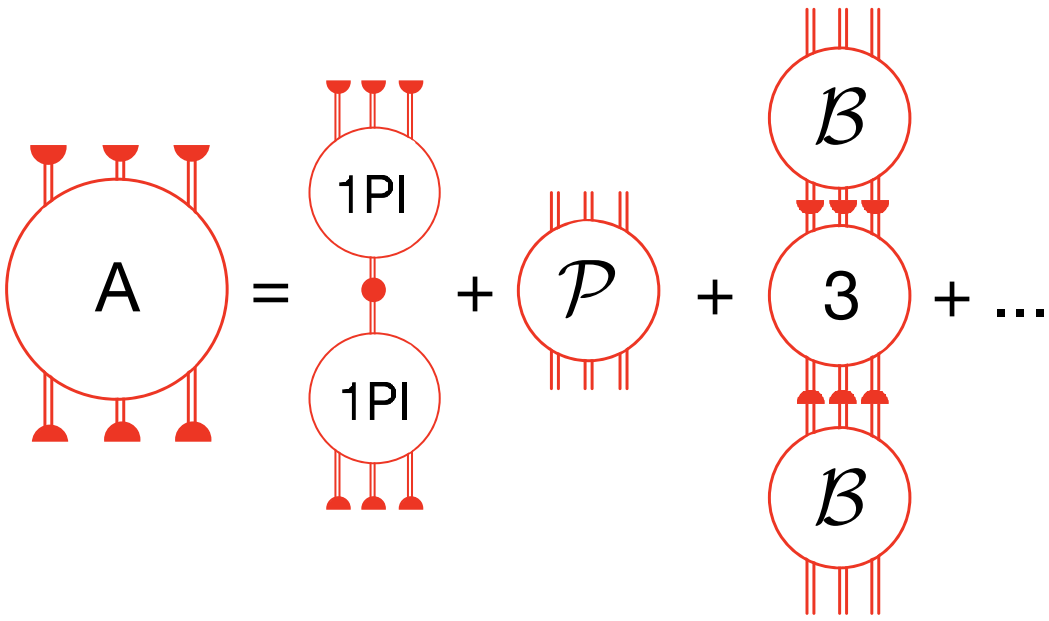}
	\caption{To solve for $\calx^{(3)}$, we systematically classify the diagrams that contribute to the amputated connected six-point function, which is contained in $\cala$ and appears on the left hand side. The diagrams that contribute to $\calf$ (first term on the right hand side) all allow for left-right paths that cross over only one double-line. The remaining diagrams on the right side have the property that any left-right path crosses over three or more double-lines. We organize these diagrams by explicitly showing all the locations where a left-right path can cross over three double-lines. The remaining parts of the diagrams are encapsulated in $\calx^{(3)}$. The blobs labeled $\calb$ and $\calp$ should be replaced by sums over products of the first two diagrams in Figure \ref{fig:threediagrams}, according to \eqref{eq:137} and \eqref{eq:138}. The $\ldots$ includes further terms in the geometric series in \eqref{eq:calaeq}. Note that $\calp$ is defined similarly to $\calb$, except that $\calp$ only contains connected diagrams. Given that the left hand side is a sum of connected diagrams, only connected diagrams may appear on the right side.}
	\label{fig:d3diagrams}
\end{figure}

To compute $\cald^{(3)}$, we consider diagrams that allow for left-right paths that cross over three $\calo$ double-lines, but we exclude those diagrams that were already counted in $\cald^{(1)}$ and $\cald^{(2)}$. It is convenient to think of each diagram as an operator that acts on functions of two energies. Before we can enumerate the diagrams that contribute to $\cald^{(3)}$, we need to define a few operators. The first two operators we consider are defined as follows:
\begin{equation}
    \left[\calx^{(2)}_{1}\right]_{a d}(s_b,s_c;s_f,s_e) = \calx^{(2)}_{a c}(s_b,s_f) \frac{\delta(s_c - s_e)}{\rho(s_c)}, \quad     \left[\calx^{(2)}_{2}\right]_{a d}(s_b,s_c;s_f,s_e) = \calx^{(2)}_{b d}(s_c,s_e) \frac{\delta(s_b - s_f)}{\rho(s_b)}.  \label{eq:135}
\end{equation}
For convenience, we will often write $\calx^{(2)}_{1}$ and $\calx^{(2)}_2$ to refer to these operators.\footnote{All of the subsequent operators we define will depend on the same set of energies, which we will no longer explicitly write unless necessary.} We depict these operators in Figure \ref{fig:threediagrams}.
The product of $\calx^{(2)}_{i}$ and $\calx^{(2)}_{j}$ (with $i,j \in \{1,2\}$) is defined as follows:
\begin{equation}
\label{eq:productdef}
    \left[\calx^{(2)}_{j} \calx^{(2)}_{i}\right]_{a d}(s_b,s_c;s_f,s_e) =
    \int_0^\infty ds_1 \rho(s_1) ds_2 \rho(s_2) \, \left[\calx^{(2)}_{j}\right]_{a d}(s_b,s_c;s_1,s_2)
    \left[\calx^{(2)}_{i}\right]_{a d}(s_1,s_2;s_f,s_e).
\end{equation}
We use this product to define the operators $\calb$ and $\calp$ as follows:
\begin{align}
\label{eq:137}
    \calb &= \sum_{n = 0}^\infty \left( \calx^{(2)}_{1} + \calx^{(2)}_{2}\right)^n
    \\
    \label{eq:138}
    \calp &= \calb - 1 -  \sum_{n = 1}^\infty\left( \left(\calx^{(2)}_{1}\right)^n + \left(\calx^{(2)}_{2}\right)^n\right).
\end{align}
The motivation for defining $\calb$ and $\calp$ in this way will be clear from Figure \ref{fig:d3diagrams}. The next two operators we introduce, $\cala$ and $\calf$, are respectively defined from the connected planar six-point and four-point functions of $\calo$. See Figure \ref{fig:AFblob}.
The last operator we need, $\calx^{(3)}$, is defined in Figure \ref{fig:blob3}.
The product of any two operators is defined as in \eqref{eq:productdef}. The integrals over $s_1$ and $s_2$ correspond to integrating over the energies that appear in the closed index loops that are formed when a diagram in the first operator is attached to a diagram in the second operator. Note that when either $\calx^{(2)}_1$ or $\calx^{(2)}_2$ appear in a product, at most one closed index loop is formed, which reflects the fact that delta functions appear in \eqref{eq:135}.
Having defined $\calx^{(2)}_1$, $\calx^{(2)}_2$, $\calb$, $\calp$, $\cala$, $\calf$, and $\calx^{(3)}$, we note that the explicit formulas for all of these operators are known {\it a priori}, except for $\calx^{(3)}$. This is because $\calx^{(2)}_1$  and $\calx^{(2)}_2$ follow from \eqref{eq:131}, while $\cala$ and $\calf$ follow directly from the connected planar correlators of $\calo$, which are known from the (regulated) gravitational Feynman rules. To determine $\calx^{(3)}$, we use the following relation:
\begin{equation}
    \cala = \calf + \calp + \calb \calx^{(3)} \calb \left[ 1 - \calx^{(3)} \calb \right]^{-1},
    \label{eq:calaeq}
\end{equation}
which is depicted and explained in Figure \ref{fig:d3diagrams}.
After using \eqref{eq:calaeq} to solve for $\calx^{(3)}$, we may directly compute the sum of all diagrams that contribute to $\cald^{(3)}$. The basic building blocks we use to enumerate these diagrams are $\calx^{(2)}_1$, $\calx^{(2)}_2$, and $\calx^{(3)}$. Note that $\calb$ counts all of the diagrams that may be enumerated using $\calx^{(2)}_1$ and $\calx^{(2)}_2$ only. The result is
\begin{align}
    \begin{split}
        \cald^{(3)}_{a d} &= \text{Tr } \sum_{n = 1}^\infty \frac{1}{n} \left(\left[\calx^{(2)}_1 + \calx^{(2)}_2\right]^n - \left[\calx^{(2)}_1\right]^n - \left[\calx^{(2)}_2\right]^n \right)
        \\
        &+ \text{Tr } \sum_{n = 1}^\infty \frac{1}{n} \left[ \calx^{(3)} \calb \right]^n.
    \end{split}
    \label{eq:d3sum}
\end{align}
The trace in \eqref{eq:d3sum} corresponds to setting $s_b = s_f$ and $s_c = s_e$ in Figure \ref{fig:threediagrams} and then integrating over these energies with measure $ds \, \rho(s)$. We are left with a function of $s_a$ and $s_d$ that corresponds to $\cald^{(3)}_{a d}$. On the first line of \eqref{eq:d3sum}, we enumerate all of the diagrams that may be built out of $\calx^{(2)}_1$ and $\calx^{(2)}_2$. The subtractions ensure that we only count connected diagrams. On the second line, we organize the diagrams by the number of $\calx^{(3)}$ factors that appear. The $\frac{1}{n}$ factors ensure that there is no overcounting and also supply the correct symmetry factors to the diagrams that have a $\mathbb{Z}_n$ symmetry.

Using \eqref{eq:calaeq}, we have that
\begin{equation}
\label{eq:141}
   \cala - \calf - \calp + \calb =  \calb \left[ 1 - \calx^{(3)} \calb \right]^{-1} := \calr,
\end{equation}
from which it follows that
\begin{equation}
    1- \calr^{-1} \calb  =   \calx^{(3)} \calb.
\end{equation}
The second line of \eqref{eq:d3sum} becomes
\begin{equation}
    \text{Tr } \sum_{n = 1}^\infty \frac{1}{n} \left[ \calx^{(3)} \calb \right]^n = -     \text{Tr } \log \calr^{-1} \calb =      \text{Tr } \log \calr - \text{Tr } \log \calb, 
\end{equation}
so that \eqref{eq:d3sum} becomes
\begin{equation}
    \cald^{(3)}_{a d} =  \text{Tr } \log (1 - \calx^{(2)}_1) + \text{Tr } \log (1 - \calx^{(2)}_2) +     \text{Tr } \log \calr,
    \label{eq:144}
\end{equation}
where we have used \eqref{eq:137} to substitute for $\calb$. Our method for computing $\cald^{(3)}$ is valid for any single-trace, two-matrix model. Given the connected planar correlators which are fixed by the disk amplitudes of the model, we may determine $\calx^{(2)}_1$, $\calx^{(2)}_2$, and $\calr$ and thus $\cald^{(3)}$.

We now obtain explicit expressions for $\cald^{(3)}$ using the $q$-deformed and Selberg regulators. We first define the operators $\cals^{q}_{1}$ and $\cals^{q}_{2}$ as follows:
\begin{align}
\left[\cals^{q}_{1}\right]_{a d}(s_b,s_c;s_f,s_e) &\equiv  \left\{\begin{array}{ccc}
		\Delta & s_b & s_c \\
		\Delta & s_f & s_a
	\end{array}\right\}_q \frac{\delta(s_c - s_e)}{\rho_q(s_c)},
	\\
	\left[\cals^{q,}_{2}\right]_{a d}(s_b,s_c;s_f,s_e) &\equiv \frac{\delta(s_b - s_f)}{\rho_q(s_c)}  \left\{\begin{array}{ccc}
		\Delta & s_c & s_d \\
		\Delta & s_e & s_b
	\end{array}\right\}_q. 
\end{align}
Note that the Yang-Baxter equation is
\begin{equation}
\label{eq:yb}
    \cals^{q}_{1} \cals^{q}_{2} \cals^{q}_{1} = \cals^{q}_{2} \cals^{q}_{1} \cals^{q}_{2}.
\end{equation}
Using the regulated gravitational Feynman rules to determine the disk amplitudes, we find that
\begin{equation}
\cala = \calf + \epsilon^2 \cals^{q}_{2}\cals^{q}_{1} + \epsilon^2 \cals^{q}_{1}\cals^{q}_{2} + \epsilon^{\#}\cals^{q}_{1}\cals^{q}_{2}\cals^{q}_{1},
\end{equation}
where $\# = 3$ for the $q$-deformed regulator and $\# = 2$ for the Selberg regulator, and
\begin{equation}
\left[\calf\right]_{a d}(s_b,s_c;s_f,s_e) = \epsilon^2 \left\{\begin{array}{ccc}
		\Delta & s_b & s_c \\
		\Delta & s_d & s_a
	\end{array}\right\}_q
	 \left\{\begin{array}{ccc}
		\Delta & s_e & s_f \\
		\Delta & s_a & s_d
	\end{array}\right\}_q.
\end{equation}
From \eqref{eq:135} and the $q$-analogue of \eqref{eq:131}, we have that
\begin{equation}
    \calx^{(2)}_1 = \frac{\epsilon ~ \cals^{q,}_{1}}{1 + \epsilon ~ \cals^{q}_{1}} , \quad     \calx^{(2)}_2 = \frac{\epsilon ~ \cals^{q}_{2}}{1 + \epsilon ~ \cals^{q}_{2}} , 
\end{equation}
and it follows using \eqref{eq:138} that
\begin{align}
    \calb - \calp &= \left[1 - \calx^{(2)}_1\right]^{-1} + \left[1 - \calx^{(2)}_{2}\right]^{-1} - 1,
    \\
    &= 1 + \epsilon ~ \cals^{q}_{1} + \epsilon ~ \cals^{q}_{2}.
\end{align}
Using \eqref{eq:141} and \eqref{eq:144}, we finally have
\begin{align}
\label{eq:154}
    \begin{split}
\cald^{(3)} &= -\text{Tr } \log (1 + \epsilon ~ \cals^{q}_{1}) -\text{Tr } \log (1 + \epsilon ~ \cals^{q}_{2}) 
\\
&+ \text{Tr } \log (1 + \epsilon~\cals^{q}_{1} + \epsilon~\cals^{q}_{2} + \epsilon^2 ~ \cals^{q}_{2}\cals^{q}_{1} + \epsilon^2 ~ \cals^{q}_{1}\cals^{q}_{2} + \epsilon^{\#} \cals^{q}_{1}\cals^{q}_{2}\cals^{q}_{1}).
    \end{split}
\end{align}
The first line of \eqref{eq:154} only depends on the spectrum of $\cals_1^{q}$ and $\cals_2^{q}$, which is independent of the two energies that $\cald^{(3)}$ could depend on. We now argue that the second line of \eqref{eq:154} is also independent of these energies.

To compute the second line of \eqref{eq:154}, we will construct two convenient orthonormal bases for the space of functions of two energies that the operators $\cals_1^{q}$ and $\cals_2^{q}$ act on. We will refer to basis vectors of the first (resp. second) basis as $\ket{n,m}_1$ (resp. $\ket{n,m}_2$), where $n,m \in \mathbb{Z}_{\ge 0}$. The basis vector $\ket{s}$ in the space of functions of one energy is normalized such that
\begin{equation}
    \braket{s_1 | s_2} = \frac{\delta(s_1 - s_2)}{\rho_q(s_1)}.
\end{equation}
Then, we define $\ket{n,m}_1$ by
\begin{equation}
    \left(\bra{s_f} \otimes \bra{s_e} \right)\ket{n,m}_1 = P^{\Delta, \Delta}_{n}(s_f;s_a,s_e|q) P^{2\Delta + n,\Delta}_{m}(s_e;s_a,s_d|q),
\end{equation}
and $\ket{n,m}_2$ is defined by
\begin{equation}
    \left(\bra{s_f} \otimes \bra{s_e} \right)\ket{n,m}_2 = P^{\Delta,2\Delta + m}_{n}(s_f;s_a,s_d|q) P^{\Delta,\Delta}_{m}(s_e;s_f,s_d|q).
\end{equation}
Using \eqref{eq:askeywilsonpolyorth}, we have
\begin{equation}
    \, _1\braket{n,m|\tilde{n},\tilde{m}}_1 = \, _2\braket{n,m|\tilde{n},\tilde{m}}_2 = \delta_{n,\tilde{n}}\delta_{m,\tilde{m}},
\end{equation}
while \eqref{eq:completeness} ensures that these bases are complete. Our two bases are eigenbases of $\cals_1^{q}$ and $\cals_2^{q}$,
\begin{equation}
    \cals_1^{q,\epsilon}\ket{n,m}_1 = (-1)^n q^{\frac{n(n-1)}{2}} q^{2 \Delta n} \ket{n,m}_1, \quad     \cals_2^{q,\epsilon}\ket{n,m}_2 = (-1)^m q^{\frac{m(m-1)}{2}} q^{2 \Delta m} \ket{n,m}_2,
\end{equation}
which shows that $\cals_1^{q}$ and $\cals_2^{q}$ are Hermitian.

The operators $\cals^{q}_{1}$ and $\cals^{q}_{2}$, subject to the relation \eqref{eq:yb}, generate an algebra. A casimir of this algebra is given by
\begin{equation}
    \calc := (\cals^{q}_{1} \cals^{q}_{2} \cals^{q}_{1})^2.
\end{equation}
Using \eqref{eq:pentagon} and \eqref{eq:diagrelation}, we find that
\begin{equation}
\calc \ket{n,m}_1 = \calc_{n,m} \ket{n,m}_1, \quad \calc \ket{n,m}_2 = \calc_{n,m} \ket{n,m}_2, \quad \calc_{n,m} := q^{(n + m)(6 \Delta  - 1 + n + m)}. 
\end{equation}
In particular, the casimir depends only on the sum of $n$ and $m$. Because $n$ and $m$ are non-negative integers, each value of the casimir corresponds to a finite-dimensional representation of the algebra. Hence, in either of the two bases we introduced, any operator in the algebra takes a block-diagonal form, where each block is finite-dimensional. To compute the second line of \eqref{eq:154}, we need to compute the log of the determinant of the operator appearing in parentheses. It suffices to compute the determinant of this operator in each individual block. The second line of \eqref{eq:154} thus takes the form of an infinite sum, where each term in the sum corresponds to a single block. Let us consider the matrix elements of $\cals^{q}_{1}$ and $\cals^{q}_{2}$ in the $\ket{n,m}_1$ basis for concreteness. The matrix elements of $\cals^{q}_{1}$ are determined by its eigenvalues, which do not depend on the two energies that $\cald^{(3)}$ could depend on. To determine the matrix elements of $\cals^{q}_{2}$, we use \eqref{eq:whatshouldicallthis}, which implies that
\begin{equation}
    \ket{n,m}_1 = \sum_{y = 0}^{n + m} S_q^{\Delta,\Delta,\Delta,n + m}(y,n) \ket{n+m-y,y}_2.
\end{equation}
Hence, the matrix elements of $\cals^{q}_{2}$ in the basis that diagonalizes $\cals^{q}_{1}$ do not depend on the two energies that $\cald^{(3)}$ could have depended on. It follows that $\cald^{(3)}_{a d}$ is independent of $s_a$, $s_d$ and does not contribute to the empty double-trumpet.

Having explicitly shown that $\cald^{(2)}$ and $\cald^{(3)}$ do not contribute to the empty double-trumpet, we conjecture that $\cald^{(n)}$ does not contribute for all $n \ge 2$ . The formula for $\cald^{(n)}$ should involve determinants of operators acting on the space of functions of $n$ energies, and we expect that these operators are part of the algebra generated by the appropriate analogues of $\cals^{q}_{1}$ and $\cals^{q}_{2}$. An argument similar to the one given above will then show that $\cald^{(n)}_{a d}$ does not depend on $s_a$ or $s_d$.

\subsection{2-point function on the double-trumpet}

\label{sec:twopointdoubletrumpetappendix}

\begin{figure}
	\centering
	\includegraphics[scale=.35]{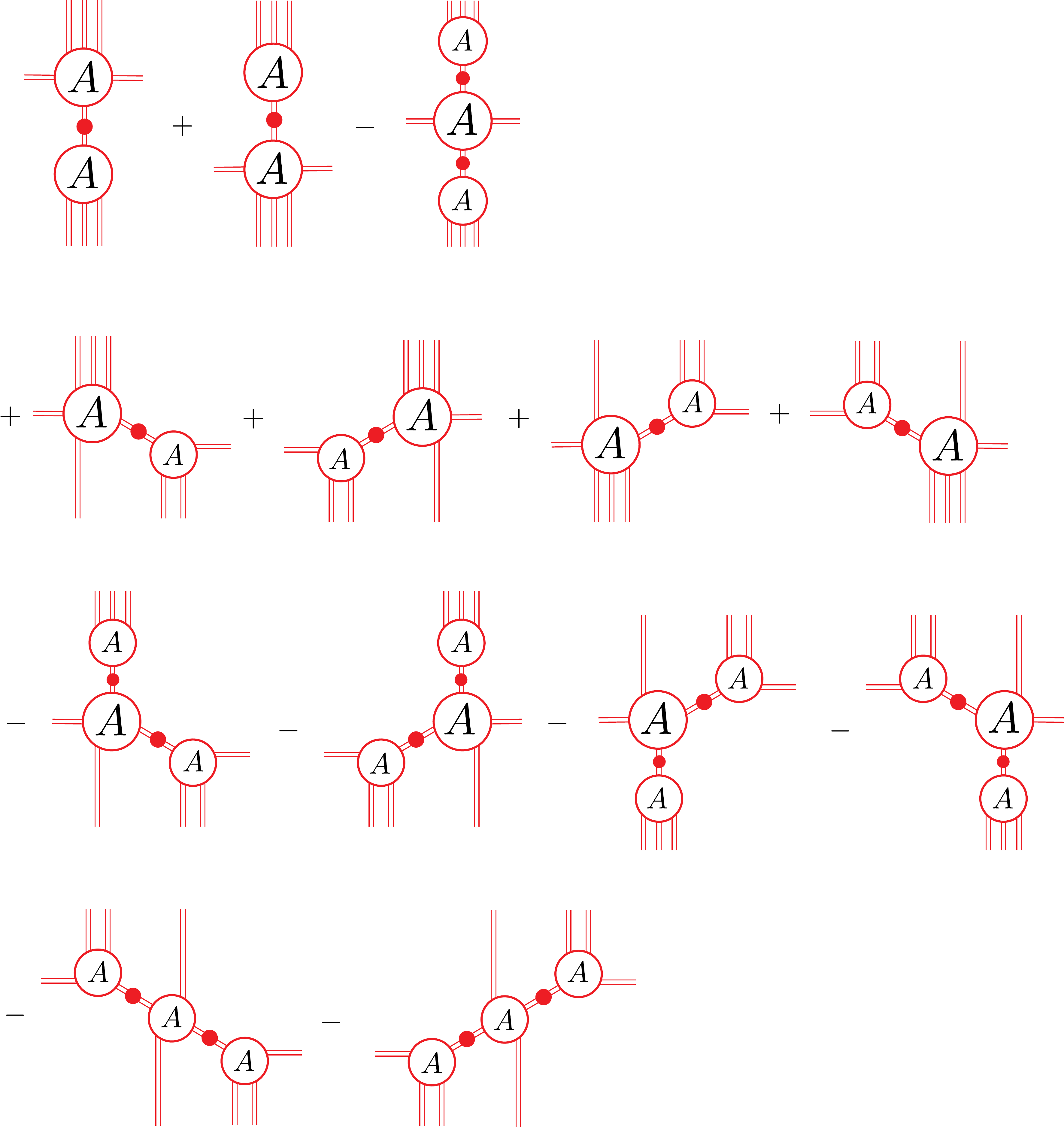}
	\caption{In this Figure, Figure \ref{fig:part2} and Figure \ref{fig:part3}, we enumerate all of the connected eight-point diagrams that admit a ``shortcut'' from the left to the right side of the diagram. A ``shortcut'' is defined as a left-right path that crosses over one or two double-lines. In keeping with the conventions in Figure \ref{fig:AFblob}, a blob with an ``A'' in it refers to a sum over connected planar amputated diagrams.}
	\label{fig:part1}
\end{figure}

\begin{figure}
	\centering
	\includegraphics[scale=.35]{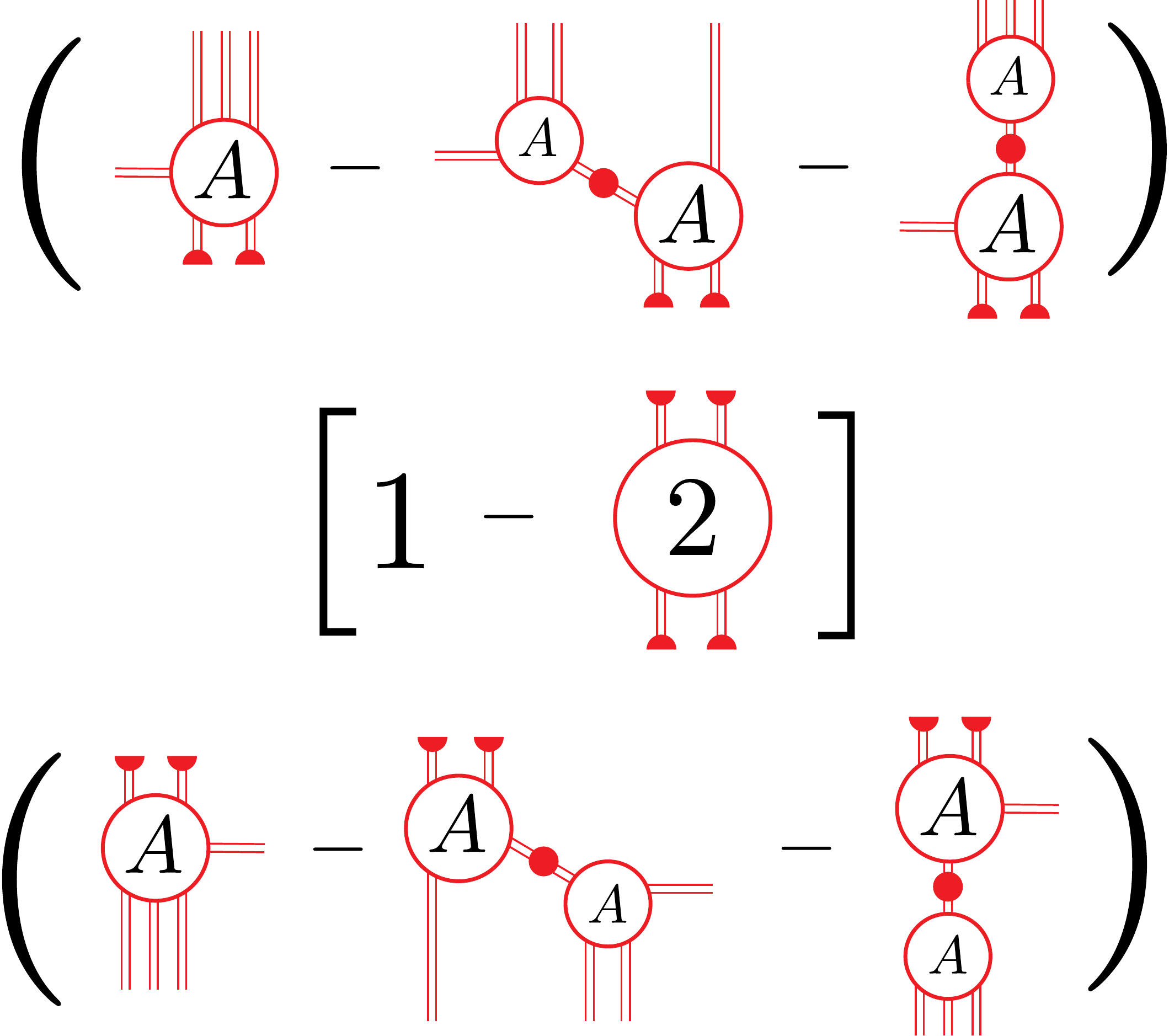}
	\caption{In this Figure, Figure \ref{fig:part1} and Figure \ref{fig:part3}, we enumerate all of the connected eight-point diagrams that admit a ``shortcut'' from the left to the right side of the diagram. A blob with an ``A'' in it refers to a sum over connected planar amputated diagrams. The three factors that are arranged vertically are meant to be multiplied. The ``2'' blob was defined in Figure \ref{fig:b2}.}
	\label{fig:part2}
\end{figure}

\begin{figure}
	\centering
	\includegraphics[scale=.35]{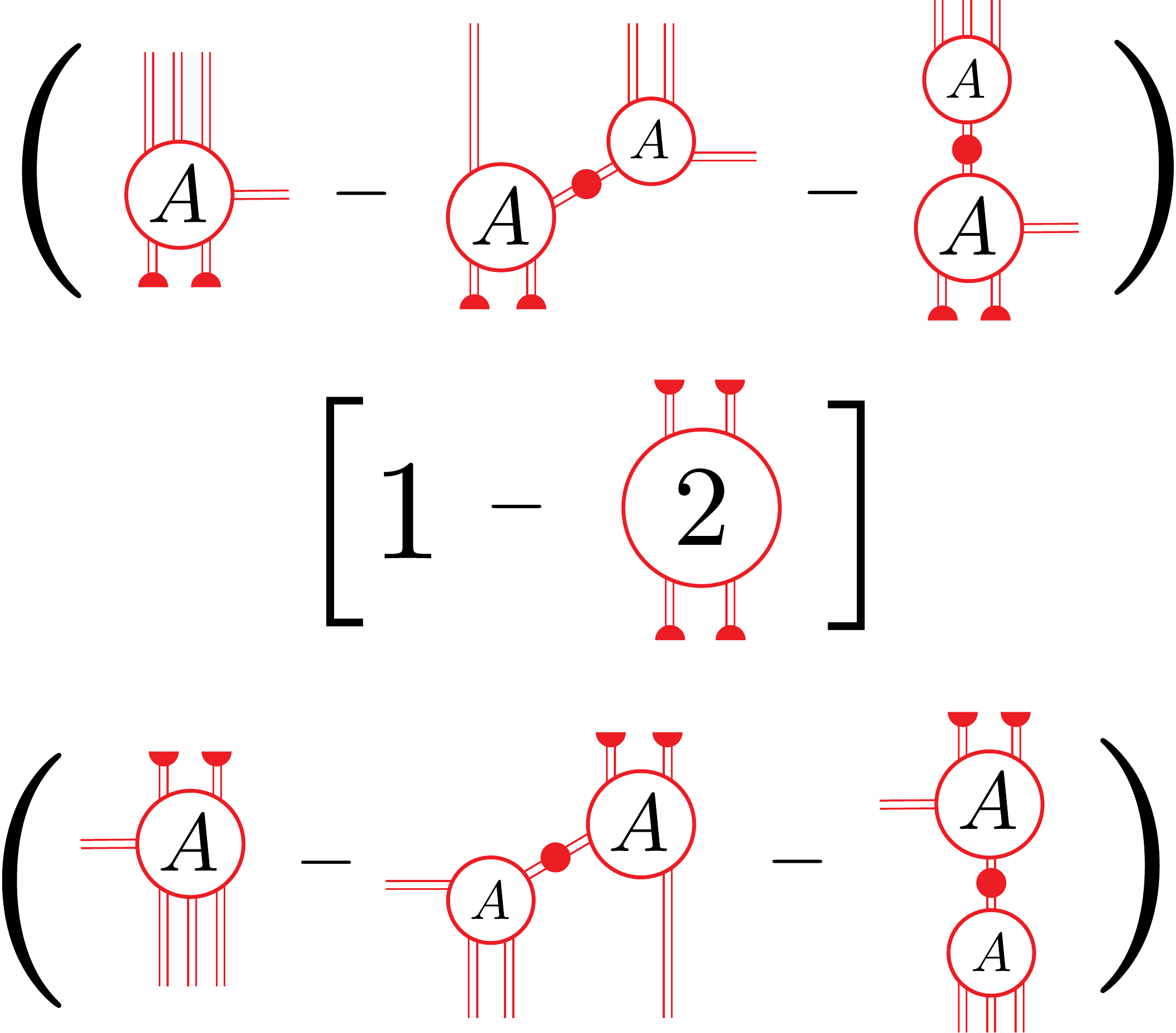}
	\caption{In this Figure, Figure \ref{fig:part1} and Figure \ref{fig:part2}, we enumerate all of the connected eight-point diagrams that admit a ``shortcut'' from the left to the right side of the diagram. A blob with an ``A'' in it refers to a sum over connected planar amputated diagrams. The three factors that are arranged vertically are meant to be multiplied. The ``2'' blob was defined in Figure \ref{fig:b2}.}
	\label{fig:part3}
\end{figure}

\begin{figure}
	\centering
	\includegraphics[scale=.4]{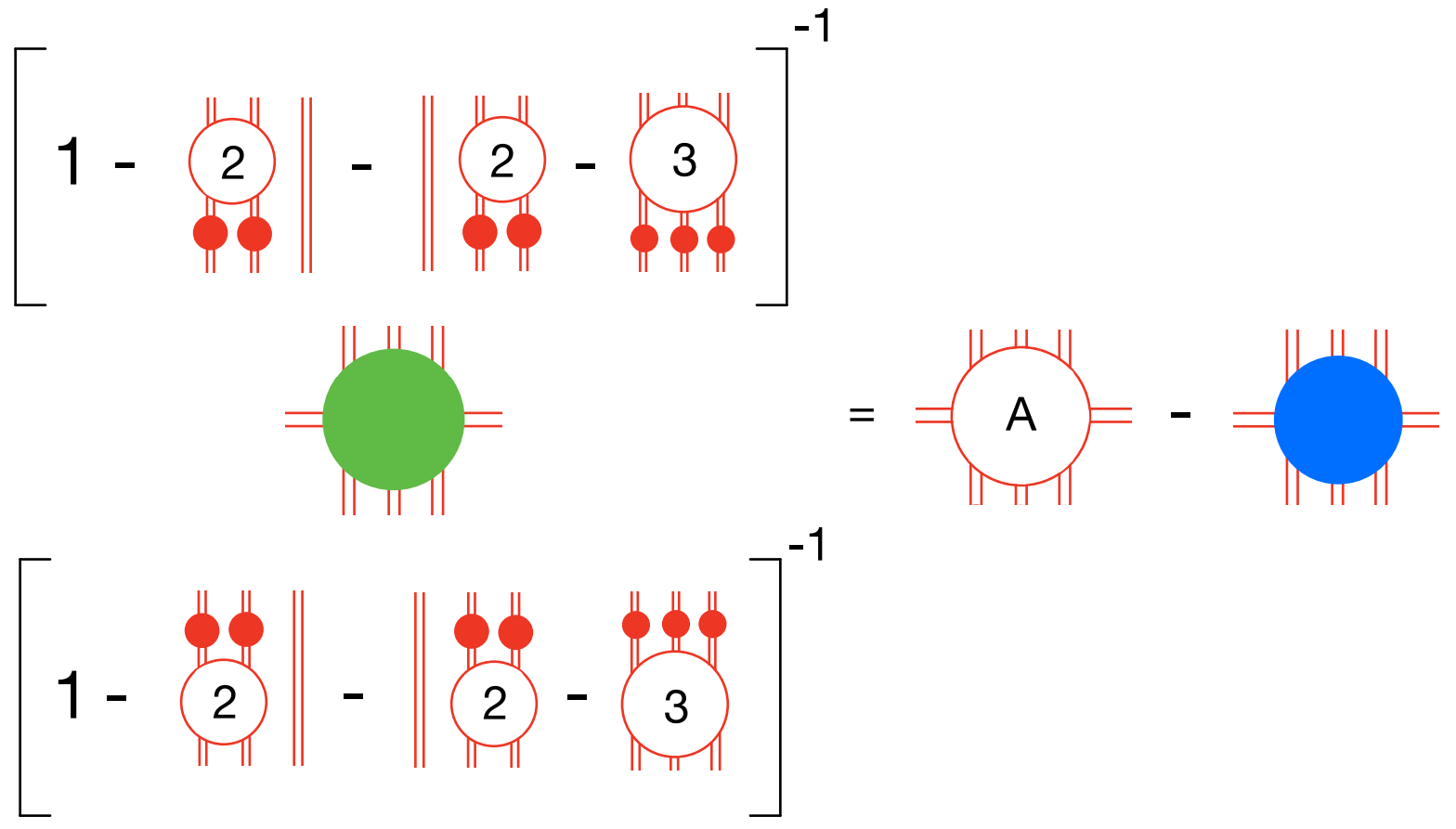}
	\caption{We define the green blob with eight external double-lines. First, we subtract all diagrams that permit shortcuts from the sum over connected amputated eight-point diagrams. The blue blob is defined to be the sum over all the diagrams in Figures \ref{fig:part1}, \ref{fig:part2}, and \ref{fig:part3}. Furthermore, we must amputate off factors of the ``2'' and ``3'' blobs from the top and bottom three legs. The ``3'' blob is defined in Figure \ref{fig:blob3}. This figure defines the green blob in analogy to Figure \ref{fig:dttwopoint2_greenblob}. The top and bottom ends of this diagram are not identified.}
	\label{fig:dttwopoint3_greenblob}
\end{figure}

\begin{figure}
	\centering
	\includegraphics[scale=.3]{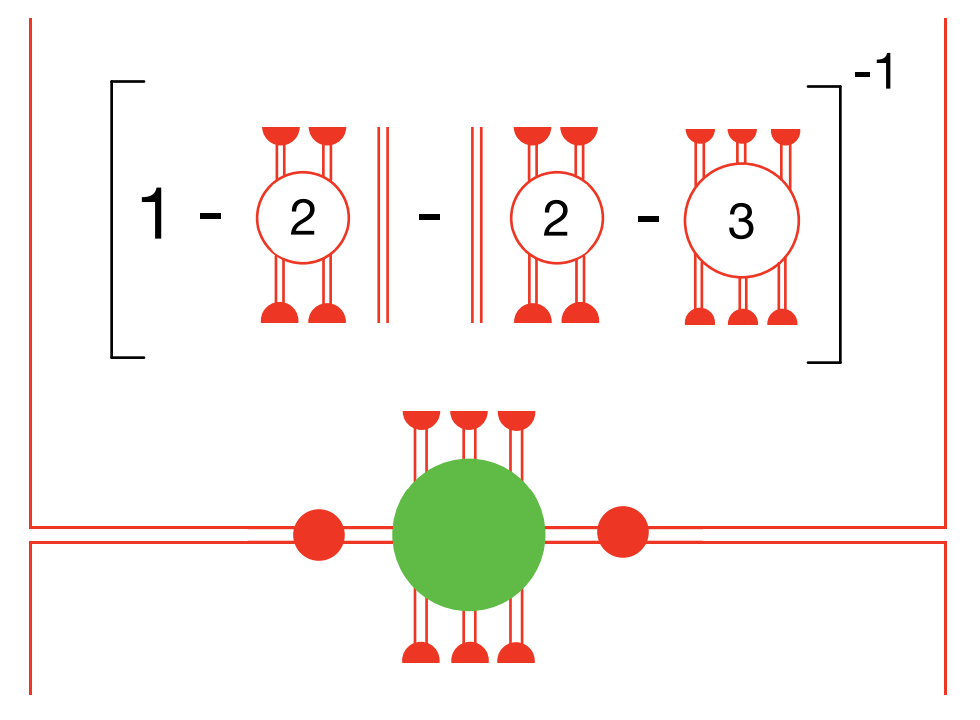}
	\caption{We enumerate the sum over all diagrams such that the minimum number of double-lines crossed by a left-right path is three. The green blob is defined in Figure \ref{fig:dttwopoint3_greenblob}. The top and bottom ends of this diagram are identified.}
	\label{fig:dttwopoint3}
\end{figure}

Continuing from the end of section \ref{sec:twopointfunctiondoubletrumpet}, we now consider diagrams such that the minimum number of $\calo$ double-lines that are crossed by a left-right path is three. This is the last class of diagrams that we explicitly compute. In this case, the analogue of the green blob in Figure \ref{fig:dttwopoint2} will have eight external double-lines. Six of the double-lines will wrap around the double-trumpet, while the other two double-lines go to the AdS boundaries where they represent an insertion of $\calo$. The green blob should only include connected diagrams. However, there are connected diagrams that allow for left-right paths that cross over only one or two double-lines. These diagrams must be subtracted from the sum over all connected eight-point diagrams (this subtraction is analogous to the subtraction on the right hand side of Figure \ref{fig:dttwopoint2}). In Figures \ref{fig:part1}, \ref{fig:part2}, and \ref{fig:part3}, we enumerate all of the connected diagrams that should be subtracted. We then define the green blob in Figure \ref{fig:dttwopoint3_greenblob}. In Figure \ref{fig:dttwopoint3}, we enumerate all the diagrams that we wish to resum.

Next, note that \eqref{eq:calaeq}, \eqref{eq:137}, and \eqref{eq:138} imply that
\begin{align}
    \cala - \calf &= \calp + \calb \left(\left[1 - \calx^{(3)} \calb\right]^{-1} - 1\right)
    \\
    &=  - 1 - \frac{\calx_1^{(2)}}{1 - \calx_1^{(2)}} - \frac{\calx_2^{(2)}}{1 - \calx_2^{(2)}} + \calb \left[1 - \calx^{(3)} \calb\right]^{-1} 
            \\
    &=  - 1 - \frac{\calx_1^{(2)}}{1 - \calx_1^{(2)}} - \frac{\calx_2^{(2)}}{1 - \calx_2^{(2)}} +  \frac{1}{\calb^{-1} - \calx^{(3)} }
            \\
    &=  - 1 - \frac{\calx_1^{(2)}}{1 - \calx_1^{(2)}} - \frac{\calx_2^{(2)}}{1 - \calx_2^{(2)}} +  \frac{1}{1 - \calx_1^{(2)} - \calx_2^{(2)} - \calx^{(3)} }    
\end{align}
from which it follows that
\begin{equation}
\frac{1}{1 - \calx_1^{(2)} - \calx_2^{(2)} - \calx^{(3)} } = 1 + \frac{\calx_1^{(2)}}{1 - \calx_1^{(2)}} + \frac{\calx_2^{(2)}}{1 - \calx_2^{(2)}} + \cala - \calf
\label{eq:202}
\end{equation}
Recalling the graphical representations of $\calx_1^{(2)}$ and $\calx_2^{(2)}$ in Figure \ref{fig:threediagrams} as well as the representation of $\calx^{(3)}$ in Figure \ref{fig:blob3}, it follows that the left hand side of \eqref{eq:202} is graphically represented by
\begin{equation}
\includegraphics[scale=0.4]{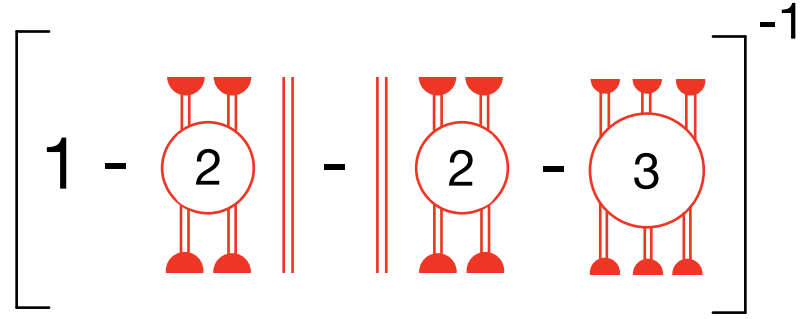}
\end{equation}
which appears in Figure \ref{fig:dttwopoint3}.

We now evaluate the right hand side of Figure \ref{fig:dttwopoint3_greenblob} using the $q$-deformed regulator. There are 27 connected chord diagrams with eight external lines. These contribute to the ``A'' blob. Of these chord diagrams, 21 are subtracted. The right hand side of Figure \ref{fig:dttwopoint3_greenblob} becomes the sum over the following 6 chord diagrams:
\begin{equation}
    \includegraphics[scale=0.15]{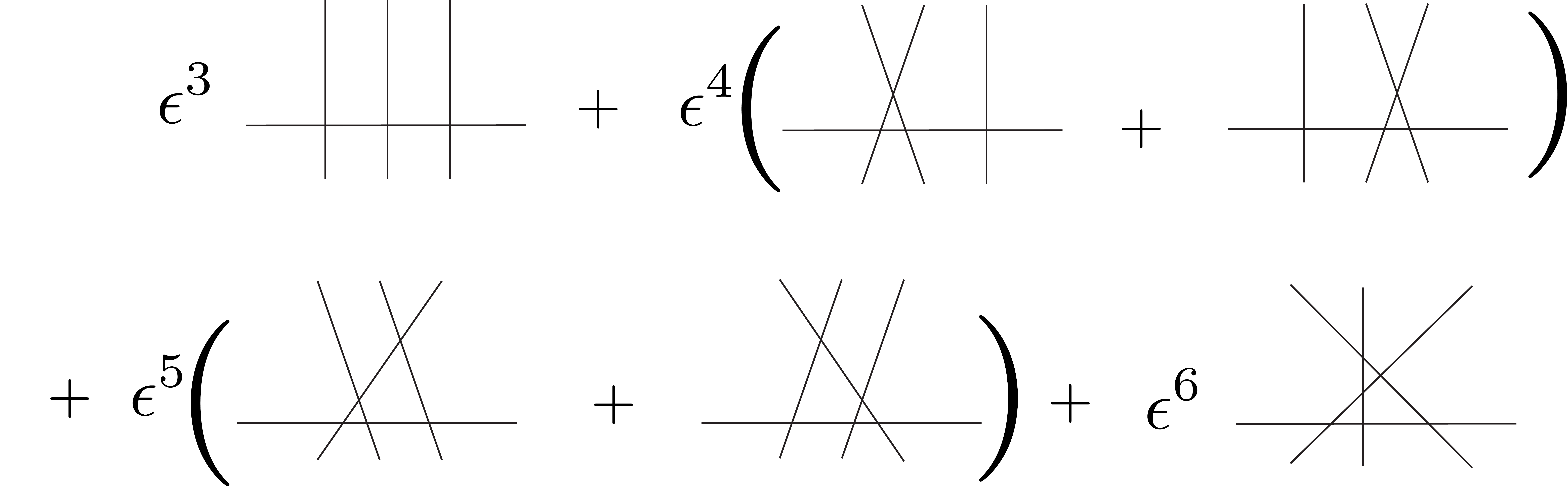}
    \label{eq:qdeformed3symmetric}
\end{equation}
This is a symmetric sum over six permutations. To obtain the sum over the diagrams in Figure \ref{fig:dttwopoint3}, we should multiply \eqref{eq:qdeformed3symmetric} by the inverse of \eqref{eq:202} and identify the top and bottom ends of the diagrams (or equivalently, take a trace). In the $q$-deformed model, \eqref{eq:202} becomes
\begin{equation}
    \includegraphics[scale=0.15]{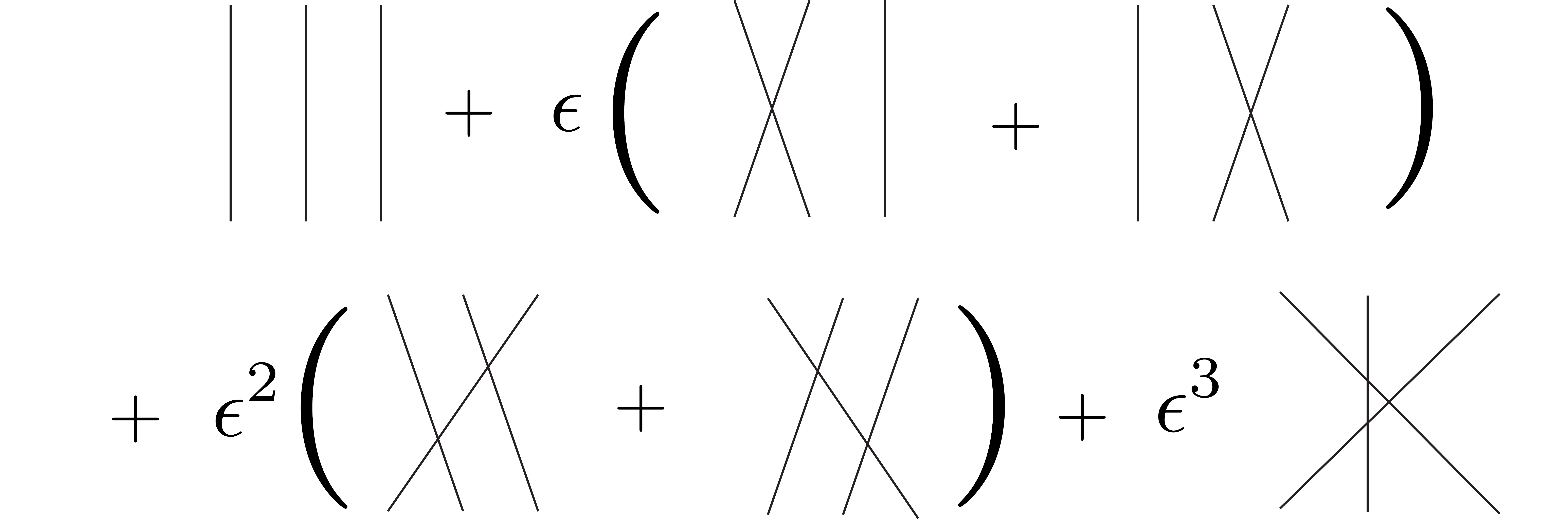}
    \label{eq:diagram_symmetrized3lines}
\end{equation}
Multiplying \eqref{eq:qdeformed3symmetric} by the inverse of \eqref{eq:diagram_symmetrized3lines} has the effect of undoing the symmetrization over the three lines in \eqref{eq:qdeformed3symmetric}. The final result for Figure \ref{fig:dttwopoint3} is
\begin{equation}
    \includegraphics[scale=0.15]{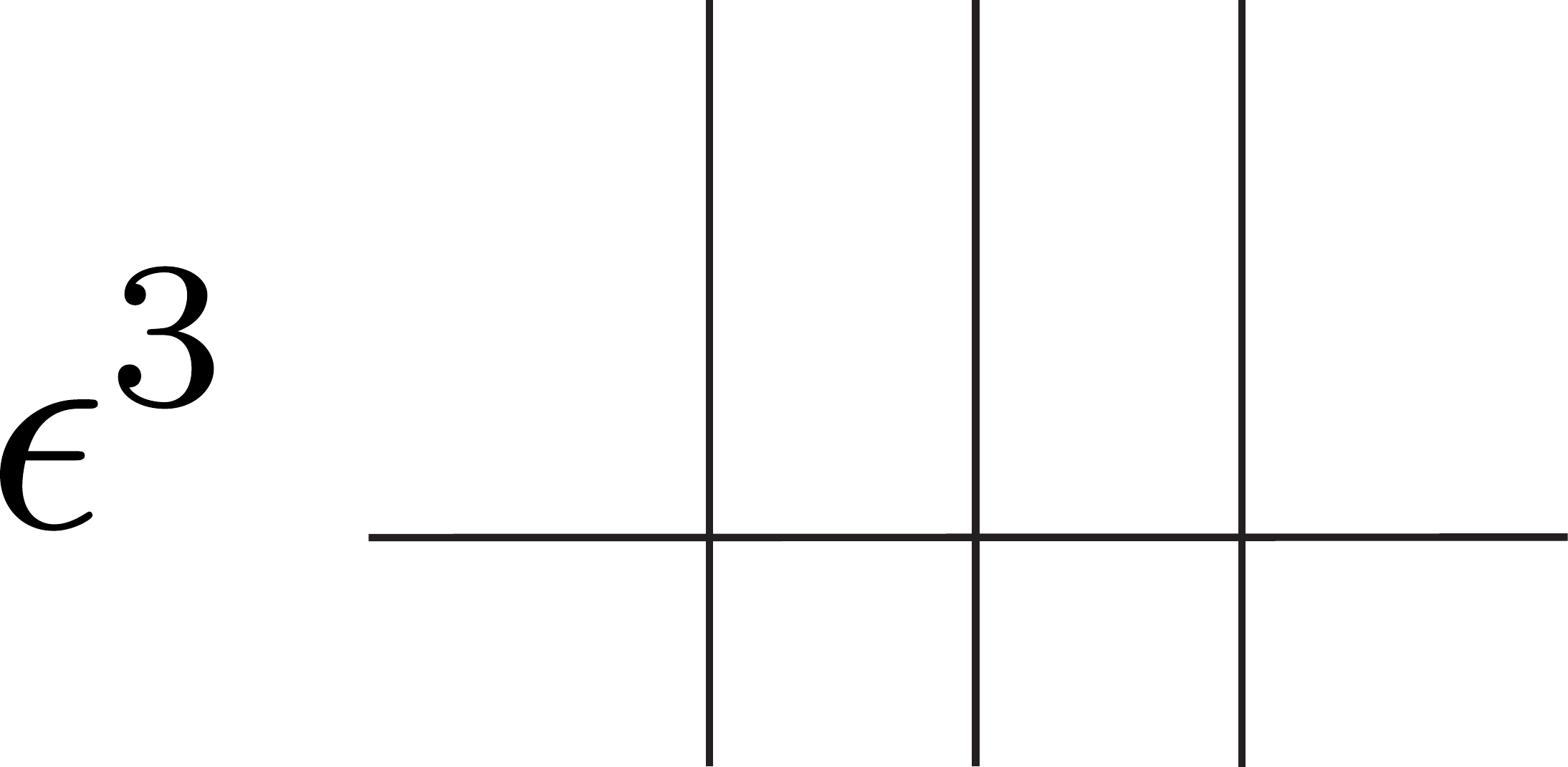}.
\end{equation}
After using the pentagon identity \eqref{eq:afterpentagon} and taking the JT limit, this becomes
\begin{equation}
    \int ds_a \rho(s_a)  \, ds_c \rho(s_c) \, e^{-\beta_L s_a^2 - \beta_R s_c^2}
    (\Gamma^\Delta_{aa}  \Gamma^\Delta_{cc})^{1/2}
     \sum_{n,m = 0}^\infty    \left\{\begin{array}{ccc}
		3 \Delta + n + m & s_a & s_c \\
		 \Delta & s_c & s_a
	\end{array}\right\}.
\end{equation}

\bigskip

Finally, we turn to the Selberg regulator. Using the Selberg regulator, \eqref{eq:202} becomes
\begin{equation}
\label{eq:208}
    \includegraphics[scale=0.15]{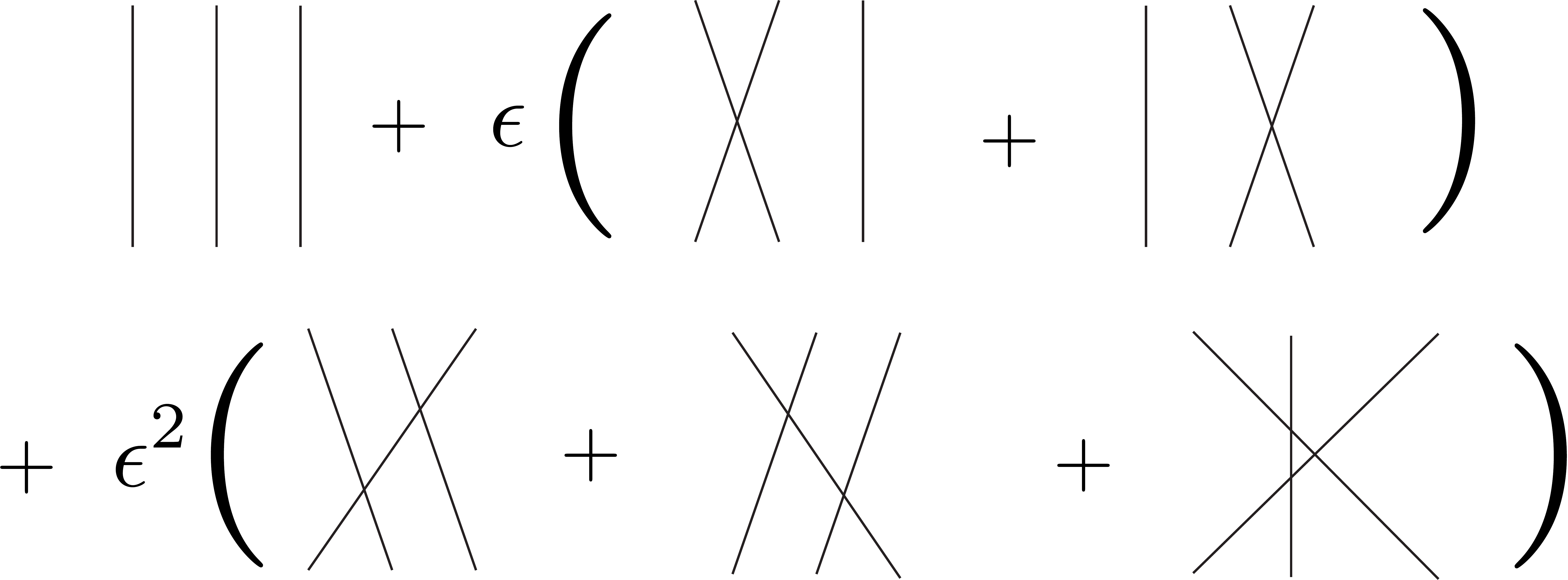}
\end{equation}
which differs from \eqref{eq:diagram_symmetrized3lines} by the weighting assigned to the last diagram.

We now consider the right hand side of Figure \ref{fig:dttwopoint3_greenblob} using the Selberg regulator. Note that each chord diagram that contributes to the ``A'' blob is weighted by $\epsilon^3$. Furthermore, every diagram in Figure \ref{fig:part1} is weighted by $\epsilon^3$. In Figures \ref{fig:part2} and \ref{fig:part3}, the terms inside the round brackets are weighted by $\epsilon^2$. The factor inside the square brackets may be evaluated from \eqref{eq:190}. To simplify the remainder of this computation, we now set $q = 1$. From \eqref{eq:190}, we have that
\begin{equation}
    \includegraphics[scale=0.1]{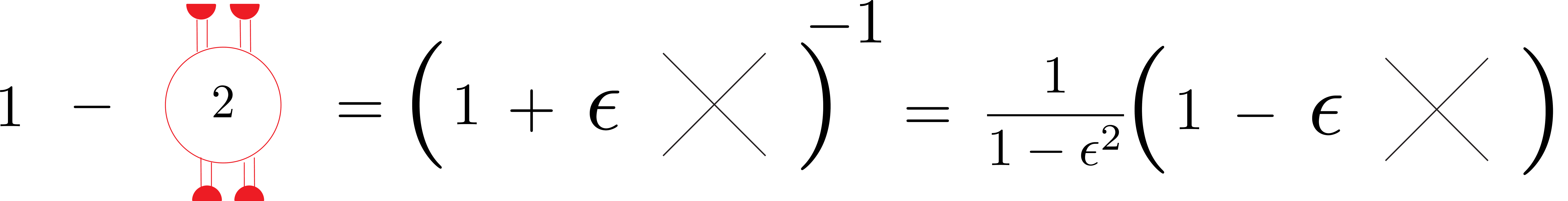}
\end{equation}
where in the last equality we used \eqref{eq:6jorthog}. It follows that the diagrams in Figure \ref{fig:part2} evaluate to
\begin{equation}
    \includegraphics[scale=0.15]{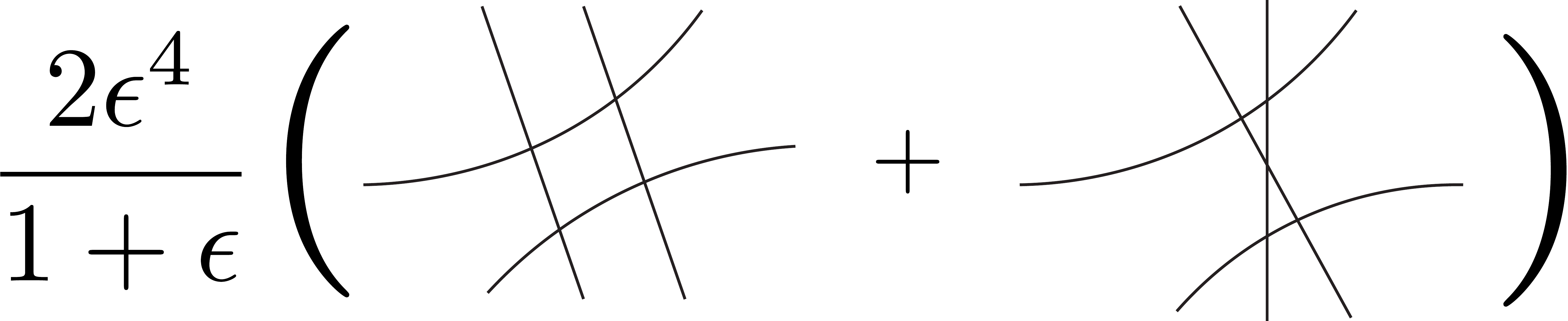}
\end{equation}
while those in Figure \ref{fig:part3} evaluate to
\begin{equation}
    \includegraphics[scale=0.15]{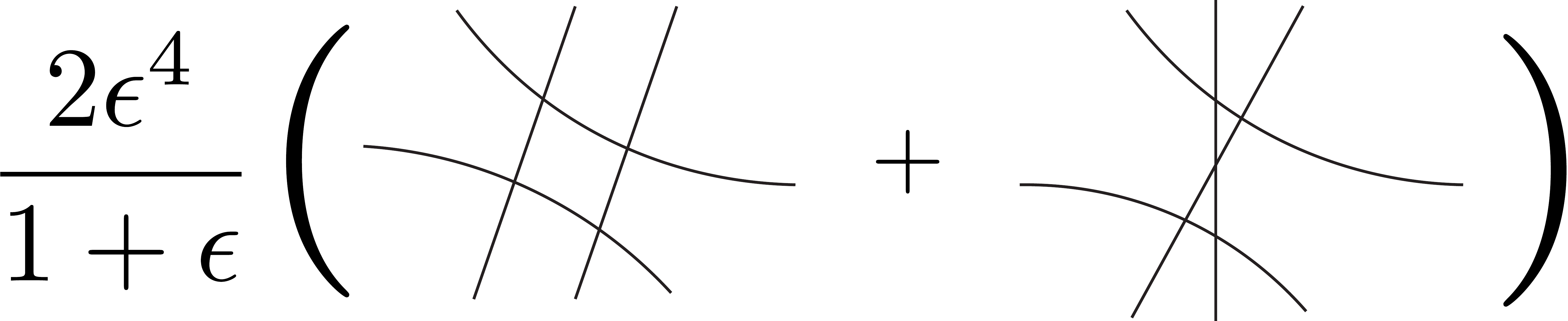}.
\end{equation}
The right hand side of Figure \ref{fig:dttwopoint3_greenblob} then becomes the sum over the following chord diagrams:
\begin{equation}
\includegraphics[scale=0.13]{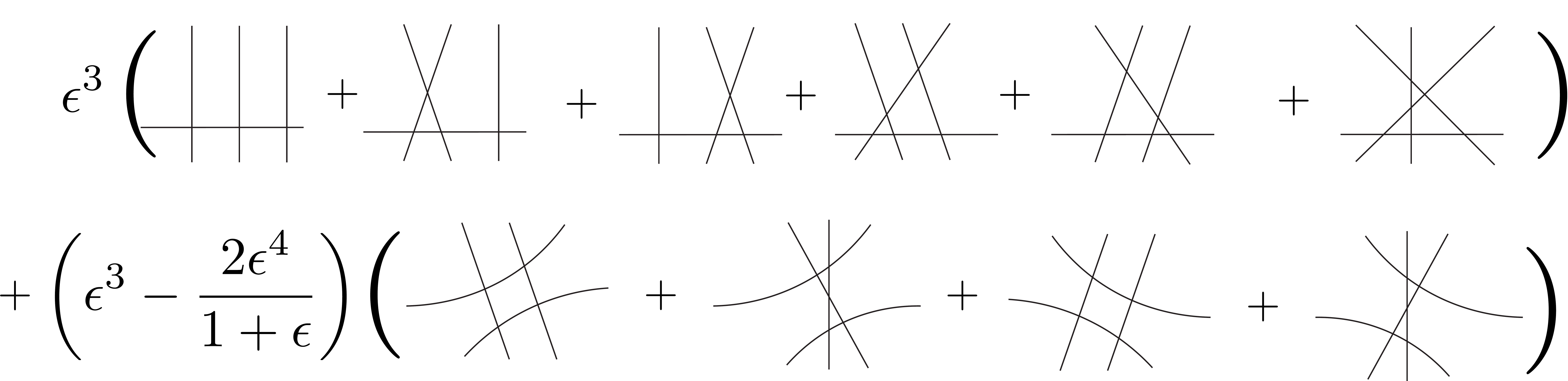}.
\label{eq:212}
\end{equation}
Next, we must multiply the inverse of \eqref{eq:208} with \eqref{eq:212} and then take a trace. Evaluating the inverse of \eqref{eq:208} is easier when $q = 1$. The inverse of \eqref{eq:208} is
\begin{equation}
    \includegraphics[scale=0.13]{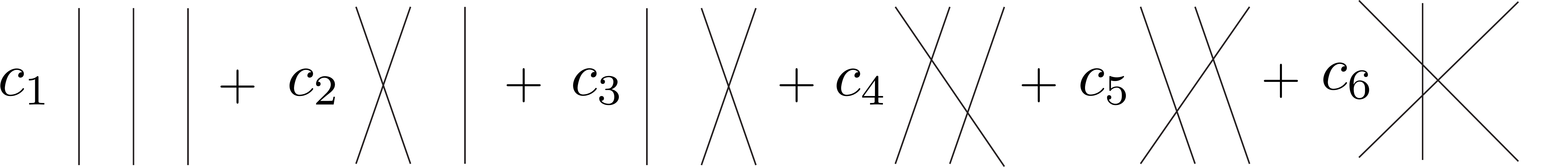}
    \label{eq:213}
\end{equation}
where the coefficients $c_j$ are
\begin{align}
\begin{split}
c_1 &= \frac{2 \epsilon  \left(-\epsilon ^3+\epsilon +1\right)+1}{(\epsilon -1)^2 (\epsilon  (\epsilon  (6 \epsilon
   +7)+4)+1)}
   \\
   c_2 &= c_3 = \frac{\epsilon  ((\epsilon -2) \epsilon  (\epsilon +1)-1)}{(\epsilon -1)^2 (\epsilon  (\epsilon  (6 \epsilon
   +7)+4)+1)}
   \\
   c_4 &= c_5 = \frac{\epsilon ^3 (\epsilon +2)}{(\epsilon -1)^2 (\epsilon  (\epsilon  (6 \epsilon +7)+4)+1)}
   \\
   c_6 &= -\frac{\epsilon ^2 \left(2 \epsilon ^2+1\right)}{(\epsilon -1)^2 (\epsilon  (\epsilon  (6 \epsilon +7)+4)+1)}.
\end{split}    
\end{align}
We now multiply \eqref{eq:213} with the final line of \eqref{eq:212} and then take the $\epsilon \rightarrow 1$ limit (we will come back to the rest of \eqref{eq:212} later). We then take a trace by identifying the top and bottom ends of the diagrams. The result is
\begin{equation}
    \includegraphics[scale=0.13]{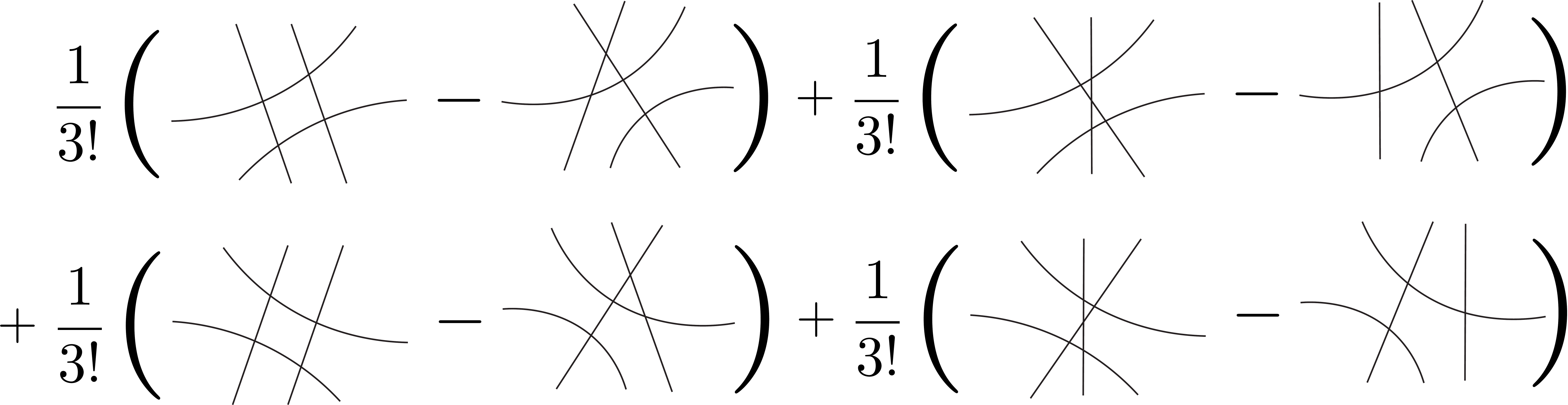}.
    \label{eq:cancelling}
\end{equation}
Because we have identified the top and bottom ends of each diagram, it turns out that \eqref{eq:cancelling} evaluates to zero. In fact, each of the bracketed terms separately vanish.

We now multiply \eqref{eq:213} with the sum over the six terms appearing in brackets in \eqref{eq:212}. In the JT limit, the result finally becomes
\begin{equation}\label{dt2ptw3Final0}
    \includegraphics[scale=0.15]{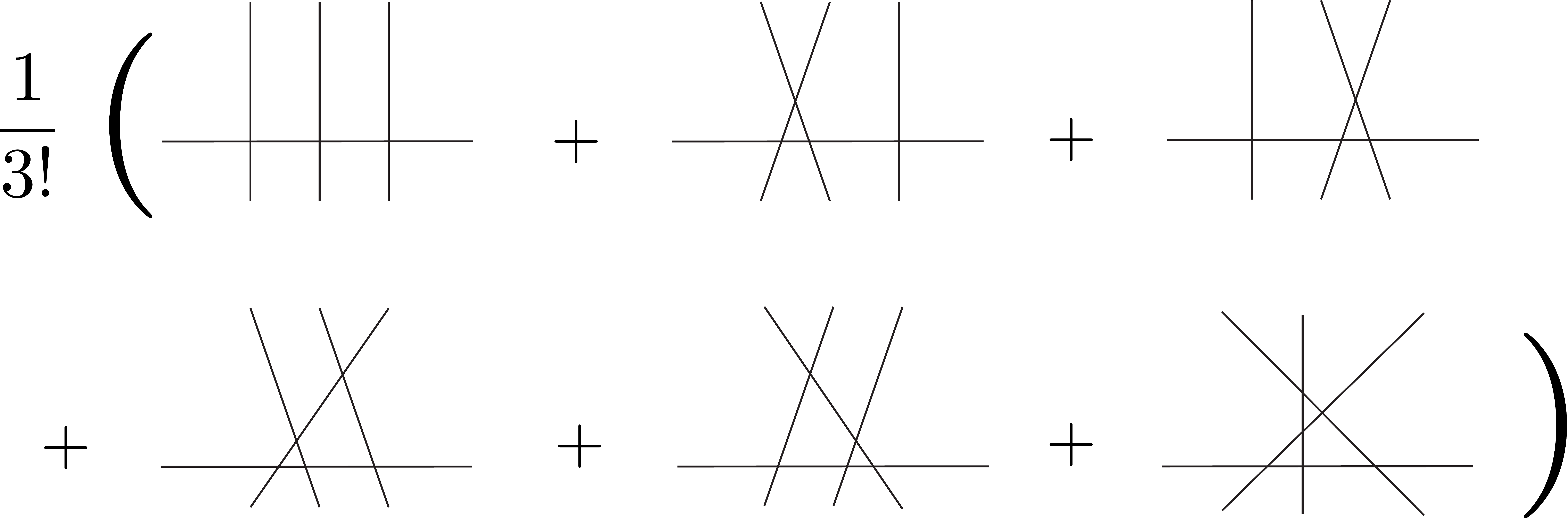}
\end{equation}

\bigskip

Now we would like to compute \eqref{dt2ptw3Final0}, where we identify the top and bottom parts of the diagram. We start with the first 3 terms 
\begin{align}
\text{tr} ~ \raisebox{-.1in}{\includegraphics[scale=.1]{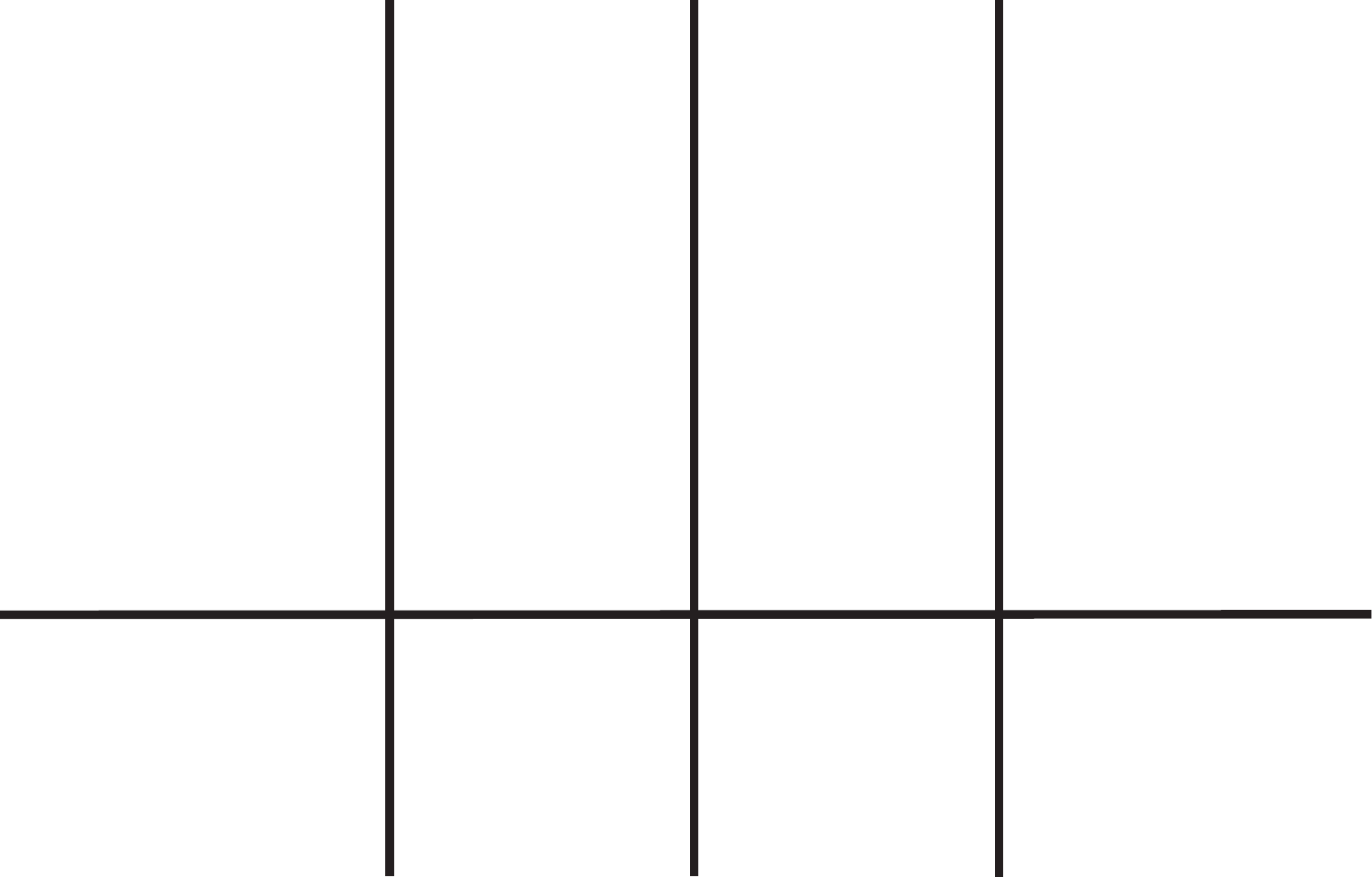}} ~&=~
\int_0^\infty dk_1 dk_2 ~ \rho(k_1) \rho(k_2) ~
\raisebox{-.1in}{\includegraphics[scale=.1]{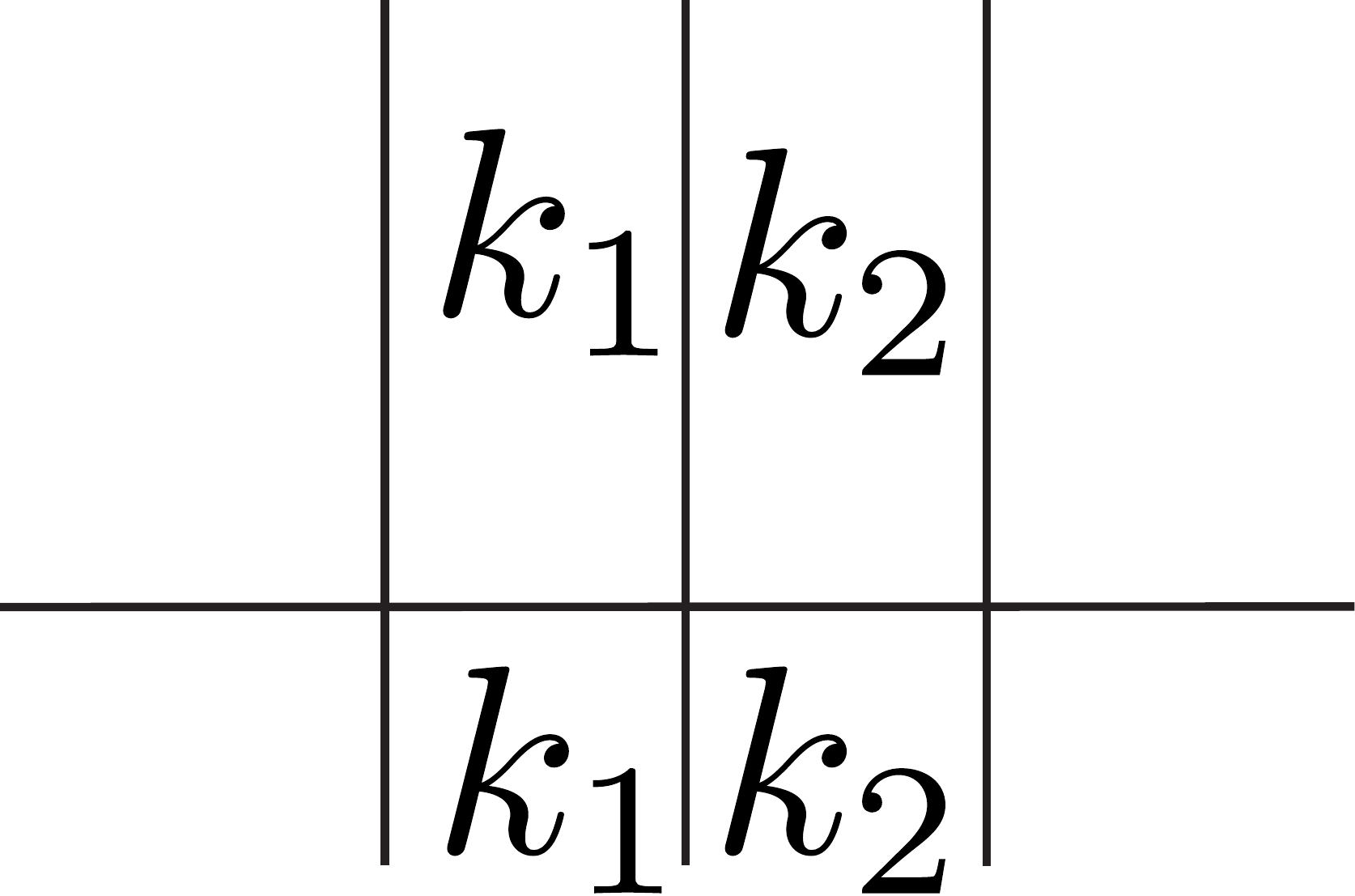}} 
 ~=~ 
\sum_{n,m=0}^\infty 
\left\{ 
\begin{matrix}
3\Delta +n +m & s_a & s_b \\
\Delta & s_b & s_a
\end{matrix}
\right\} \ , \\
\text{tr}~
\raisebox{-.1in}{\includegraphics[scale= .1]{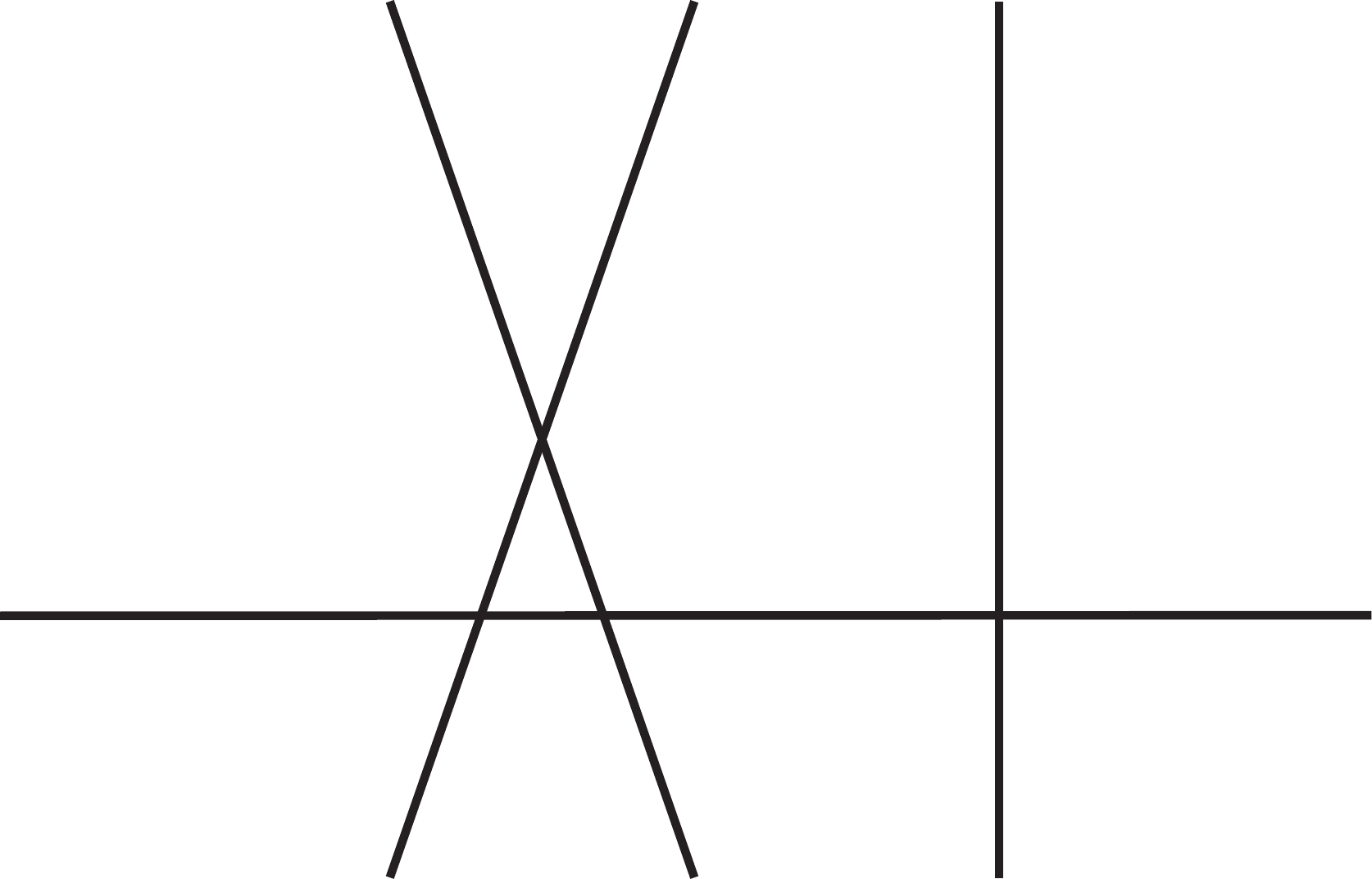}}
~&=~
\text{tr}~
\raisebox{-.1in}{\includegraphics[scale= .1]{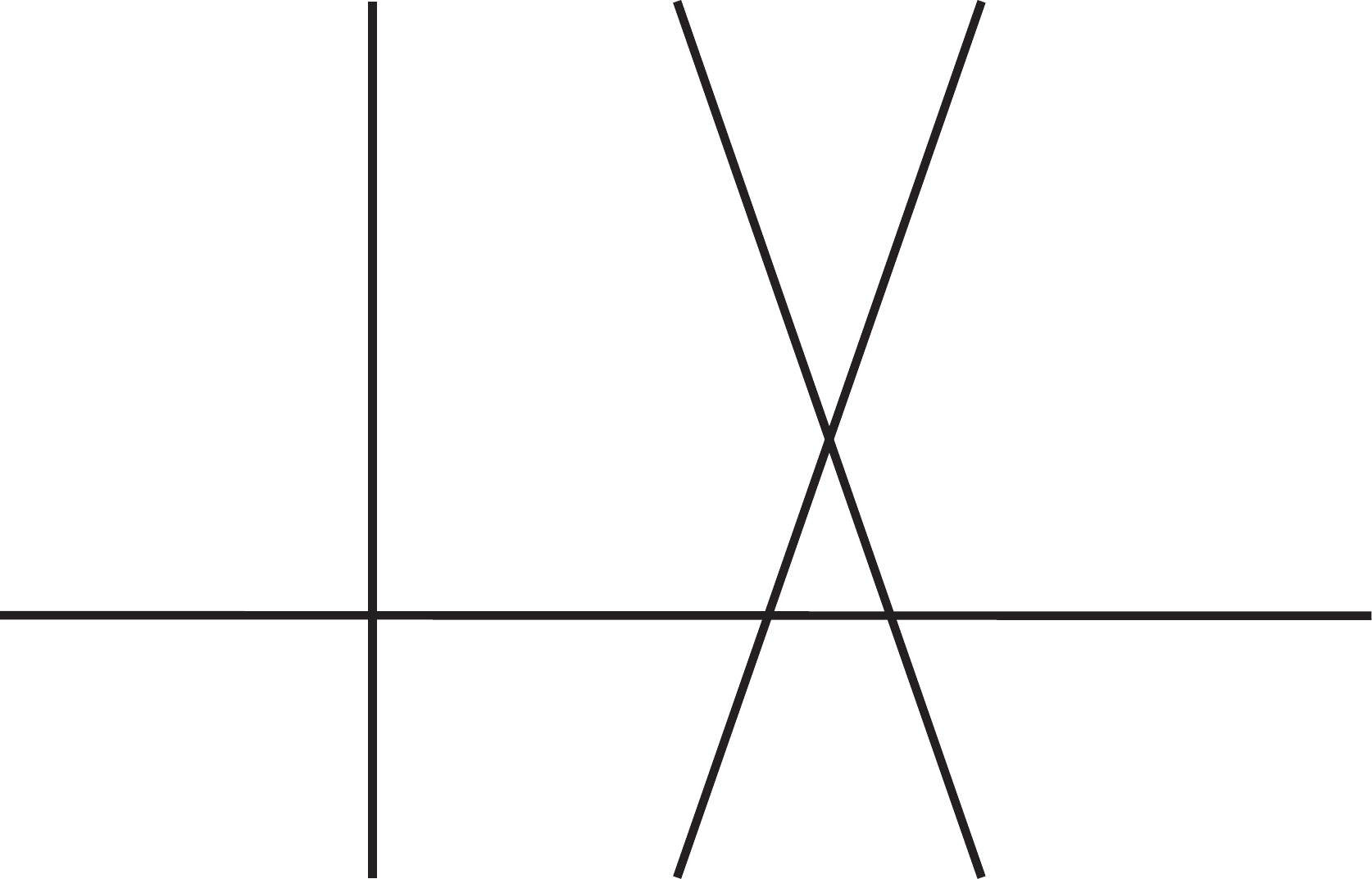}}
~ = ~
\sum_{n,m=0}^\infty (-1)^n
\left\{ 
\begin{matrix}
3\Delta +n +m & s_a & s_b \\
\Delta & s_b & s_a
\end{matrix}
\right\} \ .
\end{align}
The trace here means that we identify top and bottom parts of the diagram and integrate over $k_1, k_2$ with the density of states. In both equations we used the pentagon identity \eqref{Pent1} twice and orthogonality of Wilson polynomials. In the 2nd equation we also used \eqref{WtransW}.

The last 3 terms in \eqref{dt2ptw3Final0} are slightly more tricky to compute. The 4th term gives 
\begin{align}
&\text{tr}~
\raisebox{-.1in}{\includegraphics[scale= .15]{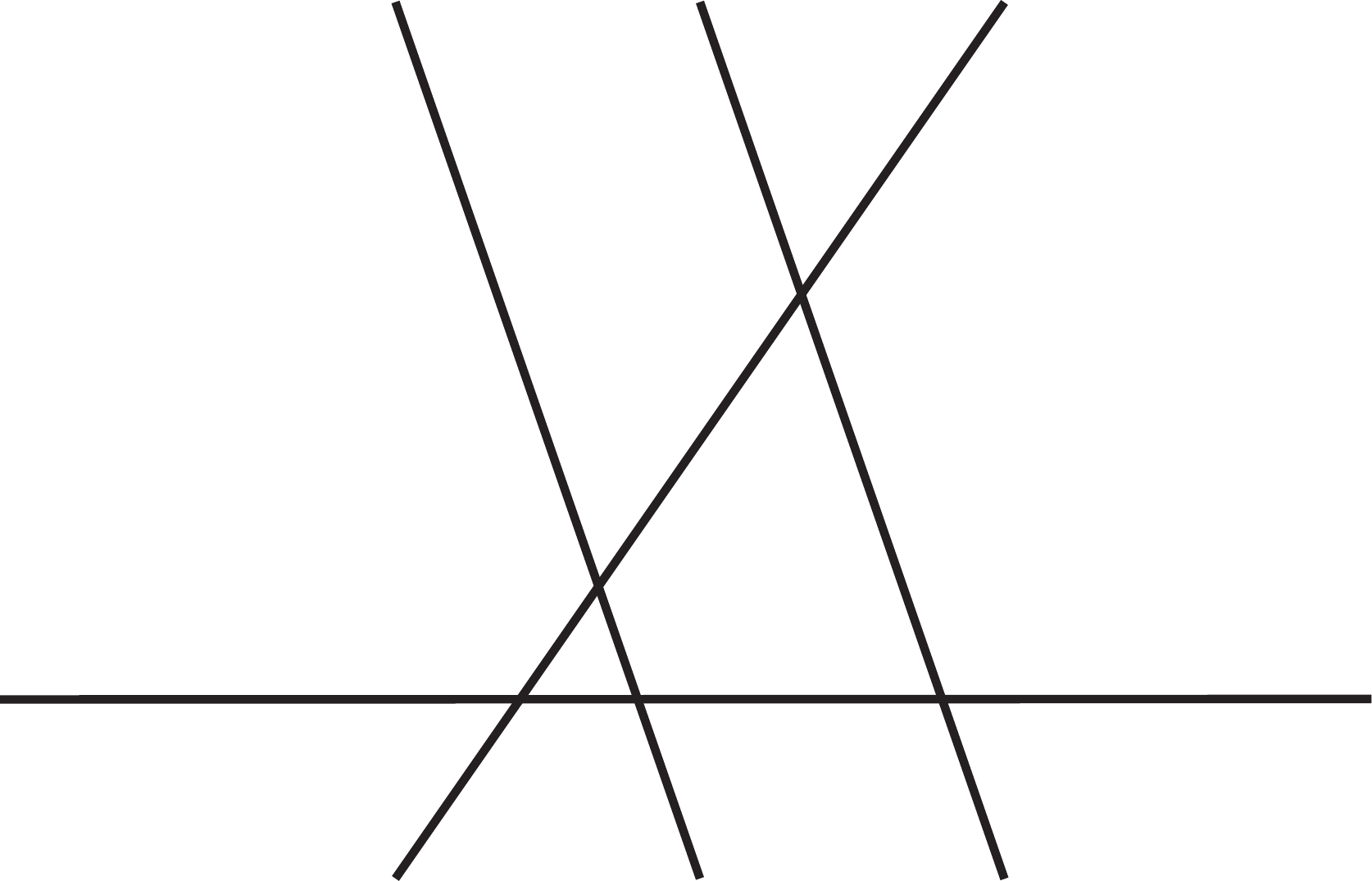}}
~ = ~ 
\sum_{n=0}^\infty 
\text{tr}~~~
\raisebox{-.35in}{\includegraphics[scale= .2]{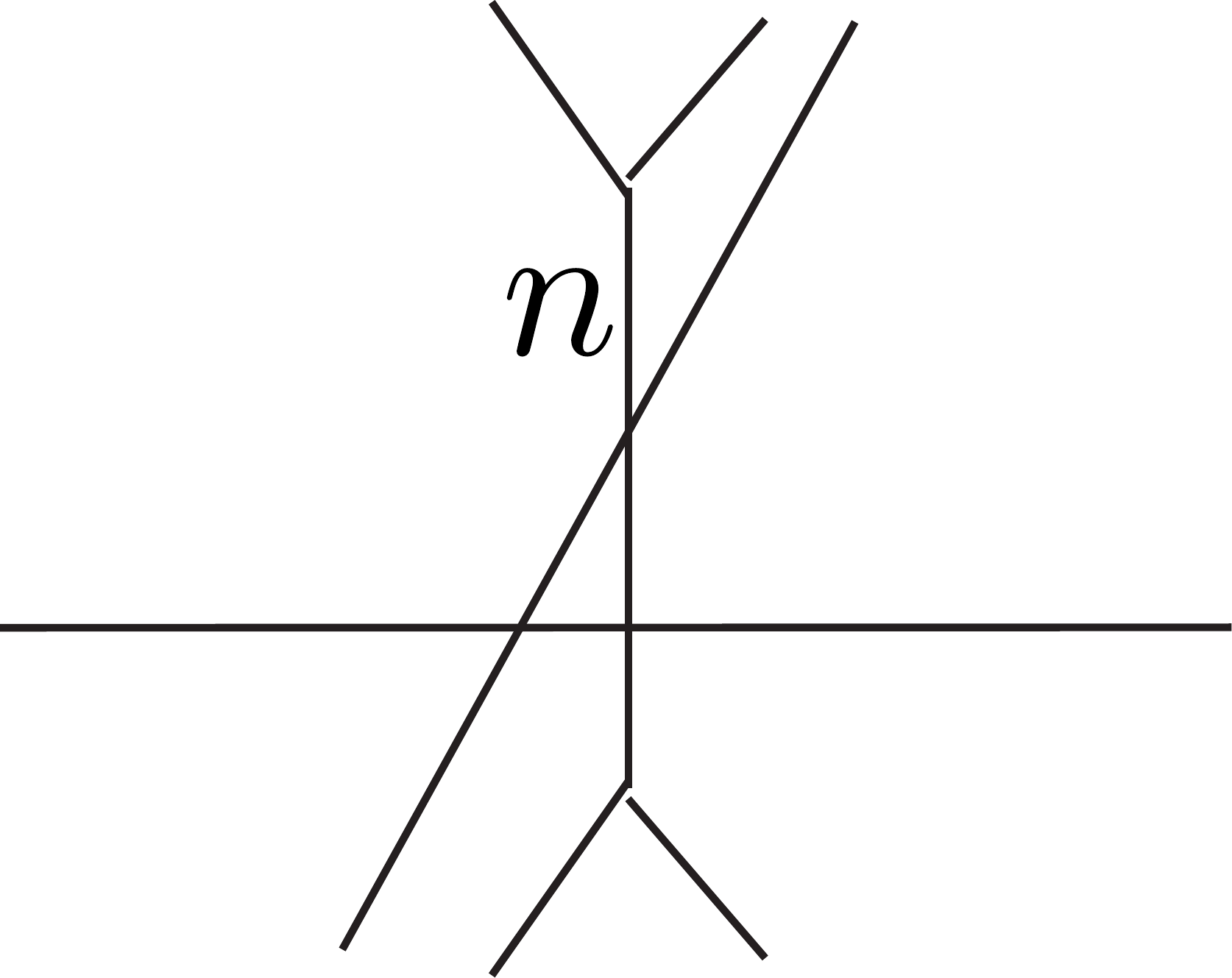}}
~ = ~
\sum_{n,m=0}^\infty (-1)^m
~~~\text{tr}~~~
\raisebox{-.6in}{\includegraphics[scale= .2]{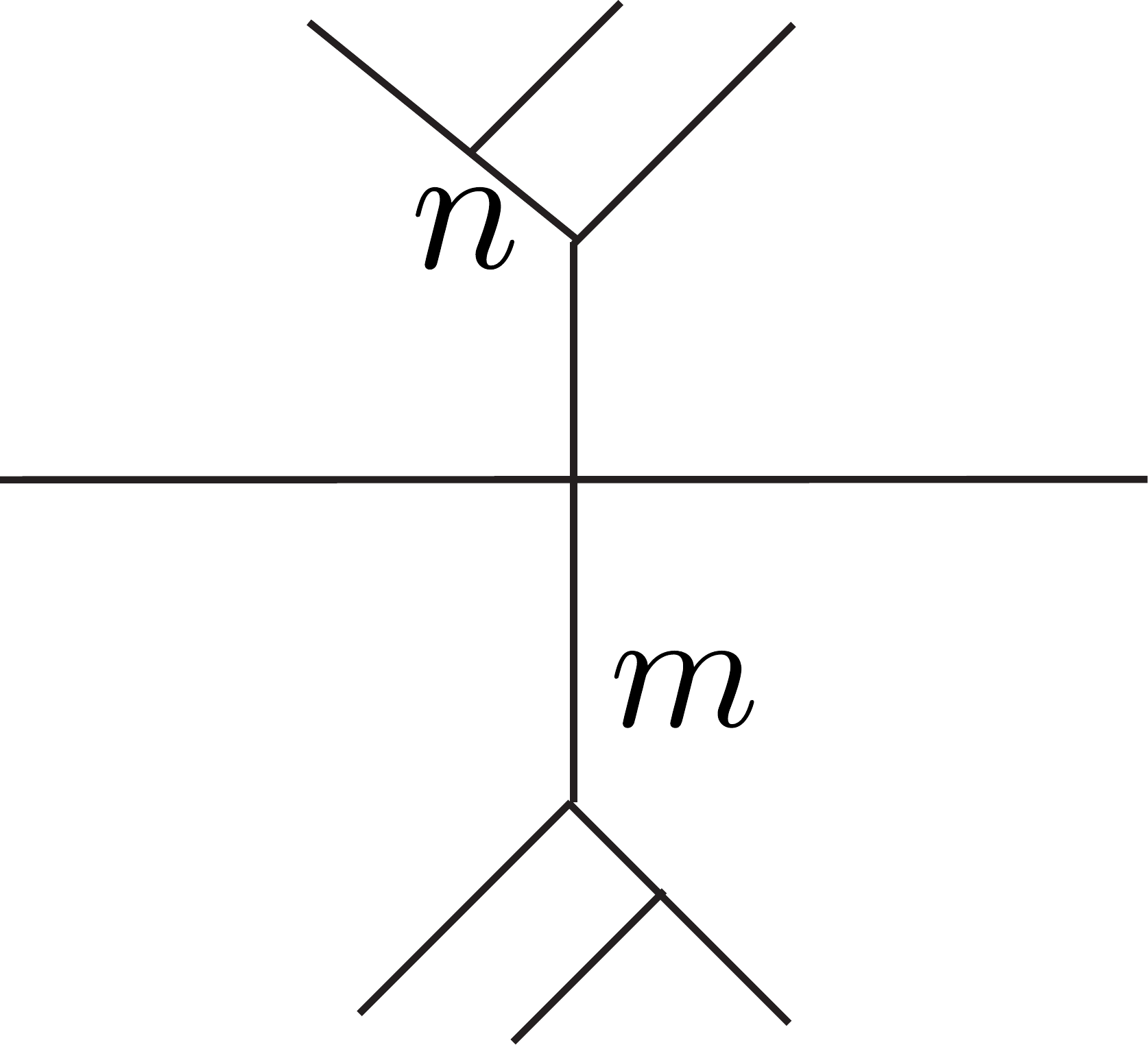}}
\\
=&\sum_{n,m=0}^\infty (-1)^m
~~~\text{tr}~~~
\raisebox{-.3in}{\includegraphics[scale= .2]{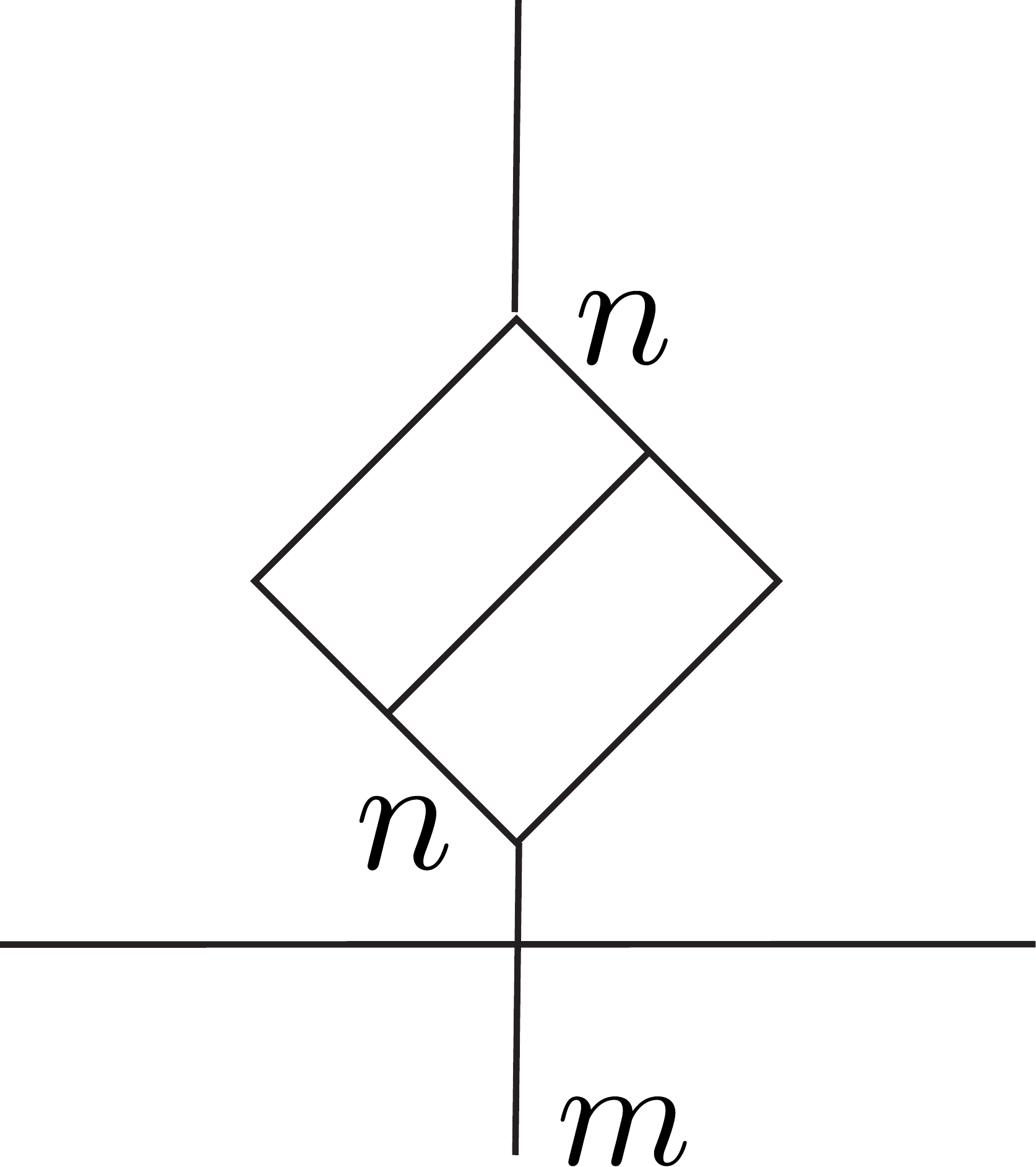}} \ .
\label{w3W2}
\end{align}
In the 1st equality we used the pentagon identity. Then we used an expansion of the 6j-symbol into Wilson polynomials \eqref{6jdiag}. And finally, we computed the trace by connecting top and bottom lines. 

To compute \eqref{w3W2} we note that it is an overlap between two products of Wilson polynomials
\begin{align}
\raisebox{-.4in}{\includegraphics[scale=.3]{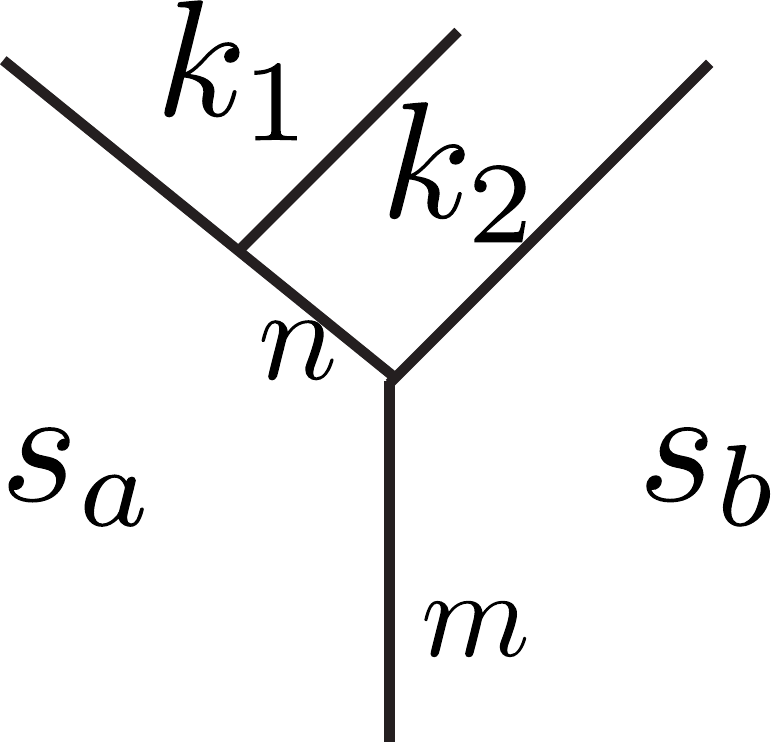}} 
~~~&=~~~
P_n(k_1; \Delta \pm i s_a, \Delta \pm i k_2)
P_m(k_2; 2\Delta +n \pm i s_a, \Delta \pm i s_b ) \ , \\
\raisebox{-.4in}{\includegraphics[scale=.3]{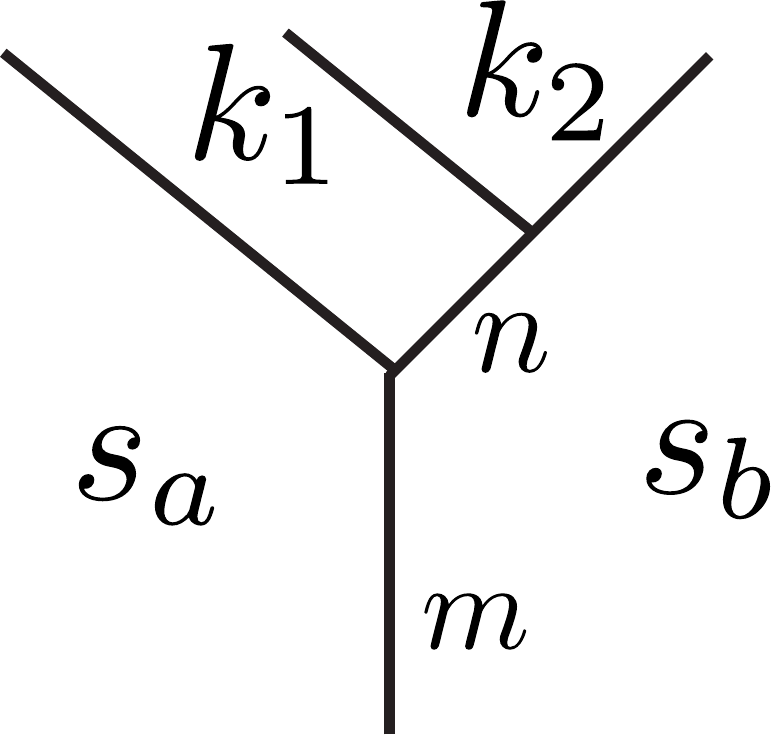}} 
~~~&=~~~
P_m(k_1; \Delta \pm i s_a, 2\Delta + n \pm i s_b)
P_n(k_2; \Delta \pm i k_1, \Delta \pm i s_b ) \ .
\end{align}
These two products can be thought of as bases in the space of functions of two variables $k_1, k_2$. Since both of them are bases, they can be related to each other by a linear transformation. Indeed, such a transformation can be derived by analytically continuing the theorem 7.6 of \cite{Groenvelt2005}
\begin{align}\label{PP1toPP2}
\raisebox{-.4in}{\includegraphics[scale=.3]{figures/PP1.pdf}} 
~~~=~~~
\sum_{j=0}^{n+m} c_j R_j ~~~
\raisebox{-.4in}{\includegraphics[scale=.3]{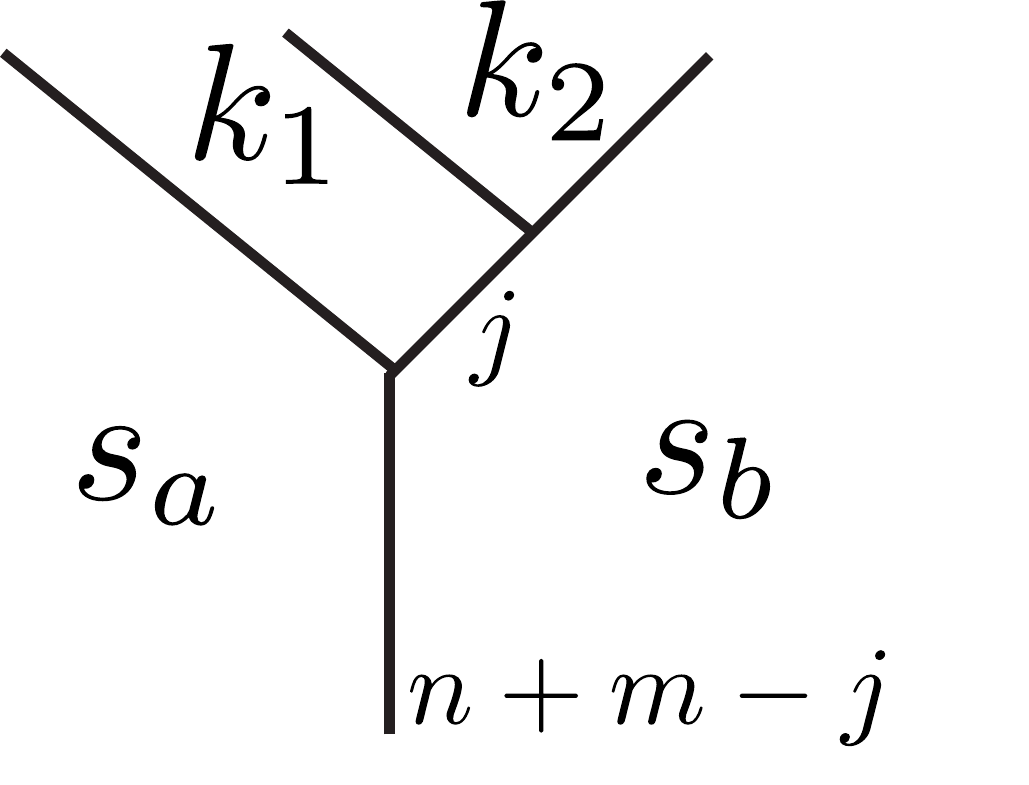}} \ ,
\end{align}
where 
\begin{align}
c_j &= \sqrt{{\cal N}_j} \left( n+m \atop j \right) {(2\Delta)_n (2\Delta)_m \over (2\Delta)_j} 
{(6\Delta +n +m -1)_j \over (4\Delta +j-1)_j (4\Delta+2j)_{n+m-j}}  \ , \\
{\cal N }_j &= {j! (n+m-j)! \over n! m!} 
{(6\Delta +n+m+j-1)_{n+m-j} \over (6\Delta +2n+m-1)_m}
{(4\Delta+ j-1)_{j} \over (4\Delta +n-1)_n} \\
&
{\Gamma(4\Delta + 2n ) \over \Gamma(4\Delta + 2j)}
{\Gamma(4\Delta + n+m+j ) \over \Gamma(4\Delta + 2n+m)}
{\Gamma(2\Delta + n+m-j ) \Gamma(2\Delta +j)^2 \over \Gamma(2\Delta + m) \Gamma(2\Delta + n)^2} 
\end{align}
and $R_j$ is the Racah polynomial defined on page 1 of \cite{Groenvelt2005} 
\begin{align}
R_j &\equiv R_j(n; 2\Delta -1, 2\Delta - 1, -n-m-1, 4\Delta +n+m-1 ) \\
&=
\tensor[_4]{F}{_3}
\left( 
{-j , -n, 4\Delta +j-1, 4\Delta +n-1 \atop 
2\Delta, -n-m, 6\Delta +n+m-1 } ; 1
\right) \ .
\end{align}
In \eqref{w3W2} the top and bottom lines are identified and carry the index $m$. This means that we are interested in the term $j=n$ in the RHS of \eqref{PP1toPP2}. In this case some of the expressions simplify. In particular, ${\cal N}_n = 1$. We find 
\begin{align}
&\raisebox{-.6in}{\includegraphics[scale= .2]{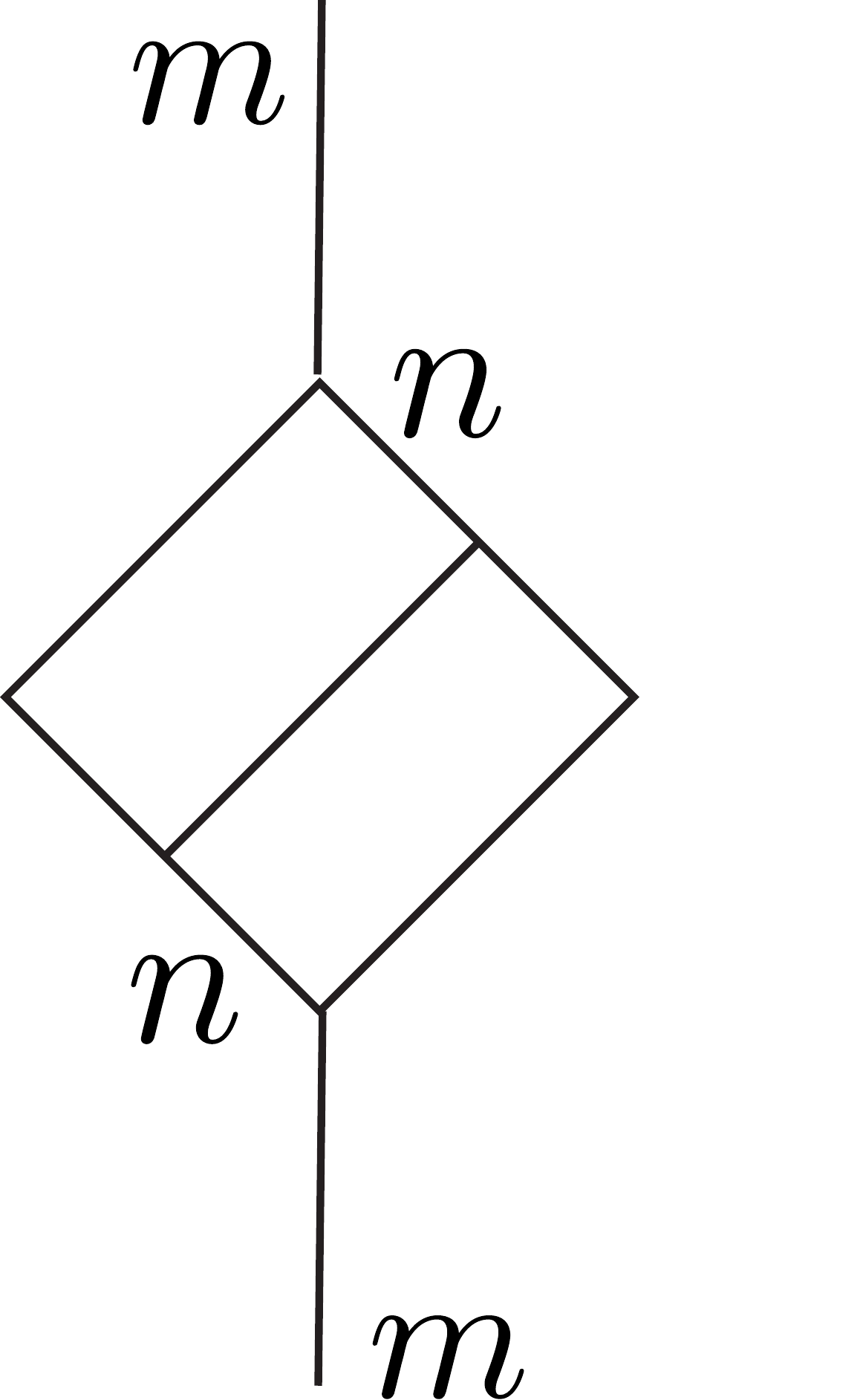}} 
~~~=~~~
(c_j R_j)|_{j=n}\\
&=
\left( n+m \atop n \right) {(2\Delta)_n (6\Delta +n+m-1)_n \over (4\Delta + 2n)_m (4\Delta +n-1)_n} 
\tensor[_4]{F}{_3}
\left( 
{-n , -n, 4\Delta +n-1, 4\Delta +n-1 \atop 
2\Delta, -n-m, 6\Delta +n+m-1 } ; 1
\right) \ .
\label{cnRn}
\end{align}
 Finally, we find 
 \begin{align}
 \text{tr}~
\raisebox{-.1in}{\includegraphics[scale= .15]{figures/w34.pdf}}
~ =& ~ 
\text{tr}~
\raisebox{-.1in}{\includegraphics[scale= .15]{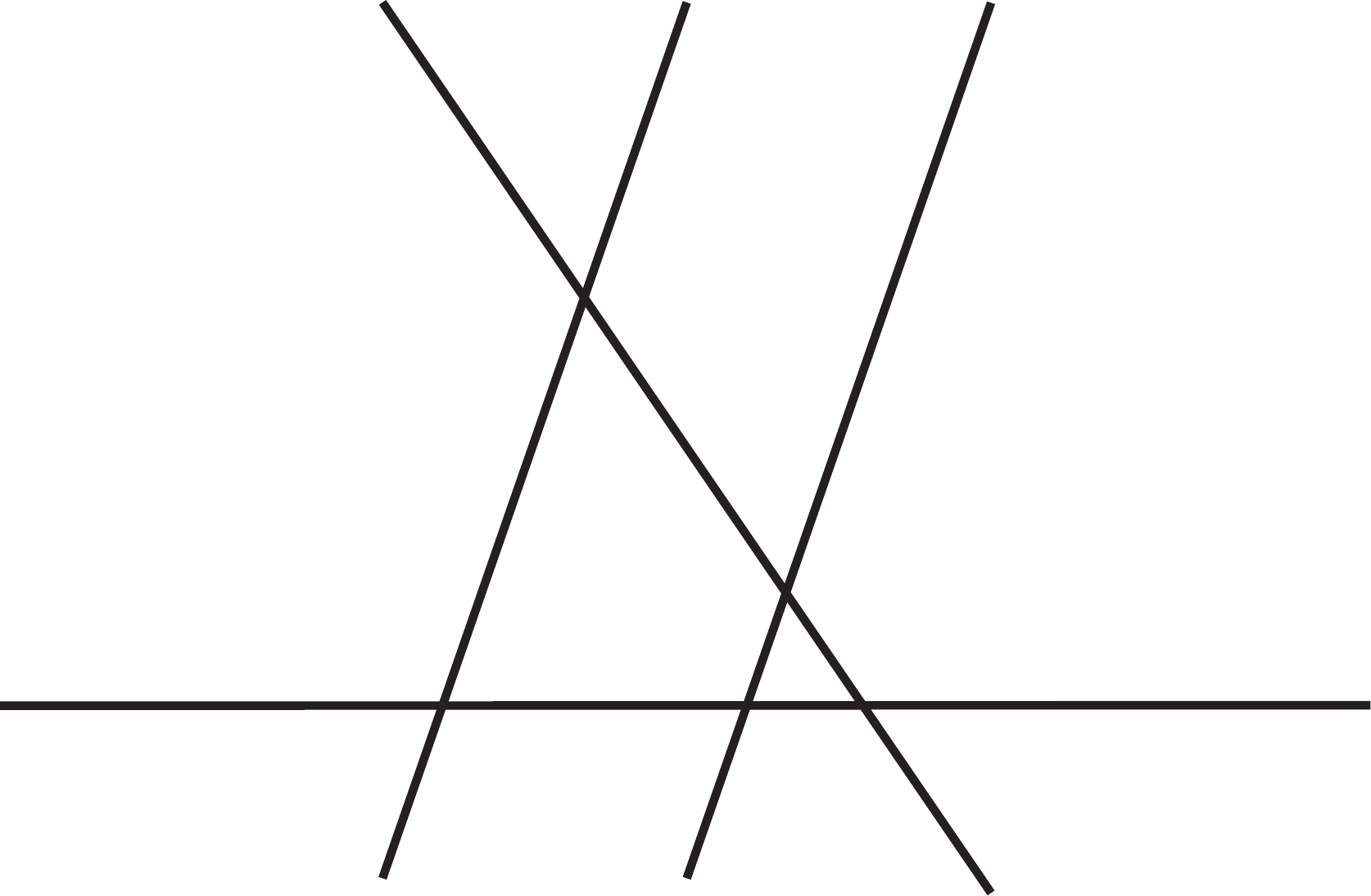}}
 = \sum_{n,m=0}^\infty (-1)^m c_n R_n ~
 \left\{ 
 \begin{matrix}
 3\Delta +n +m & s_a & s_b \\
 \Delta & s_b & s_a
 \end{matrix}
 \right\} \ .
 \end{align}
 A similar computation shows 
 \begin{align}
     \text{tr}~
\raisebox{-.1in}{\includegraphics[scale= .15]{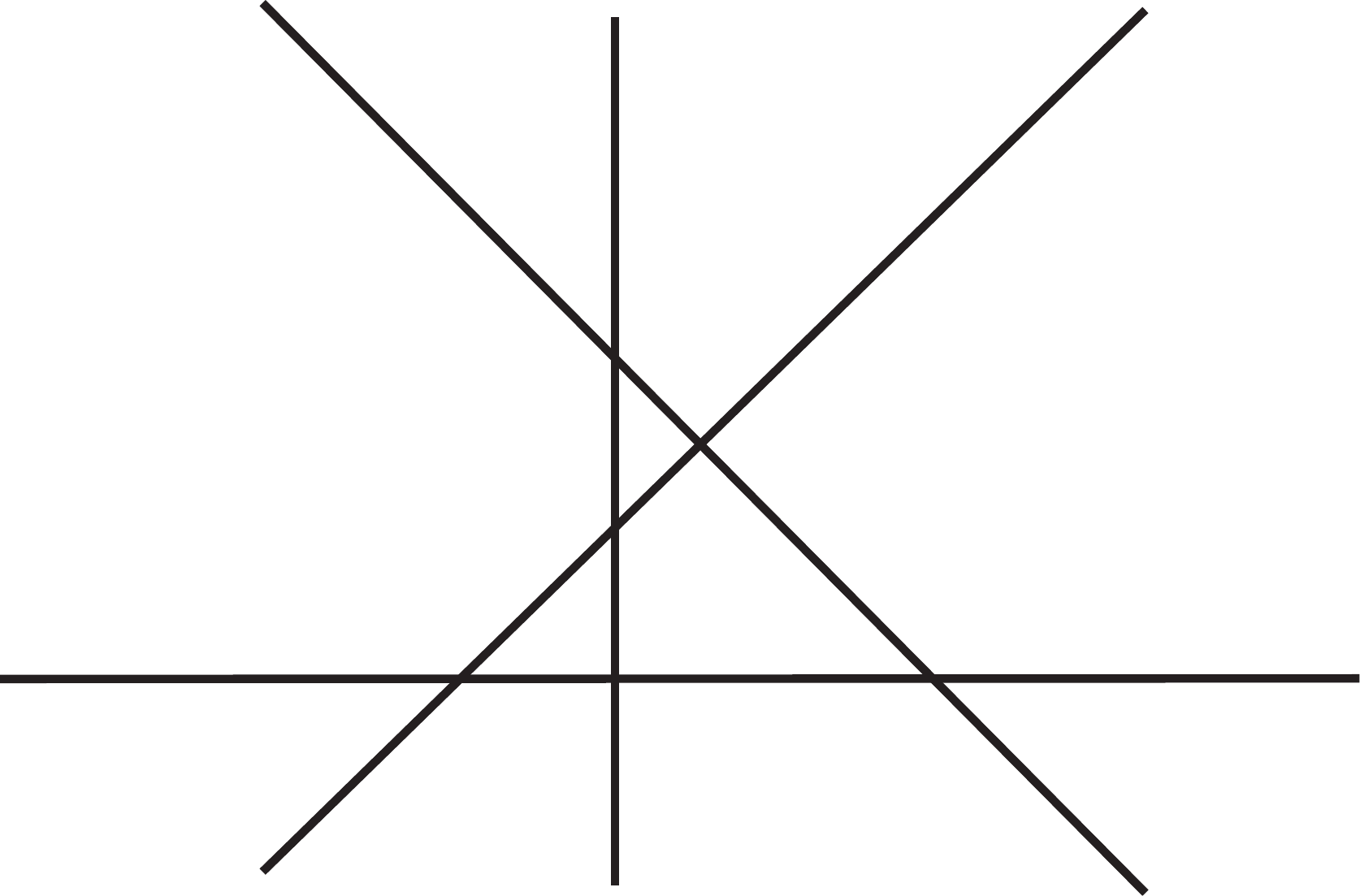}}
~ = ~ 
 \sum_{n,m=0}^\infty (-1)^{n+m} c_n R_n ~
 \left\{ 
 \begin{matrix}
 3\Delta +n +m & s_a & s_b \\
 \Delta & s_b & s_a
 \end{matrix}
 \right\} \ .
 \end{align}
Combining the above results we have
\begin{align}
&{1\over 3!}\text{tr}
\left( 
\raisebox{-.2in}{\includegraphics[scale=0.1]{figures/w31.pdf}} 
~+~ 
\raisebox{-.2in}{\includegraphics[scale=0.1]{figures/w32.pdf}} 
 ~+~ 
\raisebox{-.2in}{\includegraphics[scale=0.1]{figures/w33.pdf}} 
 ~+~ 
\raisebox{-.2in}{\includegraphics[scale=0.1]{figures/w34.pdf}} 
 ~+~ 
\raisebox{-.2in}{\includegraphics[scale=0.1]{figures/w35.pdf}} 
 ~+~ 
\raisebox{-.2in}{\includegraphics[scale=0.1]{figures/w36.pdf}} 
\right) \\
=&
{1\over 3!}\sum_{n,m=0}^\infty \left[ 1+ 2(-1)^n + 2(-1)^m c_n R_n + (-1)^{n+m}c_nR_n \right]
\left\{ 
 \begin{matrix}
 3\Delta +n +m & s_a & s_b \\
 \Delta & s_b & s_a
 \end{matrix}
 \right\} \\
 =& {1\over 3!}\sum_{N=0}^\infty 
 \left\{ 
 \begin{matrix}
 3\Delta +N & s_a & s_b \\
 \Delta & s_b & s_a
 \end{matrix}
 \right\} 
 \sum_{n=0}^N \left[ 1+ 2(-1)^n + 2(-1)^m c_n R_n + (-1)^{n+m}c_nR_n \right]\Big|_{m=N-n} \ .
 \label{w3Racah}
\end{align}
We would like to compare this result with JT gravity. The next term in \eqref{2pt-dt2} (that we didn't write there) is
\begin{align}
\sum_{n,m=0}^\infty 
\left\{ 
 \begin{matrix}
 3\Delta +2n +3m & s_a & s_b \\
 \Delta & s_b & s_a
 \end{matrix}
 \right\} 
 = \sum_{N=0}^\infty 
 \left\{ 
 \begin{matrix}
 3\Delta +N & s_a & s_b \\
 \Delta & s_b & s_a
 \end{matrix}
 \right\} 
 \sum_{n,m = 0}^\infty  \delta_{2n+3m, N} \ .
\end{align}
To match this with \eqref{w3Racah} we need
\begin{align}
{1\over 3!}\sum_{n=0}^N \left[ 1+ 2(-1)^n + 2(-1)^m c_n R_n + (-1)^{n+m}c_nR_n \right]\Big|_{m=N-n}
=
\sum_{n,m = 0}^\infty  \delta_{2n+3m, N} \ ,
\end{align}
where $c_n R_n$ is defined in \eqref{cnRn}. Incredibly, this identity is indeed true. We didn't find its derivation, but checked it numerically for several values of $N$.


\section{Representation theory of $\mathcal{U}_q(\mathfrak{su}(1,1))$}

\label{sec:reptheory}

\label{AppendixB}

In this appendix, we introduce the $\mathcal{U}_q(\mathfrak{su}(1,1))$ algebra and derive important relations involving the various 6j symbols that are encountered throughout this paper. We mostly follow the discussion in \cite{Groenvelt2005}, although our conventions differ. Any results that are not proven here are attributable to \cite{Groenvelt2005}. The main purpose of this section is to derive the identities \eqref{eq:pentagon} and \eqref{eq:whatshouldicallthis}.

\subsection{The $\mathcal{U}_q(\mathfrak{su}(1,1))$ algebra}

The $\mathcal{U}_q(\mathfrak{su}(1,1))$ algebra is a $q$-deformation of the universal enveloping algebra of $\mathfrak{su}(1,1)$. The generators $K$, $K^{-1}$, $E$, and $F$ obey the relations
\begin{equation}
	K K^{-1} = 1 = K^{-1} K, \quad KE = q^{1/2} EK, \quad KF = q^{-1/2} FK, \quad EF - FE = \frac{K^2 - K^{-2}}{q^{1/2} - q^{-1/2}}
	\label{eq:algebradef}
\end{equation}
and the Casimir is given by
\begin{equation}
	\Omega = \frac{q^{-1/2} K^2 + q^{1/2} K^{-2} - 2}{(q^{-1/2} - q^{1/2})^2} + EF = \frac{q^{-1/2} K^{-2} + q^{1/2} K^2 - 2}{(q^{-1/2} - q^{1/2})^2} + FE.
	\label{eq:casimir}
\end{equation}
This algebra is a Hopf $*$-algebra, and the comultiplication operator $\Delta$ is defined as follows:
\begin{equation}
	\Delta(K) = K \otimes K, \quad \Delta(E) = K \otimes E + E \otimes K^{-1}, \quad \Delta(K^{-1}) = K^{-1} \otimes K^{-1}, \quad \Delta(F) = K \otimes F + F \otimes K^{-1}.
	\label{eq:coproduct}
\end{equation}
Note that this algebra is not cocommutative, which means that the tensor product of two representations depends on their ordering. The adjoint operation is defined by
\begin{equation}
	K^\dagger = K, \quad E^\dagger = -F, \quad F^\dagger = - E, \quad (K^{-1})^\dagger = K^{-1}.
	\label{eq:adjoint}
\end{equation}
We may define an automorphism of the algebra as follows:
\begin{equation}
\tilde{K} = K^{-1}, \quad \tilde{K}^{-1} = K, \quad \tilde{E} = F, \quad \tilde{F} = E.
\label{eq:autom}
\end{equation}
The tilde generators also obey \eqref{eq:algebradef}, \eqref{eq:adjoint}, and \eqref{eq:casimir} for the same $\Omega$. The coproduct is not invariant under \eqref{eq:autom}. If $X$ represents a generator of the algebra such that
\begin{equation}
	\Delta(X) = \sum_i a_i \otimes b_i,
\end{equation}
then \begin{equation} \Delta(\tilde{X}) = \sum_i \tilde{b}_i \otimes \tilde{a}_i.
\end{equation}

We are interested in three unitary representations: the positive discrete series, the negative discrete series, and the principal series.

\subsubsection{Discrete series}

The positive discrete series $\pi^+_k$ is labeled by $k > 0$. The representation space is $\ell^2(\mathbb{Z}_{\ge 0})$ with orthonormal basis $\{ e_ n\}_{n \in \mathbb{Z}_{\ge 0}}$. The action is given by

\begin{align}
	\begin{split}
\pi^+_k(K) e_n &= q^{\frac{k+n}{2}} e_n, \quad \pi^+_k(K^{-1}) e_n = q^{-\frac{k+n}{2}} e_n
\\
(q^{-1/2} - q^{1/2}) \pi^+_k(E) e_n &= q^{-\frac{1}{4} - \frac{k}{2} - \frac{n}{2}} \sqrt{(1 - q^{n + 1})(1 - q^{2 k + n})} e_{n+1}
\\
(q^{-1/2} - q^{1/2}) \pi^+_k(F) e_n &= - q^{\frac{1}{4} - \frac{k}{2} - \frac{n}{2}} \sqrt{(1-q^n)(1 - q^{2k + n - 1})} e_{n-1}
\\
(q^{-1/2} - q^{1/2})^2 \pi^+_k(\Omega) e_n &= (q^{k - \frac{1}{2}} + q^{\frac{1}{2} - k} - 2)e_n.
	\end{split}
\label{eq:discseries}
\end{align}
The positive discrete series representation becomes the negative discrete series representation under the automorphism in  \eqref{eq:autom}. That is, if $X$ is a generator of the algebra, then it is represented by $\pi^-_k(X)$ in the negative discrete series representation, where
\begin{equation}
	\pi_k^-(X) = \pi_k^+(\tilde{X}).
\end{equation}

\subsubsection{Principal series}

The principal unitary series representations $\pi^P_{\rho,\epsilon}$ are labeled by $0 \leq \rho \leq \frac{\pi}{|\log q|}$ and $\epsilon \in [0,1)$, where $(\rho,\epsilon) \neq (0,\frac{1}{2})$. At times we may find it convenient to allow $\epsilon$ to take values outside of this range with the understanding that $\epsilon$ is defined modulo unit shifts. The representation space is $\ell^2(\mathbb{Z})$ with orthonormal basis $\{e_n\}_{n \in \mathbb{Z}}$. The action is given by
\begin{align}
\begin{split}
\pi^P_{\rho,\epsilon}(K) e_n &= q^{\frac{n + \epsilon}{2}} e_n, \quad \pi^P_{\rho,\epsilon}(K^{-1}) e_n = q^{- \frac{n + \epsilon}{2}} e_n,
\\
(q^{-1/2} - q^{1/2})\pi^P_{\rho,\epsilon}(E) e_n &= q^{-\frac{1}{4} - \frac{n}{2} - \frac{\epsilon}{2}} \sqrt{ (1 - q^{n + \epsilon + i \rho + \frac{1}{2}})(1- q^{n + \epsilon - i \rho + \frac{1}{2}})} e_{n+1}	
\\
(q^{-1/2} - q^{1/2}) \pi^P_{\rho,\epsilon}(F)e_n &= - q^{\frac{1}{4} - \frac{n}{2} - \frac{\epsilon}{2}} \sqrt{(1 - q^{n + \epsilon + i \rho - \frac{1}{2}})(1 - q^{n + \epsilon - i \rho - \frac{1}{2}})} e_{n-1}
\\
(q^{-1/2} - q^{1/2})^2 \pi^P_{\rho,\epsilon}(\Omega) e_n &= (q^{i \rho} + q^{- i \rho} - 2)e_n
\end{split}
\label{eq:prinseries}
\end{align}
If we define the automorphism in \eqref{eq:autom} to act on $e_n$, $\rho$, and $\epsilon$ as follows,
\begin{equation}
	\tilde{\rho} = \rho, \quad \tilde{\epsilon} = - \epsilon, \quad \tilde{e}_n = (-1)^n e_{-n},
	\
\end{equation}
then \eqref{eq:prinseries} is invariant under the automorphism.

\subsection{Clebsch-Gordan coefficients}

We now consider the Clebsch-Gordan coefficients that we will need.

\subsubsection{$\pi^+ \otimes \pi^+$}

Consider the tensor product representation $\pi^+_{k_1} \otimes \pi^+_{k_2}$, which is spanned by $e_{n_1} \otimes e_{n_2}$ for $n_1,n_2 \in \mathbb{Z}_{\ge 0}$. A state of definite $K$ weight is given by
\begin{equation}
	\ket{p} \equiv \sum_{n=0}^p v^p_n \, e_n \otimes e_{p - n},
	\label{eq:ketppippip}
\end{equation}
where $p \in \mathbb{Z}_{\ge 0}$, and we wish to choose $v^p_n$ such that $\ket{p}$ is an eigenvector of the casimir $\Omega$. The correct choice is to take

\begin{align}
	\begin{split}
 v^p_n(x,k_1,k_2) &= -(q^{\frac{1}{2} -  k_1 -  k_2})^n  q^{ p  k_1 } \sqrt{\frac{(q^{2 k_2 };q)_p}{(q^{2 k_1 + 2 k_2 };q)_p}} \sqrt{\frac{(q^{2 k_1},q^{-p};q)_n}{(q,q^{1-p-2k_2};q)_n}}
		  \\
		  &\times \sqrt{\frac{(q^{2 k_1},q^{2 k_1 + 2 k_2-1},q^{-p};q)_x}{(q,q^{2 k_1 + 2 k_2 + p},q^{2 k_2};q)_x}\frac{(q^{p x - \frac{x(x-1)}{2}})(1 - q^{2 k_1 + 2 k_2 - 1 + 2 x})}{(1 - q^{2 k_1 + 2 k_2 - 1})(-q^{2 k_1})^x}}
			\\
		 &\times R_n(\mu(x); q^{2 k_1-1},q^{2 k_2-1},p|q) 
		 \label{eq:vnp}
\end{split}
\end{align}
where $x \in\{ 0,1,\ldots,p\}$, and $R_n$ is a dual $q$-Hahn polynomial, defined on page 450 of \cite{orthpolyreference}. Note that the quantities inside the square roots are positive. With $v_n^p$ chosen as in \eqref{eq:vnp}, $\ket{p}$ in \eqref{eq:ketppippip} is a vector in the $\pi^+_{k_3}$ representation, where $k_3 = k_1 + k_2 + x$.

The Clebsch-Gordan coefficients $v^p_n(x,k_1,k_2)$ obey the following orthogonality relations:
\begin{align}
	\sum_{x = 0}^p v^p_m(x,k_1,k_2) v^p_n(x,k_1,k_2) = \delta_{nm}, \quad n,m \in \{0,1,\ldots,p\}
	\\
		\sum_{n = 0}^p v^p_n(x_1,k_1,k_2) v^p_n(x_2,k_1,k_2) = \delta_{x_1x_2}, \quad x_1,x_2 \in \{0,1,\ldots,p\},
\end{align}
and it follows that the decomposition of $\pi^+_{k_1} \otimes \pi^{+}_{k_2}$ is given by
\begin{equation}
	\pi^+_{k_1} \otimes \pi^{+}_{k_2} = \bigoplus_{j = 0}^\infty \, \pi^+_{k_1 + k_2 + j}.
\end{equation}

\subsubsection{$\pi^+ \otimes \pi^P$}

Consider the tensor product representation $\pi^+_{\Delta} \otimes \pi^P_{\rho,\epsilon}$, whose decomposition into irreps contains both principal series and positive discrete series representations \cite{Groenvelt2005}. We are only interested in the principal series representations. The space $\pi^+_{\Delta} \otimes \pi^P_{\rho,\epsilon}$ is spanned by $e_{n_1} \otimes e_{n_2}$ for $n_1 \in \mathbb{Z}_{\ge 0}$ and $n_2 \in \mathbb{Z}$. A state of definite $K$ weight is given by
\begin{equation}
\ket{\tau,p} = \sum_{n \ge 0} f^p_n(\tau,\rho,\epsilon,\Delta) \, e_{n} \otimes e_{p-n},
\end{equation}
where $p \in \mathbb{Z}$, and we wish to define $f^p_n(\tau,\rho,\epsilon,\Delta)$ such that $\ket{\tau,p}$ is an eigenvector of $\Omega$ that belongs to the principal series representation $\pi^P_{\tau,\Delta + \epsilon}$. The Clebsch-Gordan coefficient is given by
\begin{align}
	\begin{split}
f^p_n(\tau,\rho,\epsilon,\Delta) &= (-1)^n \sqrt{\frac{ |\log q|}{ 2\pi} \frac{1}{(q;q)_n (q^{2 \Delta};q)_n (q^{\frac{1}{2} -  p -  \epsilon \pm  i \rho} ; q)_n} \frac{(q^{\pm 2 i \tau}, q, q^{\frac{1}{2} - p  - \epsilon \pm i \rho } , q^{2 \Delta} ;q  )_{\infty}}{(q^{\frac{1}{2} - p - \Delta - \epsilon \pm i \tau},q^{\Delta \pm i \rho \pm i \tau};q)_\infty}}  
\\
&\times p_n\left(\frac{q^{ i \tau} + q^{-  i \tau}}{2},q^{\frac{1}{2} - p - \Delta - \epsilon},q^{\Delta \pm i \rho} | q \right),
\label{eq:fnp}
\end{split}
\end{align}
where $p_n$ is a continuous dual q-Hahn polynomial, defined on page 429 of \cite{orthpolyreference}. Because $\ket{\tau_1,p}$ and $\ket{\tau_2,p}$ must be orthogonal for $\tau_1 \neq \tau_2$, we have that
\begin{equation}
	\sum_{n \ge 0} f^p_n(\tau_1,\rho,\epsilon,\Delta) f^p_n(\tau_2,\rho,\epsilon,\Delta) = \delta(\tau_1 - \tau_2),
\label{eq:orthorelation}
\end{equation}
And the normalization of \eqref{eq:fnp} is chosen to make \eqref{eq:orthorelation} consistent with the orthogonality relation of the continuous dual $q$-Hahn polynomials provided in \cite{Groenvelt2005}.

\subsubsection{$\pi^P \otimes \pi^-$}

The decomposition of $\pi^P \otimes \pi^-$ may be deduced from the decomposition of $\pi^+ \otimes \pi^P$ and the automorphism \eqref{eq:autom}. In particular, $\pi_{\rho,\epsilon}^P \otimes \pi_{\Delta}^-$ is spanned by $e_{n_1} \otimes e_{n_2}$ with $n_1 \in \mathbb{Z}$ and $n_2 \in \mathbb{Z}_{\ge 0}$. If we define
\begin{equation}
	\ket{\tau,p} = \sum_{n \ge 0} (-1)^n f_n^{-p} (\tau,\rho,-\epsilon,\Delta) e_{p+n} \otimes e_n,
\end{equation}
then $\ket{\tau,p}$ belongs to a $\pi^P_{\tau,\epsilon - \Delta}$ representation.

\subsubsection{$\pi^- \otimes \pi^-$}

The decomposition of $\pi^- \otimes \pi^-$ may be deduced from the decomposition of $\pi^+ \otimes \pi^+$ and the automorphism \eqref{eq:autom}. In particular, $\pi^-_{k_1} \otimes \pi^-_{k_2}$ is spanned by $e_{n_1} \otimes e_{n_2}$ for $n_1,n_2 \in \mathbb{Z}_{\ge 0}$. If we define
\begin{equation}
	\ket{p} = \sum_{n = 0}^p v^p_{p-n}(x,k_2,k_1)e_n \otimes e_{p-n},
\label{eq:vtp}
\end{equation}
then $\ket{p}$ is a vector in the $\pi^-_{k_1 + k_2 + x}$ representation.

\subsection{6j symbols}

\begin{figure}
	\centering
	\includegraphics[width=0.7\linewidth]{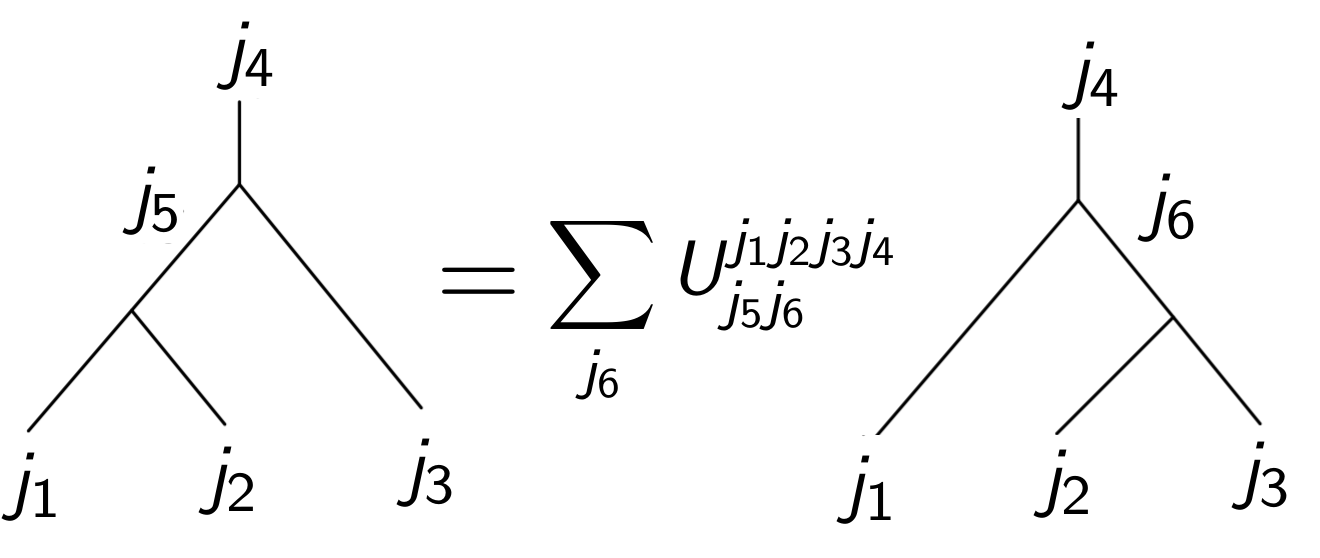}
	\caption{A diagrammatic illustration of the definition of the 6j symbol. Each junction of three lines represents a Clebsch-Gordan coefficient.}
	\label{fig:6j}
\end{figure}

In this section, we study three different 6j symbols that may be built out of the above Clebsch-Gordan coefficients. We now briefly review the general definition of a 6j symbol. Let $j_i$ denote a unitary irrep of an algebra, where $i \in \{1,2,\cdots,6\}$. To define a 6j symbol, consider the triple tensor product $j_1 \otimes j_2 \otimes j_3$. One way to decompose this into irreps is to first take the tensor product of $j_1 \otimes j_2$ and then tensor the result with $j_3$. Let us then project the result onto the space of states transforming in a $j_4$ representation (call this space $\mathcal{S}_4$). The space $\mathcal{S}_4$ may be written as a direct sum of orthogonal subspaces, where each subspace corresponds to a representation that appears in the $j_1 \otimes j_2$ tensor product (let $j_5$ refer to such a representation). Thus an orthonormal basis vector of $\mathcal{S}_4$ may be written as $\ket{j_4 \, m_4 ; j_5}_{12}$, where $m_4$ labels a basis vector in the $j_4$ irrep. Alternatively, we can repeat the entire process in the other channel by taking the $j_2 \otimes j_3$ tensor product first. Let $j_6$ refer to a representation in the tensor product of $j_2 \otimes j_3$. Thus a different orthonormal basis of $\mathcal{S}_4$ is given by vectors of the form $\ket{j_4 \, m_4 ; j_6}_{23}$. The 6j symbol is defined to be the overlap
\begin{equation}
	U_{j_{5} j_6}^{j_1 j_2 j_3 j_4} \equiv \,_{12} \braket{j_4 \, m_4 ; j_5 | j_4 \, m_4 ; j_6}_{23},
	\label{eq:6j}
\end{equation}
which does not depend on $m_4$. The 6j symbol is represented graphically in Figure \ref{fig:6j}. Equation \eqref{eq:6j} defines a unitary matrix because it is an overlap of orthonormal basis vectors. In particular,
\begin{equation}
\sum_{j_6} U^{j_1j_2j_3j_4}_{j_5j_6} \left(U^{j_1j_2j_3j_4}_{\tilde{j}_5 j_6}\right)^* = \delta_{j_5 \tilde{j}_5}, \quad \sum_{j_5}  \left(U^{j_1j_2j_3j_4}_{j_5 j_6}\right)^* U^{j_1j_2j_3j_4}_{j_5\tilde{j}_6} = \delta_{j_6 \tilde{j}_6}.
\label{eq:B.23}
\end{equation}
The 6j symbols that we discuss are real.

In general, 6j symbols may be constructed from Clebsch-Gordan coefficients. Let us define the Clebsch-Gordan coefficients $C^{j, j_1, j_2}_{m, m_1, m_2}$ such that
\begin{equation}
	\ket{j \, m} = \sum_{m_1,m_2} C^{j, j_1, j_2}_{m, m_1, m_2} \ket{j_1 \, m_1} \otimes \ket{j_2 \, m_2}.
\end{equation}
The coefficients obey an orthogonality relation:
\begin{equation}
\label{eq:cgorth}
\sum_{m_1,m_2} 	C^{j, j_1, j_2}_{m, m_1, m_2}	C^{\tilde{j}, j_1, j_2}_{\tilde{m}, m_1, m_2} = \delta_{m, \tilde{m}} \delta_{j,\tilde{j}}.
\end{equation}
The Clebsch-Gordan coefficients and 6j symbols are related as follows:
\begin{equation}
	\sum_{m_5}	C^{j_4, j_5, j_3}_{m_4, m_5, m_3} C^{j_5, j_1, j_2}_{m_5, m_1, m_2}
	=
	\sum_{j_6,m_6} U^{j_1j_2j_3j_4}_{j_5j_6} 	C^{j_4, j_1, j_6}_{m_4, m_1, m_6} C^{j_6, j_2, j_3}_{m_6, m_2, m_3},
	\label{eq:B.26}
\end{equation}
Using \eqref{eq:cgorth}, we have
\begin{equation}
	\sum_{m_5,m_2,m_3}	C^{j_4, j_5, j_3}_{m_4, m_5, m_3} C^{j_5, j_1, j_2}_{m_5, m_1, m_2} C^{j_6, j_2, j_3}_{m_6, m_2, m_3}
	=
 U^{j_1j_2j_3j_4}_{j_5j_6} 	C^{j_4, j_1, j_6}_{m_4, m_1, m_6} .
 \label{eq:B.27}
\end{equation}
We now derive another identity that we will use later. We will start with \eqref{eq:B.26} and multiply by three new Clebsch-Gordan coefficients on each side to obtain
\begin{align}
	\begin{split}
		&\sum_{m_5,m_1,m_2,m_3,m_8,m_9}	C^{j_4, j_5, j_3}_{m_4, m_5, m_3} C^{j_5, j_1, j_2}_{m_5, m_1, m_2}
		C^{j_8,j_7,j_1}_{m_8,m_7,m_1} C^{j_9,j_8,j_2}_{m_9,m_8,m_2} C^{j_{10},j_9,j_3}_{m_{10},m_9,m_3}
		\\
		=
		&\sum_{j_6,m_6,m_1,m_2,m_3,m_8,m_9} U^{j_1j_2j_3j_4}_{j_5j_6} 	C^{j_4, j_1, j_6}_{m_4, m_1, m_6} C^{j_6, j_2, j_3}_{m_6, m_2, m_3}
		C^{j_8,j_7,j_1}_{m_8,m_7,m_1}
		C^{j_9,j_8,j_2}_{m_9,m_8,m_2}
		C^{j_{10},j_9,j_3}_{m_{10},m_9,m_3}
	\end{split}
\end{align}
Next, we use \eqref{eq:B.27} to simplify the $m_1,m_2,m_8$ sum on the left and the $m_2,m_3,m_9$ sum on the right. The result is
\begin{align}
	\begin{split}
		&\sum_{m_5,m_3,m_9}		
		U^{j_7j_1j_2j_9}_{j_8j_5}
		C^{j_{10},j_9,j_3}_{m_{10},m_9,m_3}
		C^{j_9,j_7,j_5}_{m_9,m_7,m_5}
		C^{j_4, j_5, j_3}_{m_4, m_5, m_3}    
		\\
		=
		&\sum_{j_6,m_6,m_1,m_8} U^{j_1j_2j_3j_4}_{j_5j_6} 	
		U^{j_8j_2j_3j_{10}}_{j_9j_6}
		C^{j_{10},j_8,j_6}_{m_{10},m_8,m_6}
		C^{j_8,j_7,j_1}_{m_8,m_7,m_1}
		C^{j_4, j_1, j_6}_{m_4, m_1, m_6} 
	\end{split}
\end{align}
Next, we use \eqref{eq:B.27} again to simplify the $m_5,m_3,m_9$ sum on the left and the $m_6,m_1,m_8$ sum on the right. The result is
\begin{align}
	\begin{split}
		&		
		U^{j_7j_1j_2j_9}_{j_8j_5}
		U^{j_7j_5j_3j_{10}}_{j_9j_4} C^{j_{10},j_7,j_4}_{m_{10},m_7,m_4}
		\\
		=
		&\sum_{j_6} U^{j_1j_2j_3j_4}_{j_5j_6} 	
		U^{j_8j_2j_3j_{10}}_{j_9j_6}
		U^{j_7j_1j_6j_{10}}_{j_8j_4}C^{j_{10},j_7,j_4}_{m_{10},m_7,m_4}
	\end{split}
\end{align}
which becomes
\begin{equation}		
		U^{j_7j_1j_2j_9}_{j_8j_5}
		U^{j_7j_5j_3j_{10}}_{j_9j_4}
		=
		\sum_{j_6} U^{j_1j_2j_3j_4}_{j_5j_6} 	
		U^{j_8j_2j_3j_{10}}_{j_9j_6}
		U^{j_7j_1j_6j_{10}}_{j_8j_4}.
		\label{eq:special6jidentity}
\end{equation}

\subsubsection{$\pi^+ \otimes \pi^+ \otimes \pi^+$}

Consider the triple tensor product $\pi_{k_1}^+ \otimes \pi_{k_2}^+ \otimes \pi^+_{k_3}$. Our convention for the 6j symbol is given in Figure \ref{fig:6j} with the following identifications:
\begin{align}
\begin{split}
	&j_1 = \pi^+_{k_1}, \quad j_2 = \pi^+_{k_2}, \quad j_3 = \pi^+_{k_3}, \quad j_4 = \pi^+_{k_1 + k_2 + k_3 + N}, \quad j_5 = \pi^+_{k_1 + k_2 + x_{12}}, \quad j_6 = \pi^+_{k_2 + k_3 + x_{23}},
	\\
	&U^{j_1 j_2 j_3 j_4}_{j_5 j_6} = S_q^{k_1, k_2, k_3, N}(x_{12},x_{23}), \quad \sum_{j_6} = \sum_{x_{23} = 0}^N,
\end{split}
\end{align}
where the 6j symbol $S_q^{k_1, k_2, k_3, N}(x_{12},x_{23})$ is defined as follows:

\begin{align}
	\begin{split}
		S_q^{k_1, k_2, k_3, N}(x_{12},x_{23}) &=  q^{k_2(N - x_{12} - x_{23})} \left[\begin{array}{c}
	N \\
	x_{23}
\end{array}\right]_q 
\times  \,_4 \phi_3\left(\begin{array}{c}
	q^{-1 + 2 k_1 + 2 k_2 + x_{12}}, q^{-1 + 2 k_2 + 2 k_3 + x_{23}}, q^{-x_{12}},q^{-x_{23}} \\
	q^{2 k_2}, q^{-1 + 2 k_1 + 2 k_2 + 2 k_3 + N}, q^{-N}
\end{array};q,q\right) 
\\
&\times
   \frac{(q^{-1 + 2 k_1 + 2 k_2 + 2 k_3 + N};q)_{x_{23}} (q^{2 k_3};q)_{N - x_{12}} (q^{2 k_2};q)_{x_{12}} }{\sqrt{(q,q^{2 k_1},q^{2 k_2},q^{2 k_1+2 k_2+x_{12}-1};q)_{x_{12}}  (q,q^{2 k_3},q^{2 (k_1+k_2+x_{12})},q^{2 k_1+2 k_2+2 k_3+N+x_{12}-1};q)_{N-x_{12}}}} 
     \\
     &\times
      \frac{\sqrt{(q,q^{2 k_2};q)_{x_{23}}  (q,q^{2 k_1},q^{2 k_1+2 k_2+2k_3+N+x_{23}-1};q)_{N-x_{23}} } }{\sqrt{\left(q^{2 (k_2+k_3+x_{23})};q\right)_{N-x_{23}} \left(q^{2 k_3},q^{2 k_2+2 k_3+x_{23}-1};q\right)_{x_{23}}} },
      \end{split}
  \end{align}
and the $q$-binomial coefficient is defined by
\begin{equation}
	\left[\begin{array}{c}
		n \\
		k
	\end{array}\right]_q = \frac{(q;q)_n}{(q;q)_k (q;q)_{n-k}}.
\end{equation}
In particular, we have that for $n_1, n_2, n_3 \in \mathbb{Z}_{\ge 0}$ and $n_1 + n_2 + n_3 \ge N$,
\begin{align}
\begin{split}
	&v_{n_1}^{n_1 + n_2}(x_{12},k_1,k_2) v_{n_1 + n_2 - x_{12}}^{n_1 + n_2 - x_{12} + n_3}(N - x_{12},k_1 + k_2 + x_{12},k_3) 
	\\
	&= \sum_{x_{23} = 0}^{\min(N,n_2 + n_3)} S_q^{k_1,k_2,k_3,N}(x_{12},x_{23}) v_{n_2}^{n_2 + n_3}(x_{23},k_2,k_3) v_{n_1}^{n_1 + n_2 + n_3 - x_{23}}(N - x_{23},k_1,k_2 + k_3 + x_{23}).
\end{split}
\end{align}
Each Clebsch-Gordan coefficient above corresponds to a junction in Figure \ref{fig:6j}. Also, \eqref{eq:B.23} becomes
\begin{equation}
\sum_{x_{23} = 0}^N	S_q^{k_1,k_2,k_3,N}(x_{12},x_{23}) S_q^{k_3,k_2,k_1,N}(x_{23},\tilde{x}_{12}) = \delta_{x_{12}, \tilde{x}_{12}}.
\end{equation}

\subsubsection{$\pi^- \otimes \pi^- \otimes \pi^-$}
The $\pi^- \otimes \pi^- \otimes \pi^-$ 6j symbol is related to the $\pi^+ \otimes \pi^+ \otimes \pi^+$ 6j symbol via the automorphism \eqref{eq:autom}. We may make the following substitutions in Figure \ref{fig:6j},
\begin{align}
	\begin{split}
	j_1 = \pi^-_{k_1}, \quad j_2 = \pi^-_{k_2}, \quad j_3 = \pi^-_{k_3}, \quad j_4 = \pi^-_{k_1 + k_2 + k_3 + N}, \quad j_5 = \pi^-_{k_1 + k_2 + x_{12}}, \quad j_6 = \pi^-_{k_2 + k_3 + x_{23}},
	\\
	U^{j_1 j_2 j_3 j_4}_{j_5 j_6} = S_q^{k_3,k_2,k_1,N}(x_{23},x_{12}), \quad \sum_{j_6} = \sum_{x_{23} = 0}^{N}.
\end{split}
\end{align}

\subsubsection{$\pi^+ \otimes \pi^+ \otimes \pi^P$}

Consider the triple tensor product $\pi^+_{\Delta_1} \otimes \pi^+_{\Delta_2} \otimes \pi^P_{s_a,\epsilon}$. Our convention for the 6j symbol is given in Figure \ref{fig:6j} with the following identifications,
\begin{align}
	\begin{split}
	j_1 &= \pi^+_{\Delta_1}, \quad j_2 = \pi^+_{\Delta_2}, \quad j_3 = \pi^P_{s_a,\epsilon}, \quad j_4 = \pi^P_{s_c,\Delta_1 + \Delta_2 + \epsilon}, \quad j_5 = \pi_{\Delta_1 + \Delta_2 + x}^+, \quad j_6 = \pi^P_{s,\Delta_2 + \epsilon}
	\\
	U^{j_1 j_2 j_3 j_4}_{j_5 j_6} &= (-1)^{x + 1} \sqrt{\rho_q(s)} P^{\Delta_2,\Delta_1}_{x}(s;s_a,s_c|q), \quad \sum_{j_6} = \int_0^{\frac{\pi}{|\log q|}} ds.
	\end{split}
\end{align}
While the tensor product of $\pi^+_{\Delta_2}$ and $\pi^P_{s_a,\epsilon}$ does contain positive discrete series representations, the label $j_6$ only runs over principal series representations because the tensor product of a positive discrete series representation and $\pi^+_{\Delta_1}$ cannot produce $\pi^P_{s_c,\Delta_1 + \Delta_2 + \epsilon}$.

To be precise, the following identity holds:
\begin{align}
	\begin{split}
		&\sum_{x = 0}^{n_1 + n_2} v_{n_1}^{n_1 + n_2}(x,\Delta_1,\Delta_2) f_{n_1 + n_2 - x}^{n_1 + n_2 + n_3 - x}(s_c,s_a,\epsilon,\Delta_1 + \Delta_2 + x) (-1)^{1 + x} \sqrt{\rho_q(s)} P^{\Delta_2,\Delta_1}_{x}(s;s_a,s_c|q) 
		\\
		&= f_{n_2}^{n_2 + n_3}(s,s_a,\epsilon,\Delta_2) f_{n_1}^{n_1 + n_2 + n_3}(s_c,s,\epsilon + \Delta_2,\Delta_1), \quad \quad n_1,n_2 \in \mathbb{Z}_{\ge 0}, \quad n_3 \in \mathbb{Z}.
	\end{split}
\label{eq:6jcg}
\end{align}
Using \eqref{eq:askeywilsonpolyorth}, this becomes
\begin{align}
	\begin{split}
		&v_{n_1}^{n_1 + n_2}(x,\Delta_1,\Delta_2) f_{n_1 + n_2 - x}^{n_1 + n_2 + n_3 - x}(s_c,s_a,\epsilon,\Delta_1 + \Delta_2 + x) 
		\\
		&= \int_{0}^{\frac{\pi}{|\log q|}} ds (-1)^{1 + x} \sqrt{\rho_q(s)} P^{\Delta_2,\Delta_1}_{x}(s;s_a,s_c|q) \,  f_{n_2}^{n_2 + n_3}(s,s_a,\epsilon,\Delta_2) f_{n_1}^{n_1 + n_2 + n_3}(s_c,s,\epsilon + \Delta_2,\Delta_1), 
		\\
		&\quad \quad 0 \leq x \leq n_1 + n_2.
	\end{split}
\label{eq:B.25}
\end{align}
Each Clebsch-Gordan coefficient in \eqref{eq:B.25} is represented by a junction in \ref{fig:6j}. Unitary of the 6j symbol also implies the completeness relation:\footnote{This is equivalent to a special case ($\alpha = 0$) of (4.1.11) in \cite{tariq1997some}. See also equations (3.8) and (10.1) in \cite{askey1996general}.}
\begin{equation}
	\sum_{n = 0}^\infty P^{\Delta_2,\Delta_1}_{n}(s_b;s_c,s_a|q) P^{\Delta_2,\Delta_1}_{n}(s_d;s_c,s_a|q) = \frac{\delta(s_b - s_d)}{\rho_q(s_b)}.
	\label{eq:completeness}
\end{equation}
Using the graphical notation in \eqref{eq:A8}, the completeness relation is given in \eqref{IdResPic}.

\subsubsection{$\pi^P \otimes \pi^- \otimes \pi^-$}

The 6j symbol for $\pi^P_{s_a,\epsilon} \otimes \pi^-_{\Delta_1} \otimes \pi^-_{\Delta_2}$ may be inferred from the 6j symbol for $\pi^+_{\Delta_2} \otimes \pi^+_{\Delta_1} \otimes \pi^P_{s_a,-\epsilon}$. In particular, it follows from \eqref{eq:6jcg} that
\begin{align}
	\begin{split}
	&(-1)^{n_2} f_{n_2}^{n_2 - n_1}(s,s_a,-\epsilon,\Delta_1) (-1)^{n_3}f^{n_2 + n_3 - n_1}_{n_3}(s_c,s,-\epsilon + \Delta_1,\Delta_2) 
	\\
	&= \sum_{x = 0}^{n_2 + n_3} v_{n_3}^{n_2 + n_3}(x,\Delta_2,\Delta_1) (-1)^{n_2 + n_3 - x}f_{n_2 + n_3 - x}^{n_2 + n_3 - x - n_1}(s_c,s_a,-\epsilon,k_1 + k_2 + x) \left[- \sqrt{\rho_q(s)}P^{\Delta_1 \Delta_2}_{x}(s;s_a,s_c|q)\right],
	\\ &n_2,n_3 \in \mathbb{Z}_{\ge 0}, \quad n_1 \in \mathbb{Z}.
	\end{split}
\end{align}
Thus, the 6j symbol is represented by Figure \ref{fig:6j} with the following identifications,
\begin{align}
	\begin{split}
		j_1 &= \pi^P_{s_a,\epsilon}, \quad j_2 = \pi^-_{\Delta_1}, \quad j_3 = \pi^-_{\Delta_2}, \quad j_4 = \pi^P_{s_c,\epsilon -\Delta_1 - \Delta_2 }, \quad j_5 = \pi^P_{s,\epsilon - \Delta_1}, \quad j_6 = \pi^-_{\Delta_1 + \Delta_2 + x}
		\\
		U^{j_1 j_2 j_3 j_4}_{j_5 j_6} &= - \sqrt{\rho_q(s)} P^{\Delta_1,\Delta_2}_{x}(s;s_a,s_c|q), \quad \sum_{j_6} = \sum_{x \ge 0} .
	\end{split}
\end{align}

\subsubsection{$\pi^+ \otimes \pi^P \otimes \pi^-$}

\begin{figure}
	\centering
	\includegraphics[width=0.7\linewidth]{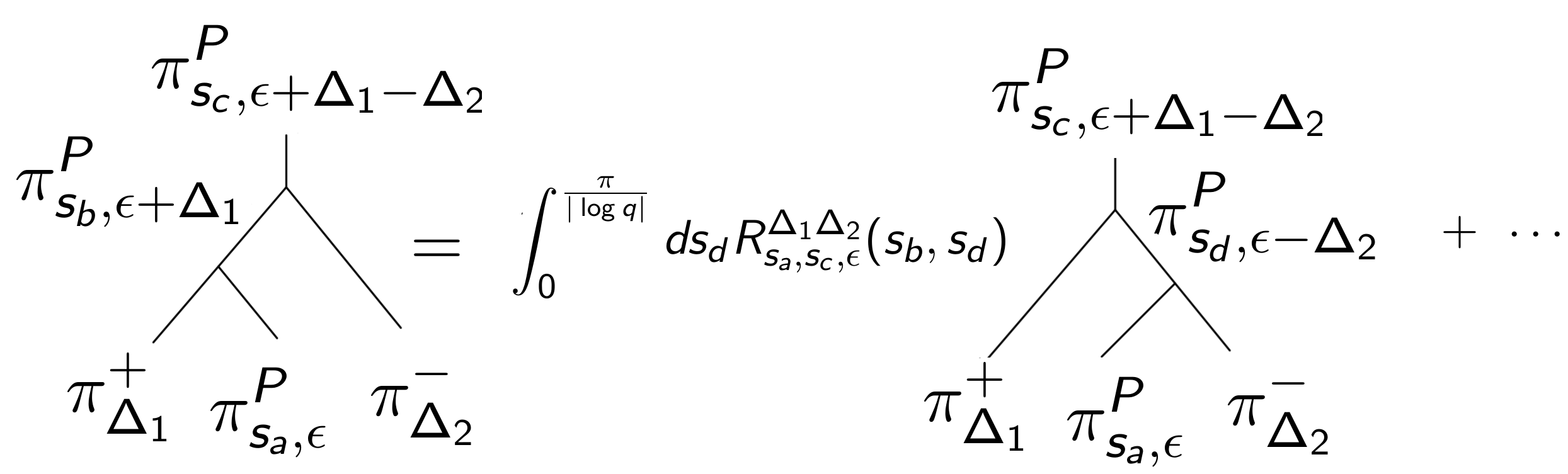}
	\caption{A diagrammatic representation of the 6j symbol in \eqref{eq:R6j}. On the right hand side, we have only depicted $\pi^P$ representations running through the channel. The $\cdots$ contain contributions from negative discrete series representations that also run in the channel.}
	\label{fig:6jpluspminus}
\end{figure}

Consider the triple tensor product $\pi^+_{\Delta_1} \otimes \pi^P_{s_a,\epsilon} \otimes \pi^-_{\Delta_2}$ projected onto the subspace of states that transform in  the $\pi^P_{s_c,\epsilon + \Delta_1 - \Delta_2}$ representation. In the two channels, we will project onto the $\pi^P_{s_b,\Delta_1 + \epsilon}$ and $\pi^P_{s_d,\epsilon - \Delta_2}$ channels. The 6j symbol is given by
\begin{equation}
	\label{eq:R6j}
	R_{s_a,s_c,\epsilon}^{\Delta_1,\Delta_2}(s_b,s_d) =   \sqrt{\frac{(q^{1/2 \pm \epsilon \pm i s_a},q^{\frac{1}{2} \pm (\Delta_1 - \Delta_2 + \epsilon) \pm i s_c};q)_\infty}{(q^{\frac{1}{2} \pm (\epsilon + \Delta_1) \pm i s_b},q^{\frac{1}{2} \pm (\Delta_2 - \epsilon) \pm i s_d};q)_\infty}}
	 \left\{\begin{array}{ccc}
		\Delta_2 & s_b & s_c \\
		\Delta_1 & s_d & s_a
	\end{array}
	\right\}_q \, \sqrt{\rho_q(s_b) \rho_q(s_d)},
\end{equation}
and is diagrammatically represented in Figure \ref{fig:6jpluspminus}. Note that the tensor product of $\pi^P \otimes \pi^-$ contains both $\pi^P$ and $\pi^-$ representations, and tensoring $\pi^+$ with either of these produces a $\pi^P$ representation. Thus, to derive an orthogonality identity analogous to \eqref{eq:B.23}, we must consider both $\pi^P$ and $\pi^-$ representations in the channel on the right hand side of Figure \ref{fig:6jpluspminus}. However, we have only explicitly considered the $\pi^P$ representations. The discrete series representations will not play a role in our derivation of the pentagon identity, which we carry out in the next section.  

\subsection{Pentagon identity}

\begin{figure}
	\centering
	\includegraphics[width=\linewidth]{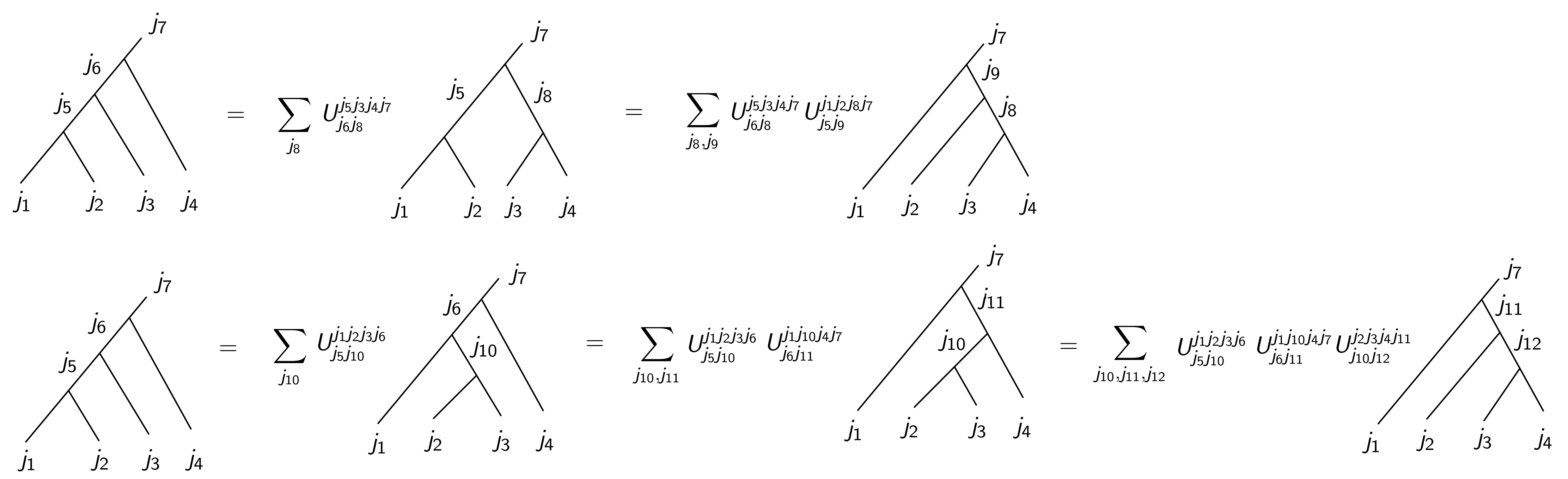}
	\caption{A graphical illustration of the pentagon identity. We may exploit the fact that there are various ways to take a tensor product of four representations to derive a nontrivial relation involving 6j symbols.}
	\label{fig:pentagon}
\end{figure}

We now derive a pentagon identity\footnote{See \cite{Baez} for an explanation of the pentagon identity, which is also called the Biedenharn-Elliot identity.} for the tensor product $\pi^+ \otimes \pi^P \otimes \pi^- \otimes \pi^-$. The pentagon identity may be derived by tensoring four representations together and applying the relation in Figure \ref{fig:6j} in different ways, as indicated in Figure \ref{fig:pentagon}. Equality of the two lines in Figure \ref{fig:pentagon} indicates that the 6j symbols must obey
\begin{equation}
	U_{j_6 j_8}^{j_5 j_3 j_4 j_7} U_{j_5 j_9}^{j_1 j_2 j_8 j_7} = \sum_{j_{10}} U_{j_5 j_{10}}^{j_1 j_2 j_3 j_6} U_{j_6 j_{9}}^{j_1 j_{10} j_4 j_7} U_{j_{10} j_{8}}^{j_2 j_3 j_4 j_{9}}.
\end{equation}

To obtain the identity of our interest, we make the substitutions
\begin{align}
\begin{split}
	j_1 &= \pi^+_{\Delta_1}, \quad j_2 = \pi^P_{s_a,\epsilon}, \quad j_3 = \pi^-_{\Delta_2}, \quad j_4 = \pi^-_{\Delta_3}
	\\
	j_5 &= \pi^P_{s_b,\epsilon + \Delta_1}, \quad j_6 = \pi^P_{s_c,\epsilon + \Delta_1 - \Delta_2}, \quad j_7 = \pi^P_{s_d,\epsilon + \Delta_1 - \Delta_2 - \Delta_3},
	\\
	j_8 &= \pi^-_{\Delta_2 + \Delta_3 + x}, \quad j_9 = \pi^P_{s_e,\epsilon - \Delta_2 - \Delta_3}, \quad j_{10} = \pi^P_{s_f,\epsilon - \Delta_2},
\end{split}
\end{align}
so the 6j symbols are
\begin{align}
\begin{split}
&U^{j_5 j_3 j_4 j_7}_{j_6 j_8} = - \sqrt{\rho_q(s_c)} P^{\Delta_2 \Delta_3}_{x}(s_c;s_b,s_d|q), \quad U^{j_1 j_2 j_8 j_7}_{j_5 j_9} = R^{\Delta_1,\Delta_2 + \Delta_3 + x}_{s_a,s_d,\epsilon}(s_b,s_e),
\\ &U^{j_1 j_2 j_3 j_6}_{j_5 j_{10}} = R^{\Delta_1 \Delta_2}_{s_a,s_c,\epsilon}(s_b,s_f), \quad U^{j_1 j_{10} j_4 j_7}_{j_6 j_9} = R^{\Delta_1 \Delta_3}_{s_f,s_d,\epsilon - \Delta_2}(s_c,s_e), \quad U^{j_2 j_3 j_4 j_9}_{j_{10} j_8} = - \sqrt{\rho_q(s_f)}P^{\Delta_2 \Delta_3}_{x}(s_f;s_a,s_e|q)
\end{split}
\end{align}
and the pentagon identity becomes
\begin{align}
	\begin{split}
	 &\sqrt{\rho_q(s_c)} P^{\Delta_2 \Delta_3}_x(s_c;s_b,s_d|q) R^{\Delta_1,\Delta_2 + \Delta_3 + x}_{s_a,s_d,\epsilon}(s_b,s_e) 
	\\
	&= \int_0^{\frac{\pi}{|\log q|}} ds_f \,  R^{\Delta_1 \Delta_2}_{s_a,s_c,\epsilon}(s_b,s_f) R^{\Delta_1 \Delta_3}_{s_f,s_d,\epsilon - \Delta_2}(s_c,s_e)  \sqrt{\rho_q(s_f)}P^{\Delta_2 \Delta_3}_{x}(s_f;s_a,s_e|q),
	\end{split}
\end{align}
and using \eqref{eq:R6j} and simplifying, this becomes
\begin{align}
	\begin{split}
		&\sqrt{\frac{(q^{\frac{1}{2} \pm (\Delta_2 + \Delta_3 - \epsilon) \pm i s_e};q)_\infty}{(q^{\frac{1}{2} \pm (\Delta_2+\Delta_3  - \epsilon + x) \pm i s_e};q)_\infty}}\sqrt{\frac{(q^{\frac{1}{2} \pm (\Delta_1 - \Delta_2 - \Delta_3  + \epsilon - x) \pm i s_d};q)_\infty}{(q^{\frac{1}{2} \pm (\Delta_1 - \Delta_2  - \Delta_3 + \epsilon ) \pm i s_d};q)_\infty}}
		\\
&\times P^{\Delta_2 \Delta_3}_{x}(s_c;s_b,s_d|q)		 \left\{\begin{array}{ccc}
			(\Delta_2+\Delta_3 + x) & s_b & s_d \\
			\Delta_1 & s_e & s_a
		\end{array}
		\right\}_q \, 
		\\
		&= \int_0^{\frac{\pi}{|\log q|}} ds_f \rho_q(s_f)\, 
		 \left\{\begin{array}{ccc}
			\Delta_2 & s_b & s_c \\
			\Delta_1 & s_f & s_a
		\end{array}
		\right\}_q \,  \left\{\begin{array}{ccc}
			\Delta_3 & s_c & s_d \\
			\Delta_1 & s_e & s_f
		\end{array}
		\right\}_q \,  P^{\Delta_2 \Delta_3}_{x}(s_f;s_a,s_e|q),
	\end{split}
\end{align}
Next, we use the identity
\begin{equation}
	\frac{(q^{\frac{1}{2} \pm (k - x) \pm i s};q)_\infty}{(q^{\frac{1}{2} \pm k \pm i s};q)_\infty} = q^{-x^2 + 2 k x}
\end{equation}
and conclude that
\begin{align}
	\begin{split}
		&q^{ x \Delta_1 } P^{\Delta_2 \Delta_3}_{x}(s_c;s_b,s_d|q)		 \left\{\begin{array}{ccc}
			(\Delta_2+\Delta_3 + x) & s_b & s_d \\
			\Delta_1 & s_e & s_a
		\end{array}
		\right\}_q \, 
		\\
		&= \int_0^{\frac{\pi}{|\log q|}} ds_f \rho_q(s_f)\, 
		 \left\{\begin{array}{ccc}
			\Delta_2 & s_b & s_c \\
			\Delta_1 & s_f & s_a
		\end{array}
		\right\}_q \,  \left\{\begin{array}{ccc}
			\Delta_3 & s_c & s_d \\
			\Delta_1 & s_e & s_f
		\end{array}
		\right\}_q \,  P^{\Delta_2 \Delta_3}_{x}(s_f;s_a,s_e|q),
	\end{split}
\end{align}

Using the completeness relation \eqref{eq:completeness}, this becomes
\begin{align}
\label{eq:pentagon}
	\begin{split}
		\sum_{x = 0}^\infty &q^{ x \Delta_1 } P^{\Delta_2 \Delta_3}_{x}(s_c;s_b,s_d|q)		 \left\{\begin{array}{ccc}
			(\Delta_2+\Delta_3 + x) & s_b & s_d \\
			\Delta_1 & s_e & s_a
		\end{array}
		\right\}_q \, P^{\Delta_2 \Delta_3}_{x}(s_f;s_a,s_e|q)
		\\
		&=  
		 \left\{\begin{array}{ccc}
			\Delta_2 & s_b & s_c \\
			\Delta_1 & s_f & s_a
		\end{array}
		\right\}_q \,  \left\{\begin{array}{ccc}
			\Delta_3 & s_c & s_d \\
			\Delta_1 & s_e & s_f
		\end{array}
		\right\}_q \,  .
	\end{split}
\end{align}

Using the graphical notation in \eqref{eq:A8}, this becomes
\begin{equation}
    \includegraphics[scale=0.4]{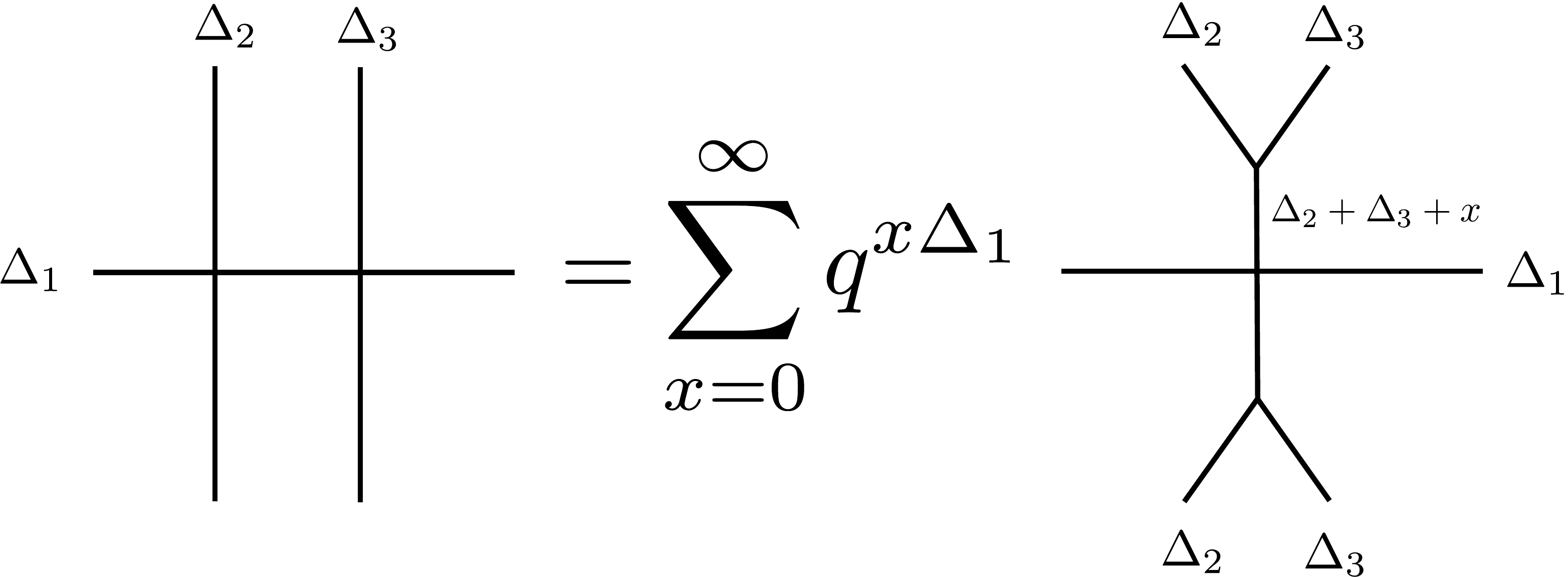}
    \label{eq:pentagongraphic}
\end{equation}

\subsection{Final identity}

The last identity that we will need follows from \eqref{eq:special6jidentity}, with the following identifications:
\begin{align}
	\begin{split}
	j_1 &= \pi^-_{\Delta_1}, \quad j_2 = \pi^-_{\Delta_2}, \quad j_3 = \pi^-_{\Delta_3}, \quad j_4 = \pi^-_{\Delta_1 + \Delta_2 + \Delta_3 + N}
	\\
	j_5 &= \pi^-_{\Delta_1 + \Delta_2 + x}, \quad j_6 = \pi^-_{\Delta_2 + \Delta_3 + y}, \quad j_7 = \pi^P_{s_a,\epsilon}, \quad j_8 = \pi^P_{s_b,\epsilon - \Delta_1},
	\\
	j_9 &= \pi^P_{s_c,\epsilon - \Delta_1 - \Delta_2}, \quad j_{10} = \pi^P_{s_d,\epsilon - \Delta_1 - \Delta_2 - \Delta_3}, \quad \sum_{j_6} = \sum_{y = 0}^{N}
	\end{split}
\end{align}
Then, \eqref{eq:special6jidentity} becomes
\begin{align}
    \begin{split}
&P^{\Delta_1,\Delta_2}_{x}(s_b;s_a,s_c|q)
P^{\Delta_1 + \Delta_2 + x,\Delta_3}_{N-x}(s_c;s_a,s_d|q)
	\\
	&=
	\sum_{y = 0}^N 
	S_q^{\Delta_3,\Delta_2,\Delta_1,N}(y,x)
 P^{\Delta_1,\Delta_2 + \Delta_3 + y}_{N-y}(s_b;s_a,s_d|q)
	 P^{\Delta_2,\Delta_3}_{y}(s_c;s_b,s_d|q)
	.
	\label{eq:whatshouldicallthis}
	\end{split}
	\end{align}

Using the graphical notation introduced in \eqref{eq:A8}, this identity becomes
\begin{equation}
    \includegraphics[scale=0.4]{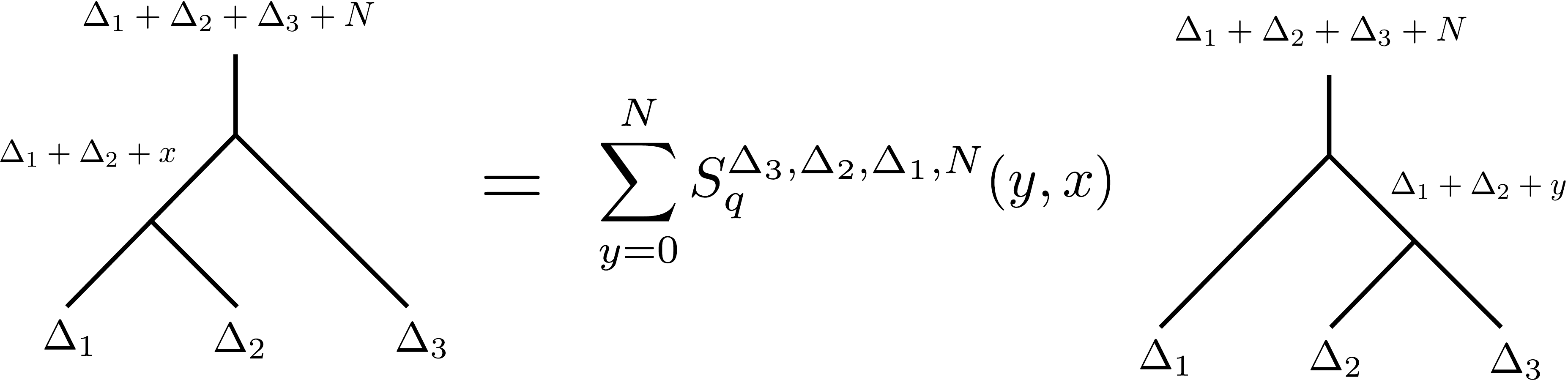}
\end{equation}

\bibliographystyle{JHEP}
\nocite{}
\bibliography{thebibliography}

\end{document}